# A Framework for Speechreading Acquisition Tools

**Benjamin Millar Gorman**

This thesis is submitted in partial fulfilment for the degree of

*Doctor of Philosophy*

at the University of Dundee

Doctor of Philosophy

University of Dundee

March 2018

# Declarations

## Candidate's Declaration

I, Benjamin Gorman, hereby declare that I am the author of this thesis; that I have consulted all references cited; that I have done all the work recorded by this thesis; and that it has not been previously accepted for a degree.

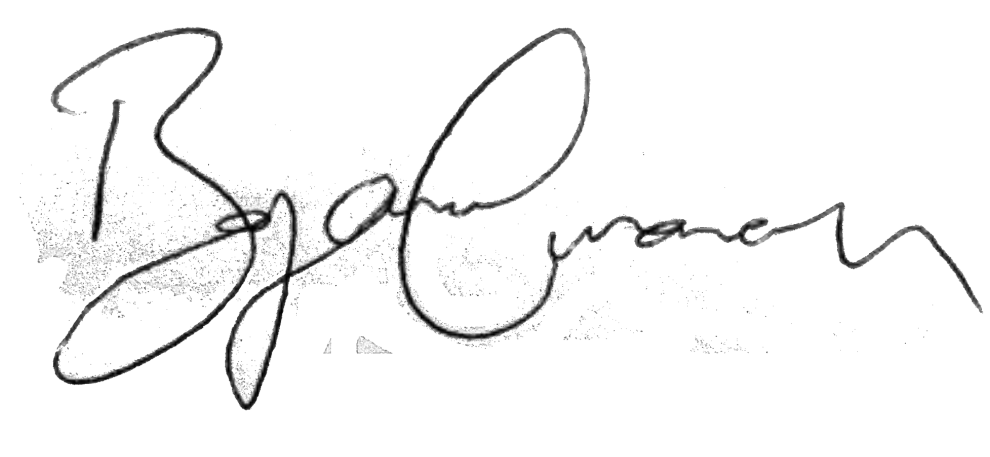

## Supervisor's Declaration

I, David Flatla, hereby declare that I am the supervisor of the candidate, and that the conditions of the relevant Ordinance and Regulations have been fulfilled.

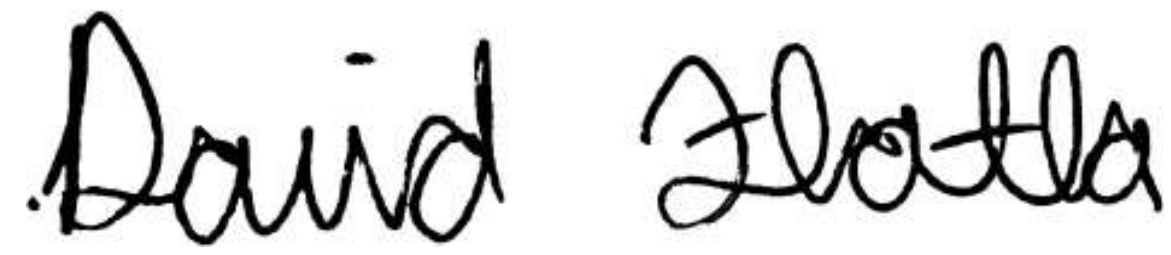

# Abstract

At least 360 million people worldwide have disabling hearing loss that frequently causes difficulties in day-to-day conversations. Hearing aids often fail to offer enough benefits and have low adoption rates. However, people with hearing loss find that speechreading can improve their understanding during conversation. Speechreading (often called lipreading) refers to using visual information about the movements of a speaker's lips, teeth, and tongue to help understand what they are saying. Speechreading is commonly used by people with all severities of hearing loss to understand speech, and people with typical hearing also speechread (albeit subconsciously) to help them understand others.

However, speechreading is a skill that takes considerable practice to acquire. Publicly-funded speechreading classes are sometimes provided, and have been shown to improve speechreading acquisition. However, classes are only provided in a handful of countries around the world and students can only practice effectively when attending class. Existing tools have been designed to help improve speechreading acquisition, but are often not effective because they have not been designed within the context of contemporary speechreading lessons or practice.

To address this, in this thesis I present a novel speechreading acquisition framework that can be used to design Speechreading Acquisition Tools (SATs) – a new type of technology to improve speechreading acquisition. I interviewed seven speechreading tutors and used thematic analysis to identify and organise the key elements of the framework. I evaluated the framework by using it to: 1) categorise every tutor-identified speechreading teaching technique, 2) critically evaluate existing Conversation Aids and SATs, and 3) design three new SATs.

I then conducted a postal survey with 59 speechreading students to understand students' perspectives on speechreading, and how their thoughts could influence future SATs. To further evaluate the framework's effectiveness I then developed and evaluated two new SATs (*PhonemeViz* and *MirrorMirror*) designed using the framework. The findings from the evaluation of these two new SATs demonstrates that using the framework can help design effective tools to improve speechreading acquisition.

# Acknowledgements

I would first like to thank my supervisor, David Flatla, who took me on as his first PhD student. You have taught me so much about writing, research and the generation of compelling ideas. Thank you for your patience, humour, corrections, faith, and encouragement.

I would also like to thank my committee for their insight and suggestions:

Professor Matt Huenerfauth, Rochester Institute of Technology (External Examiner)

Dr Nick Taylor, University of Dundee (Internal Examiner)

Professor Keith Edwards, University of Dundee (Convener)

I am indebted to the seven speechreading tutors who took part in my research interviews. Their knowledge and expertise helped shape the core ideas behind my thesis, and without their insights this would be a very different research project.

Thank you to all of the staff from the Queen Mother Building. In particular, Anne Millar who first emailed me to say that there was a PhD opportunity with someone "...from Canada who teaches HCI". Thank you for always helping me whenever I asked. Thanks to Rachel Menzies for always giving me solid advice and the opportunity to teach and contribute towards two very enjoyable modules. Thanks also to the students I have taught over the past three years, you rarely asked how my research was going – which was awesome. Thanks to all the fellow PhD students who asked thoughtful questions and advice at the annual PhD symposiums. Thank you to Mike Crabb, Alan Newell and Norman Alm who read drafts of my CHI submissions for 2017 and 2018, your feedback definitely helped to clarify and strengthen the written work within those submissions.

Thanks to Garreth Tigwell, who I first met in the labs of the Queen Mother Building during the first week of our undergrad degrees, who would later join DAPRlab in the first year of my PhD to start his MSc and then returned a year later to start his own PhD! Thanks for being my sounding board and providing your insight that improved my research rigour. More importantly, thanks for the many lunches and countless laughs discussing the the oddity of academia. Thanks to Daniel Herron, who I also met during my undergrad at Dundee. You always give the most helpful advice in a way you know I'll understand. Thanks for your encouragement and your ability to quickly turn a sour day into a funny one. Thanks also to Michael Mauderer, your ideas and approach to research really helped to wrap up my thesis. Thanks to every other past member of DAPRlab or the "Flatlets" as Anne Millar likes to call us.

Thanks to all my close friends who have encouraged me along the way, in particular Sarah for always giving me a much needed distraction over coffee discussing the smaller problems of life.

Thanks to all those who I have shared waves with over the past four years from the Dundee University Surf Club; from the cold waters of Lunan Bay, Montrose and St Andrews to the warm waters of Taghazout and Imsouane. In particular, thanks to Joseph, David, Alex, Adam and also to Eamon for the countless snowboaring trips up to the 'Shee. Thanks also to Mike for your weekly yoga classes and showing me there is fun for a researcher after submitting a thesis!

Thanks to my big brother Dave, your support whilst remote was always appreciated. Thanks to my little brother Edd, (who is a much more skilled programmer than I am) for your help and advice on those annoying little bugs I have had over the years and for the endless rounds of Injustice. Finally, thank you to my Mum and Dad. Thank you for your never-ending support, for helping to celebrate every small victory along the way, for your encouragement, and for always listening to my moans and instead making me laugh about them.

# Associated Publications

Some of the concepts and figures described in this thesis have been published before. The following list gives an overview of these publications. Each of these papers were written by me with editing advice supplied by my supervisor Dr David Flatla.

- **Benjamin M. Gorman** *and David R. Flatla. 2018. MirrorMirror: A Mobile Application to Improve Speechreading Acquisition. In Proceedings of the 2018 CHI Conference on Human Factors in Computing Systems. ACM, ACM, New York, NY.*

  This paper forms the basis for Chapters 6 and 8.

- **Benjamin M. Gorman** *and David R. Flatla. 2017. A Framework for Speechreading Acquisition Tools. In Proceedings of the 2017 CHI Conference on Human Factors in Computing Systems. ACM, ACM, New York, NY, 519–530.*

  This paper forms the basis for Chapters 4 and 5.

- **Benjamin M. Gorman** *2016. Reducing viseme confusion in speech-reading. ACM SIGACCESS Accessibility and Computing 114 (2016), 36–43.*

  This paper forms the basis for Chapter 7.

# Contents

# List of Figures

# List of Tables

# 1

# Introduction

As you read these words consider the sounds that you have encountered throughout your day: friends laughing, birds singing, a song on the radio – these sounds embellish our daily lives. However, along with these daily events our ability to hear provides us with one of our most important channels of communication. Without our hearing, it would be difficult to discuss daily events with one another, conduct business during a meeting, or make a new connection at an event.

More than 11 million people (1 in 6) in the UK have some degree of hearing loss [2] and in the USA, an estimated 30 million people (12.7%) 12 years and older have hearing loss in both ears [24]. On a global scale, the World Health Organisation estimates that 360 million people (~5%) worldwide have disabling hearing loss[a] [139]. Hearing loss prevalence increases with age [24], resulting in an anticipated growth in hearing loss in the future (e.g., 1 in 5 people are expected to have hearing loss by 2035 in the UK [2]).

Hearing loss results in difficulties understanding what others are saying during conversation [131]. Our relationships and identities are shaped through the various conversations we engage in throughout our lives [37]. As hearing loss causes difficulties during conversations it can result in social isolation [51], career stagnation [98], and a decrease in life satisfaction [131].

Hearing aids are designed to reduce these problems, but can be detrimental in noisy environments [61], and have low adoption rates (~14% people who need a hearing aid regularly wear one [2]) due to comfort issues [99], perceived social stigma [71], and expense [25].

[a]Hearing loss greater than 40 dB in the better hearing ear in adults and greater than 30 dB in the better hearing ear in children.

Speechreading (often called lipreading) refers to using visual information about the movements of a speaker's lips, teeth, and tongue to understand what they are saying [131]. Speechreading is commonly used by people with all severities of hearing loss to help understand speech [37], and people with typical hearing also speechread (albeit subconsciously) to help them understand others [64]. Speechreading has the advantage that it does not rely on the other conversation partner's knowledge of a Signed Language or a technique such as Cued Speech [45].

However, speechreading is a skill that takes considerable practice and training to acquire [81]. Publicly-funded speechreading classes are sometimes provided, and have been shown to improve speechreading acquisition [10]. Speechreading classes are primarily focused on learning how different mouth shapes are produced during speech [81], as well as how to use conversational repair strategies to gain important contextual information to help 'fill in' any gaps in understanding [81]. Classes also include information about hearing aids or other assistive listening devices, and give people a social space to meet with others who have hearing loss [131]. Classes can also improve an individual's self-confidence [20], and help attendees become more knowledgeable about their hearing loss and how they can make communication easier [106].

Within classes there are two main approaches to teaching speechreading: *synthetic* and *analytic* [57]. Synthetic methods (sometimes referred to as context-training) use a 'top-down' approach where focus is placed on understanding the topic of a conversation to determine words being spoken [131]. Analytic methods (sometimes referred to as 'eye-training' [36, 78]) use a 'bottom-up' approach where focus is placed on the visual speech pattern to identify what is being spoken [131].

Unfortunately, classes are only provided in a handful of countries around the world, and often there is an insufficient number of classes running in areas in which they are provided (e.g., only 50 of an estimated 325 required classes are currently running in Scotland [10]) and classes require mobility to attend.

To address this, there have been many attempts to design tools to improve the acquisition of speechreading. However, these previous solutions are typically not helpful to speechreaders because their designs do not align with how speechreading is currently taught or practiced within speechreading classes. Any solution that is developed to help speechreaders practice their speechreading or use their speechreading skill should be influenced by how speechreading is currently taught in speechreading classes.

## 1.1 Problem

The problem to be addressed in this thesis is: *Existing tools designed to improve speechreading acquisition are not effective because they have not been designed within the context of contemporary speechreading lessons or practice*.

In general, these previous solutions are not helpful to speechreaders because they were not influenced or based on how speechreading is currently taught. For instance, many of the analytic tools that are designed to help during speechreading provide information related to how a sound is produced but not how it appears on the lips. Furthermore, the synthetic tools for practicing speechreading do not teach students how to learn how to generalise their synthetic skills from training sessions to day-to-day life.

## 1.2 Motivation

By designing SATs that are influenced by how speechreading is currently taught, people with hearing loss will be able to augment their class-based learning, or learn on their own if no suitable classes are available. Once in the hands of people with hearing loss, appropriately-designed SATs will help enhance their speechreading capabilities, increasing their conversational confidence and reducing their social isolation.

## 1.3 Solution

To help expand speechreading training worldwide, I developed a novel framework for developing Speechreading Acquisition Tools (SATs) – a new type of technology designed specifically to improve speechreading acquisition. The framework consists of two dimensions (*Type of Skill* and *Amount of Information*), each with three levels (*Analytic/Synthetic/Hybrid* and *Low/Medium/High*, respectively).

Through the dissemination and adoption of the framework into the research community and assistive technology commercial sector, I foresee new technology being developed that is much more effective than the state-of-the-art so far.

## 1.4 Steps in the Solution

Five major steps were carried out to develop and evaluate the framework presented in this thesis:

1) Speechreading Tutor Interviews: I conducted interviews with seven practicing speechreading tutors. Using thematic analysis of the interview transcripts, I identified four main themes relevant to the future development of speechreading acquisition tools: 1) speechreading as a skill, 2) access to speechreading, 3) teaching practices, and 4) attitudes to technology.

2) Framework Development: The findings from Step 1 were used to organise the key elements of the framework, which consists of two dimensions (*Type of Skill* and *Amount of Information*), each with three levels (*Analytic*/*Synthetic*/*Hybrid* and *Low*/*Medium*/*High*, respectively)

3) Speechreading Student Questionnaire: To further inform the design of SATs, I conducted a postal questionnaire with students from speechreading classes to explore the challenges and situations they encounter while speechreading, and their approach to practice outside of class.

4) Development of PhonemeViz: The idea behind PhonemeViz was initially conceptualised during the third step of the framework evaluation (see Section 1.5). PhonemeViz is a visualisation that is positioned at the side of a speaker's face, beginning at the forehead and ending at the chin and presents textual representations of consonant speech sounds in a semi-circular arrangement, with an arrow beginning from the centre of this semi-circle pointing at the last spoken initial consonant speech sound (phoneme) to provide persistence. This design is intended to enable a speechreader to focus on the speaker's eyes and lip movements (as in traditional speechreading), while also monitoring changes in PhonemeViz's state using their peripheral vision to help disambiguate confusing lip movements.

   The design of PhonemeViz was inspired by the initial fingerspelling technique that was highlighted by four speechreading tutors during the interviews conducted during Step 1. PhonemeViz's design was further informed by the challenges reported by participants of the questionnaire conducted in Step 3.

5) Development of MirrorMirror: The idea behind MirrorMirror was initially conceptualised during the third step of the framework evaluation (see Section 1.5). MirrorMirror is a new SAT in the form of a mobile application that allows students to practice their speechreading by recording and watching videos of people they frequently speak with.

The design of MirrorMirror was inspired by the mirror training technique that was highlighted by seven speechreading tutors during the interviews conducted during Step 1. MirrorMirror's design was further informed by the positive and negative aspects of mirror training as reported by participants of the questionnaire conducted during Step 3.

## 1.5 Evaluation

To evaluate the framework, I used it to: 1) classify every existing speechreading teaching technique identified by the participants, 2) critically reflect on previously-developed solutions, and 3) design three new SATs for enhancing speechreading acquisition and proficiency.

By employing the framework in this fashion, I show that it: 1) comprehensively reflects existing speechreading teaching practice, 2) can be used to help understand the strengths and weaknesses of previously-developed solutions, and 3) can be used to identify clear opportunities for the development of new SATs to help improve speechreading skill acquisition.

As a further demonstration of the framework's ability to scaffold the development of new SATs, I then built and evaluated two of the SATs (PhonemeViz and Mirror ) designed during step 3) of the framework evaluation. To guide the development of these SATs, I conducted a postal questionnaire with 59 speechreading students to investigate if their views aligned with the speechreading tutors. Both PhonemeViz and MirrorMirror were inspired by teaching techniques that were fitted into the framework cells, and I used the results from both the interviews and the student questionnaire to guide their development.

I evaluated PhonemeViz with 14 participants against five existing visualisation techniques (plus a no visualisation control condition) in a lab-based user study. The results demonstrated that PhonemeViz allowed participants to achieve 100% word recognition (showing successful disambiguation), and PhonemeViz was well-received in subjective and qualitative feedback.

I evaluated MirrorMirror through case studies with three speechreading students. The case study evaluation of MirrorMirror was comprised of three stages: 1) a briefing, initial questionnaire, and tutorial session, 2) a week-long in-the-wild-deployment, and 3) a post-deployment discussion session. The results demonstrated that MirrorMirror enabled participants to effectively target their speechreading practice on people, words and situations they encounter during daily conversations.

## 1.6 Contributions

The central contribution of this thesis is the development of a novel framework that can be used to develop Speechreading Acquisition Tools (SATs) – a new type of technology designed specifically to improve speechreading acquisition. Through the development and release of SATs, people with hearing loss will be able to augment their class-based learning, or learn on their own if no suitable classes are available.

This thesis also presents a number of secondary contributions:

1) A critical overview of current Conversation Aids, and related approaches to improving speechreading acquisition, framed within the cells of the framework.

2) Presentation of novel interview data from seven practicing speechreading tutors with thematic analysis of that data.

3) Presentation of novel questionnaire data from a postal survey with 59 students from speechreading classes.

4) A description of the development and evaluation of PhonemeViz, a new SAT in the form of a visualisation that displays a subset of a speaker's spoken phonemes to the speechreader to reduce viseme confusion that can occur at the start of words. The design of PhonemeViz was inspired by the 'initial-letter fingerspelling' technique described by speechreading tutors during the interviews.

5) A description of the development and evaluation of MirrorMirror, a new SAT that addresses the limitations of current SATs by allowing users to capture (and practice with) videos of people they frequently speak with. The design of MirrorMirror was inspired by the 'mirror practice' technique described by speechreading tutors during the interviews.

## 1.7 Overview of thesis

This thesis contains the work described in this introductory chapter, presented in the following sequence of ten chapters:

**Chapter 1** *Introduction*

**Chapter 2** *Background*: Presents background research related to this thesis. Includes necessary background on hearing, hearing loss, speechreading and speechreading teaching techniques.

**Chapter 3** *Related Work*: Presents related work on Conversation Aids and previously developed tools to help people acquire speechreading.

**Chapter 4** *Speechreading Tutor Interviews*: Describes the motivation, method and findings of interviews conducted with seven practicing speechreading tutors.

**Chapter 5** *Speechreading Acquisition Tools Framework*: Describes the motivation, design and evaluation of a framework to help design Speechreading Acquisition Tools – a new type of technology to improve speechreading acquisition.

**Chapter 6** *Speechreading Student Questionnaire*: Describes the motivation, method and findings of a postal questionnaire with 59 speechreading students, sourced from classes taught by tutors who took part in the interviews in Chapter 4.

**Chapter 7** *PhonemeViz*: Presents a new Speechreading Acquisition Tool (SAT), which is a visualisation of initial spoken consonants to help disambiguate words that appear similar when spoken. This chapter describes the motivation, design, implementation and evaluation of PhonemeViz.

**Chapter 8** *MirrorMirror*: Presents a new Speechreading Acquisition Tool (SAT) that allows students to practice their speechreading by recording and watching videos of people they frequently speak with. This chapter describes the motivation, design, implementation and evaluation of MirrorMirror.

**Chapter 9** *Discussion*: Summarises main findings from previous chapters, and discusses implications, challenges, and limitations of the work described in this thesis.

**Chapter 10** *Conclusion and Future Work*: Briefly summarises this thesis, and outlines future directions for this research.

**Appendix A** *Ethical Approval Forms*: This appendix contains the letters of approval from the University Teaching and Research Ethics committee.

**Appendix B** *Speechreading Tutor Interviews Material*: This appendix contains study material used during the speechreading tutor interviews presented in Chapter 4.

**Appendix C** *Speechreading Student Questionnaire Material*: This appendix contains study material used during the student questionnaire presented in Chapter 6.

**Appendix D** *PhonemeViz Material*: This appendix contains study material used during PhonemeViz study presented in Chapter 7.

**Appendix E** *MirrorMirror Material*: This appendix contains material used during the MirrorMirror study presented in Chapter 8.

2

# Background

## 2.1 Introduction

The research in this thesis is based on three foundational areas: hearing, speech, and speechreading. This chapter presents the foundational knowledge required to understand how hearing works and the affect that hearing loss has on an individual's ability to perceive speech. From there, I describe how speechreading can be used to improve understanding and the theoretical approaches used to teach speechreading.

## 2.2 Hearing

Although our sense of smell, and vision allow us to perceive events at a distance, the detection of many day-to day events relies exclusively on our hearing. For both animals and humans our ability to hear serves as an important detection system. For an animal, hearing the snap of a twig, or the rustle of leaves can help prevent capture or death from a predator. For humans, we rely on our hearing to detect important signals/events such as a smoke alarm, a child crying in another room or for hearing and locating a ringing phone. However, in addition to allowing us to perceive the world of acoustic vibrations all around us, it also provides one of our most important forms of communication – speech.

### 2.2.1 Hearing Mechanism

Sound is created when a source creates vibrations within a surrounding medium whether it is a solid, liquid or gas. These vibrations propagate away from the source at the speed of sound producing a sound wave. For instance, when we speak our vocal chords create vibrations within the air that is being exhaled, which leads to the production of sound. Human ears are capable of processing vibrations within the air at frequencies between 20 Hz to 20 kHz into sound waves [104]. When a sound wave reaches our ears it is converted into a series of messages that our brains can interpret.

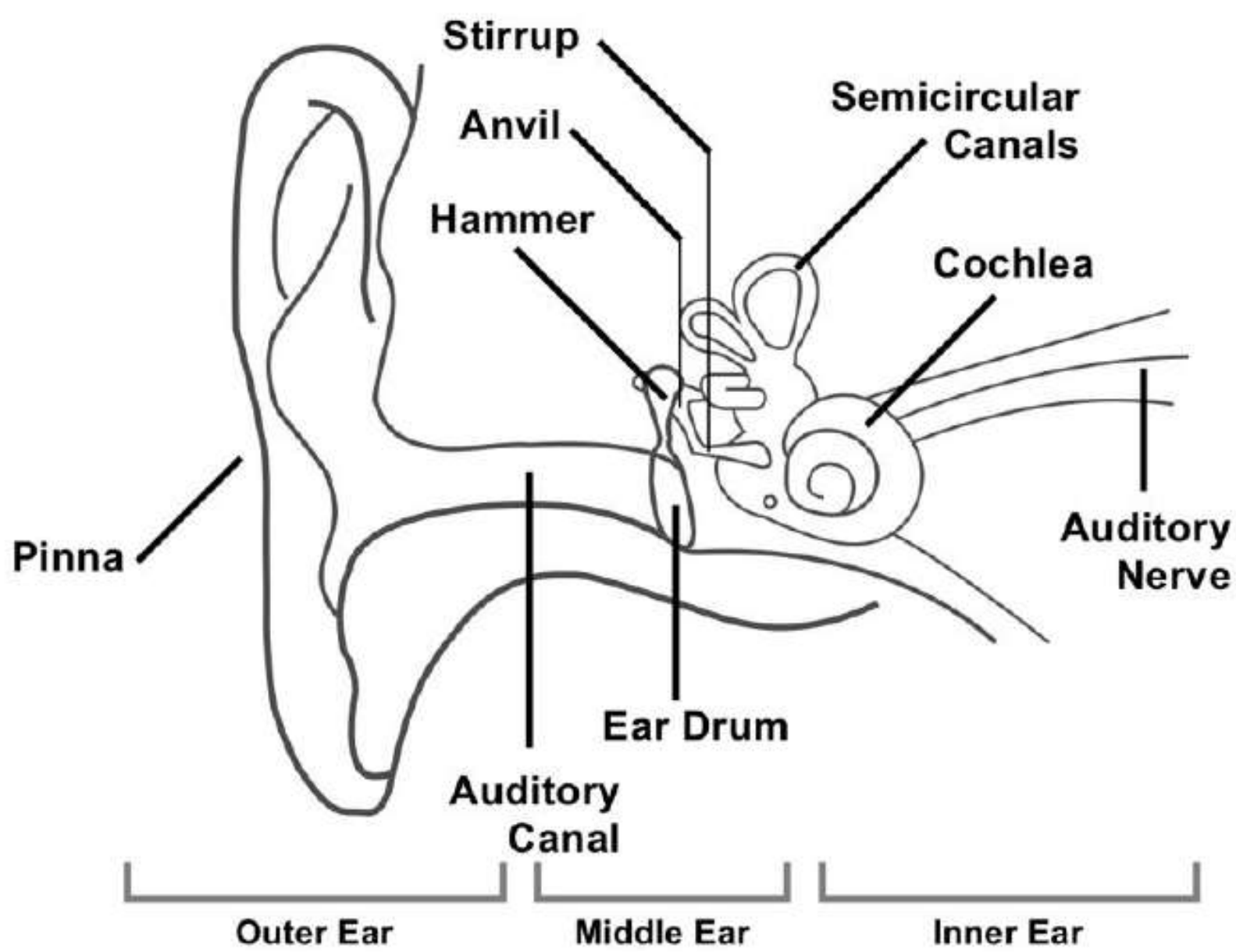

Figure 2.1: Diagram showing the structure of the human ear, detailing the parts of the outer, middle, and inner ear.

The outer part of the human ear, known as the pinna, gathers sound energy from a sound source and focuses it into the middle ear. The structure and shape of the pinna (as shown in Figure 2.1) is designed to bounce the sound in different patterns into the auditory canal, depending on whether the source is located above, below, behind, or in front of you [23].

The brain then determines the direction or angle of a sound source in relation to the head through a process known as sound localisation [23]. The brain calculates the interaural time difference (ITD) which is the difference in arrival time of a sound wave between the two ears. If a sound wave arrives from one side of the head the sound wave has a further distance to travel and thus a longer time to reach the far ear compared to the near ear [104].

The primary goal of the middle ear is to transfer sound waves in air into mechanical pressure waves that are transferred to the fluids of the inner ear. From the pinna, sound enters the ear canal and vibrates the eardrum and three small bones (the hammer, anvil and stirrup – known together as the ossicles as shown in Figure 2.1), which transfer vibrations of the eardrum into pressure waves in the fluid of the inner ear [104].

The pressure wave causes fluid to move through the inner ear (known as the cochlea) causing tiny hair cells within the cochlea to move. These hair cells convert movement into electrochemical nerve impulses which are passed to the brain via the auditory nerve, where they are interpreted by the brain as sound [104].

### Speech Perception

After the brain processes the initial sound signal, the signal is further processed to extract acoustic cues and phonetic information [64]. This speech information can then be used for higher-language processes such as word recognition [104]. I describe this process in more detail in Section 2.5.2 when discussing speechreading.

## 2.3 Hearing Loss

Hearing loss or a hearing impairment is the partial or total inability to perceive sound. Across the world it is estimated that 360 million people (~5%) have disabling hearing loss[a] [139]. People with hearing loss may be described as hard of hearing and a person described as d/Deaf[b] typically has little to no hearing.

The prevalence of hearing loss increases as we age [24], and it is expected that up to one in five people in the UK will have hearing loss by 2035 [2]. Hearing loss results in difficulties understanding what others are saying during conversation [131]. Our relationships and identities are shaped through the various conversations we engage in throughout our lives [37]. As hearing loss causes difficulties during conversations, it can result in social isolation [51], career stagnation [98], and a decrease in life satisfaction [131].

[a] Hearing loss greater than 40 dB in the better hearing ear in adults and greater than 30 dB in the better hearing ear in children.

[b] 'Deaf' with an upper case 'D' is an accepted way of denoting Deaf culture and describes people with a hearing loss who choose to identify with the Deaf community. A lower case 'd' in 'deaf' is a term used medically to refer to people with a high degree of hearing loss.

## Classifiying Hearing Loss

Hearing loss can be measured by an audiologist conducting a pure tone audiometry test, where an individual's responses to tones played at different frequencies (250Hz - 8000 Hz) and at different levels of loudness (between -10dBHL, which is extremely quiet, and 120dBHL, which is extremely loud) can be measured and plotted on an audiogram [131]. An audiogram is a chart that plots the loudness level (on the vertical axis) an individual can hear at each frequency tested (on the horizontal axis), and therefore illustrates their hearing threshold [131].

An individual's overall degree of hearing loss is usually described as the mean level of hearing loss in decibels in the better ear. The overall degree of hearing loss can be described using a classification system: Typical Hearing, and Mild, Moderate, Severe, and Profound Hearing Loss. Boundaries of the classification can vary, but the system used by the British Society of Audiology [32] and Action on Hearing Loss [3] (as shown in Table 2.1) is commonly used in the UK, and will be referred to throughout this thesis.

| Label | Range of hearing loss |
|---|---|
| Mild | 20-40 dB |
| Moderate | 41 - 70 dB |
| Severe | 71 - 95 dB |
| Profound | in excess of 95 dB |

Table 2.1: Classification descriptions of hearing loss used by the British Society of Audiology [32] and Action On Hearing Loss [3].

## Types of Hearing Loss

Hearing loss can be caused by a number of factors, including genetics, ageing, noise exposure, viral infections, birth complications, use of certain medications and trauma or physical damage to the ear itself.

There are two main types of hearing loss – *Conductive* and *Sensorineural*. When both types are present at the same time, it is known as *Mixed hearing loss*. It is also possible to acquire hearing loss resulting from damage to the auditory cortex of the brain [75].

**Conductive hearing loss:** Conductive hearing loss is the result of sounds not being able to pass freely from the outer to inner ear [104]. This may be a result of a blockage due to a build-up of wax (*cerumen*) in the auditory canal, or fluid in the middle ear caused by an ear infection.

However, it can also occur due to trauma or infection damaging the eardrum or an abnormality in the structure of the outer ear or auditory canal. Finally, a condition known as *otosclerosis* can also cause severe conductive hearing loss, where an abnormal growth of bone prevents the ossicles in the middle ear from moving freely [101].

The result of conductive hearing loss is that there is an attenuation of sound reaching the cochlea, leading to sound appearing muffled or quieter than normal [103]. The loss experienced may vary depending on the sound frequency leading to a loss in tonal quality [104].

**Sensorineural hearing loss:** This type of hearing loss is the result of damage to hair cells in the cochlea or damage to the auditory nerve (or both). This type of loss can be attributed to many different syndromes through genetic or non-genetic causes. For instance, diseases such as measles, mumps, and meningitis can cause permanent damage to varying degrees to the cochlea [103]. *Meniere's disease* can also damage the cochlea through repeated episodes of a build-up of excess fluid in the inner ear that causes pressure affecting both hearing and balance [110]. Additionally, a benign tumour (*acoustic neuroma*) can compress the auditory nerve, typically causing high frequency hearing loss resulting in *tinnitus* (ringing or buzzing in the ears) and balance problems [1]. Trauma to the ear as a result of repeated exposure to loud noise, a skull fracture or damage to the inner ear by surgical instruments during surgery can also damage the cochlea or auditory nerve [103]. Medications that are toxic to the ear (*ototoxic* medications) can cause temporary or permanent damage to the cochlea, typically affecting high frequency hearing [110]. Finally, gradual deterioration of the cochlea of one or both ears as a result of ageing is known as *presbyacusis* [30].

The result of sensorineural hearing loss is that there can be a lack of total hearing in different frequencies depending on the severity of the loss [103]. This is in spite of the outer and middle ear possibly functioning correctly; damage to the cochlea or auditory nerve can lead to some frequencies not being interpreted properly or not be reaching the brain at all [64].

## 2.4 Listening Devices

After an individual receives an audiological assessment, an audiologist may provide them with a listening device [131]. The main objectives for providing an individual with a listening device are to:

1) Improve speech intelligibility.
2) To restore a range of general sound intensity.

In most cases there are two devices that are commonly fitted to achieve the objectives listed above: *hearing aids* (HAs) and *cochlear implants* (CIs).

### 2.4.1 Hearing Aids

A hearing aid is an electroacoustic device designed to amplify sounds entering the ear canal with the general aim of making speech more intelligible [131]. Modern hearing aids contain the following components: one or more microphones to pick up the external sound, a preamplifier for each microphone, an analog-to-digital (ADC) converter for each amplified microphone signal, a digital signal processor (DSP), loudspeaker (also known as a receiver), battery, and a casing in which all of the components aforementioned are housed [114].

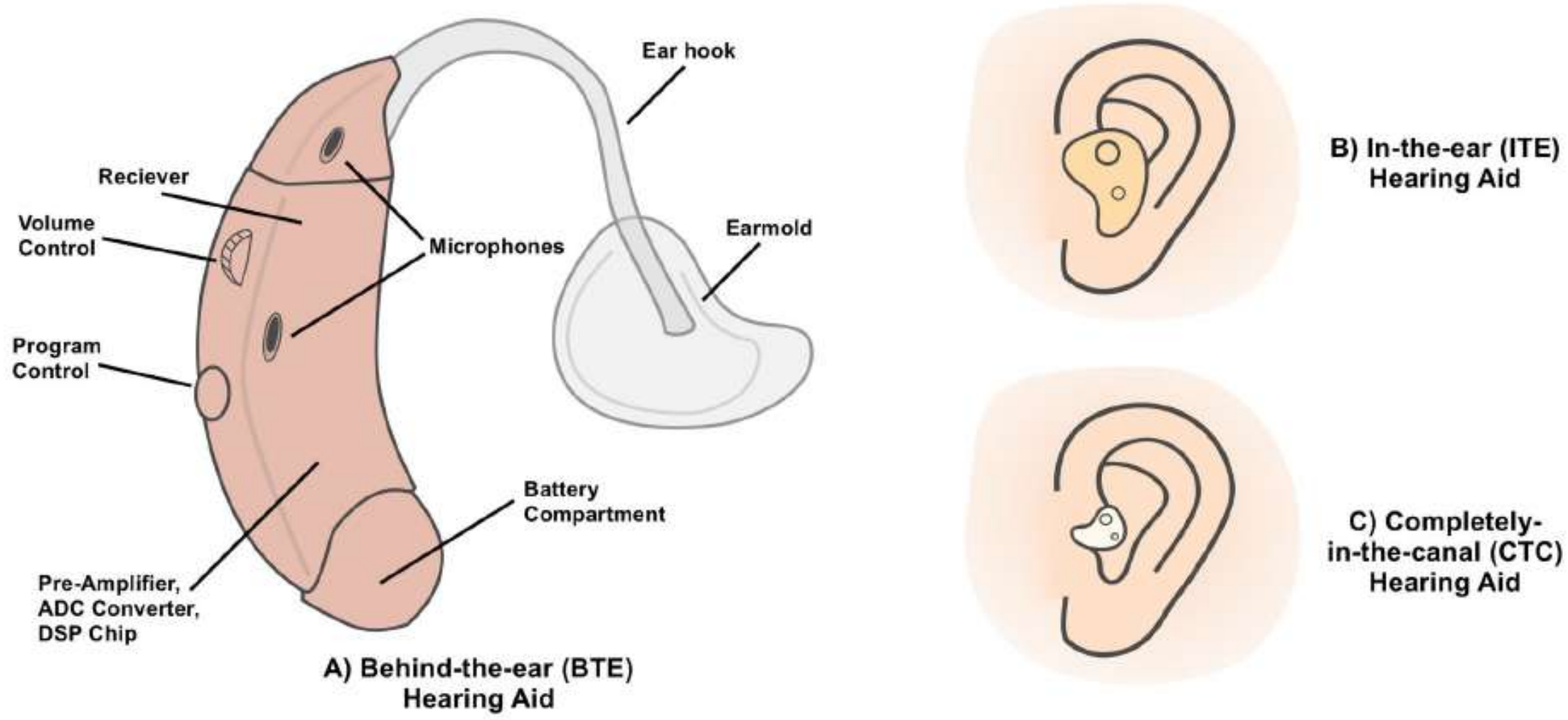

Figure 2.2: Diagram of different hearing aid styles: A) Behind the ear (BTE) hearing aid, B) In-the-ear (ITE) hearing aid, and C) Completely-in-the-canal (CTC) hearing aid.

There are three main styles of hearing aids (as shown in Figure 2.2): in the ear (ITE), completely in the canal (CIC), behind the ear (BTE), and BTE with the receiver in the ear canal with an 'open dome' (where the earmold or dome is not sealed into the canal) [114].

Hearing aids are intended to help reduce conversational difficulties, but are expensive [25], often counterproductive in noisy environments [61], and have low adoption rates (~14% people who need a hearing aid actively wear one in the UK [2]) due to a lack of comfort [99] and perceived social stigma [71].

### 2.4.2 Cochlear implants

Not all individuals who are diagnosed with hearing loss have the potential to benefit from using a hearing aid. For instance, in cases where the individual has very little residual hearing, no matter how much the sound information is processed or amplified, it will not improve their ability to hear speech [103]. In these cases, it is common for the individual to have sensorineural hearing loss, which is caused by the absence or damage to the hair cells in the cochlea. The fitting of a cochlear implant (CI) can improve the speech intelligibility for individuals with this type of hearing loss.

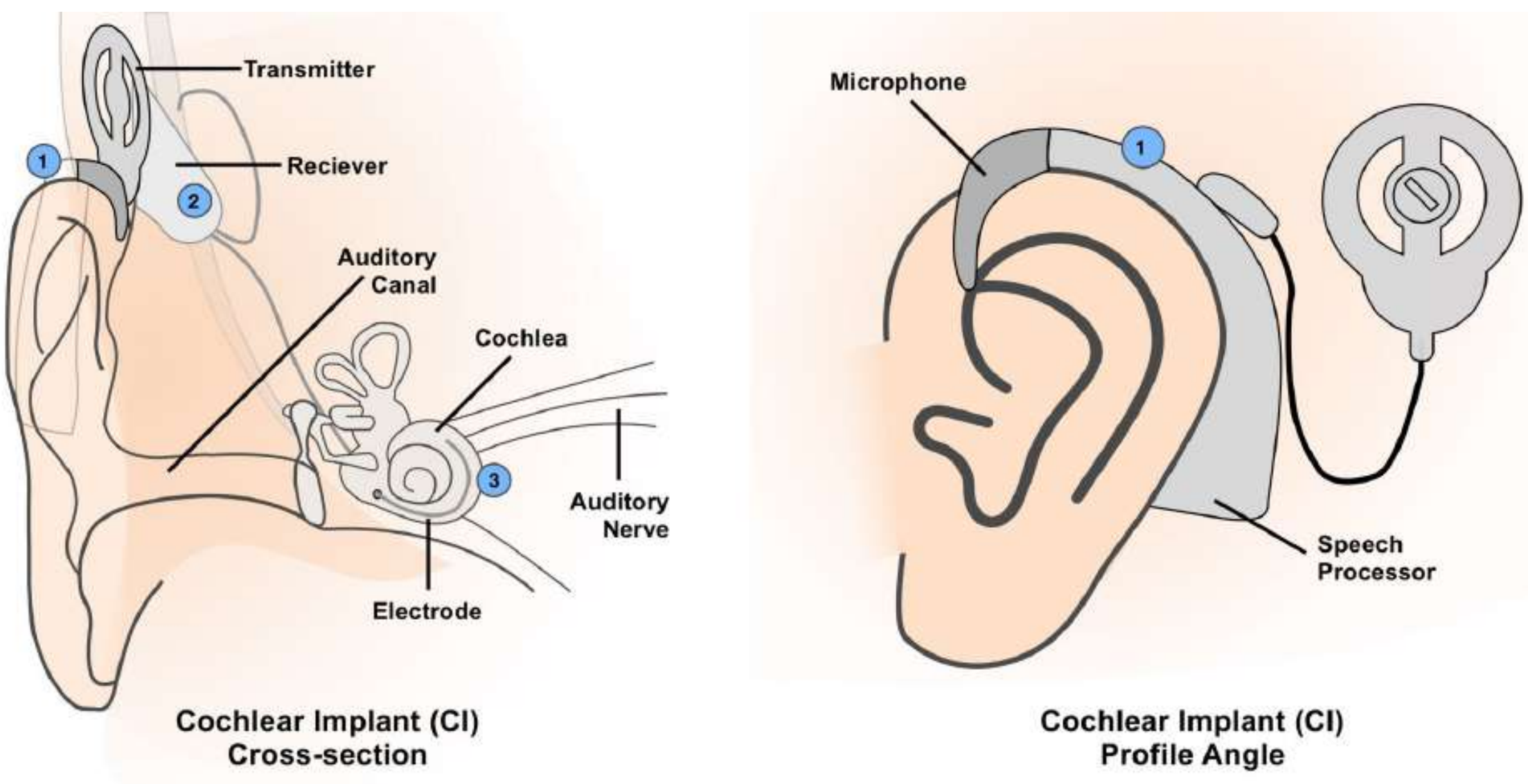

Figure 2.3: Left: Diagram showing a cross-section view of a cochlear implant along with the structure of the human ear where 1) External Speech Processor, 2) Internal implant, and 3) Electrode array inside the cochlea. Right: Diagram showing a profile angle of a cochlear implant.

A cochlear implant (CI) is a surgically implanted electronic device that replaces the need for hair cells by directly stimulating the auditory nerve. The nerve impulses are then delivered to the brain, following the typical pathways as if the cochlear was being stimulated naturally. A CI is composed of a microphone and some electronics that reside outside the skin, generally behind the ear (as shown in Figure 2.3.1), which transmits a signal to an array of electrodes placed in the cochlea (as shown in Figure 2.3.2) that directly stimulate the auditory nerve [114].

## 2.5 Speechreading

Typical human speech is produced from the combined influence that elements of the vocal tract have on air exhaled from the lungs. Using the specialised vocal fold muscles in the throat, we vibrate the air being exhaled to create sound waves. Through altering the frequency of the vibrations, we alter the pitch of the sound. The *articulators* (tongue, lips, teeth, and hard/soft palate) are then used to make specific sounds such as vowels and consonants.

As some elements of the vocal tract (namely the tongue, lips, teeth) are visible, there is a direct relationship between the perception of the auditory and visual characteristics of speech [96]. It is thought that speech is processed in a bimodal nature where the brain simultaneously processes the auditory and visual streams of speech information in order to form a comprehensive understanding of what a speaker is saying [19, 64]. This is demonstrated by the McGurk effect [100], in which conflicting auditory and visual stimuli can result in a perceived sound that is not present in the audible or visual stimuli.

Speechreading (often called lipreading) can be described as a special case of this audio-visual speech recognition where greater emphasis is placed on watching the movements of a speaker's lips, teeth, and tongue, rather than on the audible speech information. This is combined with conversation context (e.g., the speaker, the topic, the environment) and any residual hearing to understand speech. Due to the limitations of listening devices, speechreading is commonly used by people with hearing loss to improve understanding during conversation [37]. It has also been shown that those with typical hearing also speechread (albeit subconsciously) to help them understand others under noisy conditions [92].

### 2.5.1 Speechreading or Lipreading?

Both of the terms *speechreading* and *lipreading* are often used to describe the skill of improving understanding by focusing on visual aspects of a speaker, most notably their lip movements. However, to improve understanding of a speaker it is necessary for an individual with hearing loss to attend to the speaker's lip movements while also focusing on the speaker's facial expressions and gestures together with any auditory information that is available through residual hearing. In addition, research has demonstrated that visual speech information does not stem solely from lip position, as the tongue and teeth position also act as additional sources of information [127]. Furthermore, Vatikiotis-Bateson et al. [136] and Lansing and McConkie [88] both demonstrated that even though people with typical hearing tend to fixate on the mouth as noise levels increase, they also continue to gaze at a speaker's eyes more than anticipated during conditions with the highest levels of noise, suggesting that it is wrong to assume only visual cues from the mouth can influence understanding.

As the process of improving understanding by focussing on visual aspects of a speaker includes much more than just 'reading lips', the term *speechreading* is used most consistently within the literature, and will therefore be the term I will use throughout this thesis [c].

## 2.5.2 Speechreading Process

The ultimate aim of speechreading is to maximise the amount of information that can be extracted about what a speaker is saying. This is primarily achieved through watching the visual speech information, but as discussed speechreading involves combining information from three sources: 1) the visual speech stimuli, 2) the linguistic and environmental context, and 3) any residual hearing available.

### 1) Speech Stimuli

When discussing the speech stimuli, I am referring to aspects (both auditory and visually) associated with the production of speech sounds. This section briefly describes how speech is produced and how this information is extracted when speechreading.

### Phonemes

Every word in a spoken language is comprised of perceptually distinct units of sound known as *phonemes*. For instance, */b/*, */æ/*, and */t/* are the phonemes for "bat". There are 48 commonly-recognised phonemes in the English language [122].

Phonemes can be audibly distinguished by various features. One of the main features distinguishing phonemes is voice status. A *voiced* phoneme involves sound from the vocal chords [70]. For instance, F (e.g., fat, fan) and V (e.g., vat, van) are distinguishable because when the latter is produced, the vocal chords are producing sound (*voiced*), whereas in the former they are not (*unvoiced*). However, not all auditory changes affect the phoneme produced (e.g., singing words at different notes does not change what a speech sound represents phonetically). Phonemes can be divided into two main groups: *vowels* and *consonants*.

[c]There are times when I refer to speechreading as 'lipreading'. Most notably, when talking to speechreading tutors and their students who reside in the the UK – as within the UK, 'lipreading' is the commonly used term.

**Vowels:** In general, all vowel speech sounds are voiced, and the difference between vowels in English depends on the resonance level within the vocal tract that is produced by different positions of the articulators [70]. For instance, holding the tongue to the front of the mouth and varying its height produces IY as in beat, IH (bit), EH (bat), and AE (bet). With the tongue in the mid position, we get AA (ball), ER (bird), AH (but), and AO (bought). With the tongue in the back position we get UW (boot), UH (book), and OW (boat).

There is another class of vowels called *dipthongs*, which change during their duration. They can be thought of as starting with one vowel and ending with another. For example, AY (buy) can be approximated by starting with AA (ball) and ending with IY (beat).

**Consonants:** Consonants are separated into further classes, with many having both voiced and unvoiced pairs [87]. A *stop* or *plosive* involves stopping the speech sound (using the articulators) and then re-releasing a speech sound [70]. They come in unvoiced/voiced pairs: P/B, T/D, and K/G. A *fricative* involves 'hissing' sounds generated by constraining the speech sound by the lips and teeth [70]. They also come in unvoiced/voiced pairs: F/V, TH/DH (e.g., thing versus that), S/Z, and SH/ZH (e.g., shut and azure). *Nasals* are all voiced and involve moving air through the nasal cavities by blocking it with the lips and gums: M, N, and NX (sing) [70]. *Affricatives* are similar to stops but are followed by a fricative: CH (church), JH (judge) [70]. *Semi-vowels* (also referred to as *glides*), are consonants that have a continued, gliding motion of the *articulators* into the following vowel and include J, and W [87]. A *liquid* is a generic label used to classify two English approximate consonants, /r/ and /l/ [70].

## Visemes

When one of the phonemes described above is spoken, the speaker's lips, teeth, and tongue produce a visual representation known as a *viseme* [54]. Hearing loss causes some phonemes to be lost or difficult to perceive (depending on the type and severity of the hearing loss), but the viseme is still available. For example, */l/* and */r/* are acoustically similar in English (especially when following another consonant, such as 'grass' vs. 'glass'), but are generated using distinct visemes, so this visual difference can be used to determine if a speaker has said */l/* or */r/*. When discussing making use of visual speech stimuli during speechreading, typically this refers to the process of trying to map visemes to phonemes to help understand what a speaker is saying [21].

However, the viseme-to-phoneme mapping is often a 'one-to-many' relationship, in which a single viseme can be mapped to a number of phonemes [90]. For example, */v/* is a voiced phoneme, which is audibly distinct from */f/*, which is not voiced. However, the viseme for */v/* is very similar to the viseme for */f/*, making the words *'fan'* and *'van'* difficult to distinguish visually.[d]

In theory because each vowel is produced with a distinct oral cavity shape, it is thought that vowels are easier to distinguish than consonants [77, 93]. However, when we speak the articulators (tongue, lips, teeth) are constantly moving and this results in the shape of the viseme for one phoneme being influenced by the shape of the viseme for the preceding and following phoneme. This blending of speech shapes based on neighbouring phonemes is known as *co-articulation*. For example, even though the words *ball* and *boot* begin with the same initial /b/ sound, when our lips form the word *boot* they are rounded but not rounded when they form the word *ball*.

There is some disagreement among researchers as to how many viseme classes exist [39]. Lucey et al. produced a table that effectively maps all possible phonemes to the generally accepted 14 viseme classes [90]. This table is reproduced in Table 2.2, and shows the typical 48 phonemes [122] for the English language, grouped into their 14 viseme classes.

### Homophenes

As discussed above, many phonemes can be represented by the same viseme, leading to some phonemes being difficult to visually disambiguate from one another. However, in addition to individual visemes, it is estimated that between 40-60% [14] of words in English appear visually similar when spoken. Words that appear similar on the mouth are known as homophenes. For example, even though words such as *grade* and *yes* are audibly distinct they appear visually similar when spoken. Therefore, in cases where the noise level reduces the use of residual hearing, the speechreader may have difficulty telling words that are homophenous apart.

[d] I encourage readers to make *'ffffff* and *'vvvvvv'* sounds to hear the difference, and to make these sounds plus the words 'fan' and 'van' in front of a mirror (quietly or without using your voice) to see the lack of visual difference.

| Phoneme | Viseme | Phoneme | Viseme |
|---|---|---|---|
| P (p) | /p/ | K (k) | /k/ |
| B (b) | | G (g) | |
| M (m) | | N (n) | |
| EM (m̩) | | L (l) | |
| F (f) | /f/ | NX (r̩) | |
| V (v) | | HH (h) | |
| T (t) | /t/ | Y (y) | |
| D (d) | | EL (l̩) | |
| S (s) | | EN (n̩) | |
| Z (z) | | NG (ŋ) | |
| TH (θ) | | IH (ɪ) | /iy/ |
| DH (ð) | | IY (i) | |
| DX (ɾ) | | AH (ʌ) | /ah/ |
| W (w) | /w/ | AX (ə) | |
| WH (w) | | AY (aɪ) | |
| R (r) | | ER (ɝ) | /er/ |
| CH (tʃ) | /ch/ | AO (ɔ) | |
| JH (dʒ) | | OY (ɔɪ) | |
| SH (ʃ) | | IX (ɨ) | |
| ZH (ʒ) | | OW (oʊ) | |
| EH (ɛ) | /ey/ | UH (ʊ) | /uh/ |
| EY (eɪ) | | UW (u) | |
| AE (æ) | | AA (ɑ) | /aa/ |
| AW (aʊ) | | SIL & SP | /sp/ |

Table 2.2: Typical 48 ARPABET phonemes (with their IPA symbol – `https://www.internationalphoneticassociation.org/`) used in the English language including silence (SIL) and short pause (SP), grouped into their 14 viseme classes adapted from [90].

## 2) Context

### Environmental and Situational Cues

Extracting contextual information from the location in which speechreading is taking place can help improve speechreading success [93]. For example, the initial questions from a barista in a coffee shop can often be anticipated, which helps to build overall understanding [81]. Furthermore, a speaker's body posture, hand/arm gestures, and facial expressions can influence meaning; a pause and a tilt of the head can indicate a question, an extended hand towards a door means 'after you', a nod of the head means yes. Facial expressions can also serve as clues to the emotion of an utterance – a smile might indicate happiness and agreement versus a furrowed brow showing displeasure or disagreement. Interpreting the context around the speaker can help to clarify the words being speechread [93]. In addition, knowing the topic of a conversation can also aid the speechreading process. For example, Gagné et al. [59] found that when testing speechreading proficiency, embedding target words in semantically related sentences increased speechreading accuracy relative to target words embedded in unrelated sentences.

### Linguistic Context

Due to the redundancy of language, the *linguistic context* within a sentence can give clues to what a speaker might be saying [57]. Boothroyd refers to aspects of linguistic context as 'constraints' [26]. For example, imagine a speechreader has determined a speaker has said the following sentence (where each letter is represented by an underscore):

"You _ _ _ _ _ _ _ _ _ _ _ _ umb _ _ _ _ _."

The *lexical* constraints of this sentence limits the number of possibilities for the final word, as there are very few words that start with the letters 'umb' and only one word that is commonly said in conversation. Furthermore as the preceding word is only two letters the *syntactical* constraints of this sentence suggest that it must be 'an'. Therefore the last two words are likely 'an umbrella'. Finally, if the sentence was uttered as you were to leave your office with a co-worker for lunch, *topical* and *pragmatic* constraints gives us enough knowledge to say that the full sentence is probably "You should take an umbrella". Words such as "umbrealla" are easier to speechead because there are few other words that are visually similar. In contrast, words like "bat" are difficult to speechread because they have several visually-similar neighbours (e.g., mat, pat, bet, bit). However, words that are unusually used within conversations (such as highly specific technical words) will be more difficult to speechread than words that are used commonly [52, 131].

### 3) Residual Hearing

Finally, the individual's residual hearing either alone or amplified through the use of a hearing aid or cochlear implant can provide substantial benefit when speechreading [57, 124]. Even though not all phonemes will be audible, residual hearing can help to supply *suprasegmental* patterns within speech [64]. For instance, to provide intonation (which can clarify the speaker's feelings or intent), stress patterns (indicating a question or an emphasised word) or indicate word boundaries and pauses. Even being able to tell a voiced versus an unvoiced pair can help disambiguate a consonant viseme pair [131].

## 2.5.3 Speechreading Ability

It has been reported that there is a wide variability in speechreading ability between individuals. These estimates are thought to vary between zero and close to ninety percent words correct in sentences [13, 18, 92]. However, typically these estimates are based on results from study tasks that restrict the participant to use "visual-only" speechreading, in which the individual speechreads the talker without the use of residual hearing. Therefore, in typical conversation where the speechreader can make use of contextual cues and their residual hearing, this ability will likely be greater.

Furthermore, it has been demonstrated that individuals who are deaf are more successful at speechreading [13]. This could be because these individuals have relied on the visual speech signal for communication longer than those who have acquired hearing loss in later life [37].

## 2.5.4 Factors Affecting Speechreading

The differences in ability described above can be explained by the many aspects that affect the success of speechreading. Aside from the difficulty of extracting information from visual speech stimuli and the context surrounding the speaker discussed above, there are three main areas that affect speechreading success: 1) the speaker, 2) the environment, and 3) the speechreader.

### 1) The Speaker

The major speaker-related factors influencing speechreading success are the degree of lip movement, the rate of speech, familiarity with the speaker, and the presence of distractions. In addition, shouting, mumbling, turning away, speaking quickly, covering the mouth and smiling while talking, all make speechreading more difficult [131].

**Lip Movement** A speaker who speaks naturally with precise lip movements is thought to be easier to speechread [85]. In fact, a *lipspeaker*[e] is a person with typical hearing who is trained to repeat a speaker's message to speechreaders accurately, without using their voice. They clearly reproduce the shape of words, flow, rhythm and phrasing of natural speech and repeat the stress pattern as used by the speaker, all in a manner that makes speechreading easier.

Additionally, if the speaker has an accent that is not native to the speechreader they may find the speaker to be less intelligible. Irwin et al. found that even when provided with a contextual cue, participants found that speechreading a nonnative accent (Glaswegian from Glasgow, Scotland) was more difficult than speechreading their own native (East Midlands from Nottingham, England) accent [76].

**Rate of Speech** A speaker's rate of speech depends on their age, gender, and current psychological state [128]. Nitchie [105] estimated that during typical speech, a speaker may produce 13 speech sounds per second, yet estimated that the eye was only capable of consciously seeing around 8-10 speech movements per second (although it is thought that for most speechreaders recognising speech movements individually may not be a fully conscious process [78]). Regardless, it is unlikely that the speechreader is able to perceive every movement of a visual speech signal, so when a speaker talks quickly (such as in presentations, or in conversational speech) it can be difficult to determine when one word ends and another begins; the connected speech boundaries between words may not be obvious [131].

**Familiarity** If the speechreader is familiar with the speaker (such as a family member, or their speechreading teacher), they will find the task of speechreading easier because they will have become accustomed to the speaker's particular speech movements [93]. The opposite is also true, if the speechreader encounters someone they have never met, it will take some time for them to be able to speechread them effectively [93].

**Distractions** The presence of facial hair is reported by speechreaders to be distracting and is thought to enhance the difficulty of speechreading some speakers, due to it potentially obscuring or changing the appearance of some speech movements [81] (but disputed by Kitano et al. [84]). Furthermore, wearing sunglasses or reflective eyeglasses can make eye contact difficult, which is thought to be important during speechreading [88, 136]. Items such as dangling earrings and other articles of distracting clothing are also thought to increase the difficulty for the speechreader [93].

[e] `http://www.lipspeaker.co.uk/lipspeaking/`

### 2) The Environment

The environmental factors that affect speechreading are concerned with the ability to see or hear the speaker clearly. For instance, the distance between the speechreader and speaker is vital, because increased distance will make the more subtle movements of the articulators less visible [93]. The angle of the speaker also influences the difficulty of the task, with some individuals finding it easier to speechread when viewing the speaker at a frontal viewing angle versus a profile angle [53, 131]. The lighting on the speaker is also important – if the face is in shadow, then the visual cues may be difficult to see. Erber [53] found that speechreading under conditions of high background brightness results in a significant reduction in visual-only speechreading performance.

As discussed before, residual hearing can help the process of speechreading. As a result however, noisy situations make speechreading more difficult as there may be less information available to the speechreader in the auditory channel.

### 3) The Speechreader

Finally, factors related to the speechreader themselves affect the outcomes of the speechreading process. Tye-Murray et al. [132] found that younger adults speechread better than older adults, and suggested that the difference between older and younger adults was comparable across gender.

In addition, the visual acuity of a speechreader is very important to the task of speechreading [80]; if an individual has limited vision then their speechreading ability will be reduced.

As mentioned above, assessing the lexical context of speech is part of the speechreading process, therefore the greater lexical knowledge a speechreader possesses the better they will be able to speechread [93]. The individual's level of hearing loss, and how long they have had a hearing loss also affects the speechreading process [124].

Finally, the amount of time the individual has been speechreading positively influences their speechreading ability. For instance, it has been demonstrated that individuals with early-onset hearing loss have enhanced speechreading ability versus those with typical hearing because they have relied on the visual speech signal throughout their lives particularly to acquire spoken language [13].

## 2.6 Speechreading Teaching

As a result of the factors discussed above, speechreading can be a difficult skill to acquire, and can therefore take considerable practice and training [81]. Novice speechreaders (e.g., someone who has received no formal speechreading training) often find it difficult to fully understand what a speaker is saying, resulting in confusion, frustration, and reduced conversational confidence [37]. Individuals with hearing loss have been informally learning speechreading since at least the early sixteenth century, where it was acquired as a by-product of learning speech production [57]. Through development of speech production skills, it was hoped that the individual would gain experience in observing the visual information of their instructor's facial movements and thus acquire the skill of speechreading. The explicit teaching of speechreading with methodologies that were not concerned primarily with speech production did not occur until the nineteenth century [78].

Publicly-funded speechreading classes are sometimes provided, and have been shown to improve speechreading acquisition [10]. However, classes are only provided in a handful of countries around the world and when provided, there is often an insufficient number of classes running (e.g., only 50 of an estimated 325 required classes are currently running in Scotland [10]) and classes typically require mobility to attend.

Speechreading classes primarily focus on learning how different mouth shapes are produced during speech [81], as well as how to use conversational repair strategies to gain important contextual information to help 'fill in' any gaps in understanding [81]. Classes also include information about hearing aids or other assistive listening devices, and give people a social space to meet with others who have a hearing loss [131]. Classes can also improve an individual's self-confidence [20], and help attendees become more knowledgeable about their hearing loss and how they can make communication easier.

### 2.6.1 Teaching Approaches

Within classes there are two main approaches to teaching speechreading: *synthetic* and *analytic* [57].

#### Synthetic Approach

Synthetic methods use a 'top-down' approach in which the speechreader is encouraged to focus on the gist or the topic of a conversation to help them determine individual words being spoken [78].

Synthetic methods are often referred to as ‘mind-training’ or ‘context-training’, as they focus on teaching students to use situational cues and lexical ability to help understand the topic of the conversation. Synthetic methods consider the sentence to be the basic unit of speech [131].

### Analytic Approach

Analytic methods use a ‘bottom-up’ approach in which the speechreader is encouraged to focus on individual speech movements (visemes) to identify the word, phrase or sentence being spoken [131]. Analytic methods are often referred to as ‘eye-training’ as the speechreader focusses on the visual aspects of a speaker to disambiguate visual speech patterns [36, 78]. Analytic methods hold that the syllable or phoneme is the basic unit of speech [78].

## 2.6.2 Teaching Methods

Formal speechreading teaching methods are traditionally divided into the two approaches described above. However, when comparing each method, there are many elements of each that are common. Therefore, this distinction within the methods is perhaps unnecessary, and it is the specific techniques used within each method that should be categorised as analytic or synthetic.

### Nitchie Method

The Nitchie Method [105] was initially developed using an analytic approach, however it shifted towards a synthetic approach in later years. Nitchie is credited with establishing the foundations of modern speechreading training as well as the synthetic approach to speechreading. His method stresses that eye-training materials and those based around association or context be separated. He also believed that context training was more important than training for visual disambiguation [22]. However, his materials do not strictly follow his method. The context materials are a series of unrelated sentences and short stories that are written for reading and not as a conversation (so the context surrounding the speaker cannot be extracted). Sentence materials are based around speech movements that come from eye-training materials rather than standing alone.

### Mueller-Walle Method

The Mueller-Walle Method (also known as the Bruhn method [22]) was introduced by Martha Emma Bruhn, who studied speechreading with Julius Mueller-Walle in Hamburg, Germany and then introduced the method into the United States [34]. The Mueller-Walle Method is an analytic approach that emphasises eye-training through syllable drills, which are rhythmic drills consisting of contrasting syllables that are restricted to sound combinations found in the English language [34]. Syllable drills are spoken as quickly as possible by the tutor.

### Kinzie Method

The Kinzie Method [83], was introduced by the Kinzie sisters, who studied with Bruhn and then Nitchie and then combined features from both methods in order to form their own method. The Kinzie Method uses a synthetic approach to speechreading and includes mirror practice (where the student talks before a mirror to learn visual differences between speech movements) and the use of voice. Materials in this method are organised into graded lessons for both children and adults, with sentences forming the basis of instruction [83].

### Jena Method

The Jena Method [36], was developed by Karl Brauckmanin Jena, from Germany and was introduced into the United States by Anna Bunger. The Jena Method is an analytic approach that emphasises syllable drills and stresses kinaesthetic awareness during speech production as well as eye-training [22]. Eye-training is accomplished through syllable drills in a similar manner as the Mueller-Walle and early Nitchie Methods. During the drills, the speaker is expected to speak in unison with the instructor and imitate their lip and jaw movements thereby concentrating on the kinaesthetic sensations experienced. It has been called ‘the talking way to speechreading’ [11].

## 2.7 Conclusion

This chapter presented the foundational knowledge required to understand how hearing works and the affect that hearing loss has on an individual’s ability to perceive speech. From there, I described how speechreading can be used to improve understanding and the theoretical approaches used to teach speechreading. In the next chapter, I will provide a review of previously designed Conversation Aids, and introduce the concept of Speechreading Acquisition Tools (SATs), which are a subset of Conversation Aids.

# 3

# Related Work

## 3.1 Introduction

This chapter presents a literature review of related work in this area of research. In particular, I discuss *Conversation Aids* and *Speechreading Acquisition Tools (SATs)*. I define a *Conversation Aid* as any technique or technology that could enable or be used to support face-to-face conversation for people with hearing loss. I define a *Speechreading Acquisition Tools* (SATs) as any technique or technology designed specifically to improve speechreading acquisition. At this stage it is important to recognise that *Conversation Aids* are a superset of *Speechreading Acquisition Tools*, as the latter have been designed specifically to be used by within the context of practicing or using speechreading, which is a special case of human conversation.

## 3.2 Conversation Aids

A number of techniques have been developed to overcome the challenges faced by people with hearing loss during conversation. I refer to these techniques as *Conversation Aids*, as they could be used by somebody with hearing loss to help improve understanding during conversation.

*Signed Languages* (e.g., American Sign Language [126], British Sign Language [47]) are natural languages with fully developed linguistic systems [49]. There are many different signed languages used internationally, yet all utterances are produced using two activities; the *manual activity* produced with the hands/arms and the *non-manual* activity which is produced by the shoulders, head and facial expressions [49]. Signed languages are often the preferred language of communication by people who are deaf or those with profound hearing loss [131].

*Cued Speech* [45] is a visual system of communication that uses eight hand-shapes, known as cues (representing consonants), placed in four positions around the mouth (representing vowels) that aim to clarify speech movements during speechreading. Using cued speech, it is possible to be able to determine phonemes purely through the visual signal, as each phoneme has a distinct combination of cue and position, and lipshape.

In spite of their benefits and the importance of signed languages to Deaf culture [118], a signed language (or cued speech) needs to be known by both conversation partners in order to help; they do not help during conversations with people who do not know the language/technique.

However, there has been work that investigates generating animations of signed languages using transcripts [73, 74]. Similar work has also shown the possibility of using automatic speech recognition to generate cued speech [48]. Therefore, it may be possible in the future to generate animations of both signed languages and cued speech using automatic speech recognition, which could be displayed to users during conversation through the use of glanceable displays (e.g., Google Glass[a]), or head mounted displays (e.g., Microsoft Hololens[b], Epson Moverio Glasses[c]).

Speech can be visualised by showing the intensity of sound at different frequencies over time. This can be shown graphically in a *spectrogram*, where time is on the X axis, frequency is on the Y axis, and the colour of the area corresponds to intensity. Spectrograms are used by linguists to identify words phonetically, although becoming competent can take considerable training (e.g., after 22 hours of training Greene et al. demonstrated that participants could accurately identify words they had been trained on [65]).

Watanabe et al. [138] introduced a visualisation that creates readable patterns by integrating different speech features into a single image, with the final image resembling a spectrogram enhanced with additional patterns, colours and labels (as shown in Figure 3.1). However, the evaluation of thus visualisation used participants with extensive spectrogram reading experience, so the technique's generalisability to people with hearing loss, and those who have limited experience with spectrograms, is unknown.

*VocSyl* [66] is a software system that provides real-time visual feedback in response to vocal pitch, loudness, duration, and syllables to allow speakers to gain new insights into their speech (as shown in 3.2). Pietrowicz and Karahalios [111] built upon this work (renaming it *SonicShapes*) by adding additional colour information to represent the classes of phonemes uttered by the speaker (as shown in Figure 3.3). Colour mappings in the visualisation represent phonological detail with distinct colours for high-closed vowels, mid vowels, low-open vowels, diphthongs, liquids, nasals/glides, and fricatives/affricates/stops [111].

[a] `https://www.x.company/glass/`
[b] `https://www.microsoft.com/en-gb/hololens`
[c] `https://epson.com/moverio-augmented-reality`

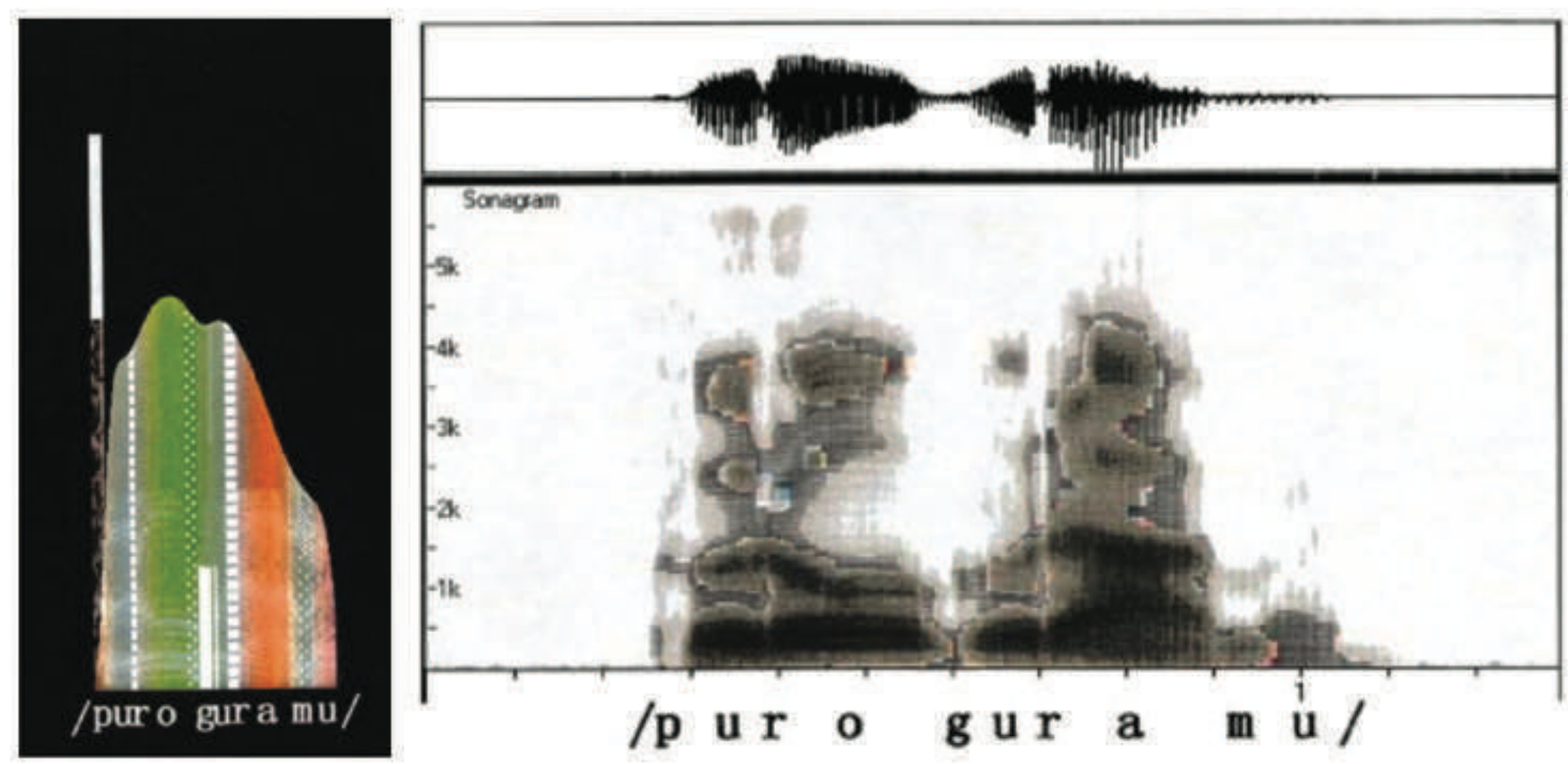

Figure 3.1: Example of Watanabe et al.'s visualisation (left) in comparison with a sonogram (top, right) and spectrogram (bottom, right) for the Japanese word /puroguramu/ ("program"). Image from [138].

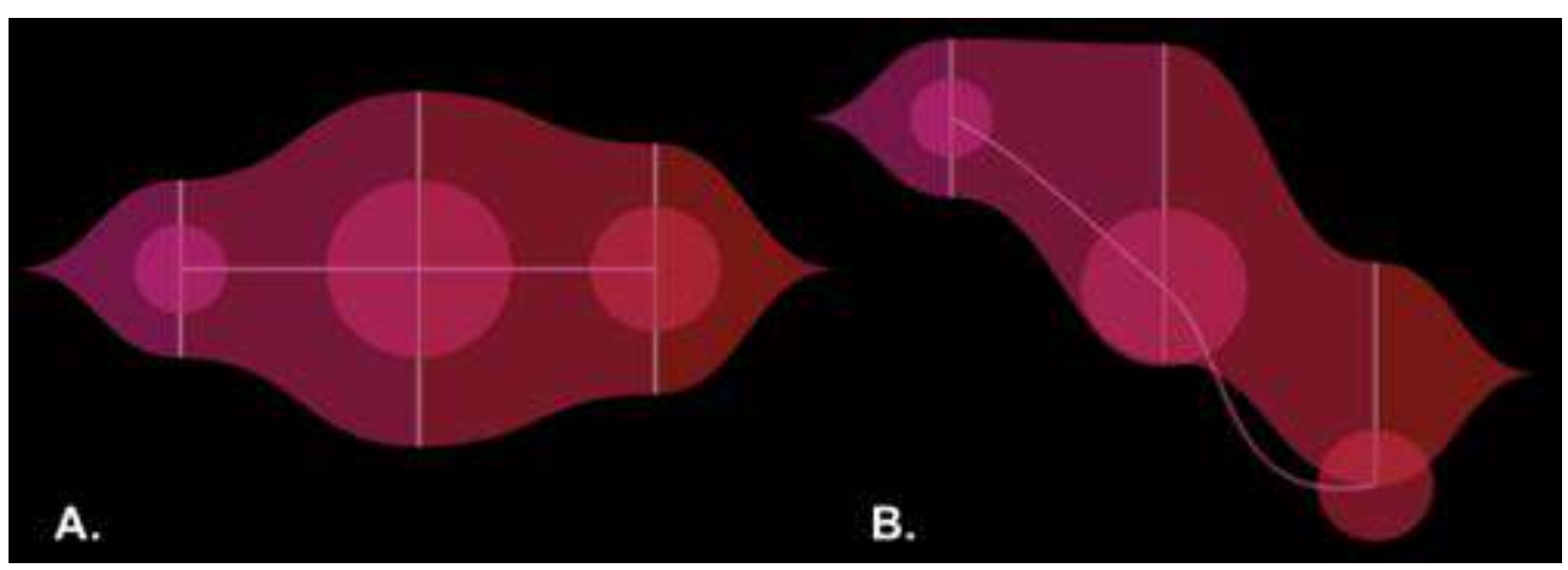

Figure 3.2: Example image of VocSyl, showing visualisation for 'Hello World' without voice pitch in A), and with visualised voice pitch in B). Image created using VocSyl [66] (downloaded from `http://social.cs.uiuc.edu/projects/vocsyl/vocsyl.html`).

VocSyl and SonicShapes were not designed to supplement speechreading, therefore both can lead to multiple words having similar visual representations; words within the same viseme groups such as 'fan' and 'van' are coded with the same colour and resulting visualisation because /f/ and /v/ both have the same phoneme class (fricative) and VocSyl does not provide voiced/unvoiced distinction.

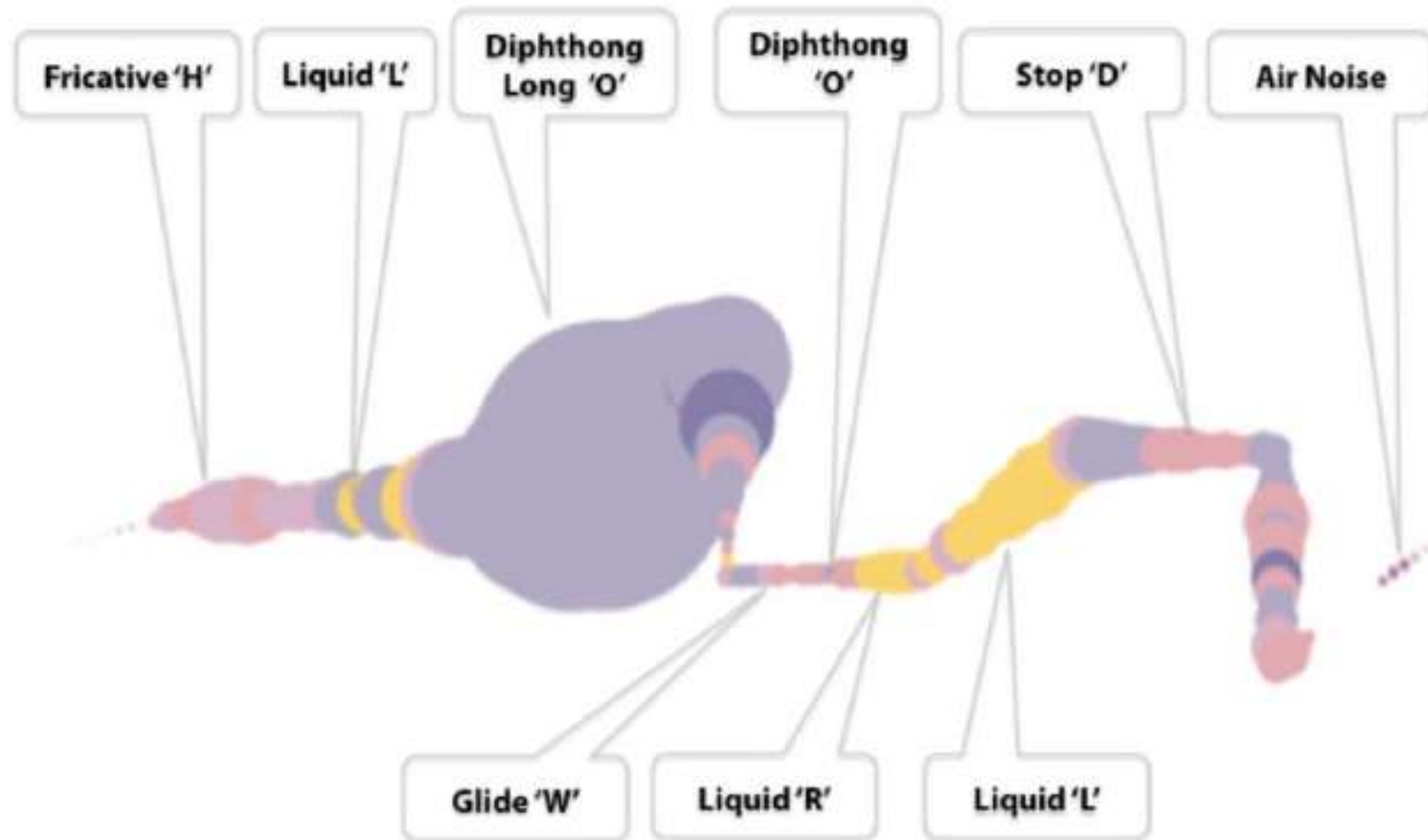

Figure 3.3: Example image of SonicShapes, showing visualisation for 'Hello World'. Image from [111].

*Subtitles* (captions, closed-captioning) displays the audio of a television programme as text on the TV screen, providing access to the speech and sound effects to individuals with hearing loss. As the accuracy of Automatic Speech Recognition (ASR) has been shown to be poor in some situations [89], subtitle creation typically relies to some extent on human transcription, which typically introduce delays. Subtitles also require the viewer to split their attention between reading and watching the video content (or the speaker's face); one eye-tracking study found that participants spent ~84% of their viewing time focussed exclusively on captions [79].

## 3.3 Speechreading Acquisition Tools (SATs)

A number of tools have been developed to help support people with hearing loss acquire speechreading, either through supporting practice or supporting their use of speechreading (which can aid acquisition). I define these tools as *Speechreading Acquisition Tools* (SATs) – a new type of technology specifically designed to support speechreading.

### SATs for Speechreading Support

The earliest example of technology designed specifically to support speechreading can be seen in Upton's *Wearable Eyeglass Speechreading Aid* [134]. This SAT used a clip-on microphone (Figure 3.5, left) to detect speech components, and processed the signal via high-and low-pass filters to classify spoken phoneme components. The first version of this SAT used five Light Emitting Diodes (LEDs) embedded within the lens of the left side of a pair of eyeglasses (as shown in Figure 3.4), but was later replaced with an LED matrix positioned at the side of a pair of modified eyeglasses (as shown in Figure 3.5, middle). The light output from the LED matrix was channeled using a mirror so that it appeared at the centre of that side's lens (as shown in Figure 3.5, right), enabling an early augmented reality system (e.g., the bottom LED illuminated when a phoneme was voiced, making it appear as if the speaker's throat was glowing).

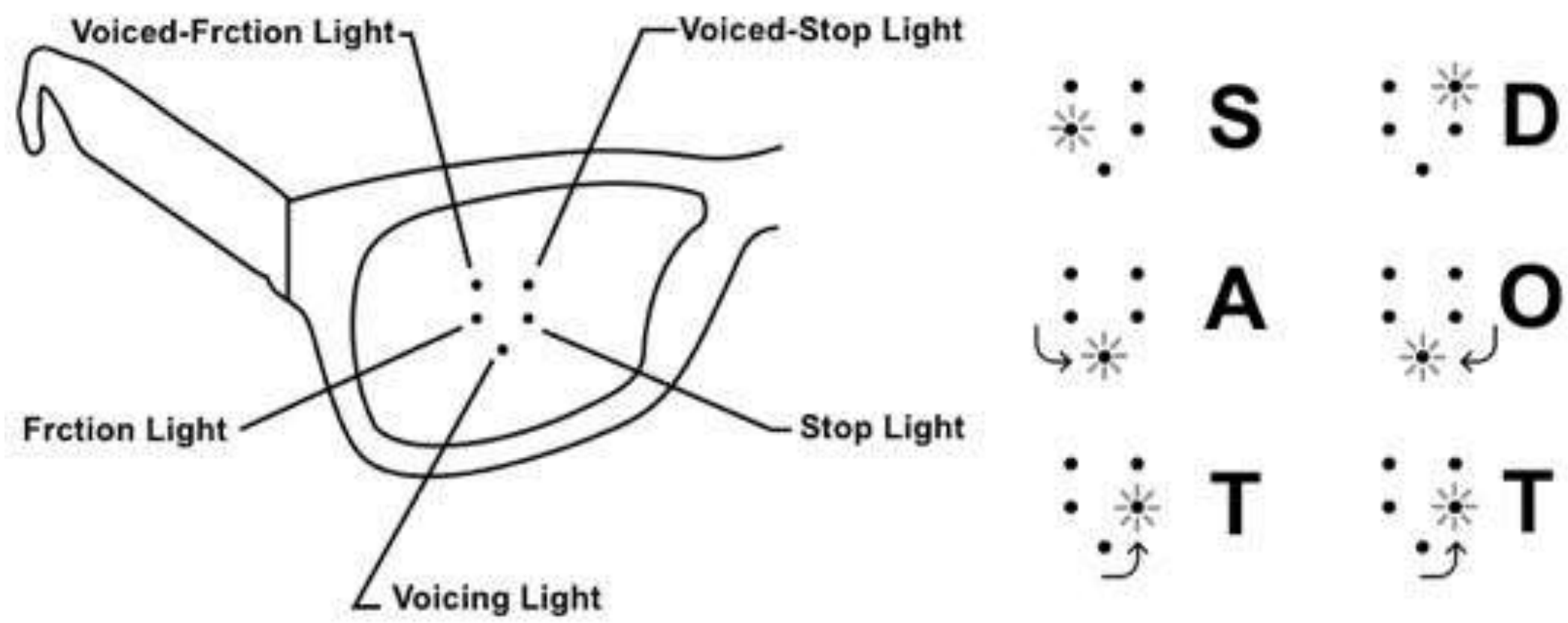

Figure 3.4: Display arrangement of Upton's 'Wearable Eyeglass Speechreading Aid' (Left). The small circles represent LEDs mounted on the eyeglass lens, which light up based on what is processed by the analyser. Pattern examples for 'SAT' and 'BAT' (right). Based off drawings from [134].

Pickett et al. [109] evaluated the matrix version of the eyeglasses using six participants with moderate to profound sensorineural hearing loss. During six one hour evaluation sessions across consecutive days, participants were asked to recognise initial or final consonants in monosyllabic words (e.g., pit, mit, bit) with and without using the eyeglasses on a male and female speaker. The participants responded by circling potential responses printed on a form. After each response the speaker fingerspelled the correct answer to the participant. In general, the results showed that performance with the eyeglasses was, on average, higher than speechreading alone (~60% with versus ~50% without the eyeglasses) although there was considerable individual variability between days [109].

Much later, a similar approach (albeit with a peripheral display) was taken by *Ebrahimi* et al. [50]. This SAT was also a modified pair of eyeglasses, with a microphone for input and a two-dimensional 5x7 red LED matrix mounted in the periphery of the right eyeglass lens. Selected speech features (voicing, plosion, and friction) were processed and then encoded as visual patterns to be used in conjunction with speechreading (as shown in Figure 3.6). An evaluation was conducted with participants with and without hearing loss. The evaluation was around four hours long divided into three sessions; half an hour speechreading the speaker without the SAT, half an hour introduction to the SAT and then two hours of training using the SAT, and then a final hour consisting of six 10 minute test sessions. The test sessions consisted of the participant viewing a speaker saying 60 vowel-consonant-vowel 'nonsense syllables' without audio. Participants responded by marking on an answer sheet consisting of 60 numbered rows, where each row contained the 'nonsense syllables' in a random order. The participant was to choose one answer in each row. Between sessions the participants were also allowed to practice with the SAT and learn the test material. The results showed that performance with this SAT was around 76% using the SAT and 41% without [50]. However, as the evaluation description demonstrates, for the final session participants were familiar with both the speaker and the material.

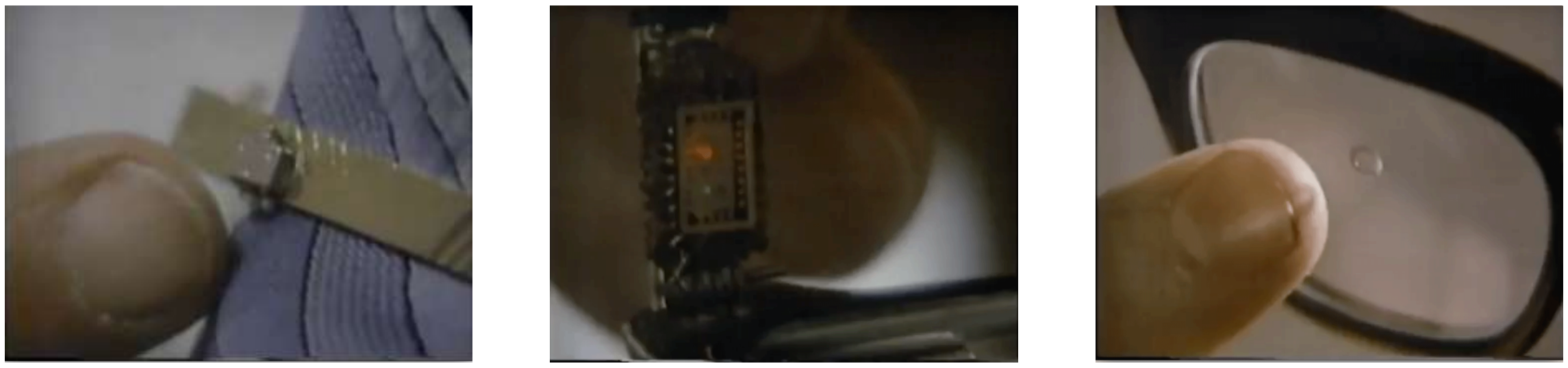

Figure 3.5: The three components of Upton's 'Wearable Eyeglass Speechreading Aid': microphone on tie-lapel pin (left), voice analyser and LED output device (middle), and the mirror fitted to the centre of the eyeglass lens (right). Frames of the video were captured from [135].

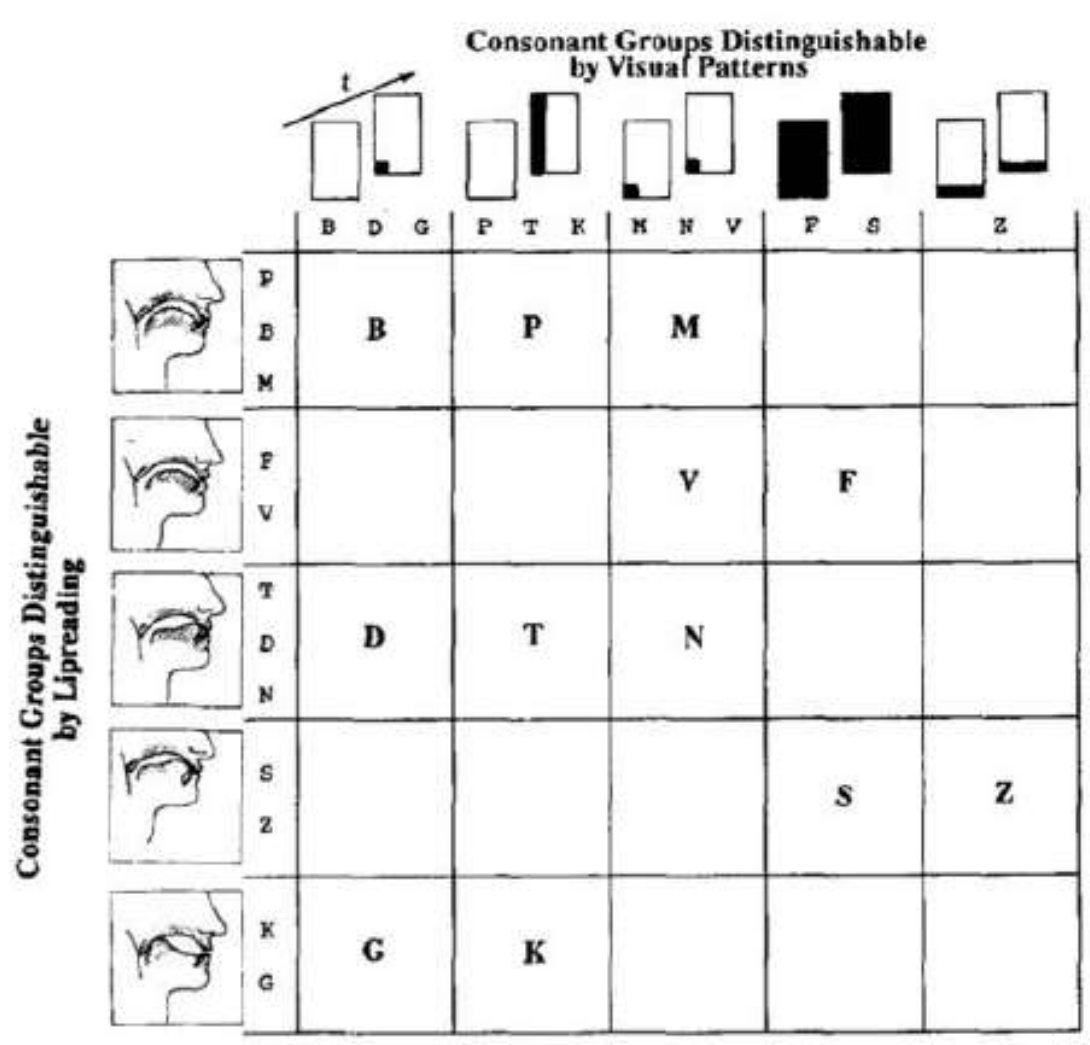

Figure 3.6: Patterns presented by Ebrahimi's peripheral display. The presented patterns are shown in a time sequence consisting of two patterns to illustrate the difference between stops and the other consonants. Diagram from [50].

Massaro et al. [95] designed an SAT called *iGlasses*, which uses a pair of modified eyeglasses with two microphones for input and three LEDs mounted in the periphery of the right eyeglass lens for output. The input was analysed (using an iPhone) for low frequency voicing information, high frequency friction energy, and nasal resonance. The authors chose these speech properties as they are relatively easy to process in real-time and they wanted to help disambiguate visemes. These properties were then transformed into output using the three LEDs, with a blue LED showing voicing, a white LED for friction, and a red LED for nasal sounds.

There has also been work that involves delivering vibrotactile feedback through the users skin to improve speechreading (*Tactile SATs*), which typically provide spoken phoneme information using a vibrotactile or electrotactile array [112]. Tactile SATs may be placed against the chest or located around the wrists [131]. When sound occurs, the signal is transduced into an electrical signal and then delivered to the vibrotactile or electrotactile array, which stimulates the skin. Some of these devices also use a spectral (frequency by vibration intensity) display, which are capable of providing information about the spectral characteristics of the signal. One such display of voice fundamental frequency showed a 10% improvement in a speech discrimination task [28], however a later study found positive results in terms of identifying voicing and for consonant identification, but no benefit for speechreading words in sentences [142]. However, with the advent of cochlear implants, Tactile SATs are not used by many people today [131].

SATs that focus on helping identify components of speech based on how they are produced (Upton, Ebrahimi, Tactile SATs, iGlasses) contradict typical speechreading approaches by training the speechreader to focus on *auditory* aspects of speech (e.g., voiced vs unvoiced phoneme, frication vs frictionless phoneme), rather than *visual*. Even though these SATs can provide rich information, this information is of limited value to someone whose understanding of speech is improved through the visual signal rather than through auditory information. For them to be more effective they would require exclusive training on speech production aspects and concepts.

### SATs for Practice

Currently, there are a limited number of SATs that support speechreading practice. However, it is unclear to what extent these are used by speechreaders.

*lipreading.org* is a website-based SAT that provides practice with vowels, consonants, syllables and words. There are practice sessions based on topics such as going to a restaurant, or a doctors appointment. However, there is a limited number of speakers and amount of content available. The website offers what is called “live lipreading” that supposedly connects you with another user via a webcam, however this is actually a video with a set of pre-recorded responses.

Another limitation is that they employ professional lip-speakers[d], so might not provide examples of typical human speech production.

*lipreadingpractice.co.uk* is a website-based SAT offering subtitled videos of consonants, vowels, and passages. The speaker says these with and without voice, shown from the front and from a profile angle, and repeats each a number of times. Words and phrases are provided as written exercises.

The Dynamic Audio Vision Interactive Device or *DAVID* [123] is an SAT offering videos of sentences on everyday topics. The student watches and responds by typing the complete sentence or content words, or via multiple choice. DAVID also provides repair strategies such as repeating the sentence, or presenting words in isolation.

*ConversationMadeEasy* [130] is an SAT comprised of three programs, with each program presenting videos of speakers with or without audio. The programs increase in complexity: Program A is for analytic practice, where students learn to discriminate and identify phonemes. Program B is for practicing unrelated sentences, where students respond by selecting a picture that best represents the sentence. Program C is for synthetic scenario-based training with commands or questions based on the scenarios given again within a closed response set of four pictured options.

Overall, the above SATs have three limitations: 1) a limited selection of content, 2) a limited selection of speakers, and 3) the user has no way to customise the content with particular words, situations or people they encounter on a daily basis.

In addition to the video-based SATs described above, there have also been attempts at training speechreaders by showing computer-generated facial models typically supplemented with additional cues.

*Lip Assistant* [140] is an SAT that generates magnified realistic animations of a speaker's lips that are superimposed on the bottom left of a video. The rendered mouth animations are superimposed to the bottom left corner of the original video (as shown in 3.7). Lip assistant was evaluated with eight participants with typical hearing. They were asked to transcribe sentences spoken by various speakers (under various noise levels), with synthetic lips, original magnified lips and the original video. The results show that both the addition of magnified lips and synthetic lips improved participants' ability to transcribe the sentences.

[d]A lip-speaker is trained to speak or repeat a speaker's message to speechreaders accurately, without using their voice.

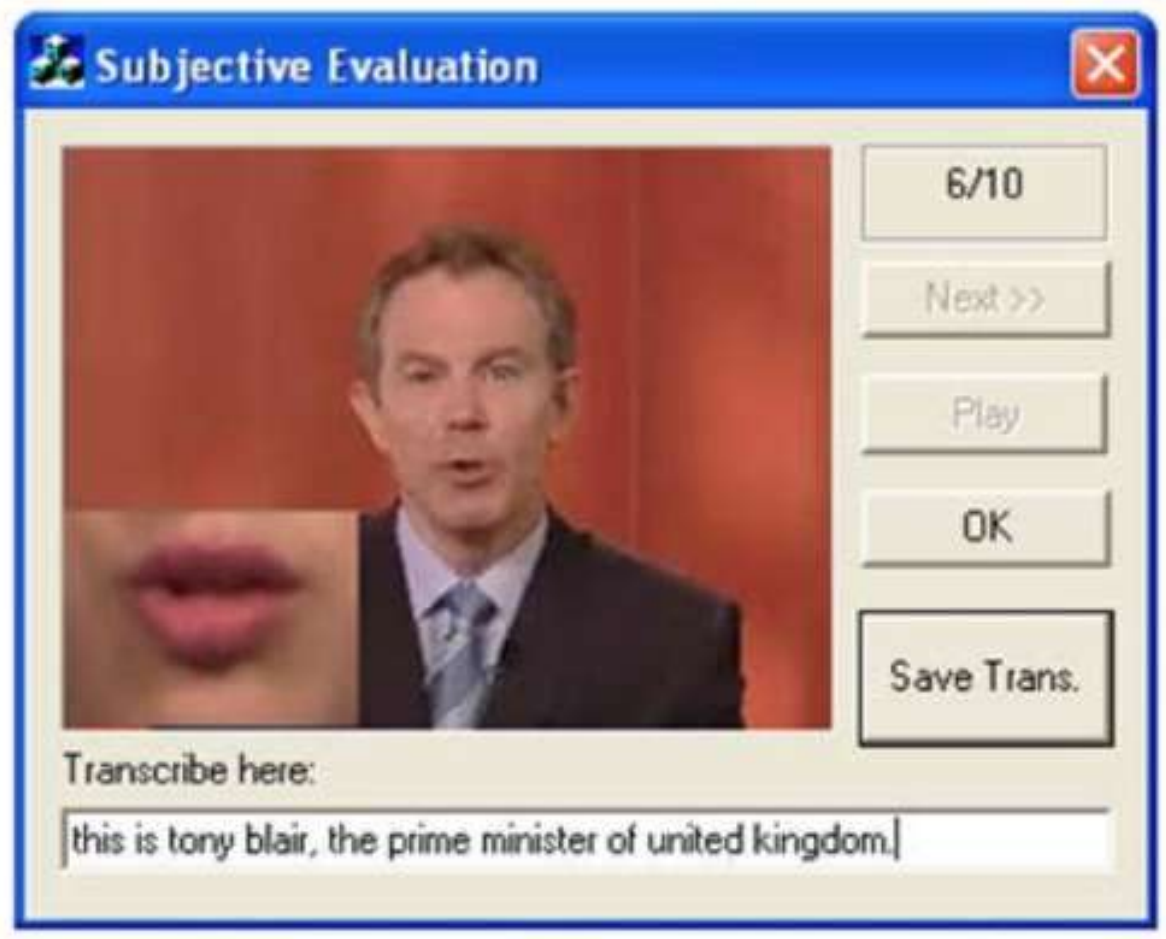

Figure 3.7: A screenshot of the evaluation program used for Lip Assistant, with the synthesised mouth sequence superimposed over the video of a speaker. Image from [140].

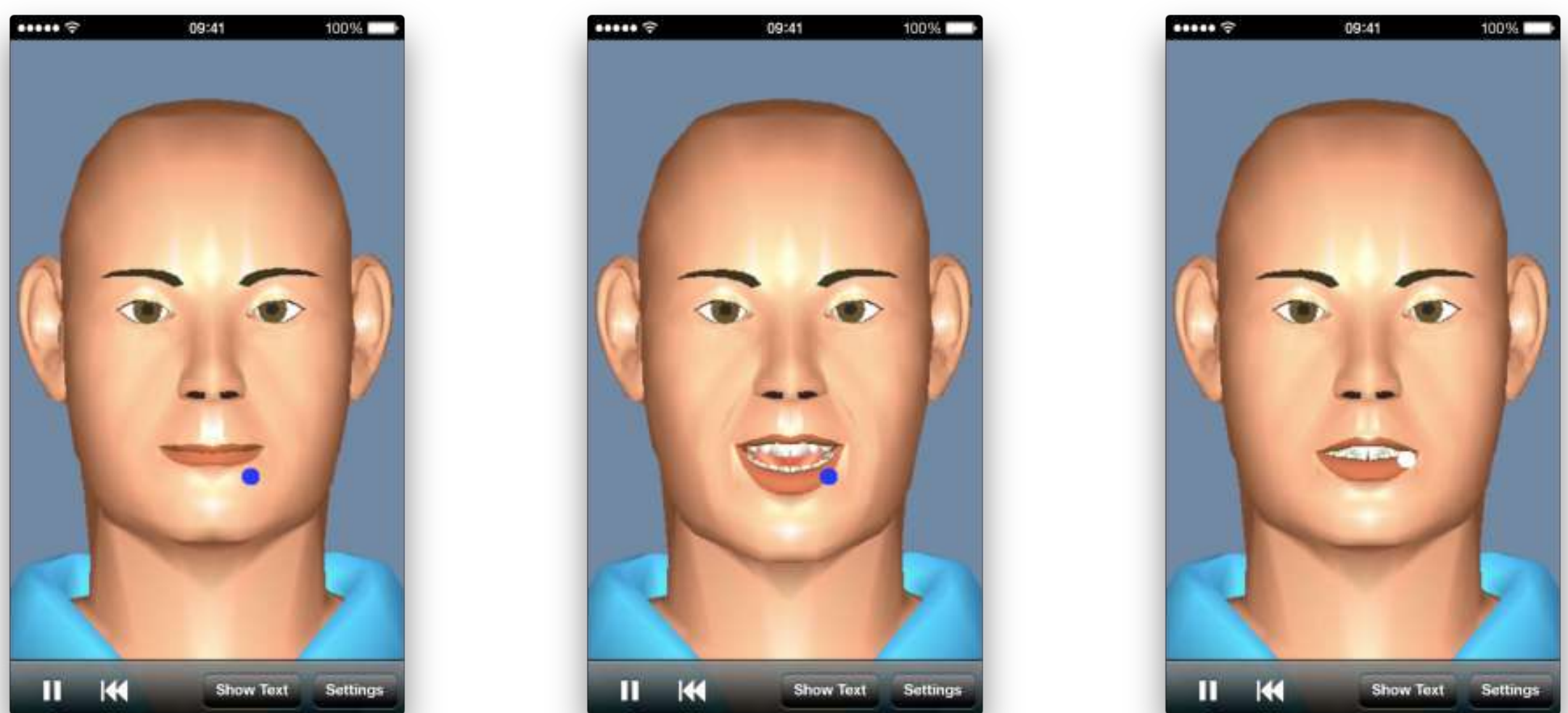

Figure 3.8: Screenshots of the iBaldi Lite application (downloaded from `https://itunes.apple.com/gb/app/ibaldi-lite/id680429104?mt=8`) running on an iPhone. In the screenshots, Baldi has been configured to speak the word ‘Bat’. The left image shows the voicing disc for /b/, the middle image shows the voicing disc for /ae/, and the right image shows fiction disc for /t/.

To improve training of the *iGlasses* SAT (described above), Massaro et al. also developed iBaldi [97], an iOS application that shows a computer animated face and transforms speech into visual cues to supplement speechreading. The cues are three coloured discs, showing nasality (red), friction (white), and voicing (blue), which appear when a phoneme from a corresponding group is presented. The cues are located near the computer generated face’s mouth.

There have also been attempts at enhancing the visual speech signal of a speaker in order to improve understanding. For instance, by adding colour to the lips of a speaker to enhance visibility of speech movements [6], or by making speech movements more exaggerated without changing their meaning [5].

## 3.4 Conclusion

This chapter provided a review of previously designed Conversation Aids, and introduced the concept of Speechreading Acquisition Tools (SATs), which are a subset of Conversation Aids. In general, these previous solutions are not helpful to speechreaders because their designs do not align with the theoretical understanding of the approaches to speechreading training discussed in Chapter 2. Any solution that is developed to help speechreaders during speechreading, or to help practice their speechreading should be influenced by how speechreading is currently taught in speechreading classes.

Although the speechreading teaching methods described in Chapter 2 outline the basic approaches to speechreading teaching, it is still necessary to investigate current practice (which is most likely influenced by these theories) to find the state-of-the-art. The teaching techniques within classes provide us with the best opportunities for design, as they will provide insight into how the two speechreading approaches (Analytic and Synthetic) can be applied in different ways to speechreading practice.

In the next chapter I present interviews conducted with seven practicing speechreading tutors that provide the foundation needed to design improved Speechreading Acquisition Tools (SATs).

4

# Speechreading Tutor Interviews

## 4.1 Introduction

This chapter presents interviews with seven practicing speechreading tutors. I analysed the interview transcripts using thematic analysis. The motivation for conducting the interviews is given first. Following this is the methodology and presentation of the findings from the thematic analysis.

## 4.2 Motivation

In Chapter 3, I presented a review of currently available Speechreading Acquisition Tools (SAT) and Conversation Aids. In general, these previous solutions are not helpful to speechreaders because their designs do not align with the theoretical understanding of the approaches to speechreading training discussed in Chapter 2. Any solution that is developed to help people practice speechreading, or to help during speechreading should be influenced by how speechreading is currently taught in speechreading classes. Although the speechreading teaching methods described in Chapter 2 outline the basic approaches to speechreading teaching, it was necessary to investigate current practice (which is most likely influenced by these theories). Therefore, I conducted interviews with practicing speechreading tutors to generate the dataset needed to inform the design of new Speechreading Acquisition Tools (SATs).

## 4.3 Method

I conducted individual interviews with seven of the 21[a] reported Scottish speechreading tutors using a list of their contact details provided by the Scottish Course to Train Tutors of Lipreading (SCTTL) website[b]. The goal of the interviews was to explore each tutor's background, approach to teaching, current use of technology, and thoughts on how speechreading can be improved. After obtaining informed consent from the participant, I audio recorded each interview, then transcribed and thematically analysed the transcripts using the approach outlined by Braun and Clarke [29].

### Aims

I had five main aims (phrased as questions) guiding the interviews:

1) Do speechreading tutors primarily employ analytic or synthetic training approaches?
2) What do speechreading tutors consider to be the basic unit of speechreading?
3) What technology (if any) do speechreading tutors currently use to teach speechreading?
4) Do speechreading tutors feel that speechreading training could be improved with new technology or training techniques?
5) How do speechreading students continue to learn when not in class?

The interview guide is included in Appendix B.7.

### Participants

As discussed in Chapter 2, Scotland is the one of the few countries that provides formal training for speechreading tutors. As such, all participants reside in Scotland, and offer classes throughout central Scotland. Participants were recruited through direct emails via contact details obtained from the SCTTL website[f], along with word-of-mouth. Participant details are summarised in Table 4.1.

All participants were female and aged from 42 to 78. The 21 tutors listed on the SCTTL website[f] were all female, therefore gender could not be balanced. The range of teaching experience was from six months to 32 years. Participants self-reported details about their hearing, which I classified into five levels using the textual descriptions of hearing loss identified by Action On Hearing Loss [3]: Typical Hearing, and Mild, Moderate, Severe, and Profound Hearing Loss.

[a] At the time of conducting the interviews there were only 21 reported tutors practicing in Scotland [10].
[b] `http://www.scotlipreading.org.uk/index.php/classes/`

Four participants reported having moderate to profound hearing loss, with one of the remaining three participants reporting that she had previously experienced temporary severe hearing loss. The participants' backgrounds, teaching history, and experience levels were varied:

**P1, 78:** P1 has had profound hearing loss from a young age. She has been teaching speechreading for 32 years after being asked to take over a class by a friend. She has no formal teacher or speechreading training. P1 teaches three classes a week.

**P2, 68:** P2 has severe hearing loss. She has been teaching speechreading for 12 years. She was a high school science teacher who retired early due to her hearing loss. She had attended speechreading classes for around two years before her tutor convinced her to take the SCTTL so that she could teach classes herself. P2 teaches two classes a week.

**P3, 61:** P3 has typical hearing. She has been teaching speechreading for 12 years. She previously worked as a subtitler for the BBC. Her father had a hearing loss which motivated her to become a speechreading tutor. She undertook the SCTTL at the same time as P2. P3 teaches two classes a week.

**P4, 57:** P4 has typical hearing. She has been teaching speechreading for 18 years. She also works as a Speech and Language Therapist for the Scottish National Health Service (NHS). P4 teaches one class per week and also trains tutors as part of the SCTTL.

**P5, 42:** P5 has typical hearing, however she experienced severe hearing loss for eight months after a viral infection. She has been teaching speechreading for six months. Her motivation for teaching started after experiencing hearing loss. P5 initially wanted to learn to sign to increase access for Deaf individuals at her community centre where she works. However, upon realising that the Deaf community in her area instead required a speechreading tutor she undertook the SCTTL. P5 teaches two classes a week.

**P6, 67:** P6 has moderate hearing loss. She has been teaching speechreading for seven years. She temporarily lost her hearing due to a viral infection 10 years ago, so her daughter encouraged her to learn speechreading to help her cope. After a year within the class, her tutor encouraged her to take the SCTTL. P6 teaches four classes a week.

**P7, 66:** P7 has severe hearing loss. She has been teaching speechreading for one year. Her motivation for becoming a speechreading tutor was to help individuals who experience hearing loss later in life; she worked in a library and noticed that these individuals seemed prone to isolation. She also took the SCTTL. P7 teaches three classes a week.

| ID | Gender | Age | Hearing Loss | Cause | HA/CI Use | Years Signing | Years Teaching |
|---|---|---|---|---|---|---|---|
| P1 | F | 78 | Profound Hearing Loss | Childhood Illness | Cochlear Implant | 32 | 32 |
| P2 | F | 68 | Severe Hearing Loss | Childhood Illness | One Hearing Aid | None | 14 |
| P3 | F | 61 | Typical Hearing | - | None | None | 12 |
| P4 | F | 57 | Typical Hearing | - | None | None | 18 |
| P5 | F | 42 | Typical Hearing* | - | None | 3 | 6 Months |
| P6 | F | 67 | Moderate Hearing Loss | Not Specified | One Hearing Aid | None | 7 |
| P7 | F | 66 | Severe Hearing Loss | Congenital Hearing Loss | Two Hearing Aids | 19 | 1 |

Table 4.1: Summary of participant demographics. HA/CI = Hearing Aid/Cochlear Implant. Years Signing = Number of Years Using Sign Language (BSL). Years Teaching = Number of Years Teaching Speechreading. ‘*’ indicates severe hearing loss for eight months due to viral infection.

## 4.3.1 Approach

After obtaining ethical approval from the School of Computing's ethics committee, I conducted semi-structured one-to-one interviews. Interviews took place in mutually-convenient and quiet locations. The mean interview time was 40 minutes (max 74 minutes, min 28 minutes); some interviews took longer due to participants discussing past students and their experiences interacting with them. The interviews consisted of questions that explored their background, general teaching approach, assessment, current use of technology and their thoughts on where speechreading could be improved in the future (the interview guide can be seen in Appendix B.7). The interview questions were open-ended (for example, "Why did you decide to become a lipreading tutor?" [c]) and I let the participant lead whenever possible (following [17]), encouraging elaboration by asking probing follow-up questions when necessary (following [17], e.g., "And do they find that helpful?"). Audio recordings were gathered during the interviews. Transcripts were coded and analysed using thematic analysis [29], grouping similar experiences together in order to identify themes across all participant interviews. The themes were refined through an iterative thematic analysis process to generate a final, distinct set of themes.

## 4.3.2 Phases Of Analysis

### Phase 1: Becoming Familiar With the Data

I manually transcribed the interviews using custom-built software. I strove for a verbatim account of all verbal and nonverbal (e.g., laughs) utterances. I added punctuation where necessary to indicate pauses, full sentences, and questions. I formatted the transcripts using 'Interviewer:' to indicate interview statements and 'P1:' to indicate statements by Participant1. All names and locations were removed to maintain anonymity.

[c] In the UK, 'speechreading' is referred to as 'lipreading', therefore in discussions with participants I used the term 'lipreading'.

## Phase 2: Generating Initial Codes

I started by reading paper copies of the transcripts and manually highlighting all interesting extracts. I then transferred the highlighted extracts into MAXQDA12 [d], resulting in 502 extracted elements from the original transcripts.

Using the 'coded segment' feature of MAXQDA12, I then systematically processed the original 502 elements iteratively using a data-driven approach to ensure that the final themes emerged exclusively from the interview data.

During this process, I coded for maximum diversity of potential themes and patterns, and did not discard data unless it was clearly not relevant to the research (e.g., an unrelated anecdote). I also kept enough of the text around each coded segment to retain the segment's context, because a common criticism of coding is that the context is often lost [35].

By giving equal attention to each extract, I further segmented the extracts and coded them using iteratively-shaped codes, resulting in 944 coded segments under 116 unique codes.

As I followed Braun and Clarke's [29] 15-point checklist of criteria for good thematic analysis, I did not conduct inter-rater coding. Inter-rater reliability checks are not always used in thematic analysis since there is scepticism regarding such tests [9, 113, 133]; it can be argued that one researcher merely trains another to think as she or he does when looking at a fragment of text and so the reliability check does not establish that the codes are objective but merely that two people can apply the same subjective perspective to the text [94].

## Phase 3: Searching for Themes

In this phase, I created a short definition for each code that described when that code would be used and what it represented across the entire data set. I then printed each code plus its definition on individual strips of paper, and iteratively organised the strips into 'theme-piles' on a whiteboard (as shown in Figure 4.1). Using the resulting 'theme-piles', I produced an initial thematic map (as shown in Figure 4.2).

[d] A qualitative analysis software package, `http://www.maxqda.com/`

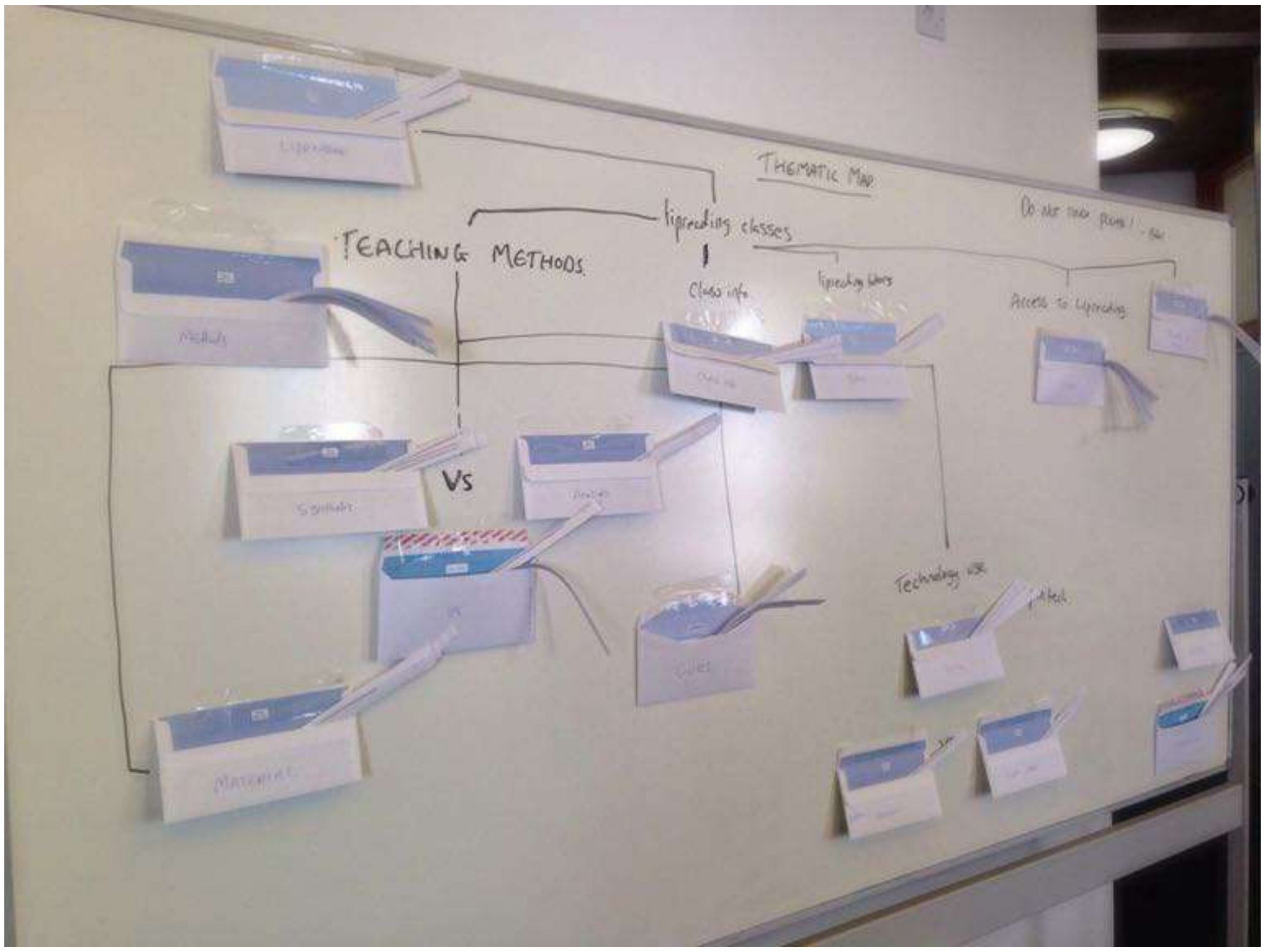

Figure 4.1: Photograph of the whiteboard analysis of the 'theme-piles' conducted during Phase 3.

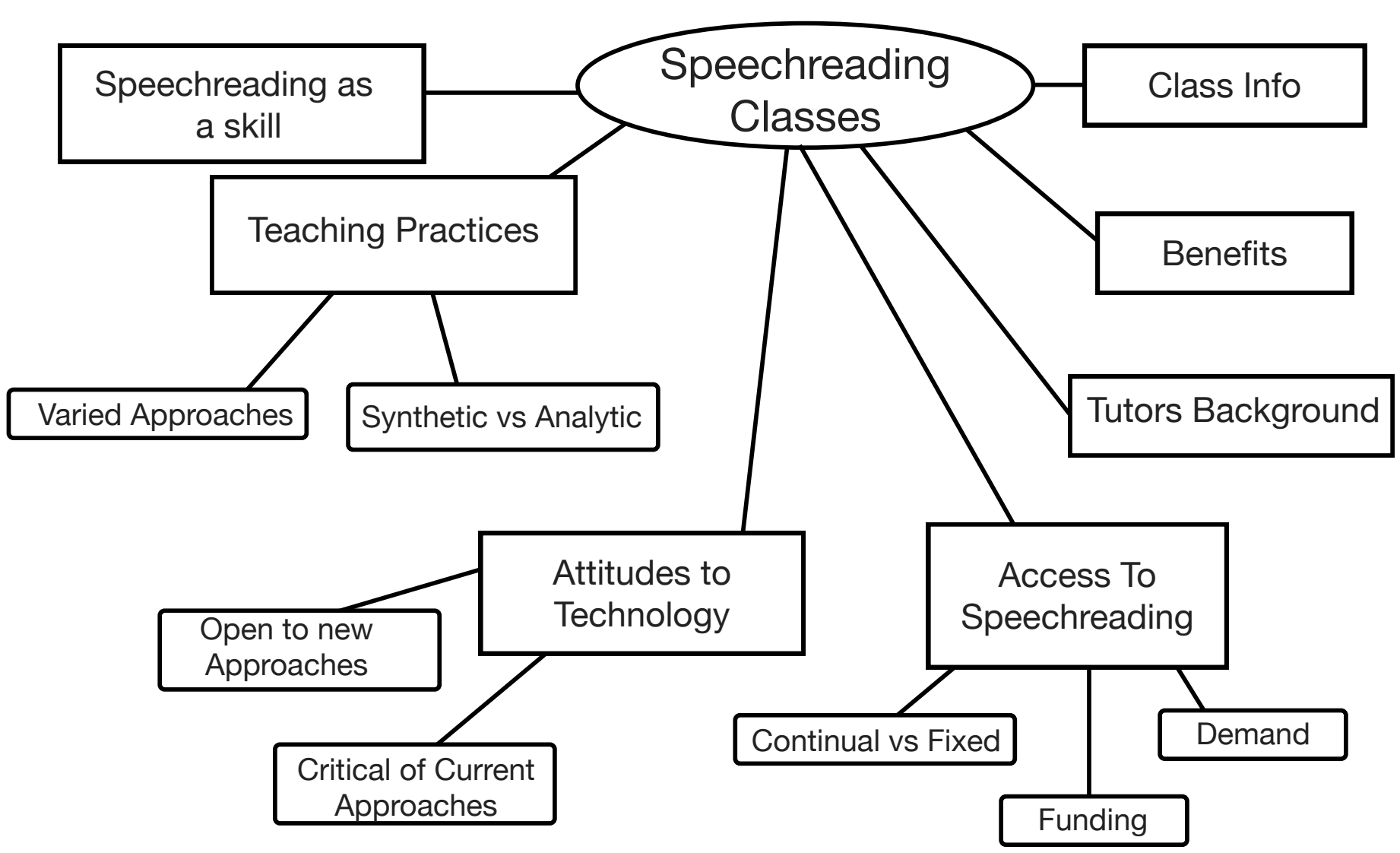

Figure 4.2: First thematic map showing seven main themes and their sub-themes.

### Phase 4: Reviewing Themes

Starting with the initial thematic map from Phase 3 (as shown in Figure 4.2), I removed themes that did not directly relate to the study aims outlined above, these were the themes called 'Class Info' and 'Tutors Background'. 'Benefits' was collated under 'Speechreading as a skill' and two subthemes of 'Access to Speechreading'; 'Funding' and 'Demand' were collated into 'Continual vs Fixed'.

I then reviewed the collection of coded extracts for each remaining theme to ensure that they formed a coherent pattern. Finally, I re-read each original transcripts to check that the revised themes provided suitable coverage.

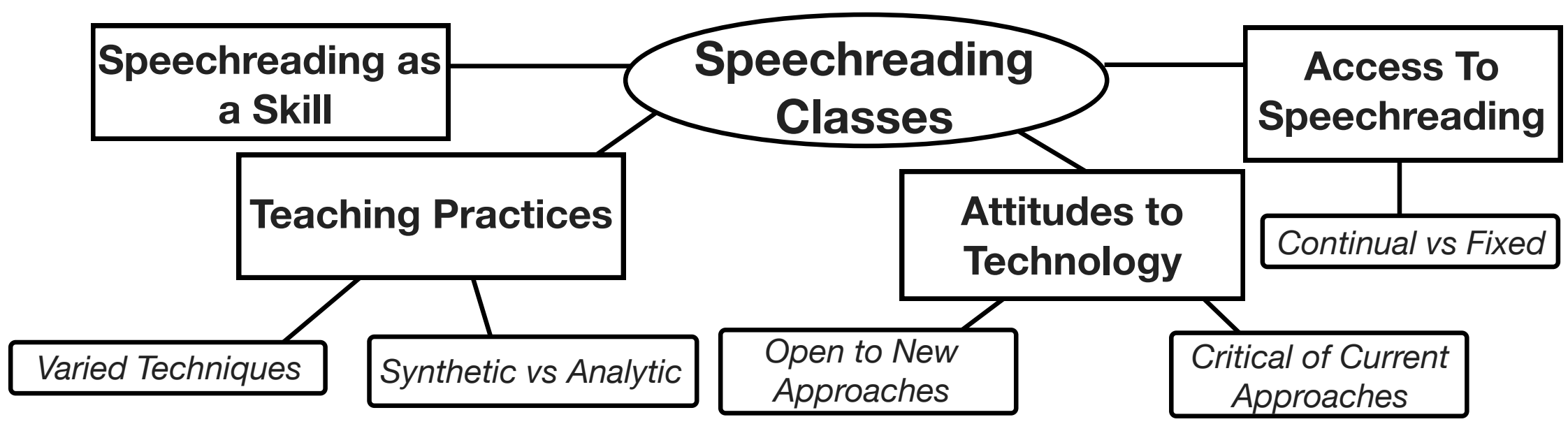

Figure 4.3: Final thematic map of four main themes and their subthemes.

### Phase 5: Defining and Naming Themes

Once the thematic map was finalised (Figure 4.3), I defined each theme by examining its collection of coded extracts to determine the main aspect of the data captured by the theme. I then revisited the collection of coded extracts for each theme, refining the story told by each theme. Finally, I produced internally-consistent accounts, with an accompanying narrative for each theme, which are presented in the next section.

## 4.4 Findings

I identified four themes through the thematic analysis (that are shown on the final thematic map in Figure 4.3): 1) Speechreading as a Skill, 2) Access to Speechreading, 3) Teaching Practices, and 4) Attitudes to Technology. I now explore each theme in detail using quotes from participants to scaffold the narrative of each theme.

### 4.4.1 Speechreading as a Skill

Participants saw speechreading as a skill, one that requires long periods of concentration and focus to learn. Classes are typically two hours long with a short break, so students focus for around an hour at a time. All seven participants discussed the need for concentration and focus within classes:

> P4: *"For the person themselves they need that confidence, that assertiveness, to do that, they also need to concentrate and pay attention for that length of time."*

> P6: *"…not looking around the room and trying to listen but actually just focusing…the amount of concentration these people give, it's amazing…it's as almost as they are drilling holes in you…it's excellent, it's really good."*

As illustrated by these comments, a student's ability to focus and concentrate is of paramount importance to learning speechreading. A high level of concentration is necessary due to the limitations of speechreading – many aspects of speech (e.g., voicing) are mostly audible instead of visible:

> P2: *"I warn them beforehand, that only 1/3 of speech is lip-readable…they are aware that there is a limitation to what we are doing but it's an added help to everything they do…it's useful but it doesn't solve all your problems."*

In addition to the level of concentration required and the limitations of speechreading, participants also described additional factors that pose difficulties for students, such as particular words or speech movements having little visual difference:

> P6: *"Knowing that…some of these skills are very subtle, you are not going to see huge [differences]. Some speech movements are very clear, [but] when you get to others there are some sounds that are so subtle you can hardly see them…vowel at the beginning of the word, like 'ahead'…it's difficult to spot. Sometimes if you can't lipread, [it] could be that there is a vowel in there that you are not aware of."*

P1: *"Certain words and sounds they find really difficult. I have to repeat them…I go over things quite a lot."*

Many participants identified how different accents can affect individual speech movements as well:

P4: *"Sometimes you've got…English [as opposed to Scottish] vowel sounds coming at you and therefore sometimes you look and think 'I don't know what that is'. Because it's got the accent it kind of changes things."*

The challenges of speechreading, plus the level of concentration required, often lead to fatigue during and after class:

P2: *"And there are times when you get tired and your lipreading is absolutely rubbish."*

P4: *"As to how tiring it is to sit and watch somebody for…two hours, we build in breaks but it's still a lot of effort and concentration. It's very tiring and I think that's what comes after two or three weeks…somebody will come up to you and say I went home and I was absolutely exhausted. I didn't realise how tiring it was."*

### 4.4.2 Access to Speechreading

Access to speechreading classes was discussed by all participants. In particular, participants focussed on issues surrounding funding of classes, the length of classes and how students beginning classes (and the general public) have little awareness of what is involved in learning speechreading:

P4: *"…it's a very difficult one, lipreading is…kinda like a Cinderella Service [ignored or treated as less important]. People don't recognise that actually everybody lipreads to a certain extent. I think what could be improved with lipreading is general awareness of the fact that everybody lipreads, so…more people would be aware of it and more people would therefore come to the classes."*

Participants also highlighted that local governments can view speechreading classes as recreational, causing funding issues:

P7: *"I would prefer it to come under university rather than sitting under [local governments], because it is a life changing skill…rather than a hobby or a job."*

P2: *"Some authorities regard it as a recreational thing."*

### Continual vs. Fixed

A subtheme within access reflects whether speechreading classes should be continual or fixed length. All tutors agreed that classes should be continual.

> P1: *"I do think that judging by this and my [other] class that it's important for the class to be ongoing unless they for individual reasons want to drop out."*

> P2: *"It's a continual practice thing…you could really do with a little practice every week…for the rest of your life."*

However, many of the tutors teach fixed-length classes. This was generally due to funding issues, with many local governments only offering two years of speechreading classes:

> P6: *"I know, two years is my maximum and then you have to go on to a paid class."*

> P2: *"So I'm paid by [anonymised] city council. They provide two years free lipreading, I am the only tutor in [anonymised] that does it. I did one on the Tuesday, two on a Thursday, but the budget is decreasing every year. We don't know how long this will go on for. I only teach for thirty weeks…that's all the council will pay you for."*

> P5: *"This is a problem for me especially because I have got funding from [anonymised], and they will not perpetually fund something. So I have got funding for one year of lipreading classes, which is a 30 week course."*

In some cases, local governments offer no funding for classes, and students living in these areas pay for classes.

> P3: *"They do pay in the [anonymised] groups. In the other groups no they didn't pay."*

Students also appear to be willing to attend continual classes rather than fixed length classes. This is supported by participants reporting that students only stop attending classes due to becoming ill or passing away.

> Interviewer: *"So why do people stop coming to a class?"*
> P7: *"I haven't had any experience of that, they have all been very faithful including [over] the holidays."*

P1: *"Some people do drop away. Usually they either die or [grow] too old. Some people died in this class and they died in my [anonymised] class too. Not many have just dropped out."*

P6: *"People who come tend to stay. A gentleman stopped coming to this class about a year ago. He is…very old, in his 90's. He lives about 10 miles down the road and has to come by bus and therefore it's very difficult for him."*

### 4.4.3 Teaching Practices

Teaching practices varied widely, with all tutors using a variety of approaches and techniques to teach speechreading.

#### Synthetic vs. Analytic

Synthetic and analytic are the two main approaches to teach speechreading. Although the teaching methods described in Chapter 2 typically emphasise one over the other, all of the tutors draw from both approaches as needed:

P4: *"We tend to have a general topic for the class…if I am doing something on a movement, 'f'…I might choose a topic that begins with that sound…I might talk about 'fireworks' or 'fire'. The whole lesson will not be around that particular speech sound…but a certain chunk of it will…if I taught that sound, I would do a follow up story or exercise with that sound appearing in it regularly."*

P6: *"If you know the context, you can make a really good guess, which is a lot of lipreading anyway. If they are talking about horse racing then there is not going to be anybody talking about ballet dancers, it's unlikely. Knowing that it's very subtle, that some of these skills are very subtle…some speech movements are very clear, when you get to others there are some sounds that are so subtle you can hardly see them."*

#### Varied Techniques

Within the teaching approaches outlined above, tutors also reported many individual techniques that they used within classes. I now briefly describe how each technique is used within classes using quotes from the participants. Participant descriptions of techniques were often fragmented throughout an interview, resulting in many relevant quotes, I have selected a quote from each participant that best represents the technique. An overview of these techniques is presented in Table 4.2.

| Technique | Tutors | Type |
|---|---|---|
| Mirror Practice | 7 | Analytic |
| Cue Recognition | 7 | Hybrid |
| Speech Movements (Lipshapes) | 7 | Analytic |
| Pair work | 7 | Hybrid |
| Quick Recognition Exercises | 6 | Analytic |
| Stories | 6 | Synthetic |
| Finger Spelling | 4 | Analytic |
| Scenarios | 4 | Synthetic |
| Word Quizzes | 3 | Hybrid |
| Framed Sentence Exercise | 2 | Synthetic |
| Mystery Object | 2 | Synthetic |

Table 4.2: Summary of the 11 teaching techniques used by the participants.

**Mirror Practice:** Mirror Practice involves students looking at a mirror to learn their own mouth shapes and the differences between mouth shapes when certain speech sounds or words are spoken. Mirror practice is an **Analytic** technique as the student focuses on visual disambiguation. Mirror practice was used by all seven tutors during their classes:

> Interviewer: *"Do you use mirror practice?"*
> P1: *"I tell them to go home and look in the mirror."*
> Interviewer: *"Ok, and that helps them remember what it looks like?"*
> P1: *"Yeah."*
>
> P2: *"We only use mirror practice for the speech movements and for seeing [lipshapes] and they watch it on their own face and they watch their partners face."*
>
> P3: *"Certainly at the beginning I give them a mirror and I introduce the idea of it and some people are too self conscious, fine, other people find it quite useful and get used to [watching] themselves and then in future pairs work. This is in the beginners class, I'll say 'practice these on your mirror' you know."*
>
> Interviewer: *"Do you use mirror practice in your class?"*
> P4: *"When I'm doing speech movements or QRE's then yes. . . We would start with*

*mirrors and then go onto doing it in pairs, doing it with other members of the class."*
Interviewer: *"Ok…do they have mirror practice at home as well?"*
P4: *"Yes, yes, so we'd recommend that, so I would recommend that they stand or sit in front of a mirror…And seeing what their own mouth does."*

Interviewer: *"So you already said you use mirror practice?"*
P5: *"We do, and I encourage them to do it at home as well, so any of the sheets that I give them out, any of the pairs exercises or anything, they go home [to be practiced in front of a mirror]."*

P6: *"We do mirror practice often, everytime we do a speech movement."*
…
Interviewer: *"Ok, so you find it quite helpful in terms of teaching then?"*
P6: *"Yes definitely, definitely."*

Interviewer: *"So do you use mirror practice then?"*
P7: *"Yes always."*

**Cue Recognition:** Cue recognition encompasses looking for body language, facial emotion, and hand gestures. Used by all seven tutors. Analytic and synthetic – **Hybrid**.

P1: *"Yes. Well I always try to go over with new people the basics of what they should be doing and looking for. Like taking in body language and the whole person, things like that…I try to make them aware if somebody is saying 'what's the time?' or points to the watch or their clock…I teach them to look for things like if a person is sad or a person is happy or angry but I also teach them not to jump to conclusions. Because sometimes people think oh they are talking about [something] and they are not."*

P2: *"I do teach gestures…I would say that they know when I'm saying something sad or I'll say [mouths a sentence with a smile], and they know."*

P3: *"Yes I do a bit on [gestures] as well, facial expressions and body language and gestures, I sometimes get them to work in pairs giving them you know emotions, a phrase with emotions."*

Interviewer: *"Do you teach them how to read facial expressions and what that does to lipshape?"*
P4: *"Again I wouldn't actually teach it but I would tell them to be looking out for it as part of the kind of coping strategies in terms of how we manage."*

Interviewer: *"Do you teach hand gestures as well so that might help them with context? So if they are going to the supermarket and somebody is pointing at something?"*
P5: *"Yeah we do that I suppose as part of the gesture thing, yeah."*

Interviewer: *"Do you teach them how to read maybe gestures in terms of hand gestures to kind of help them build any context?"*
P6: *"Generally facial expressions, we do a lot of eyebrows up or eyebrows down and that is how you tell an angry face to a pleasant face."*
Interviewer: *"And does that help them with the lipreading then?"*
P6: *"I think they would probably do that anyway wouldn't they…People just automatically read body language and facial expressions."*

Interviewer: *"Do you teach them how to read gestures? Such as hand gestures or face gestures?"*
P7: *"Oh we do that in the class, what was the last one we were talking about. Something with a naval hero I go, 'I don't think you are going to get this because the morning class didn't get it' so I go [mouths phrase] and he was a naval hero, and I pointed towards my naval."*
Interviewer: *"Right yeah (laughs)."*
P7: *"And they fell about (laughs) So yes we do mime, I call it mime."*
Interviewer: *"Mimes ok. And then facial expressions if someone is sad?"*
P7: *"Yes uh huh. I say [mouths phrase with a smile]…so yes."*

**Speech Movements and Lipshapes:** Speech movements or lipshapes is a technique where the student is told to focus on the visual representation of a single speech sound isolated or within a word and is therefore an **Analytic** teaching technique. Tutors inform the student of the target speech movement and the student has to identify the word or words spoken. Used by all seven tutors.

P1: *"At the start of a class I might emphasise a mouth or a different lip pattern and how unless you can hear you can't see some of the sounds on your lips because like with p, b, m it's hidden. So it's difficult so if maybe you are deaf or very hard of hearing it's difficult.…In the very beginning I do things like show them the lip patterns on a chart and then go over."*

P2: *"Do I teach lipshapes? Yes that's the speech movement part of this."*
Interviewer: *"How do you individually teach the lipshapes then, do you just say*

*them?"*

P2: *"Yes, and I ask them to tell me what shape my lips were and I write up what shape are my lips…Is my mouth open, closed, where is my tongue."*

Interviewer: *"So you do that without voice as well?"*

P2: *"I do that with all of my lip, my speech movements."*

P3: *"I'll think what speech movement is useful within the material, so you think Shakespeare you think sh, yeah you know that's a bit obvious but you know what I mean…So then I'll try and get a speech movement based on the material I'm working on.…When you are doing speech movement work you are inclined to do random sentences because you are looking for a demonstration of the speech movement and that is more difficult because you don't have this context thing that you know. Which is useful for them."*

P4: *"I do teach lipshapes, and just by demonstrating by putting them up, getting them to describe what they see on my lips when I do the sound on their own, so they are [saying] 'this is what I can see' and I will write that up on the flip chart, they will then have a look at themselves doing it in a mirror, I will then get them to look at each other to see if there are any variations of lipshape on different people and does it look any different with a beard does it look any different you know, with one speaker than another speaker. And from the side as well, face on from the side, what does it look like, what does it look like when it's followed by the different vowel sounds as well. Because that will change how the lip shape looks at the beginning and in the middle of a word and at the end of a word and that's where they would do a kind of speech movement exercise of saying words where they know what the word is so they are not actually having to think what the word is, they are purely looking at what's the mouth doing, what's the lip shape doing."*

P5: *"Then it's just about how they look, it's about picking it up right ok so that could have been a p/b or an m at the beginning of them because they went, clearly made that shape, right so that must have been a p/b or an m so it's trying to get them to absorb that and go, ok I saw that mouth just opening and closing, that must be a p/b or an m."*

P6 : *"I go with the speech movement with them, look at the different sounds and use those sounds in sentences."*

Interviewer: *"Do you teach a lip shape then on it's own? Without being in a word or a sentence?"*

P6: *"I have to tell them [we are] going to look at say it's f/v we are going to look at these, so I get them to look at me so it's the same. Every speech movement is done the same, watch me, watch my lips, when I say f/v and I do it three times round the room. And then point to the next one v and they can see 'oh they look just the same, just the same', then we move on, [get the] mirrors out so I say watch your own lips, use your mirrors, watch your own lips, when you are watching be aware of what you are seeing because you are going to talk about it afterwards. So they watch themselves and then they watch their partners say these. Ok so what did you see what is actually happening? What are the lips doing? The teeth doing? The tongue? Air being expelled? All of those things."*

P7: *"The speech movement would be maybe a sentence with f or v. Maybe 'Florence Nightingale' or f and n, n you can't see but f and v are a pair and recognisable. So we put them into sentences, maybe four or five."*

**Pair Work:** Pair Work is when the class members work in pairs. Pair Work is a **Hybrid** technique, as it involves Analytic and Synthetic skills. During pair work the student may be familiar with their partner's facial movements having trained with them before, and also may have the ideas about the topic of conversation. Used by all seven tutors.

P1: *"Yeah I have now and again, done pairs."*

P2: *"It teaches, you to see the shape on the mouth…and they do a pairs exercise to look at it on someone else's face, because all of them can lipread me after about 15 weeks, but they have to practice with other people, so I quite like doing pairs exercises when they are working not facing me."*

P3: *"I'll explain what the movement we are looking at and then I'll do examples on the board of the sentences and then we will practice that and then I'll give out work for them to practice with each other."*
Interviewer: *"Right that makes sense so pair work then?"*
P3: *"Yes. Pair work."*

P4: *"Depending on the size of the class, I would do it in pairs…I could do a conversation, a kind of mockup conversation between say a waiter and a customer."*

P5: *"So they say it with me, first and then they say it themselves in their mirror and then they say it with each other."*
Interviewer: *"So it's like pair work then."*

P5: *"Yep."*
Interviewer: *"So do the students find it quite useful as well then?"*
P5: *"Yes they do, another part of it, is getting students to change seats every week."*
Interviewer: *"Right so they are kinda getting different people to look at."*
P5: *"So they are getting different people and in the class when I started it I said to them you have to sit, and they were like really? And I was like you know 'it will be good for you' and you will find out later on why it's gonna be good for you, and explain it to them, but now they actually realise and they come in now and go 'did I sit here last week?, I'll move'."*

P6: *"We do pairs exercises, but speech movement for me is what I start with."*

P7: *"[It's important] to not be uncomfortable with each other which is why I say look at your partner see what they are doing and then look in the mirror and see what you are doing and then look at me and that way they loose that embarrassment, you know [to] stop being self conscious, [everyone] does become very comfortable in the class."*

**Quick Recognition Exercises (QREs):** QREs or *Syllable Drills* are rhythmic drills consisting of contrasting syllables or words spoken as quickly as possible in different orders by the tutor. Students repeat back the order. This is an **Analytic** technique. Used by P1, P2, P4, P5, P6, P7.

Interviewer: *"Do you use syllable drills?"*
P1: *"Could you give me an example?"*
Interviewer: *"So you would say something like Mo, Po, Mo."*
P1: *"Yes, I used them in the past."*
Interviewer: *"…Do you find they are quite helpful?"*
P1: *"Yes. Because they are difficult."*

Interviewer: *"Have you heard of syllable drills before? So you might say so, mo doe, and then mo so doe."*
P2: *"I call that a Quick Recognition Exercise, where you say words that look the same but have got a letter different in them…like will, bill, mill?"*

Interviewer: *"Do you use syllable drills? Or I think they are called Quick Recognition Drills?"*
P4: *"I use Quick Recognition Exercises…with words, minimal pair words, you know a word with the same sound at the beginning."*

Interviewer: *"Right like Mat, Pat or Bat or something like that?"*

P4: *"Yes. So you are looking at those kind of things, but you are doing your QRE and you are doing between three different sounds."*

Interviewer: *"And that would be syllables?"*

P4: *"That would be words as well, we would start off with just the sound on it's own and then put it into a word…Because we tend not to lipread in syllables, we lipread in words."*

P5: *"[Reading Question] Do I use syllable drills?"*

Interviewer: *"I think they are also called Quick Recognition Exercises?"*

P5: *"We do."*

Interviewer: *"Ok right, do you find they are quite useful or?"*

P5: *"Yes, very useful, and I think the thing about the repetitive thing as well it works well in a class…And if you get them to do it with me first and then they do it with each other."*

P6: *"[Do I use] QREs? – yes."*

Interviewer: *"So do you find them quite helpful?"*

P6: *"I think they are, it's almost subconsciously that people do recognise these more quickly and we are working to build up speed so I always tell them that and I think they do [find QREs helpful]."*

Interviewer: *"Do you use syllable drills?"*

P7: *"Always."*

Interviewer: *"…and do you think that's quite useful for them?"*

P7: *"Yes…it gets them into recognising the shape."*

**Stories:** Stories are based around one topic and may be a number of sentences long. Used by P2, P3, P4, P5, P6 and P7. Synthetic.

P2: *"But if I was giving a story, it's all in lines, I should have brought you samples of my work…But it would be like that, and I would say the first line, say it was Robert Louis Stevenson[e], I would say something like Robert Louis Stevenson, was born in Edinburgh…I would have written the heading Robert Louis Stevenson, names are very difficult, so you have to clue them, so I would have written Robert Louis*

[e]Robert Louis Balfour Stevenson was a Scottish novelist, poet, essayist, and travel writer. His most famous works are Treasure Island, Kidnapped, and the Strange Case of Jekyll and Mr Hyde.

*Stevenson, I would have said it three times without voice before I started the story proper, [along with] any words that are difficult."*

P2: *"Aye, you look at, you know a wee story which is 10-12 sentences long. It's not stretching some people."*
Interviewer: *"Right ok so it's not challenging them."*
P2: *"When I do a story I try to have short sentences and long sentences…So that the people who are not lipreading so well will get the short sentences and those who are smart alecs will manage to get a big long sentence because they can hold it. I'm teaching younger people."*
Interviewer: *"Yeah, right."*
P2: *"And you know, saying a sentence, nobody speaks in sentences that are 6-8 words long and especially you know with young people nowadays, I mean."*

Interviewer: *"Ok, ok that's good. So then overall is your material a big story or just sentences that make up a story?"*
P3: *"I do a story yes, not in the case of speech movements I do sentences to demonstrate the speech movement you know."*

P4: *"I mean there is a mixture within the class, we do stories, but the stories tend to be broken down into manageable sentences."*
Interviewer: *"But they all follow the same topic?"*
P4: *"They follow the same topic. The story would be about something, today is Earth Day so it could be about Earth Day, but you would present it in sentences."*

P5: *"Most of the things I do I just prepare on my own, anything interesting, or something happens or like for example Paul Daniels*[f] *died and I thought that would be a great story, Paul Daniels life, or you know magicians in general, so you do kind of get things from the media."*

P6: *"It starts with the speech movement and then everything grows from there, the sentences we work on will have the speech movement in it, the story will [focused around] the speech movement."*

P7: *"Yes, they love a story. They are all grandparents and I think they do it with the grandchildren, they love a story. If I'm telling a history lesson they like stories, I*

[f] Newton Edward "Paul" Daniels was an English magician and television presenter. Daniels achieved international fame through his television series The Paul Daniels Magic Show, which ran on the BBC from 1979 to 1994. He had died the week before the interview with P5.

*sometimes do Asian stories, or Russian stories...And they love it, you can see they have got it, you can see they enjoy it."*
Interviewer: *"Because it makes them more interested and that keeps them more focused?"*
P7: *"Yes, it's something different."*

**Fingerspelling:** Finger spelling is a component of signed languages, where a combination of hand positions can represent letters. In this technique the tutor signs the first letter of a word but the rest has to be identified. As this technique focuses on visual disambiguation it is an **Analytic** technique. Used by P2, P4, P5 and P7.

P2: *"I'll tell you what I do use, is fingerspelling...Because k is hard to see, h is hard to see, g is hard to see, t, d, m, these, if it's at the beginning of the word it helps you along the road, although these people are not slow to tell you, you are distracting me from watching your lips...So I might only fingerspell it once, when I say it without voice 3 times, I'll fingerspell it once and then I'll put my hands away, and leave them to struggle."*

P4: *"We point, we do teach fingerspelling as an addition to it, but I would point out gestures of things but I wouldn't actually go into teach it."*

P5: *"Because we give hints with fingerspelling, so say Houdini or whatever the Paul Daniels lesson or whatever, so people can't get it if it's an invisible sign on the lips (a viseme) then we give the letter of each word."*
Interviewer: *"Kind of the initial letter?"*
P5: *"The initial letter and that kind of gives them a hint, and you use that all the time, so lets say [the sentence] is 'How do people keep learning at home' you know that's the sentence but folk are struggling with it and I can see they are struggling with it, so I go 'Have you got it?' And they'll go 'No'. So you would say *fingerspells each initial letter of the following* 'How do people'. You would actually break it down and just give them the first letter of each word."*

P7: *"I've been to other classes and they don't do the fingerspelling and it's obvious they struggle."*

**Scenarios:** Scenarios is a technique where the tutor bases a lesson around pretending students are in a specific place such as the dentist, so all material is based on that scenario. It is a **Synthetic** technique. Used by P1, P2, P4 and P6.

P1: *"I have now and again, done pairs…like a plan of the street. And asked them to talk about the directions to shopping and things like that."*

P2: *"So we are going to the supermarket and we are buying a loaf of bread and potatoes."*

P4: *"If they were sitting in a cafe, what kind of things would they be asked.You know what's the waiter going to say, when they first come up to you in a cafe. That sort of information, so it makes learning things an awful lot easier and quicker…We can do some of that building in what we call coping strategies. You know we'll build in the scenario of the cafe to something. Or you know going to a dentist, or going to the doctors, what kind of things are you likely to be asked."*
Interviewer: *"And would they do that in pairs as well?"*
P4: *"They can do it [in pairs], we do some of that in small groups. Depending on the size of the class, I would do it in pairs. I could do a conversation, a kind of mockup conversation between say a waiter and a customer. So they are getting this, you know, what am I expecting to hear and what are my answers going to be."*

P6: *"I spend quite a lot of time working on context on what's being said, the individual sounds are there in their minds somewhere but getting the general idea I think is more important so we do quite a lot of work on that."*
Interviewer: *"Ok so maybe you do kind of scenario based practice?"*
P6: *"Yes, we can do."*

**Word Quizzes:** Word Quizzes are based around a topic such as 'animals' and the student has to watch for an animal for each letter of the alphabet (e.g., word starts with 'Z' and is an animal). Used by P2, P5 and P7. **Hybrid**.

P2: *"That they have to lipread the answers to the quiz, say birds…They'd have to lipread the names of the birds, but they would know…"*
Interviewer: *"That it's birds?"*
P2: *"Yeah."*

P7: *"Last thing [in the lesson] is always a quiz, they love the quizzes and so we usually end up with some kind of quiz."*

Interviewer: *"How do you make sure they are learning at home or do you not think it's important?"*
P5: *"I think it's important, it could be something as simple as next week we are*

> *gonna be doing the a to z of animals.…So go home and have a think this week when you are brushing your teeth or doing your hair about ok about an a, aardvark, b and you know just work your way [through].”*

**Framed Sentences:** A framed sentence exercise includes saying a topic-based sentence such as “In my garden I will find _____” where the gap is filled with a word of an item found in the garden such as flowers or grass. It is a **Synthetic** technique. Used by P2 and P5.

> P2: *“Where I would say, what kind of hats…or something like that, and I’d write, you can wear a _____ , and then, sombrero, homburg*[g]*, tammy*[h]*, beret, and I write these down.”*

> P5: *“I’ll give kind of a framed sentence, so for example in my garden I find apples, in my garden I find dog toys, in my garden I find birds…and I find flowers, so they have the basic thing every time, in my garden I find _____”.*
> Interviewer: *“So that’s the kind of context stuff?”*
> P5: *“Yes, and then they know then that it’s stuff they can find in their garden, so they are then thinking if I had a garden, some of them don’t have gardens, so you know that kind of thing.”*
> Interviewer: *“So it kind of narrows their search space for what word it could be?”*
> P5: *“Yeah you’ve got to give them context.”*

**Mystery Object:** Mystery Object is when you have an unknown object hidden using paper or cloth. The tutor speaks a number of ‘hint’ sentences to aid identification. It is a **Synthetic** technique. Used by P2 and P3.

> P2: *“Hot cross bun is not lip-readable, watch (mouths hot cross bun).”*
> Interviewer: *“Ok yeah, it’s not easy.”*
> P2: *“So, you just hand them it in a polythene bag and they pass it round and then you give them a set of about nine clues. And you give them all without voice…normally in a story, it’s a line without voice, [then a line] with voice, next line the same. Mystery object, no voice.”*
> Interviewer: *“Oh ok. Isn’t that really difficult then? Sounds really difficult.”*
> P2: *“Yes. A lot of them get it by about clue three or four and they love it, because they are having to work so hard, to try and determine what you are saying, when I’m*

[g] A homburg is a formal felt hat characterised by a single dent running down the center of the crown.
[h] A tammy or a tam o’ shanter is a name given to the traditional Scottish bonnet worn by men.

*finished that I give them what I said with voice and again without voice in each of the lines."*
Interviewer: *"So then is it mostly the same kind of object, so they kind of know what it might be or just completely?"*
P2: *"I'll tell you the kind of things I do…pine cone for Christmas…hot cross bun for Easter, a bar of soap…an apple."*
Interviewer: *"So they just don't know what's coming?"*
P2: *"They have no idea what's coming and I always put it in bubble wrap."*

P3: *"And sometimes you do that at the beginning and say with a mystery object or a mystery person, so I'm gonna do 8 sentences. I'm not gonna give you any clues lets just go through it and just relax and if you don't get it first don't worry try and pick it up later you know."*

### 4.4.4 Attitudes To Technology

Tutors appear to use little technology in their classes, and discussed only using hearing aids, loop systems, and in some cases Microsoft PowerPoint or videos. Five tutors reported informing students of lipreadingpractice.co.uk, a website containing videos of some of the exercises used within classes.

#### Critical Of Current Approaches

Participants were critical of current approaches to learning speechreading using technology. Subtitles were praised for allowing individuals to enjoy videos, however tutors also reported that subtitles do not improve speechreading as students have to either watch the subtitles or watch the video:

P4: *"I never advocate watching television to practice lipreading. If you have ever tried…it's horrendous!"*

lipreadingpractice.co.uk was also criticised for having distracting videos and limited training material:

P2: *"The trouble with a lipreading site like that is there are only so many stories. Eventually, if…you've got a good memory, you are just gonna know them all."*

### Open To New Approaches

Participants were open to new approaches involving technology and several discussed how mobile apps could be developed to help students practice outside of classes:

> Interviewer: *"Do you think that people who are looking into learning lipreading would benefit from additional kinds of new technology to help?"*
> P3: *"I would think so yes. But I don't have the technological ability so I can't tell you what it would be. But I'm sure because there is so much out there, iPhones, iPads and you know tablets and all that. Surely to goodness there must be something that we can do that will help."*

> P5: *"I mean I don't know if there is kind of lipreading apps…I think a lipreading app would be good. Just for people to practice. You know apps are the way forward aren't they?…How you would develop it I'm not quite sure…it would just need to be about the shapes and practicing the words and having a bit of context or whatever."*

> P7: *"I think anyone who is looking into learning to lipread would benefit from anything that encourages them to lipread, anything at all."*

## 4.5 Discussion

From my thematic analysis, I found that speechreading is a difficult skill to learn and that classes help facilitate learning. However, there is a lack of funding for classes resulting in limited longterm access. I also found that tutors employ different teaching techniques and approaches, and use little technology when teaching, but are open to new technology.

### 4.5.1 Summary of Findings

The first aim of the interviews was to investigate whether tutors employ analytic or synthetic training approaches. Although the teaching methods described in Chapter 2 typically emphasise one over the other, all of the tutors draw from both approaches as needed. This is especially apparent when seeing the different teaching techniques that are employed by tutors. With all tutors using a variety of analytic and synthetic teaching techniques.

The second aim of the interviews was to investigate what tutors consider to be the basic unit of speechreading. In general, tutors did not feel that there was one way to practice or teach speechreading, with speech movements, words, sentences and even general topics being discussed as ways to practice and teach from.

The third aim of the interviews was to investigate what technology tutors currently use in classes to teach speechreading. Tutors appear to use little technology in their classes, and reported only using hearing aids, loop systems, and in some cases Microsoft PowerPoint or videos. Five tutors did report informing students of *lipreadingpractice.co.uk* (a website containing videos of some of the exercises used within classes), however, participants were generally critical of current approaches to learning speechreading using technology. Subtitles were praised for allowing individuals to enjoy videos, however tutors also reported that subtitles do not improve speechreading as students have to either watch the subtitles or watch the video reflecting previous research [79]. Websites such as lipreadingpractice.co.uk were criticised for a lack of content.

The fourth aim of the interviews was to investigate how tutors feel that speechreading students practice outside of class. Tutors reported that they felt students practice outside of class by using a mirror (mirror practice), watching television with and without subtitles, using exercises from class, observing speakers during conversations and using websites such as *lipreadingpractice.co.uk* (in spite of its shortcomings).

The fifth, and final, aim of the interviews was to investigate tutors opinions on whether new technology or training techniques could improve speechreading teaching. Tutors were open to new approaches involving technology, and several discussed how mobile apps could be developed to help students practice outside of classes.

### 4.5.2 Limitations

First, this data is obtained exclusively from Scottish speechreading tutors, many of whom were trained on the same course – it is possible that these tutors use outdated techniques, reducing the value of these findings. However, Scotland is one of the few countries in the world that provides accredited speechreading tutor training – most other countries provide no formal training. As such, it is reasonable to assume that Scottish tutors can give us a reasonable representation of speechreading training techniques due to the lack of formal training elsewhere.

Second, all of the tutors who took part in the interviews were female. This introduces a gender bias to these findings, however at the time of conducting the interviews all of the 21 tutors listed on the Scottish Course to Train Tutors Of Lipreading (SCTTL) website[i] were female. This has been highlighted as an issue and the Scottish government, the SCTTL, and local hearing charities are aiming to address this gender bias by seeking to train a more diverse group of tutors in the future [10].

[i] (`http://www.scotlipreading.org.uk/index.php/classes/`)

## 4.6 Conclusion

In Chapter 3, I presented a review of currently available Speechreading Acquisition Tools (SATs) and Conversation Aids. In general, these previous solutions are not helpful to speechreaders because their designs do not align with the theoretical understanding of the approaches to speechreading training discussed in Chapter 2. Any solution that is developed to help speechreaders during speechreading, or to help practice their speechreading should be influenced by how speechreading is currently taught in speechreading classes. Although the speechreading teaching methods described in Chapter 2 outline the basic approaches to speechreading teaching, it was necessary to investigate current practice (which is most likely influenced by these theories).

Therefore, in this chapter, I presented the methodology and findings of in-depth interviews conducted with seven practicing Scottish speechreading tutors to explore their background, approach to teaching, current use of technology, and thoughts on how speechreading can be improved. Through thematic analysis of the interview transcripts, I identified four main themes relevant to the future development of speechreading acquisition tools: 1) Speechreading as a Skill, 2) Access to Speechreading, 3) Teaching Practices, and 4) Attitudes to Technology.

In the next chapter, I will use the findings from the thematic analysis to develop a novel framework to help design new Speechreading Acquisition Tools (SATs). To evaluate the framework, I will demonstrate that it can accommodate current teaching techniques discussed by the tutors during the interviews, as well as existing solutions from related work discussed in Chapter 3. I will also discuss how the framework can be used to help identify and design three new SATs.

5

# Speechreading Acquisition Tools Framework

## 5.1 Introduction

This chapter presents a novel framework that can be used to design Speechreading Acquisition Tools (SATs) - a new type of technology designed specifically to improve speechreading acquisition. I used the thematic analysis findings from Chapter 4 to identify and organise key elements of the framework. The motivation behind the framework is given first, followed by a justification of design decisions used in the implementation of the framework. Following this is an evaluation of the framework by using it to: 1) categorise every teaching technique identified by speechreading tutors (during the interviews presented in Chapter 4), 2) critically evaluate existing Conversation Aids and existing SATs (discussed in Chapter 3), and 3) design three new SATs: PhonemeViz, MirrorMirror and ContextCueView (where the first two are developed and evaluated in Chapter 8 and 9 respectively).

## 5.2 Motivation

In Chapter 3, I presented a review of currently available Speechreading Acquisition Tools (SATs) and Conversation Aids. In general, these previous solutions are not helpful to speechreaders because their designs do not align with the theoretical understanding of the approaches to speechreading training discussed in Chapter 2. Any solution that is developed to help use or practice speechreading, should be influenced by how speechreading is currently taught in speechreading classes.

Although the teaching methods described in Chapter 2 outline the basic theoretical approaches to speechreading teaching, it was also necessary to investigate current practice (which is most likely influenced by these theories). Therefore, I conducted interviews with practicing speechreading tutors to generate the dataset needed to inform the design of new Speechreading Acquisition Tools (SATs).

From my thematic analysis of the interviews presented in Chapter 4, I found that speechreading is a difficult skill to learn and that classes help facilitate learning. However, there is a lack of funding for classes, threatening the long-term availability of classes and these classes are only available in a handful of countries in the world. I also found that tutors employ different teaching techniques and approaches, and use little technology when teaching, but are open to new technology. These themes highlight how speechreading acquisition can be enhanced through the development of new assistive tools that will help resolve issues regarding the lack of access, and can also be specialised to different speechreading teaching approaches.

I call these tools Speechreading Acquisition Tools (SATs) – a new type of technology designed specifically to improve speechreading acquisition. I believe that through the development and release of SATs, people with hearing loss will be able to augment their class-based learning, or learn on their own if they are unable to attend speechreading classes.

However, it is currently unclear for the research community, AT commercial sector, and AT enthusiasts how to design SATs within the context of contemporary speechreading teaching and practice. To facilitate the transition of knowledge from the thematic analysis to the research community, Assistive Technology (AT) commercial sector, and AT enthusiasts, I developed a *speechreading acquisition framework* that can be employed when designing SATs.

## 5.3 Framework Design

The framework should describe the space of speechreading teaching by framing it using the techniques reported by speechreading tutors during the interviews presented in Chapter 4. The teaching techniques provide us with the best opportunities for design, as they give us insight into how the two speechreading approaches (Analytic and Synthetic) can be applied in different ways to speechreading practice. Therefore, when designing new technology to support speechreading, we can borrow elements from these techniques because they show the issues and challenges behind using and practicing speechreading.

### 5.3.1 Dimensions

To scaffold the framework, I looked at the two dimensions that can be used to describe each of the teaching techniques:

**Type of Skill** All of the teaching techniques can be classified as Analytic or Synthetic – the two approaches to speechreading discussed in Chapter 2 and reported by speechreading tutors during the interviews presented in Chapter 4. However, this dimension is likely continuous as some techniques borrow aspects from both approaches.

**Additional information available** Each teaching technique provides a different amount of information supplied about the training material to the student. From a low amount, such as when practicing *speech movements* (as only the speech movement is given and the student has to speechread the rest of the word), compared to a technique such as *framed sentences* that provide a large amount of information (e.g., “In my garden I find _____”). During the interviews, tutors noted that they would use techniques that provide a high amount of information in beginner classes (e.g., framed sentences, mirror practice) and reserve techniques providing low information (e.g., speech movements, stories) for later classes as the students’ speechreading skills increase.

### 5.3.2 Initial Version

The initial version of the framework was represented as a two-dimensional space with two continuous dimensions: *Type of Skill* in the x-axis from *Analytic* to *Synthetic* and *Amount of Information* from *High* to *Low* as shown in Figure 5.1.

Although this version could describe each of the teaching techniques, in order to improve framework accessibility I felt that the dimensions should be further discretised.

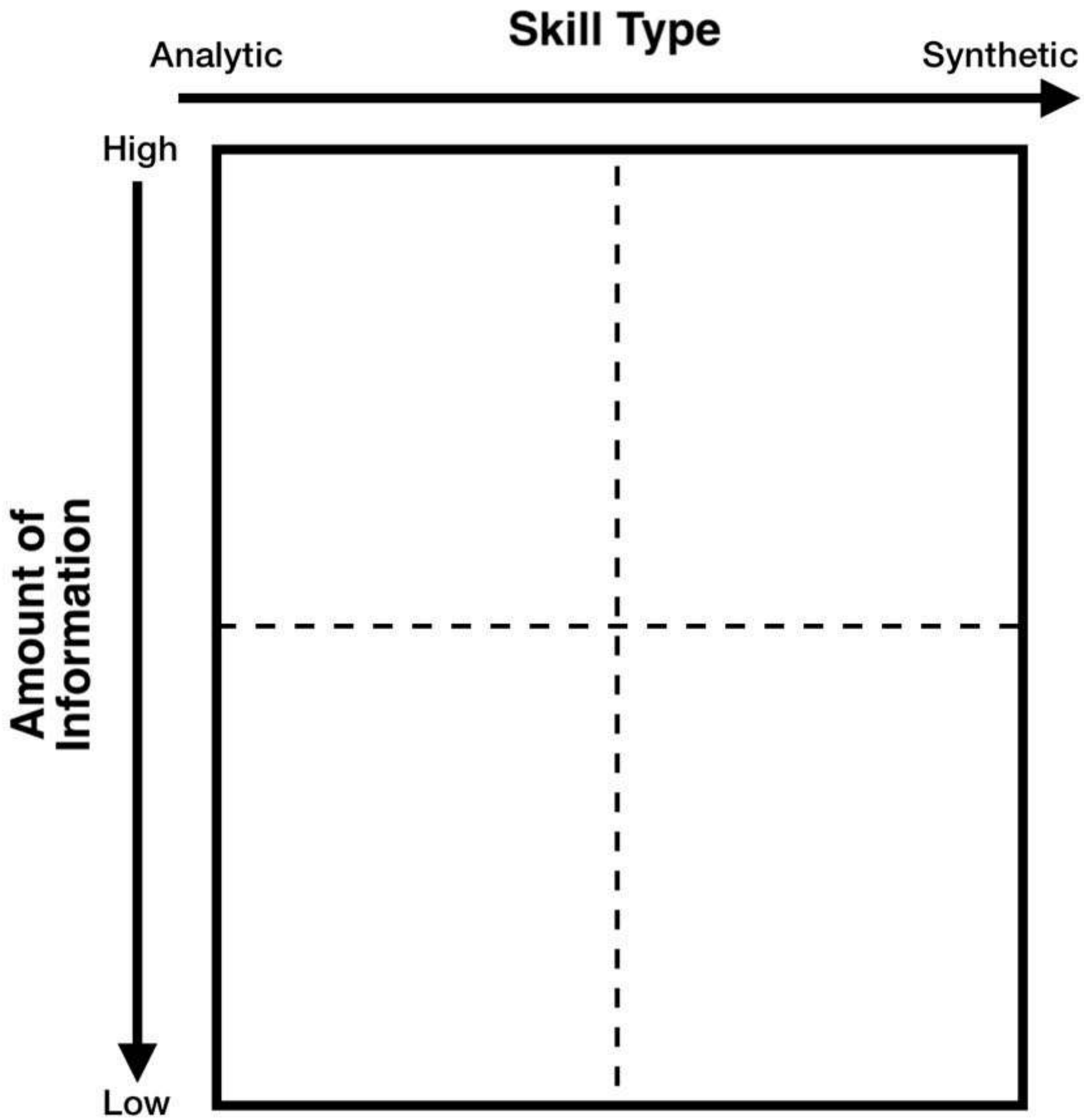

Figure 5.1: The initial version of framework, with two continuous dimensions: 1) *Type of Skill* ranging between *Analytic* and *Synthetic* and 2) *Amount of Information*, ranging from *High* to *Low*).

### 5.3.3 Second Version

In the second version of the framework, I discretised each of the continuous dimensions: *Type of Skill* and *Amount of Information*, each split into three levels (*Analytic/Hybrid/Synthetic* and *Low/Medium/High*.

The initial version as shown in Figure 5.1 was represented as a two-dimensional space, however as a result of discretising the dimensions the second version was represented as 3x3 grid (as shown in Figure 5.2). I also added a baseline and a ceiling label to attempt to explain the rationale behind each technique providing a different amount of information:

**Visual-only Speechreading** This baseline represents the amount of information supplied by the visual only channel in terms of speech recognition. It represents the least assisted case for a speechreader in which they cannot hear the speaker due to limited residual hearing or noisy conditions.

**Typical Hearing** This ceiling level is taken as the total possible information to be gained from audio-visual speech recognition by a conversant with typical hearing in a quiet room with adequate lighting and a good speaker. It represents the hypothetical maximum level of benefit that can be provided by speechreading.

However, upon reflection I decided that these labels were unnecessary because it would be beneficial for the final framework to be able to describe teaching techniques along with the Conversation Aids and SATs discussed in Chapter 3, and the labels would inhibit this due to their conversational focus.

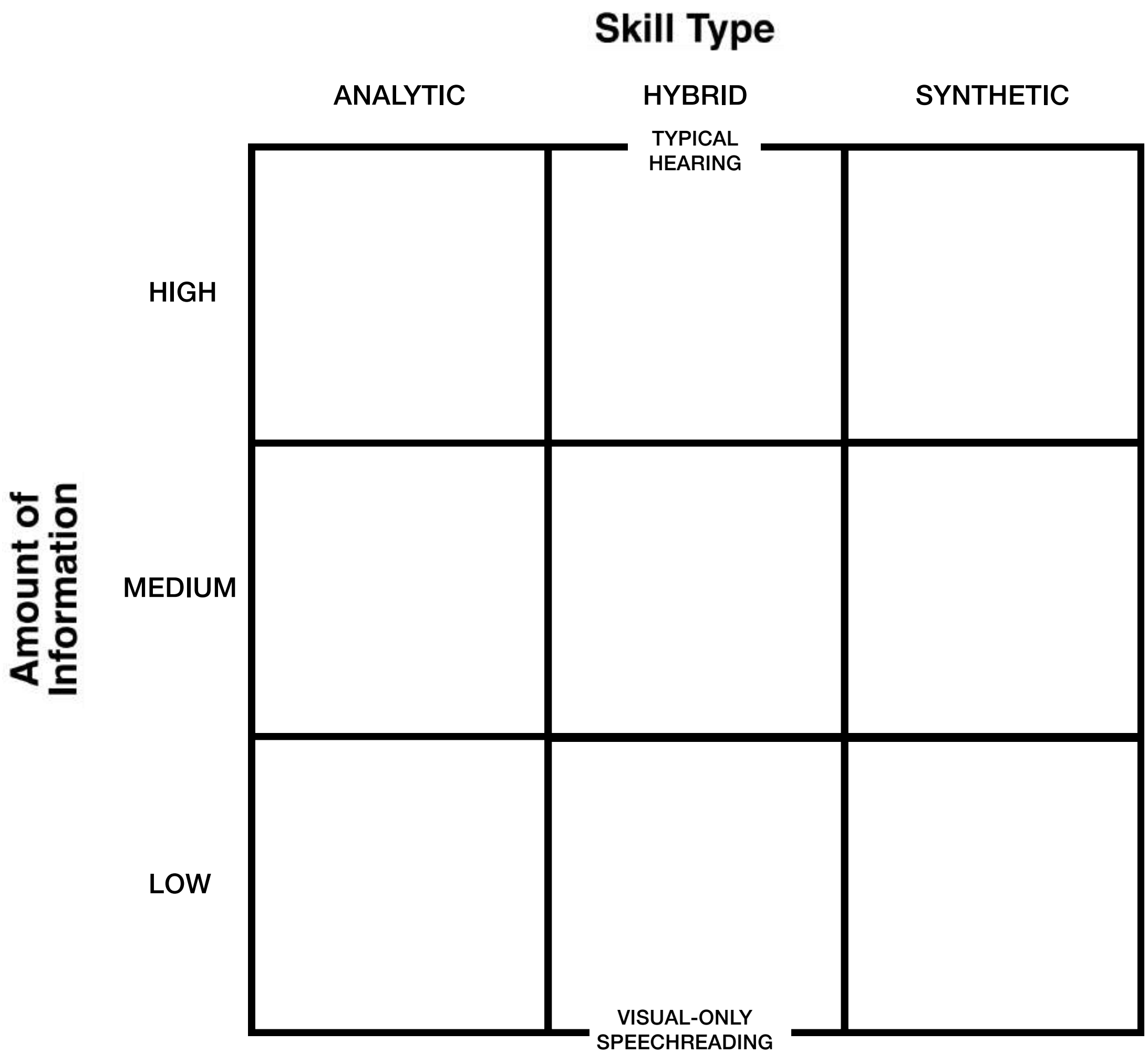

Figure 5.2: The second version of the framework, with the same two dimensions as the initial version: *Type of Skill* and *Amount of Information*, but now split into three levels (*Analytic/Hybrid/Synthetic* and *Low/Medium/High*, respectively). The label 'Visual Only Speechreading' is a baseline taken as the amount of information supplied by visual-only speechreading. It represents the least assisted case for a speechreader in which they cannot hear the speaker due to limited residual hearing or noisy conditions. The label 'Typical Hearing' represents the total possible information to be gained from audio-visual speech recognition by a conversant with typical hearing in a quiet room with adequate lighting and a good speaker. It represents the hypothetical maximum level of benefit that can be provided by speechreading.

### 5.3.4 Final Version

As discussed, because both 'skill type' and 'information amount' apply to each teaching technique, I used these as the base dimensions for the final implementation of the framework: *Type of Skill* and *Amount of Information*. The dimensions are continuous in nature, but to improve framework accessibility I discretised each into three levels (*Analytic/Hybrid/Synthetic* and *Low/Medium/High*, respectively), resulting in a 3x3 cell-based grid as shown in Figure 5.3.

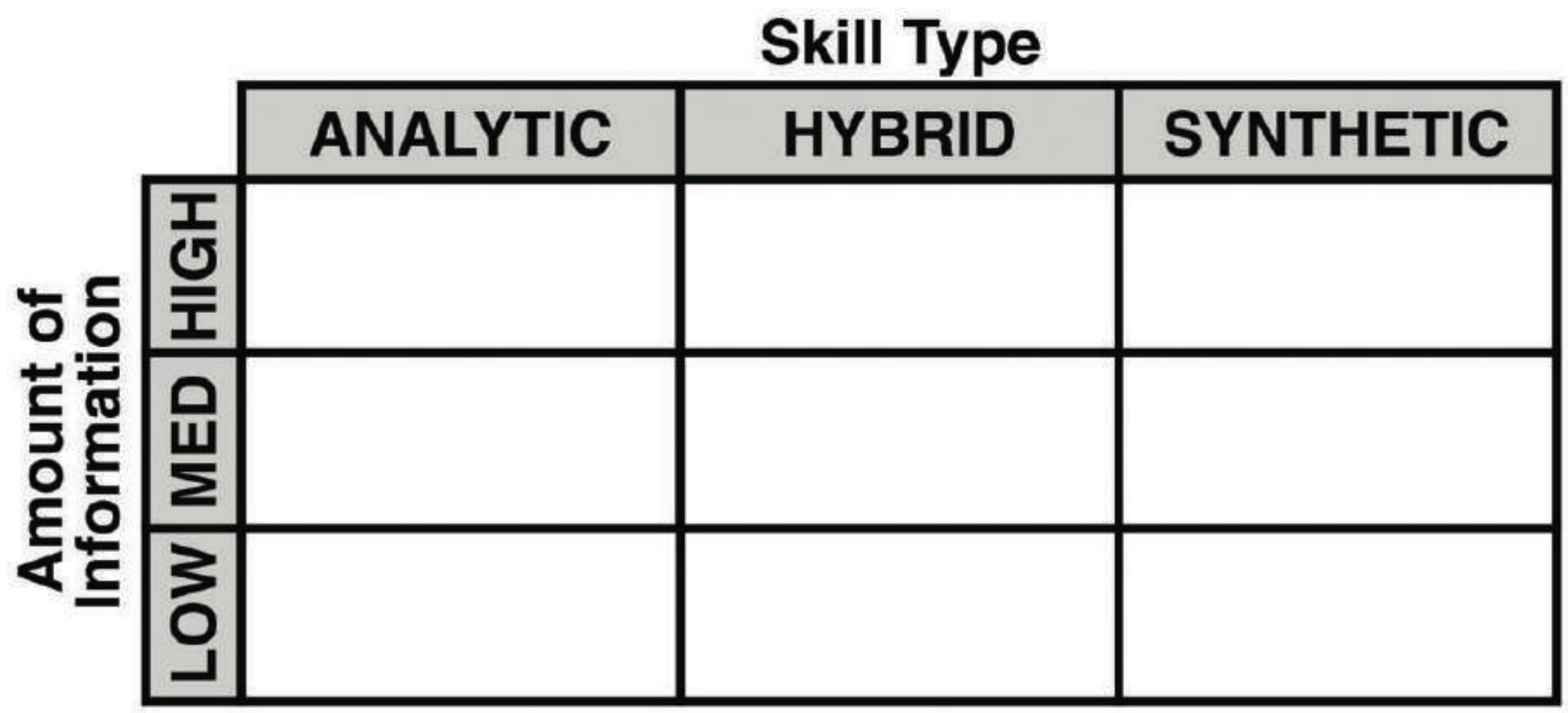

Figure 5.3: The final speechreading acquisition framework, with two dimensions: *Type of Skill* and *Amount of Information*, each split into three levels (*Analytic/Hybrid/Synthetic* and *Low/Medium/High*, respectively).

The final version does not have the baseline and ceiling labels present in the second version because they would inhibit the framework's ability to describe both teaching techniques and previously designed Conversation Aids and SATs.

A technique (Conversation Aid, SAT or teaching technique) is classified as *Analytic* if it focusses on visual disambiguation. A technique that focusses on leveraging the context is classified as *Synthetic*. *Hybrid* techniques focus on both analytic and synthetic approaches to speechreading.

As the basic unit of analytic teaching methods is the phoneme, analytic techniques that provide individual phonemes are classified as *Medium*, techniques that provide non-phonemic information (e.g., speech production properties) are *Low*, and techniques that provide more than phonemes (e.g., whole words) are *High*.

The basic unit of synthetic teaching methods is the sentence, synthetic techniques that provide the topic of a specific sentence are classified as *Medium*, techniques that provide less (e.g., the topic of a conversation) are *Low*, and techniques that provide more information (e.g., the topic and context of a sentence) are *High*.

## 5.4 Framework Evaluation

To evaluate the framework, I now use it to 1) classify existing teaching techniques, 2) critically reflect on previously-developed SATs and Conversation Aids discussed in Chapter 3, and 3) identify and describe three new technologies for enhancing speechreading acquisition and proficiency. By employing the framework in this fashion, I show that it: 1) comprehensively reflects existing speechreading teaching practice, 2) can be used to help understand the strengths and weaknesses of previously-developed solutions, and 3) can be used to identify clear opportunities for the development of new SATs to improve speechreading acquisition.

### 5.4.1 Teaching Techniques

To evaluate the framework, I first fit existing teaching techniques (reported by tutors during the interviews presented in Chapter 4) within the framework cells. The goal of this evaluation is to assess the framework's coverage; accommodating every teaching technique identified by the interview participants indicates good coverage, any teaching techniques not fitting within the framework indicate framework incompleteness.

In this evaluation, 'Amount of Information' is the amount that is supplied to the student by the tutor or technique. This information helps the student understand what the tutor is saying, thereby giving the student feedback on his/her speechreading.

#### Fitting Identified Teaching Techniques

I now describe the classification rationale for each teaching technique described in Section 4.4.3 of Chapter 4. The classifications can be seen in Figure 5.4.

**Speech Movements and Lipshapes:** Speech movements or lipshape is a teaching technique where the student is told to focus on the visual representation of a single speech sound isolated or within a word and is therefore an **analytic** teaching technique. Tutors inform the student of the target speech-movement and the student has to identify the word or words spoken. As the student is only provided with the target speech-movement this is a **low amount of information** as the student has to identify the initial letter (e.g., /p/ in /p,b,m/) and the rest of the word (e.g., pat).

**Fingerspelling:** Fingerspelling is a teaching technique which borrows components of signed languages, where a combination of hand positions can represent letters. As this technique focuses on visual disambiguation it is an **analytic** technique. In this technique the tutor signs the initial letter of a word but the rest has to be identified. This is classed as a **medium amount of information** as there is a unique visual representation for each speech sound, which provides the initial letter so that the student only has to speechread the rest of the word.

| Amount of Information \ Skill Type | ANALYTIC | HYBRID | SYNTHETIC |
|---|---|---|---|
| HIGH | Mirror Practice, QREs | Word Quizzes (name the animals) | Framed Sentence Exercises |
| MED | Finger Spelling (initial letter) | Cue Recognition (body language) | Scenarios |
| LOW | Speech Movements, Lip Shapes | Pair Work | Mystery Object*, Stories |

Figure 5.4: Placement of teaching techniques into the framework. The '*' indicates the starting amount of information provided by this technique (Mystery Object), but this level increases as more clues are given.

**Quick Recognition Exercises (QREs):** QREs or *Syllable Drills* are rhythmic drills consisting of contrasting syllables or words spoken as quickly as possible in different orders by the tutor. Students repeat back the order. This is an **analytic** technique and provides a **high amount of information** as the student knows the words or syllables spoken, so only has to work out the order of them.

**Mirror Practice:** Mirror Practice involves students looking at a mirror to learn their own mouth shapes and the differences between mouth shapes when certain speech sounds or words are spoken. Mirror practice is an **analytic** technique as the student focuses on visual disambiguation and provides a **high amount of information** as the student knows the words and movements they are speaking before the mirror.

**Pair Work:** Pair Work is when the class members work in pairs. Pair Work is a **hybrid** technique, as it involves Analytic and Synthetic skills. Pair work provides a **low amount of information** as the student may be familiar with their partner's facial movements (having trained with them before).

**Cue Recognition:** Cue recognition encompasses looking for body language, facial emotion, and hand gestures. It is a **hybrid** technique and provides **medium amount of information** as facial expressions and hand gestures provide a substantial amount of context to an utterance.

**Word Quizzes:** Word Quizzes are based around a topic such as 'animals' and the student has to watch for an animal for each letter of the alphabet (e.g., word starts with 'Z' and is an animal). It is a **hybrid** technique and it provides a **high amount of information**. For instance in this case, there are only a limited amount of animals which are commonly known for each letter so this reduces the search space for the word.

**Stories:** Stories are based around one topic and may be a number of sentences long. It is a **synthetic** technique and provides a **low amount of information** as only the topic of the story is given to the students.

**Mystery Object:** Mystery Object is a teaching technique where an object is hidden using paper or cloth. The tutor speaks a number of 'hint' sentences to aid identification. It is a **synthetic** technique and provides a **low amount of information**. However, this amount of information is variable, as the more 'hints' that are given reduces what the object can be.

**Scenarios:** Scenarios is a teaching technique when the tutor bases a lesson around pretending students are in a specific place such as the dentist, so all material is based on that scenario. It is a **synthetic** technique and provides a **medium amount of information**.

**Framed Sentence Exercises:** A framed sentence exercise includes saying a topic-based sentence such as "in my garden I will find _____" where the gap is filled with a word of an item found in the garden such as flowers or grass. It is a **synthetic** technique providing a **high amount of information** as the student is told the sentence and guesses one word; the sentence provides a high degree of contextual information.

As shown in Figure 5.4, the framework accommodates all of the identified teaching techniques with no gaps, thereby increasing the confidence in the coverage of the framework.

### 5.4.2 Existing Conversation Aids & SATs

In the second stage of the evaluation, I fit existing Conversation Aids and Speechreading Acquisition Tools (SATs) into the framework to get a deeper sense of the coverage of the framework. Through this process I also critically reflect on the design of existing Conversation Aids and SATs as well as identify where new opportunities lie.

In this evaluation, 'Amount of Information' is the amount of information that is supplied by the Conversation Aid or SAT to the person requiring conversation support.

#### Fitting Existing Conversation Aids & SATs

I will now briefly describe existing conversational aids and SATs (discussed in more detail in Chapter 3), and provide the reasoning for where I fit each in the framework (placements in Figure 5.5).

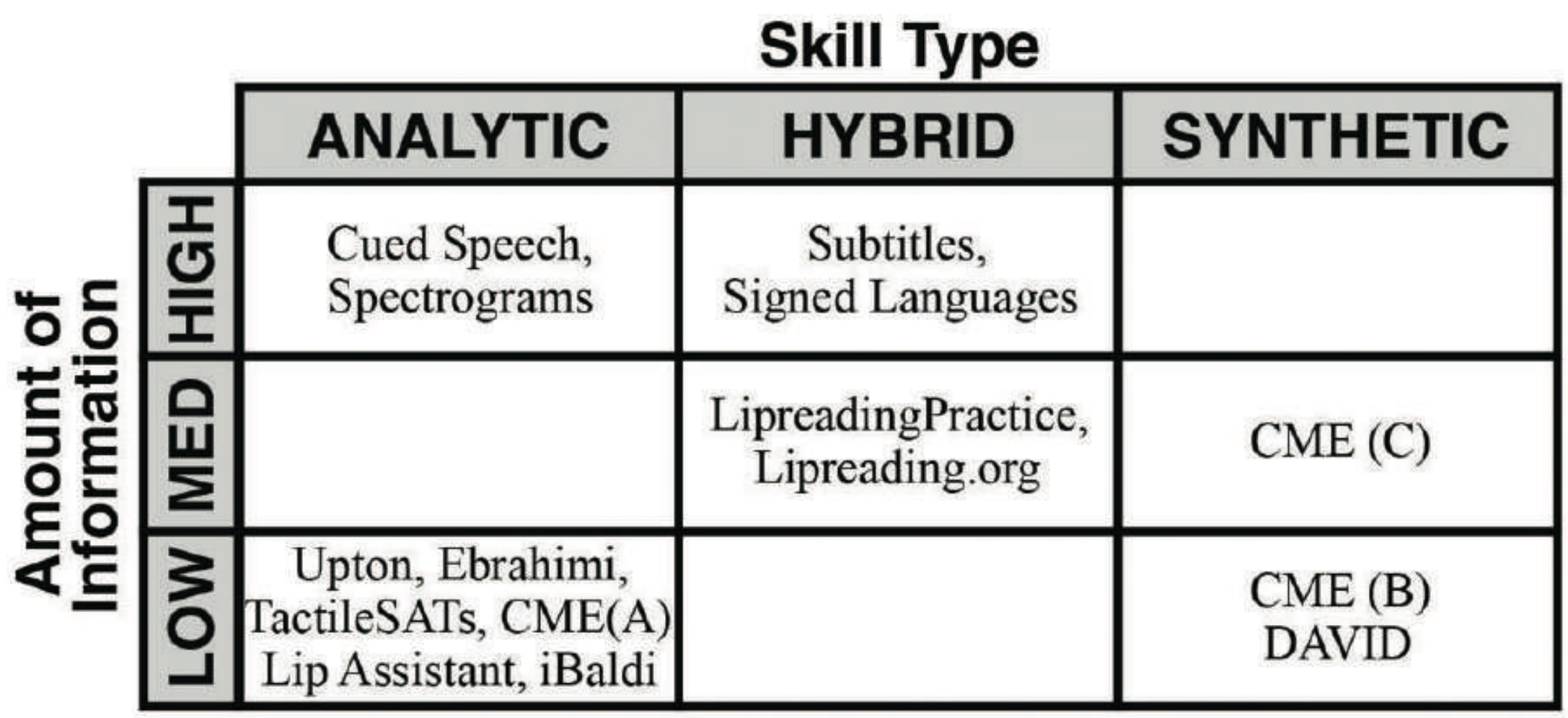

Figure 5.5: Placement of Conversation Aids and existing SATs into the framework.

**Signed Languages:** *Signed Languages* (e.g., American [126] and British Sign Language [47]) are natural languages that use hand, arm, and facial gestures to facilitate communication. A signed language provides **high information** and is a **hybrid** approach as it relies on analytic skills to understand unfamiliar names and words (typically communicated using fingerspelling) and synthetic skills (e.g., facial expressions) to understand particular aspects of a conversation (e.g., identifying a question).

**Cued Speech:** *Cued Speech* [45] is a system of eight hand-shapes placed in four positions around the mouth that aim to clarify lip-patterns during speechreading. As such, Cued Speech is **analytic** and provides **high information** as its cues disambiguate all phonemes.

In spite of their benefits and the importance of signed languages to Deaf culture [118], signed languages and/or cued speech need to be known by both conversation parters in order to help; they do not help conversations with people who do not know the language.

**Upton's Eyeglasses:** The earliest example of technology aiding speechreading can be seen in *Upton's Eyeglasses* [134]. This SAT used a clip-on microphone to detect speech, and processed the signal via high- and low-pass filters to classify spoken phoneme components. An LED matrix was positioned at the side of a pair of modified eyeglasses, and its light output was channeled so that it appeared at the centre of that side's lens, enabling an early augmented reality system (e.g., the bottom LED illuminated when a phoneme was voiced, making it appear as if the speaker's throat was glowing). Due to their focus on speech components, Upton's Eyeglasses are **analytic** and provide **low information**, as the wearer is only provided information about how a sound is produced. Much later, a similar peripheral-display approach was taken by *Ebrahimi* et al. [50].

**Tactile SATs:** Similar to Upton's Eyeglasses, *Tactile SATs* provide spoken phoneme information using tactile feedback. One such display of voice fundamental frequency showed a 10% improvement in a speech discrimination task [28], however a later study found positive results in terms of identifying voicing and for consonant identification, but no benefit for speechreading words in sentences [142]. As they provide information similar to Upton's SAT, these *Tactile SATs* are **analytic** and provide **low information**.

**iBaldi:** *iBaldi* [97] is an iOS application that overlays a visualisation of speech components onto an animated talking head. The visualisation shows one of three coloured discs (nasality in red, friction in white, voicing in blue) at the side of the head's mouth when it makes the corresponding sound. *iBaldi* provides similar information as Upton, so it is **analytic** and **low information**.

SATs that focus on helping identify components of speech based on how they are produced (Upton, Ebrahimi, Tactile, iBaldi) contradict typical speechreading approaches by training the speechreader to focus on *auditory* aspects of speech, rather than *visual*. Even though these aids can provide rich information, this information is of limited value to someone whose understanding of speech is primarily visual, not audible.

**Spectrograms:** *Spectrograms* visualise frequency (Y-axis) over time (X-axis), with intensity mapped to colour. Linguists can use them to identify words, but this requires extensive training [65]. *Watanabe* et al. [138] improved spectrograms by integrating different speech features into a single image, but the evaluation used participants with extensive spectrogram-reading experience, so the technique's generalisability to speechreading is unknown. Both examples of *Spectrograms* are **analytic** aids, and provide **high information** due to the richness of the visualised data, even if it is difficult to access for the non-expert.

**Lip Assistant:** *Lip Assistant* [140] is an SAT that generates magnified realistic animations of a speaker's lips that are superimposed on the bottom left of a video. *Lip Assistant* is **analytic** as it focusses exclusively on lipshape and it provides **low information**.

**Subtitles:** *Subtitles* (captions) present the speech (analytic) and sound effects (synthetic) of video as on-screen text. As such, they are a **hybrid** approach and provide **high information**, however they also require the viewer to split their attention between reading the subtitles and watching the video content; one eye-tracking study found that participants spent ~84% of their viewing time focussed exclusively on subtitles [79].

**ConversationMadeEasy:** *ConversationMadeEasy* [130] is an SAT comprised of three programs, each presenting videos of speakers with or without audio. The programs increase in complexity: Program A is for **analytic** training with **low information**, and program B is for **synthetic** sentence training providing **low information**. Program C is for **synthetic** scenario-based training with commands or questions based on the scenarios given within a closed response set of four pictured options. As such, Program C provides **medium information**.

**DAVID:** *DAVID* [123] is an SAT offering videos of sentences on everyday topics, such as 'going shopping'. The student watches and responds by typing the complete sentence or content words, or via multiple choice. DAVID also provides repair strategies such as repeating the sentence, or presenting words in isolation. DAVID is a **synthetic** SAT providing **low information**.

***lipreadingpractice.co.uk* and *lipreading.org*:** These are both website-based SATs offering videos (with or without subtitles) of consonants, vowels, and passages. They both offer are practice sessions based on topics such as going to a restaurant, or a doctors appointment. *lipreadingpractice.co.uk* and *lipreading.org* are a **hybrid** SATs providing **medium information**.

As shown in Figure 5.5, the framework can accommodate existing Conversation Aids and SATs, increasing the confidence in its coverage. More importantly, I also located gaps in the previous work. I next use the framework to design three new SATs that address these gaps, but also reflect the knowledge gained via the thematic analysis presented earlier.

### 5.4.3 Framework-Inspired SAT Examples

To further evaluate the framework I now demonstrate how it can be used to design new SATs that are influenced by contemporary speechreading practices. Below, I explain how I used the framework to design, prototype, and evaluate two speechreading acquisition tools, as well as to propose an additional SAT that I plan to build and evaluate as future work.

#### PhonemeViz

Four of the speechreading tutors interviewed in Chapter 4 reported the value of initial-letter fingerspelling during teaching as it helps to disambiguate words that are visually similar on the lips (due to being classed under the same viseme category). Even though fingerspelling is a powerful teaching technique, for successful use outside of classes it requires the speaker to know how to fingerspell. The concept of providing supplementary phoneme-based information to speechreaders could be borrowed from fingerspelling to inspire the design of a new SAT that could provide this type of **medium information**.

Recently, Augmented Reality smart glasses such as the Google Glass[a], Microsoft Hololens[b], and the Epson Moverio[c] have become popular. Using such devices it may be possible for us to be able to augment the speechreading process with additional information akin to fingerspelling. This would help to decrease one of the challenges of speechreading and also results in no additional knowledge required for someone speaking with a person who is speechreading.

Using the framework I designed a new SAT called *PhonemeViz*, which displays a subset of a speaker's phonemes to the speechreader. PhonemeViz focuses on reducing viseme confusion that occurs at the start of words by providing a similar amount of information as initial-letter fingerspelling. PhonemeViz places consonant phonemes in a semi-circular arrangement, with an arrow beginning from the centre of this semi-circle pointing at the last heard consonant phoneme to provide persistence. PhonemeViz is positioned at the side of a speaker's face, beginning at the forehead and ending at the chin. By combining the visualisation and their ability to speechread, speechreaders should be able to attend to the speaker's face while being able to disambiguate confusing viseme-to-phoneme mappings, therefore improving understanding during conversation. In my evaluation of a PhonemeViz prototype (discussed in Chapter 7), PhonemeViz enabled all participants to achieve 100% word recognition (showing successful viseme disambiguation), and participants lauded PhonemeViz in subjective and qualitative feedback. The results demonstrate that visualising a **medium** amount of **analytic** information can improve visual-only speechreading in constrained word recognition tasks.

[a]https://www.x.company/glass/
[b]https://www.microsoft.com/en-gb/hololens
[c]https://epson.com/moverio-augmented-reality

These results suggest that PhonemeViz can be overlaid onto video or displayed on a transparent head mounted display (as shown in Figure 5.6) to augment natural speechreading and enhance speechreading acquisition. Although, for any real-time applications, new automated speech recognition algorithms tuned to initial phonemes are needed. The motivation, design and evaluation of PhonemeViz is discussed in more detail in Chapter 7.

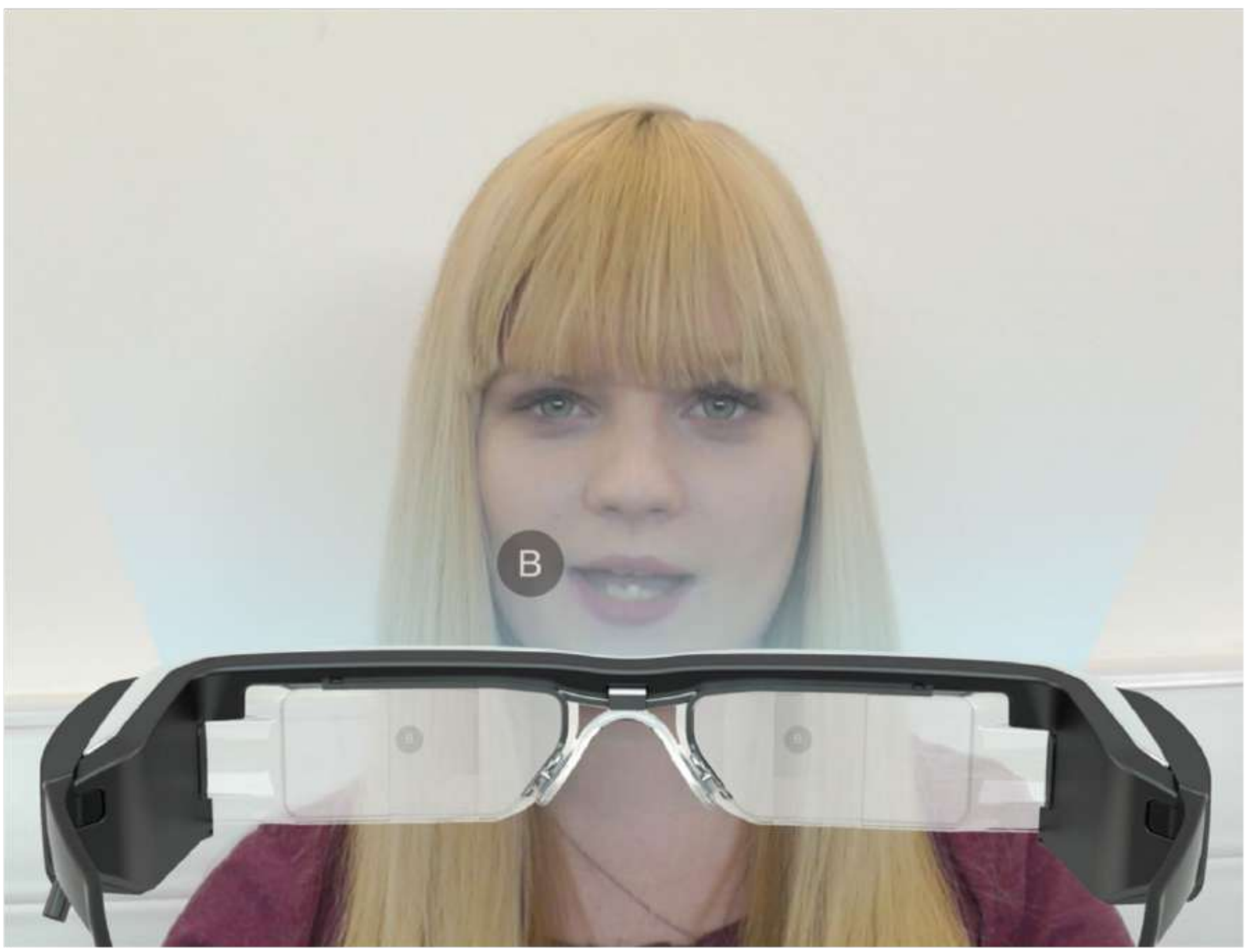

Figure 5.6: A mockup of the next iteration of PhonemeViz viewed through Epson Moverio glasses (`http://www.epson.com/moverio`) for 'bat'.

## MirrorMirror

During the interviews presented in Chapter 4, tutors reported that mirror practice plays a key role in speechreading training, and is also recommended by Action on Hearing Loss [4, 106] for practice, as it may develop visual cue integration skills needed during speechreading [7]. However, traditional mirror training does not fully develop speechreading skills as students cannot assess themselves (because they have full knowledge of what they are saying), and the technique trains them to read their own speech (instead of other people).

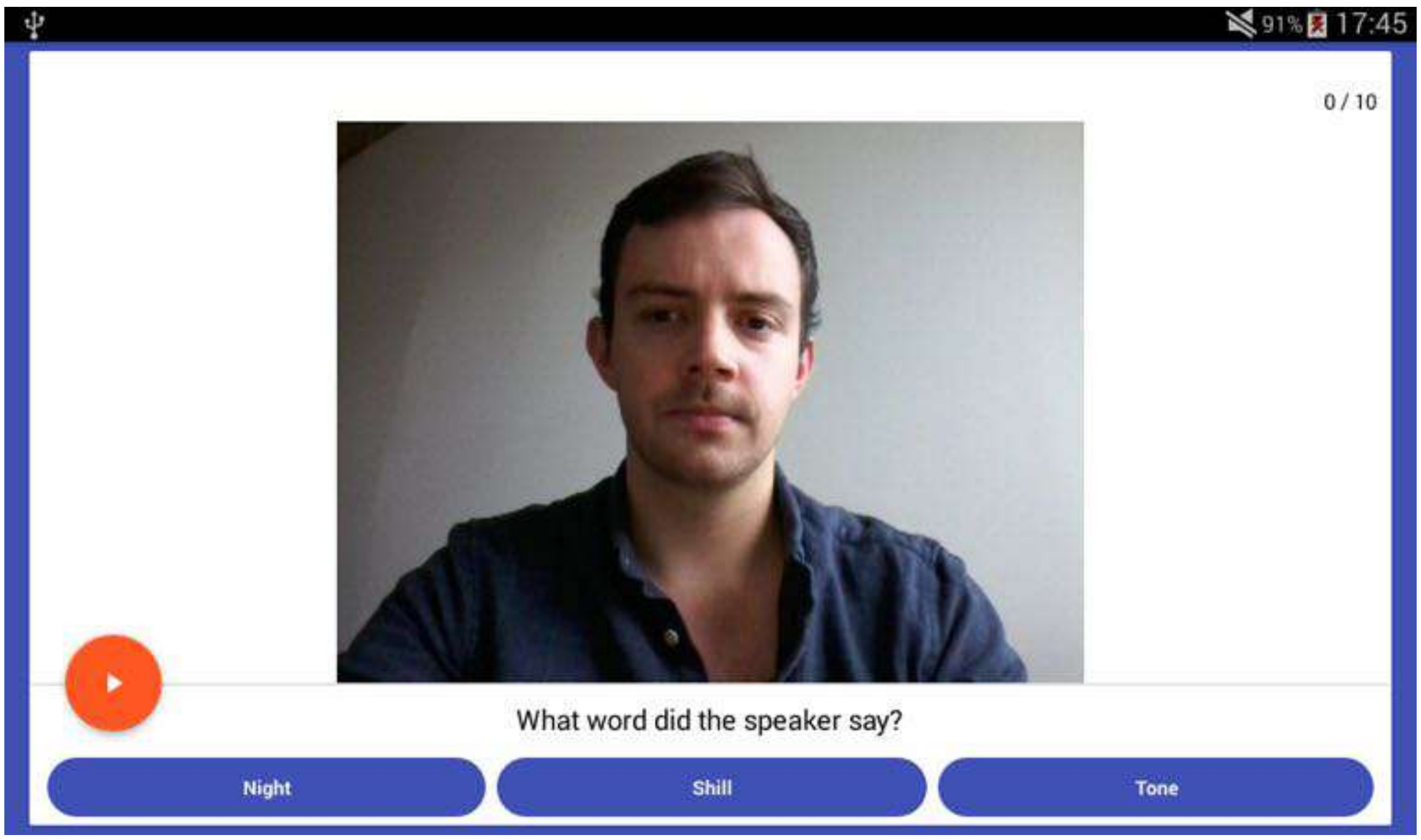

Figure 5.7: Screenshot of MirrorMirror's lipshape practice session that displaying a video of a speaker recorded by a user. At the bottom of the video are three buttons displaying three words, one of which is the correct answer.

The concept of watching speech movements and words in isolation to improve analytic speechreading skills could be borrowed from mirror training. I used this concept to inspire the design of a new SAT that overcomes the limitations of traditional mirror training. This SAT is a mobile application called MirrorMirror that allows speechreaders to practice lipshapes and words by recording videos of people they frequently talk to. MirrorMirror provides a multiple choice quiz game where the user selects the word they think the speaker has spoken (as shown in Figure 5.7). MirrorMirror provides feedback on whether they are speechreading correctly, and allows them to target specific challenges and situations.

Through practicing with MirrorMirror's endless repository of videos, it is possible to overcome the ‘full knowledge’ limitation of current mirror practice. Additionally, users could also share their speech movement libraries with each other, overcoming the ‘self-training’ limitation of traditional mirror training. MirrorMirror provides a variable (**low** to **high**) amount of information. Unlike mirror practice which provides **analytic** information, MirrorMirror allows users to practice **hybrid skills** because the user has some knowledge about the person, words and situations they have recorded. The motivation, design and evaluation of MirrorMirror is discussed in more detail in Chapter 8.

### ContextCueView

In Chapter 4, four tutors reported using scenarios in speechreading classes and they emphasised that scenarios provide a rich context for speechreading training. Within scenario training the speechreader is taught that particular phrases and topics can be pre-associated with a given location or situation. These associations can be shown using a constellation diagram introduced by Kaplan [81] (as shown in 5.8), in which a text label for the situation is placed in the middle while related topics and phrases radiate out from the ‘situation label’. Constellation diagrams help by prompting the speechreader to consider potential phrases and topics in advance of a given situation.

The concept of preparing for a situation through scenario-training or by completing a constellation diagram could be used to inspire an SAT that could help speechreaders to better prepare for potential phrases they will have to speechread in a given situation.

*ContextCueView* is an SAT that would gather contextual data (e.g., GPS, date/time) to determine a user’s situation. Using this contextual data, ContextCueView would load a matching previously-generated constellation diagram. ContextCueView constellation diagrams would be stored in a central repository, and collectively curated to rapidly provide constellation diagrams for a variety of situations. ContextCueView could run on a mobile device and operate like a ‘contextual phrase book’, but it is also well-suited for a glanceable display such as the Google Glass (as shown in Figure 5.9).

This last stage of the evaluation has demonstrated that the framework can be used to identify clear opportunities for the development of new SATs that could improve speechreading acquisition. The placement of these framework-inspired SAT examples into the framework are shown in Figure 5.10.

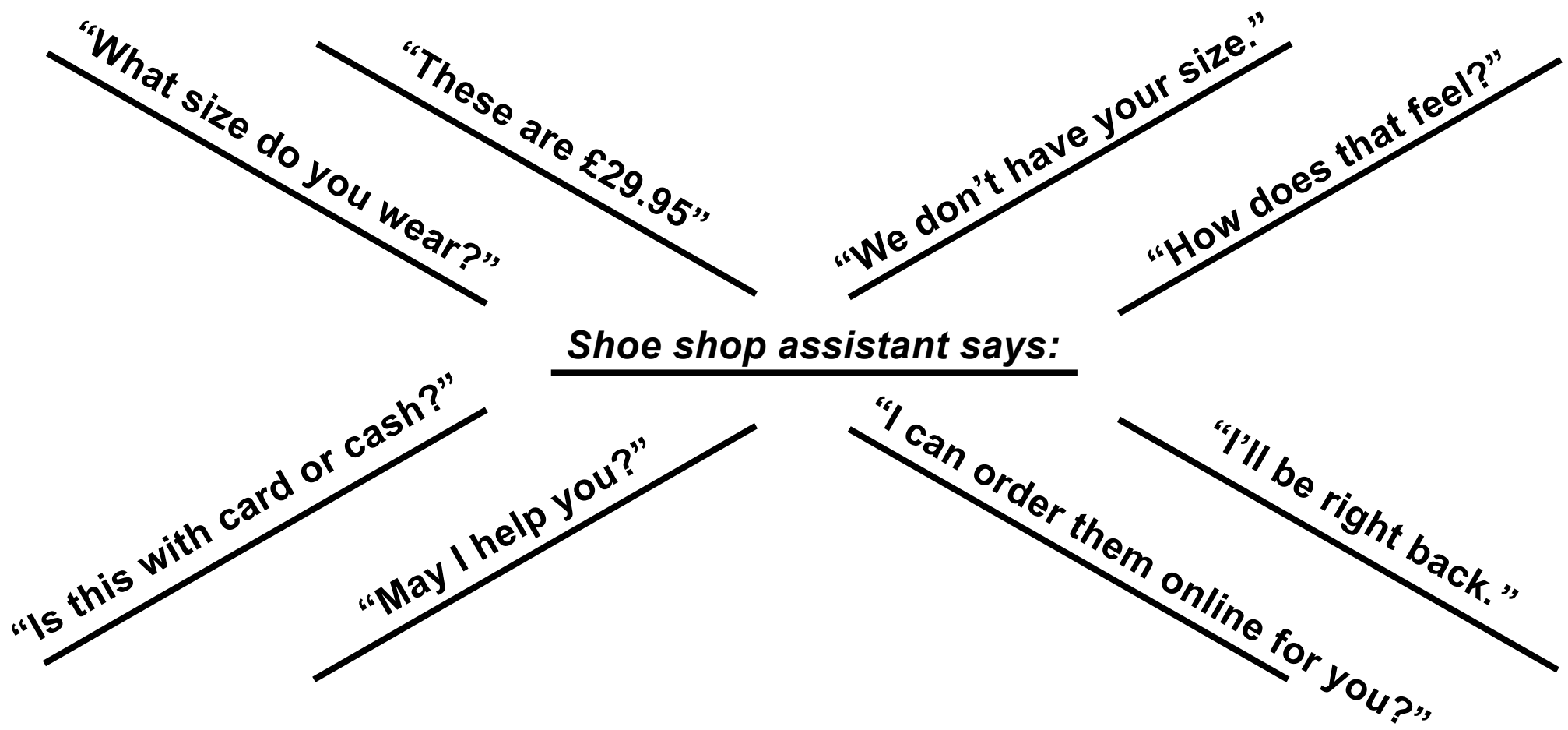

Figure 5.8: Phrases and topics that can be pre-associated with buying shoes in a shop, represented on a constellation diagram introduced by Kaplan [81].

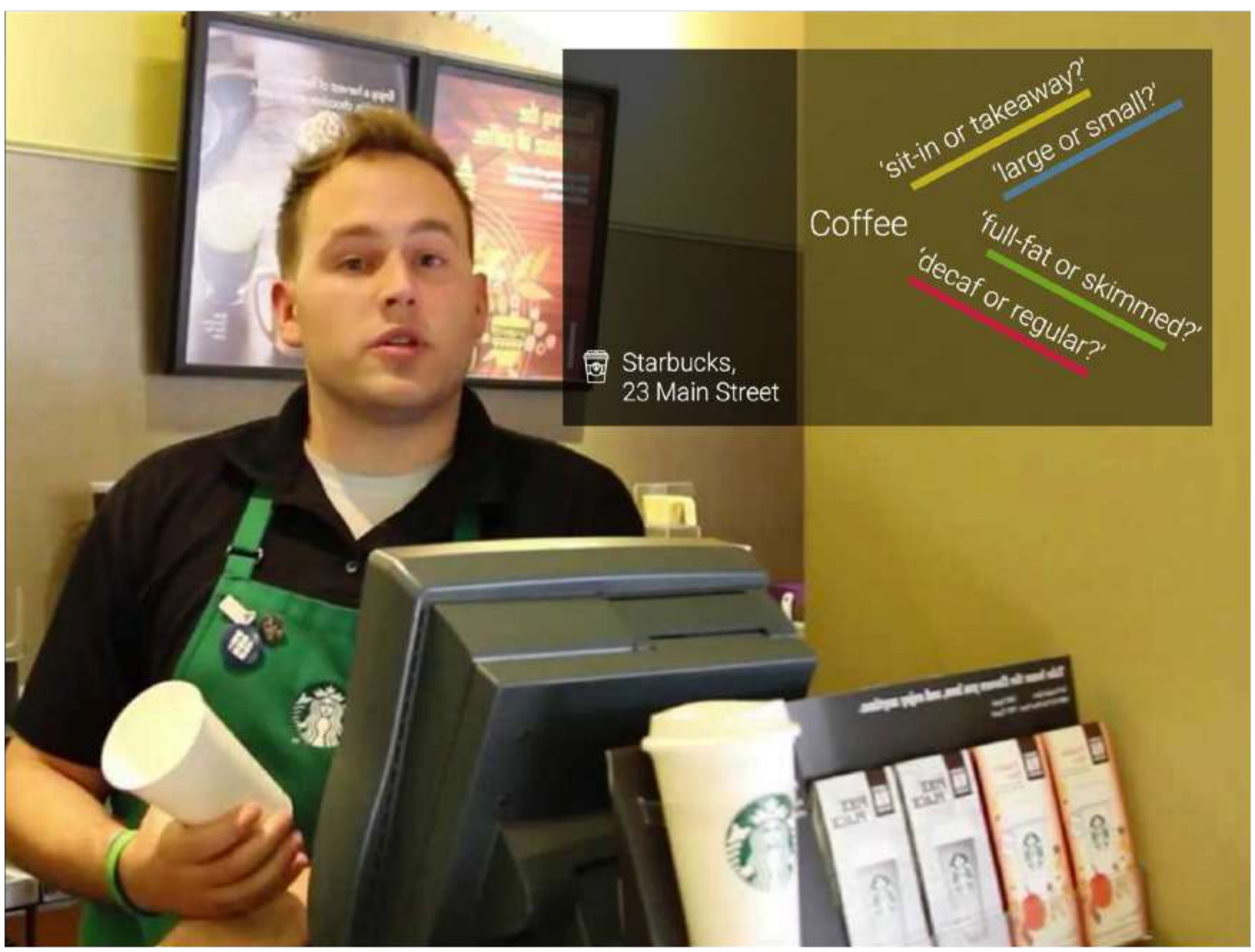

Figure 5.9: Mockup of ContextCueView showing synthetic conversation cues for a coffee shop interaction using Google Glass (`https://developers.google.com/glass/`).

| Amount of Information \ Skill Type | ANALYTIC | HYBRID | SYNTHETIC |
|---|---|---|---|
| HIGH | Cued Speech, Spectrograms | Subtitles, Signed Languages | **ContextCueView** |
| MED | **PhonemeViz** | LipreadingPractice, Lipreading.org | CME (C) |
| LOW | Upton, Ebrahimi, TactileSATs, CME(A) Lip Assistant, iBaldi | **MirrorMirror*** | CME (B) DAVID |

Figure 5.10: Placement of framework-Inspired SAT examples (in bold) into the framework. The '*' indicates the starting amount of information provided by this technique (MirrorMirror), but this level increases depending on familiarity with the content.

## 5.5 Discussion

### 5.5.1 Summary

In Chatper 4, using thematic analysis of interviews with speechreading tutors, I identified four main themes relevant to the future development of speechreading: *speechreading as a skill*, limited *access to speechreading*, a broad range of *teaching practices*, and mixed *attitudes to technology*. Using the themes, I developed a novel framework to help design new Speechreading Acquisition Tools (SATs). In evaluating the framework, I demonstrated that it can: 1) accommodate every teaching technique identified by speechreading tutors (during the interviews presented in Chapter 4), 2) accommodate existing Conversation Aids and existing SATs (discussed in 3), and 3) be used to design three new SATs: PhonemeViz, MirrorMirror and ContextCueView (where the first two are described in more detail in Chapters 8 and 9 respectively).

### 5.5.2 Limitations

First, I based the framework on data obtained exclusively from Scottish speechreading tutors, many of whom were trained on the same course – it is possible that these tutors use outdated techniques, reducing the value of the framework. However, Scotland is one of the few countries in the world that provides accredited speechreading tutor training – most other countries provide no formal training. As such, it is arguable that Scotland's training course is based on best practices.

Second, fitting a technology into the framework does not guarantee that it will be useful for speechreading acquisition. For example, spectrograms provide a high amount of analytic information (which should make them very helpful), but this information is difficult to utilise without considerable training [65]. Likewise, Upton's Eyeglasses and iBaldi provide a low amount of analytic information, but focus on speech production instead of appearance (one of the tutors – a Speech and Language Therapist – indicated that she *"...[switches] off [her] speech therapy brain because [she] would go with sounds but [indicated that] in lipreading it is very much the shape and presentation on the lips."*). The framework provides substantial guidance for the design of new SATs, but any resulting SATs still need to be evaluated using speechreaders. I argue that this is a strength of the framework not a limitation, as the framework serves to complement and enhance existing best-practice participatory design approaches, not attempt to replace them.

Third, the framework does not distinguish between SATs focussing on speechreading training versus assisting 'live' speechreading. However, as discussed, PhonemeViz and ContextCueView can be extended to 'live' speechreading, suggesting that some SATs can be adapted for both. Therefore, we leave this flexibility within the framework but will consider future refinements to make this distinction more explicit in the future.

### 5.5.3 Generalisations & Extensions

The framework currently focusses on English speechreading acquisition, but can be extended to support speechreading in other languages. French [16], German [31], Korean [41], and Japanese [72] each have their own confusing viseme-phoneme mappings, however their speechreading techniques for distinguishing between mouth-shapes (analytic), and conversational repair strategies (synthetic) are similar to English, so the framework should generalise to developing SATs for other languages.

The framework can also be extended to help develop technology for other skill-based speech domains. For example, speech therapy uses a variety of approaches [46], and already features a number of speech production aids [67, 117]. Likewise, the approach can be extended to language learning, as understanding a foreign language is analytic (e.g., pronunciation) and synthetic (e.g., using context to distinguish homonyms). In particular, ContextCueView (described above) might easily extend to supporting in-situ foreign language conversations.

## 5.6 Conclusion

From my thematic analysis of interviews presented in Chapter 4, I found that speechreading is a difficult skill to learn and that classes help facilitate learning. However, there is a lack of funding for classes which means there is limited longterm access. I also found tutors employ different teaching techniques and approaches, and use little technology when teaching, but are open to new technology.

These themes from the thematic analysis highlight how speechreading acquisition can be enhanced through the development of new assistive tools that will help resolve issues regarding the lack of access, and can also be specialised to different speechreading teaching approaches. I call these tools Speechreading Acquisition Tools (SATs) – a new type of technology designed specifically to improve speechreading acquisition. I believe that through the development and release of SATs, people with hearing loss will be able to augment their class-based learning, or learn on their own if they are unable to attend speechreading classes.

However, it is currently unclear for the research community, AT commercial sector, and AT enthusiasts how to design SATs influenced by contemporary speechreading classes. To facilitate the transition of knowledge from the thematic analysis to the research community, Assistive Technology (AT) commercial sector, and AT enthusiasts, in this chapter I introduced a *speechreading acquisition framework* that can be employed when designing SATs. An evaluation of the framework was then presented by using it to classify identified teaching techniques, criticise existing solutions, and to demonstrate how to use the framework to design three new SATs.

However, the framework was only designed using insight from speechreading tutors. To develop SATs that can improve speechreading students' ability to learn, practice and use speechreading, it was necessary to understand students' daily experience of using speechreading. In addition, it was also vital to investigate how students currently practice outside of classes, as the limitations and benefits of current techniques can inform the features of future SATs. Therefore, in the next chapter I will discuss findings from a postal questionnaire with 59 speechreading students, sourced from classes taught by tutors who took part in the interviews from Chapter 4. Some of the results of this questionnaire should help confirm if new SATs designed using the framework (such as PhonemeViz, MirrorMirror, and ContextCueView) will really fit the needs and requirements of people learning to speechread.

# 6

# Speechreading Student Questionnaire

## 6.1 Introduction

This chapter presents findings from a questionnaire completed by 59 speechreading students, sourced from classes taught by four of the speechreading tutors who took part in the interviews presented in Chapter 4. The motivation for conducting the questionnaire is given first. Following this is a description of the methodology and presentation of the findings. Finally, there is a discussion on the implication of the main findings on the development of new Speechreading Acquisition Tools.

## 6.2 Motivation

In Chapter 5 I used findings from the interviews with speechreading tutors presented in Chapter 4 to develop a framework that can be used to design Speechreading Acquisition Tools (SATs). During the evaluation of the framework, I categorised every speechreading teaching technique that tutors reported using in classes.

During the interviews, tutors also reported that they felt students practiced at home using these techniques. However I did not interview or ask students directly to determine how, or how often they practiced outside of class. In order to develop SATs that can improve speechreading students' ability to learn, practice and use speechreading, it was necessary to understand participants' daily experience of using speechreading. In addition, it was also vital to investigate how students currently practice outside of classes, as the limitations and benefits of current techniques can inform the features of future SATs.

To address this, I conducted a postal questionnaire with students from speechreading classes to explore the challenges and situations they encounter while speechreading, and their approach to practice outside of class.

# 6.3 Questionnaire

## 6.3.1 Aims

There were four questions framing the questionnaire:

1) Do speechreading students practice outside of class?
2) For those that do, how do speechreading students practice outside of class?
3) What technology do speechreading students use to practice outside of class?
4) What situations and challenges do students encounter when speechreading outside of class?

## 6.3.2 Questions

During the interviews presented in Chapter 4, tutors reported that they felt students practiced outside of class by using a mirror, watching television, using exercises from class, observing speakers during conversations and using websites such as *lipreadingpractice.co.uk*. I used these findings to design the questionnaire (shown in Appendix C.2).

The questionnaire included 25 questions across two sections. The first section contained nine questions that were used to gather the following demographic: age, sex, highest level of education, level of computer literacy, and details surrounding the participants' hearing – including the duration and cause of their hearing loss, along with their use of assistive technology.

The second section contained 16 questions and focused on participants' daily experience of speechreading: "Please rate your lipreading ability"[a], "How long have you been in lipreading classes?", "Do you practice lipreading outside of classes?", "If yes, how do you practice lipreading at home?", "Do you use mirror practice outside of class?" "If yes how often do you use mirror practice at home?”, “What do you like about mirror practice?”, “What do you dislike about mirror practice?”, “Do you use videos to practice lipreading outside of class?”, “How often do you use videos or watch television to practice lipreading outside of class with subtitles turned on?”, “How often do you use videos or watch television to practice lipreading outside of class with subtitles turned off?”, “In what situations do you find lipreading challenging? (Tick all that apply)”, “What do you find challenging when lipreading?”, “Do you rehearse/anticipate possible phrases or words that you may have to lipread before being in a situation?”, “If yes, describe how”, “Do you own a mobile device?”.

[a]In the UK, 'speechreading' is referred to as 'lipreading', therefore in discussions with participants I used the term 'lipreading'.

## 6.4 Method

I posted an information pack containing questionnaire forms, information sheets, envelopes and stamps to each speechreading tutor who had agreed to take part in the study (shown in Appendix C). Tutors were asked to distribute these to their students during or before class.

Potential participants had to be above the age of 18 and be currently enrolled in a speechreading class. Once students had completed the questionnaire they were asked to place it in a provided envelope that they could either post back to me directly, or hand to their tutor (who would place them in a larger envelope to be posted to me).

## 6.5 Participants

In total, 59 participants completed the questionnaire. Participants were sourced from four tutors, and were between 45 and 92 years old (M=73.9, SD=10.1) about 3/4 (76%) were female.

Participants reported on their highest level of education: University (27 participants), College (14), High School (13) and Other (5). Participants reported on their level of computer literacy: Excellent (3 participants), Good (17), Fair (29) and Poor (9).

### 6.5.1 Hearing Loss

#### Classification

All participants self reported having a hearing loss and were asked to classify their hearing loss using the textual descriptions of hearing loss identified by Action On Hearing Loss [3]: Mild, Moderate, Severe, and Profound Hearing Loss. Results are shown in Table 6.1.

| Hearing Loss Classification | No. of Participants |
|---|---|
| Profound | 5 |
| Severe | 19 |
| Moderate | 28 |
| Mild | 5 |

Table 6.1: How participants described their hearing loss, using the textual descriptions of hearing loss identified by Action On Hearing Loss [3].

## Cause of Hearing Loss

Participants were also asked to report the cause of their hearing loss. This was presented as checkboxes with an 'Other' field. The results of this question are summarised in Table 6.2.

| Cause | No. of Participants |
|---|---|
| Ageing | 34 |
| Congenital | 10 |
| Viral Infection | 9 |
| Exposure to loud noise | 9 |
| 'Other' – Unknown | 5 |
| 'Other' – Surgery Complication | 3 |
| Head Trauma | 3 |
| Disease | 3 |
| 'Other' – Acoustic Neuroma | 1 |
| 'Other' – Severe Shock | 1 |
| 'Other' – Medication side-effect | 1 |
| 'Other' – Tinnitus | 1 |

Table 6.2: Participants' reported causes of hearing loss.

## Duration of Hearing Loss

Participants were also asked to self-report how long they had a hearing loss. This was an open text field, that I later categorised into '0-5 Years', '5-10 Years', '10-15 Years','15-20 Years' and '20 Years +'. The results of this question are summarised in Table 6.3.

| Number of Years | No. of Participants |
|---|---|
| 0-5 Years | 8 |
| 5-10 Years | 10 |
| 10-15 Years | 5 |
| 15 - 20 Years | 10 |
| 20 Years + | 22 |
| Not Given | 4 |

Table 6.3: The length of time participants reported having a hearing loss.

### Assistive Technology

Finally participants were asked to report if they used any assistive technology. Overall 56 participants reported using hearing aids, one participant used cochlear implants and two participants stated they did not use any assistive technology. The number of participants who reported using a hearing aid does not match the current adoption rate reported by Action on Hearing Loss [2], but the reasons behind this are unclear.

## 6.6 Questionnaire Findings

### 6.6.1 Speechreading

Participants were asked to report on their level of speechreading: Excellent (0 participants), Good (19), Fair (28) and Poor (10). Participants were asked to report on their length of time in classes: Less than 6 months (4), 6-12 months (3), 1-2 years (12), 2-5 years (19), 5-10 years (3) and over 10 years (17).

### Challenges

Participants were asked to report challenges that affect their ability to speechread. This question was presented as checkboxes, with the options: ‘Words looking the same on the lips’, ‘People talking quickly’, ‘People covering their mouths’, ‘People turning away from you’, ‘Accents’, ‘Beards/Facial hair’, ‘Quiet Speakers’, ‘Concentration’, ‘Fatigue’ and an option for ‘Other’ (summarised in Table 6.4). ‘People turning away’ and ‘People covering their mouths’ were two of the most common challenges reported by participants as these pose direct problems to speechreading because you cannot see the face. ‘Words looking the same on the lips’ was reported as a challenge by 43 participants, and this is likely to be caused by the prescence of some visemes [54].

| **Challenge** | **No. of Participants** |
|---|---|
| People turning away | 54 |
| People talking quickly | 53 |
| People covering mouths | 49 |
| Quiet speakers | 44 |
| Words looking the same on the lips | 43 |
| Accents | 34 |
| Fatigue | 33 |
| Beards | 31 |
| Concentration | 25 |
| Other | 6 |

Table 6.4: Frequency of speechreading challenges reported by participants.

### 6.6.2 Situations

Participants were asked to report situations where they found speechreading difficult. This question was presented as checkboxes, with the options: 'Home', 'Dentist', 'Shopping', 'Coffee Shops', 'Transport (Bus/Taxi/Train/Plane)', 'Group Conversations', 'Doctors', 'Opticians', 'Restaurants', and an option for 'Other' (shown in Table 6.5). Group conversations were reported as the most challenging situation participants face. Restaurants and coffee shops were the next most reported as these locations often have a high amount of background noise that limits the use of residual hearing [91].

### 6.6.3 Open-ended Question Analysis

As the responses to the second set of questions were free text, I used thematic analysis [29] to understand participants' responses. However, because each question was largely independent I analysed each question separately and therefore did not produce any thematic maps.

| Situation | No. of Participants |
|---|---|
| Group Conversations | 54 |
| Restaurants | 46 |
| Coffee Shops | 32 |
| Transport | 33 |
| Dentist | 26 |
| Shopping | 23 |
| Doctors | 17 |
| Opticians | 16 |
| Home | 10 |
| General | 4 |
| Noisy Places | 3 |
| Work | 2 |
| Classes | 1 |

Table 6.5: Situations participants found speechreading to be challenging.

#### Step 1: Becoming familiar with the data

First, I read through the responses to each question to become familiar with the data set, and if necessary, split single responses into multiple rows.

#### Step 2: Generating and collating initial codes

Next, I read through all of the responses again, making a note of initial codes. The initial codes were generated using a data-driven approach and then collated and collapsed.

#### Step 3: Defining themes

Finally, I reviewed the coding of the dataset and identified patterns that could be grouped into themes within each question.

### 6.6.4 Practice

When asked if they practiced at home, 39 students reported that they did and 20 said they did not. If participants responded that they practiced at home they were asked to describe how they practiced. There were three themes within the data: Students reported they would practice through 1) Observation, 2) Watching Television, and 3) using Techniques From Classes.

### Observation

Observation refers to when participants would practice by watching faces, speakers or taking part in conversations. In total there were 34 mentions of observation taking place in many different situations. For instance, P34 reported that she practices speechreading by observing when shopping, on public transport and during social gatherings, whereas P17 reported observing the song leader during choir practice:

> P34: *"In every situation I find myself in, so in shops, on buses, in social gatherings."*

> P17: *"…using opportunities to practice in community choir, lipreading words from song leader."*

Furthermore, P50 reported practicing by taking part in question and answer sessions or by trying to follow the plot when attending the theatre:

> P50: *"By taking part in group conversations…e.g., question and answer sessions after talks or lectures. Also going to the theatre and trying to follow the plot!"*

### Watching Television

Participants reported using television to practice speechreading. In total, there were 13 mentions of using television to practice speechreading albeit with different factors. For instance, P5 and P4 reported practicing by simply watching the news:

> P4: *"Watching TV."*

> P5: *"News on TV sometimes."*

Whereas P1, P14 and P57 reported trying to reduce their reliance on their residual hearing by either turning off the sound or taking out their hearing aids:

> P1: *"TV without sound (but not as often as I should to be helpful or make a difference)."*

> P57: *"I take my hearing aids out sometimes when watching TV to see if I can lipread. Not really that successful."*

P14: *"TV (muted)"*

Participants also reported watching TV with subtitles to practice. Unfortunately, subtitles likely detract from speechreading practice, as one study found that participants with hearing loss spent around 84% of their viewing time focussed exclusively on subtitles [79]:

P2: *"I also try to watch programmes with subtitles."*

### Techniques From Classes

Finally, participants reported practicing using techniques from class such as fingerspelling or watching a DVD produced by the Association of Teachers of Lipreading to Adults [121]:

P40: *"Practice fingerspelling."*

P49: *"Listen to DVD '[look] hear'."*

## 6.6.5 Mirror Practice

Participants reported a varying frequency of using mirror practice as shown in Table 6.6.

| **Frequency** | **No. of Participants** |
|---|---|
| Daily | 0 |
| 2-3 times a week | 0 |
| Once a week | 4 |
| 1-2 times per month | 5 |
| 1-2 time per year | 4 |
| Never | 17 |
| Not Given | 29 |

Table 6.6: Participants' reported frequency of mirror practice.

### What do students like about Mirror Practice?

There were three themes describing what students liked about mirror practice. Mirror practice allows them to: 1) Learn lip-shapes, 2) Compare these lip-shapes with others, and 3) Perceive a lack of difference between certain words when spoken.

### Lipshapes

The most common part of mirror practice that students reported liking was that it helped them learn lipshapes. For instance, P34 described how mirror practice helps her notice small movements of the lips and this is due to the focussed analytic nature of mirror practice [131].

> P34: *"When we do this in class it shows very small, subtle movements of lips, tongue and teeth. Very interesting."*

P36 and P24 reported how mirror practice shows different shapes on their own lips with P24 also mentioning that it shows how their lips form shapes.

> P36: *"It helps to see the different shapes on the mouth."*

> P24: *"Seeing how good or bad my lips form shapes."*

### Compare with others

Participants also reported that looking at their own mouth shapes allows them to compare against others. P2 and P23 both reported how they can see the difference with others, likely from within their speechreading classes.

> P2: *"I see the difference between my movements [and] others."*

> P23: *"Seeing my speech pattern, sometimes different from others."*

### Lack of visual difference

Finally, participants also reported that mirror practice helps highlight how some words do not appear visually distinct. P44 and P20 both reported how some words are difficult to differentiate, likely due to the words in question being grouped under the same viseme.

> P44: *"It is [a] good way to [demonstrate] how few words can actually be seen on the lips."*

> P20: *"Seeing how some 'sounds' look the same."*

## What do students dislike about Mirror Practice?

There were four themes describing what students disliked about mirror practice. These themes were that they: 1) Dislike watching themselves, 2) That they have full knowledge of what they are saying, 3) That it was not akin to speechreading, and 4) That they over-emphasise words.

### Dislike Watching Self

The most commonly reported negative aspect of mirror practice was that participants did not like having to focus on their own appearance in the mirror:

> P24: *"Having to look at myself in a mirror."*

> P5: *"Don't like watching my own face."*

Additionally, P2 and P13 both disliked seeing the condition of their teeth and P2 mentioned hating seeing her wrinkles

> P13: *"Seeing the condition of my teeth."*

> P2: *"I hate seeing my teeth [and] wrinkles."*

### Full Knowledge

Participants also disliked that they know the words they are saying as they speak into the mirror. 'Knowing the answer' reduces the opportunity for formative learning to take place and does not represent regular speechreading. P14 reported they know what they are saying, with P5 saying that this is not helpful:

> P14: *"The fact that I know what I'm saying."*

> P5: *"Not helpful as my brain knows what I am saying so not really lipreading."*

### Not Akin To Speechreading

Participants reported that mirror practice is not similar enough to speechreading to be an effective form of practice. This is likely due to the problem of full knowledge reported above, plus a lack of naturalness:

> P3: *"Feel it is not quite [a] 'natural' situation."*

> P38: *"Not totally true to real life experience."*

This could be because mirror practice shows their own mouth shapes, with P20 saying that it does not help with understanding other people's lipshape (a core aspect of speechreading):

> P20: *"Helpful for me and my lip movement but not for seeing how others move their lips"*

**Over-Emphasis**

Finally participants reported that during mirror practice they would over-emphasise or exaggerate words when speaking into the mirror, making practice less useful, and less representative of everyday speech.

P32: *"Perhaps I over emphasise."*

P44: *"One tends to exaggerate too much."*

### 6.6.6 Video and TV Practice

To get a clearer picture of their use of subtitles to practice, I also asked participants to report the frequency they practiced with subtitles turned on and off as shown in Table 6.7. Responses to this question appear mixed, with some participants reporting never using Video/TV to practice speechreading, which may echo attitudes tutors reported towards practicing using television in Section 4.4.4.

| Frequency | Subtitles Off | Subtitles On |
|---|---|---|
| Daily | 8 | 9 |
| Once a Week | 5 | 4 |
| 2-3 Times a Week | 2 | 4 |
| 1-2 times per Month | 7 | 1 |
| 1-2 times per Year | 2 | 2 |
| Never | 20 | 19 |
| Not Given | 15 | 20 |

Table 6.7: Participants' frequency of practice with subtitles on and off.

### 6.6.7 Context Practice

The majority of participants (44/59) reported that they did not rehearse or anticipate phrases (14 Yes, 1 Not Given). If participants responded that they did rehearse or anticipate phrases, they were asked to describe how. There were two main themes, 1) Anticipating Questions and Answers, and 2) Researching Situation and Potential Topics.

**Anticipating Questions and Answers**

Participants reported that they would anticipate the questions they may be asked in situations, such as when in a restaurant or on public transport.

P24: *"Restaurants, waiters asking what you would like to drink or eat."*

P4: *"I try to think what answers I will get to the questions I intend to ask."*

P5: *"In restaurants anticipate soup choices / new soup choices."*

P24: *"Try to anticipate the bus driver's remarks."*

Participants would also constrain responses by asking specific questions. For instance when P21 asked the time of a train, they were aware that they would only have to speechread words about time.

P21: *"You know what you have said if asked, e.g., train time."*

Whereas, P6 would keep track of how much her shopping would cost, so that she would be aware of which numbers she would have to speechread on the cashier's lips.

P6: *"In shops have a rough idea how much the shopping will cost."*

**Researching Situation and Potential Topics**

Participants reported that they research or rehearse the situation and its potential topics before facing it:

P2: *"I rehearse time and dates when making appointments."*

P18: *"Find out as much information as I can about the situation I am about to face."*

P34 and P44 also reported that they try and think about the general topic of a conversation versus individual words or phrases.

P34: *"I think about what is likely to be asked. So I know the 'context' of the conversation. This preparation is helpful."*

P44: *"Knowing the subject matter being reported or the probable topics in social situations"*

### 6.6.8 Mobile Devices

Finally, participants reported owning a wide range of mobile devices as shown in Table 6.8. I found the results to be higher than expected for this demographic [108], but higher than-expected device use makes mobile devices a viable platform when considering new Speechreading Acquisition Tools.

| Device | No. of Participants |
|---|---|
| iPad | 20 |
| iPhone | 9 |
| Amazon Kindle | 14 |
| Amazon Kindle Fire | 2 |
| Tablet | 5 |
| Android Smartphone | 6 |
| Mobile phone | 21 |
| Windows Phone | 1 |
| Laptop | 2 |

Table 6.8: Participants' reported ownership of mobile devices.

## 6.7 Discussion

### 6.7.1 Summary of Findings

The findings from the questionnaire results report that over 50% of questionnaire participants have been in speechreading classes for over two years, yet less than a third (32%) of participants rated their speechreading ability as 'Good' and nobody rated it 'Excellent'. This surprising finding suggests that the development of new SATs could improve participants' ability to learn, practice and use speechreading.

The first aim of the questionnaire was to investigate whether speechreading students practiced outside of classes. During the interviews with speechreading tutors presented in Chapter 4, tutors reported that they think students practice outside of class. The questionnaire findings suggest that this is correct; 66% of participants reported that they practiced at home. However, tutors reported that they thought students practiced using techniques from classes, whereas students reported that they primarily practiced by observing speakers in daily life or on television.

The second aim of the questionnaire was to investigate further into how speechreading students practiced outside of class. Some of the participants reported a high frequency of watching television (with subtitles on and off) to practice speechreading. Together with the reported use of observation, it can be argued that these techniques are used to provide the speechreader with an endless supply of practice material. However, with television and observation, it is difficult for speechreaders to verify whether they are understanding the speaker correctly, limiting their feedback for learning.

During the interviews presented in Chapter 4, speechreading tutors reported that mirror practice plays a key role in speechreading training, and is also recommended by Action on Hearing Loss [4, 106] for practice, as it may develop visual cue integration skills needed during speechreading [7]. Participants reported that mirror practice allowed them to learn lipshapes, compare them with others, and show visual differences between words. However, they disliked watching themselves, that they have full knowledge of what they were saying, that they would over-emphasise words, and that mirror practice was not akin to genuine speechreading. This resulted in a low frequency (e.g., 28% of participants reported 'Never' using Mirror Practice and 49% did not respond at all) of usage by participants.

Finally, during the interviews tutors reported that they place a strong emphasis on synthetic based techniques during teaching, to teach students how to grasp the general idea behind what the speaker is saying. Only 14 participants reported that they practiced synthetic skills by rehearsing or anticipating phrases before facing certain situations. In general, participants reported that they would anticipate questions they may be asked or answers they would have to give or they would research the situation along with potential topics that may arise in order to improve their speechreading.

The third aim of the questionnaire was to investigate what technology students use to practice outside of class. Although tutors in Chapter 4 reported that students often use some of the SATs discussed in Chapter 3, none of the questionnaire participants reported using any of those SATs.

The fourth, and final, aim of the questionnaire was to investigate the situations and challenges students encounter when speechreading outside of class. Participants were asked to report situations where they found speechreading difficult. Group conversations were reported as the most challenging situation participants face. Restaurants and coffee shops were the next most reported as these locations often have a high amount of background noise that limits the use of residual hearing [91].

Questionnaire participants also reported that 'People turning away' and 'People covering their mouths' were two of the most common challenges they face when speechreading, as they cannot see the speaker's face. Both of these challenges can be addressed through informing the speaker that they need to see their face during conversations. Whereas the other challenges reported by participants such as ‘People talking quickly’ (53/59 participants), ‘Quiet speakers’ (44/59) and ‘Words looking the same on the lips’ (43/59) show the limitations of speechreading, and where the speechreader’s residual hearing fails to help.

These findings suggest that quality of speechreading practice outside of class is of limited value as: none of the participants reported using current SATs, TV and observation provide limited feedback and mirror practice provides too much feedback. Additionally participants reported a wide variety of challenges and situations they face whilst speechreading and these could be used to inform future development of new Speechreading Acquisition Tools (SATs); as new SATs could be specifically designed to help address these challenges. This is supported by the higher than-expected mobile device use by participants, as it suggests that mobile devices are a viable platform for new SATs.

### 6.7.2 Limitations

A limitation of the data presented in this chapter, is that it was obtained exclusively from students within Scottish speechreading classes, which are taught by tutors trained on the same course. It is therefore possible that these tutors promote outdated practice techniques. However, Scotland is one of the few countries in the world that actually provides accredited speechreading tutor training [10] – most other countries provide no formal training for speechreading tutors. It is arguable that Scotland is therefore at the forefront of speechreading training, suggesting that the tutors are utilising up to date training methods.

A second limitation is that the participants who completed the survey all had a high level of education and were computer literate. It is possible that this is not a representative demographic of those who have hearing loss and would attend speechreading classes around the world.

## 6.8 Conclusion

During the interviews presented in Chapter 4, speechreading tutors reported that they felt students practiced at home, however I did not interview or ask students directly to determine how or how often they practiced outside of class. In order to develop SATs that can improve speechreading students’ ability to learn, practice and use speechreading, it was necessary to understand how they currently practice. To address this, in this chapter I presented analysis of data from a postal questionnaire conducted with 59 students from speechreading classes to explore the challenges and situations they encounter while speechreading, and their approach to practice outside of class. The findings suggest that students are not currently well supported for practicing outside of class, and that even though 67% of participants had been in speechreading classes for over two years they still felt that their speechreading ability could be improved. In the following two chapters, I will use the findings of this chapter to influence the design of two SATs; one that focuses on improving speechreading by reducing the difficulty of disambiguating phonemes within the same viseme class, and one that aims to improve speechreading practice outside of class.

# 7

# PhonemeViz

## 7.1 Introduction

This chapter presents a speechreading Acquisition Tool (SAT) called *PhonemeViz*, which has been designed using the framework presented in Chapter 5. PhonemeViz is a visualisation that is positioned at the side of a speaker's face, beginning at the forehead and ending at the chin and presents textual representations of consonant phonemes in a semi-circular arrangement (as shown in Figure 7.1), with an arrow beginning from the centre of this semi-circle pointing at the initial consonant phoneme of a word to provide persistence. This design is intended to enable a speechreader to focus on the speaker's eyes and lip movements (as in traditional speechreading), while also monitoring changes in PhonemeViz's state using their peripheral vision to help disambiguate confusing visemes. The design of PhonemeViz was inspired by the initial-letter fingerspelling technique that was highlighted by four speechreading tutors during the interviews presented in Chapter 4. PhonemeViz's design was further informed by the challenges reported by participants of the student questionnaire presented in Chapter 6. The motivation and rationale behind the design of PhonemeViz is given first, followed by an evaluation process.

## 7.2 Motivation

The findings from the speechreading student questionnaire presented in Chapter 6 report that 66% of participants have been in speechreading classes for over two years, yet less than a third (32%) of participants rated their speechreading ability as 'Good' and nobody rated it 'Excellent', suggesting that even with extensive training, sometimes speechreading does not provide enough information.

Questionnaire participants also reported that 'People turning away' and 'People covering their mouths' were two of the most common challenges they face when speechreading, as they cannot see the speaker's face. Both of these challenges can be addressed through informing the speaker that they need to see their face during conversations. Whereas the other challenges reported by participants such as 'People talking quickly' (53/59 participants), 'Quiet speakers' (44/59) and 'Words looking the same on the lips' (43/59) show the limitations of speechreading, and where the speechreader's residual hearing fails to help. All three of these challenges are likely to be caused by visemes [54].

During the interviews presented in Chapter 4, four of the tutors highlighted the value of initial-letter fingerspelling during teaching as it helps to disambiguate words that are visually similar on the lips (due to being classed under the same viseme category). Even though initial-letter fingerspelling is a powerful teaching technique, for successful use outside of classes it requires the speaker to know how to fingerspell.

Recently, Augmented Reality smart glasses such as the Google Glass[a], Microsoft Hololens[b], and the Epson Moverio Glasses[c] have become popular. These smart glasses project light over the user's vision and allow users to see the real world with their natural vision [143], but also allow for the incorporation of virtual elements into the physical world. It is possible that in the near future this technology could be further miniaturised into smart eyeglasses or even smart contact lenses [119, 141]. Using such devices it may be possible for us to be able to augment the speechreading process with additional information akin to fingerspelling. This would allow for decreasing some of the speechreading challenges discussed above and also results in no additional training for someone speaking with a person who is speechreading.

The concept of providing supplementary phoneme-based information to speechreaders could be borrowed from fingerspelling to inspire the design of a new SAT that could provide this type of **medium information**. To address this, I used the framework introduced in Chapter 5 to develop a new SAT called *PhonemeViz*, which displays a subset of a speaker's phonemes to the speechreader. PhonemeViz focuses on reducing viseme confusion that occurs at the start of words in a similar manner to initial-letter fingerspelling. PhonemeViz places consonant phonemes in a semi-circular arrangement (as shown in Figure 7.1), with an arrow beginning from the centre of this semi-circle pointing at the last heard consonant phoneme to provide persistence.

[a] `https://www.x.company/glass/`
[b] `https://www.microsoft.com/en-gb/hololens`
[c] `https://epson.com/moverio-augmented-reality`

PhonemeViz is positioned at the side of a speaker's face, beginning at the forehead and ending at the chin (as shown in Figure 7.2). By looking at the visualisation in combination with their ability to speechread, speechreaders should be able to attend to the speaker's face while being able to disambiguate confusing viseme-to-phoneme mappings, therefore improving understanding during conversation. Currently, PhonemeViz is at the visualisation evaluation phase, where the end goal would be to display the visualisation on a transparent head mounted display, such as the Epson Moverio glasses or the Microsoft Hololens, as a visual augmentation of speech during typical conversations.

## 7.3 PhonemeViz Design

### 7.3.1 Design Requirements

Given the characteristics of the existing tools described in Chapter 3, I outline three key requirements necessary to design an SAT that visualises effective phoneme information to support speechreading.

**1) Persistence:** Spectrograms [65] demonstrate that persistence is important, as it allows speechreaders to 'catch up' on what has been said. In contrast, iBaldi has no persistence in its visualisation – the three coloured discs disappear as soon as the phoneme finishes. This may contribute to iBaldi's poor performance [97].

**2) Short Training:** Many of the previously discussed tools require a large amount of training in order to be effective (e.g., spectrograms [65], cued speech [45]). Any reduction in training time will benefit speechreaders as they learn to use a new tool, as well as encourage adoption. As a result, visualisations to support speechreading should have carefully-designed visual encodings that are as intuitive as possible. Arbitrary colour-coding can be difficult to learn [137] and confusing for people with impaired colour vision [56], so these should be avoided. Low training requirements should manifest as participants achieving high accuracy after relatively short exposure to the new technique. Coupled with this is the perceived workload for a technique – lower workloads will allow participants to use a technique for longer. During the interviews with speechreading tutors presented in Chapter 4, tutors reported that speechreading requires a lot of focus and concentration, suggesting that it is a high work load task. Participants of the postal questionnaire presented in Chapter 6 also support this by reporting that concentration and fatigue were among the challenges they faced when speechreading. Therefore it is clear that an effective visualisation should have a low work load and a low amount of training in order for a user to become competent quickly.

**3) Peripheral Position:** It is also important that the visualisation technique be positioned such that it does not obstruct the user's view of the facial features of the speaker. Being able to see the speakers mouth and eyes is a natural component of speechreading [88]. Having the visualisation on the periphery or off to the side of the speaker's face would ensure this. Placing the visualisation off to the side could result in the background colours from the surroundings of the speaker interfering with the colours used in the visualisation [58]. Solutions to this have been proposed (e.g., [125]), but to ensure maximum visibility, any new visualisation should use black and white. Furthermore, peripheral positions would allow the speech reader to more naturally look around with their eyes knowing that they have some time to look back at the visualisation, as their vision does not have to stray too far from the speaker's lips and eyes.

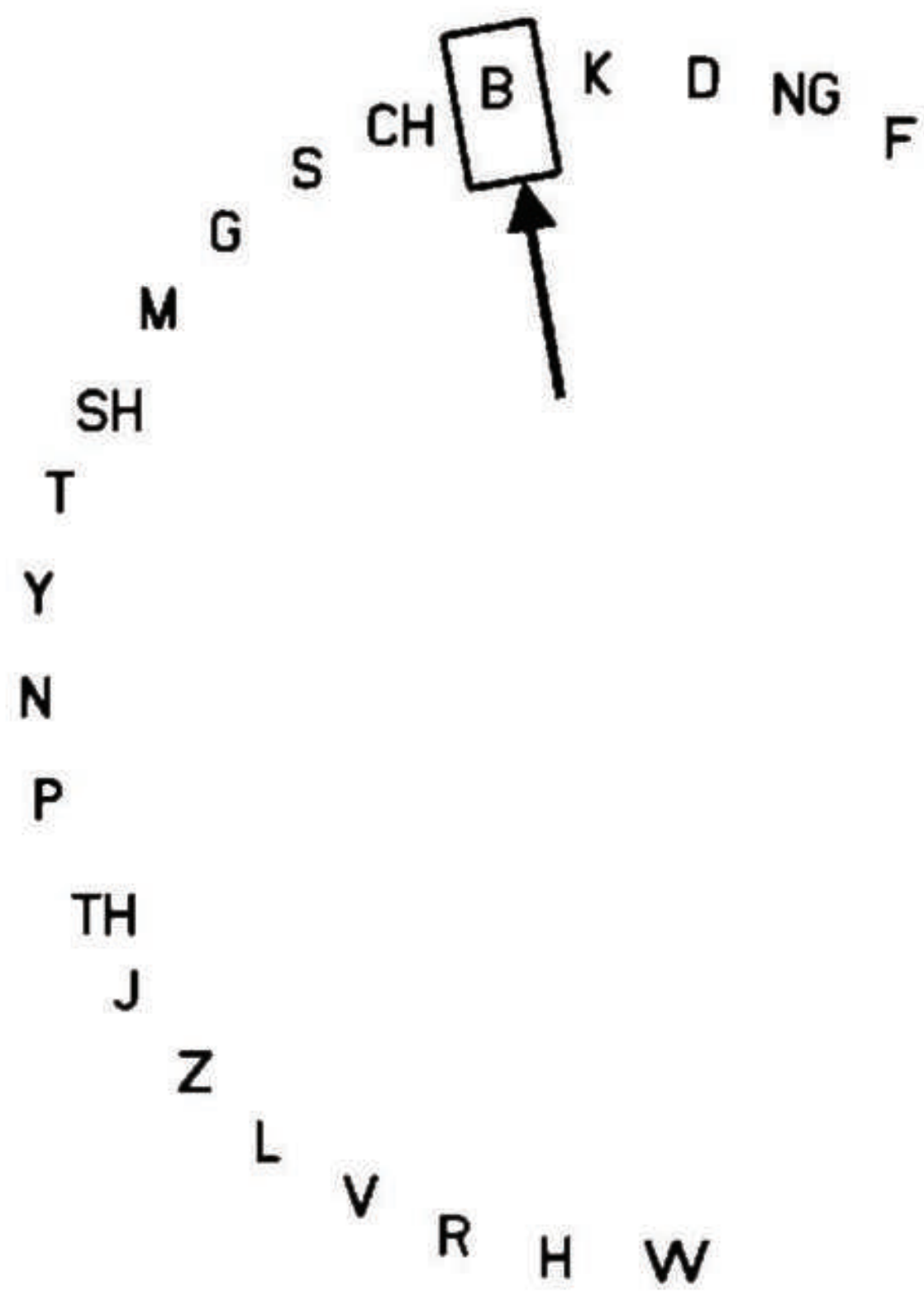

Figure 7.1: PhonemeViz: A simplified set of consonant phonemes is displayed in a semi-circular arrangement. When the system has detected the initial consonant phoneme, the arrow points to that phoneme's position. Phonemes within the same viseme group are dispersed around the semi-circle, and arranged top to bottom in alphabetic order.

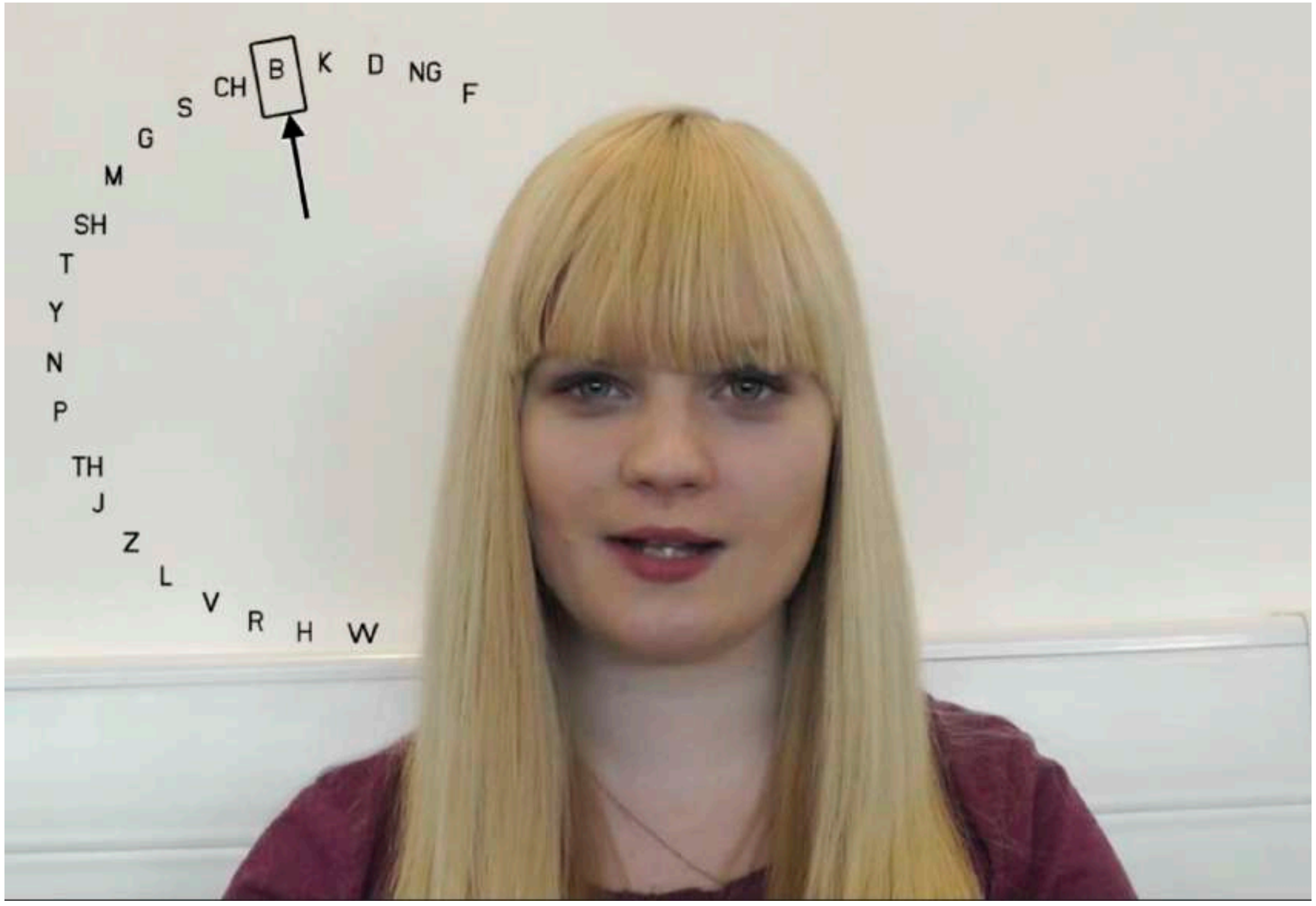

Figure 7.2: PhonemeViz shown for the word ‘Bat’.

### 7.3.2 Prototype Design

The initial prototype visualisation of PhonemeViz focuses on reducing viseme confusion when it occurs at the start of words. PhonemeViz places consonant phonemes in a semi-circular arrangement (as shown in Figure 7.1), with an arrow beginning from the centre of this semi-circle pointing at the last spoken initial consonant phoneme to provide persistence. PhonemeViz was designed to be positioned at the side of a speaker’s face, beginning at the forehead and ending at the chin (as shown in Figure 7.2). This should allow the speechreader to focus on the speaker’s eyes and lip movements while monitoring changes in PhonemeViz’s arrow using his/her peripheral vision.

The International Phonetic Alphabet (IPA) is presented using a special notation that could prove difficult to learn. To reduce this, I started with the ARPABET written form for each of the 48 phonemes [90]. I then removed all of the vowel phonemes (as vowels ‘a, e, i, o and u’ make very distinct ’open mouth’ shapes, which are easier to lipread [77, 93]) leaving 28 remaining consonant phonemes. I reduced this to a final set of 22 phonemes by simplifying similar phonemes into one representation by condensing DX, DH and TH to *TH*; WH and W to *W*; ZH and SH to *SH*; and NX, EN and NG to *NG*.

To decide letter position, I used the viseme categories shown in Table 7.1 to identify locations within the semi-circle that were spatially distributed for each phoneme within a single viseme category. To facilitate learning each phoneme representation within a viseme is alphabetically-ordered from the top of the semi-circle to the bottom to allow users to quickly learn the ordering of phonemes for a particular viseme, and hence reduce their focus on the visualisation.

When designing the visualisation, I initially thought that displaying the last three phonemes spoken would be useful, however in pilot-tests I found that it diverted attention away from the speaker's lips, which would be detrimental to speechreading. Therefore, PhonemeViz shows only the last spoken initial consonant phoneme. PhonemeViz has less persistence than captions, however it does allow more persistence than typical speechreading, as the initial phoneme of the current word is visible until the next word is spoken.

Finally, many of the visualisations discussed previously (e.g, spectrograms [65], iBaldi [97]) also rely on colour information. However, for the visualisation to be usable by as many users as possible colours are not the best approach, as users with colour vision deficiency may not be able to discriminate the subtle differences [43, 58]. Although there are suggestions on how to overcome this problem [56], for the purposes of this SAT, I chose to give the visualisation a self-contained contrast and used black text with a white outline on the arrow for maximum distinguishability for all users.

| Phoneme | Viseme | Phoneme | Viseme |
|---|---|---|---|
| P (p) | /p/ | K (k) | /k/ |
| B (b) | | G (g) | |
| M (m) | | N (n) | |
| EM (m̩) | | L (l) | |
| F (f) | /f/ | NX (r̩) | |
| V (v) | | HH (h) | |
| T (t) | /t/ | Y (y) | |
| D (d) | | EL (l̩) | |
| S (s) | | EN (n̩) | |
| Z (z) | | NG (ŋ) | |
| TH (θ) | | IH (ɪ) | /iy/ |
| DH (ð) | | IY (i) | |
| DX (ɾ) | | AH (ʌ) | /ah/ |
| W (w) | /w/ | AX (ə) | |
| WH (w) | | AY (aɪ) | |
| R (r) | | ER (ɝ) | /er/ |
| CH (tʃ) | /ch/ | AO (ɔ) | |
| JH (dʒ) | | OY (ɔɪ) | |
| SH (ʃ) | | IX (ɨ) | |
| ZH (ʒ) | | OW (oʊ) | |
| EH (ɛ) | /ey/ | UH (ʊ) | /uh/ |
| EY (eɪ) | | UW (u) | |
| AE (æ) | | AA (ɑ) | /aa/ |
| AW (aʊ) | | SIL & SP | /sp/ |

Table 7.1: Typical 48 ARPABET phonemes (with their IPA symbol – `https://www.internationalphoneticassociation.org/`) used in the English language including silence (SIL) and short pause (SP), grouped into their 14 viseme classes adapted from [90].

## 7.4 Prototype Evaluation

I evaluated PhonemeViz in-lab against five other visualisation techniques (Spectrogram, VocSyl [66], Captions, Lip-Magnification [140], and iBaldi [97]), as well as a no visualisation control (None). The main task was for participants to use each technique to help them identify a target word within a group of words that all begin with phonemes from the same viseme group. Video stimuli employing each techniques were presented without audio to control for differences in participants' hearing ability as used in previous studies [13, 18]. Participants were told to press the spacebar each time they thought the speaker had said the target word. The order of each technique and word group were counterbalanced.

### 7.4.1 Procedure, Apparatus & Design

This study used a repeated measures design and all participants attended two study sessions over consecutive days. The time between sessions allowed sufficient rest for the participants, yet was kept to a minimum to ensure sustained familiarity.

During the first session, participants filled out the demographic questionnaire, then completed a Speechreading Proficiency Test (discussed in Section 7.4.5). The participants then completed the experiment trials, in which participants had to use each technique to identify the target word within a word group (discussed in Section 7.4.3). The order of using each technique was counterbalanced using a Williams Balanced Latin Square, as was which word group the user received per technique. Before the experiment trials for each technique, participants would be shown how the technique they were about to use worked. They were trained using the words 'fan' and 'van', which were not used in the experimental trials. After training, participants completed 36 trials in which they had to identify the target word by pressing the spacebar. The 36 trials comprised four occurrences for each individual video (3 words x 3 variations = 9 individual videos). The first nine trials were a randomised ordering of all nine individual videos for a technique. The remaining 27 trials were then presented in random order. This procedure was the same for all six techniques and the control condition. Participants were given a target word for a given technique, and were instructed to press the space bar whenever they thought that word was spoken by the speaker. Participants were instructed that the target word would occur in one third of the trials.

During the second session, there was no need for participants to complete the demographic or SPT again, so went straight into experiment trials, but were given a different counterbalanced order compared to the first session. Before each evaluation, participants were again provided with the same training as the first session to refresh their memory. After finishing each technique, the participants completed a NASA Task Load Index (NASA-TLX) [69]; which is a subjective, multidimensional assessment tool that rates perceived workload in order to assess a system's perceived performance. At the end of the session, participants completed a closing questionnaire where they ranked each technique (including None) in order of preference, and provided reasons for their ordering.

### 7.4.2 Techniques

Each technique was implemented as an openFrameworks[d] application and made use of an open source add-on called ofxTimeline (`github.com/YCAMInterlab/ofxTimeline`). This add-on allowed me to add time-based triggers to each video. Using a list of phonemes and their features, I loaded each word video into an application for each technique, overlaid the technique's visualisation onto the image sequence and exported the video of the composited technique using an additional openFrameworks add-on called ofxVideoRecorder (`github.com/timscaffidi/ofxVideoRecorder`). There was no such modification made to the original videos for the no visualisation control condition.

**PhonemeViz:** PhonemeViz was implemented as outlined in Section 7.3.2 and was presented at the left hand side of the speakers' lips. The font used was Helvetica at font size 15pt.

**Spectrogram:** The spectrogram was drawn from the right side to left side of the screen and displayed a rainbow colour palette (as shown in Figure 7.3), as this is the standard or default colour chart used on many freely available spectrograms. The implementation was adapted from open source code (`github.com/Venetian/ofxSpectrogramAudioInput`) and used AccelerateFFT (native library on OSX/iOS) to perform the Fourier transform (with FFT window size of 512 and hop size of 256). The resulting spectrogram's range was from 20Hz to 20KHz.

**VocSyl:** The implementation of VocSyl followed the description outlined by Hailpern et al. [66], and used the colour palette introduced by Pietrowicze et al. [111] (as shown in Figure 7.4, left), I did not implement differences in pitch as this does not alter a word phonetically. The phoneme circles indicating volume (as can be seen in Figure 7.5, first row right) were opaque and the outlined envelope shapes were partially transparent (alpha value = 100). The enveloped shapes

[d] openFrameworks (http://openframeworks.cc/) is an open source toolkit designed for "creative coding", written in C++ and built on top of OpenGL.

are outlined with a white border in order to increase contrast between the background of the video.

**Captions:** Captions were presented as white text on a black bar at the bottom of the screen following Ofcom's guidelines [107]. The text was centred, set in Helvetica at font size 25pt with a line height of 34pt and letter spacing of 1pt. As only single words were employed, line breaks were not required.

**Lip-Magnification:** Lip-Magnification is a modified version of the technique introduced by Xie et al. [140]; instead of animating a set of computer generated lips, a magnification of the speakers's actual lips was presented at the right-hand side of the speaker's face. The magnified image is double the size of the original.

**iBaldi:** The implementation of iBaldi's cues was created followed the description and colour palette colour palette introduced by Massaro et al. [97] (as shown in Figure 7.4, right).

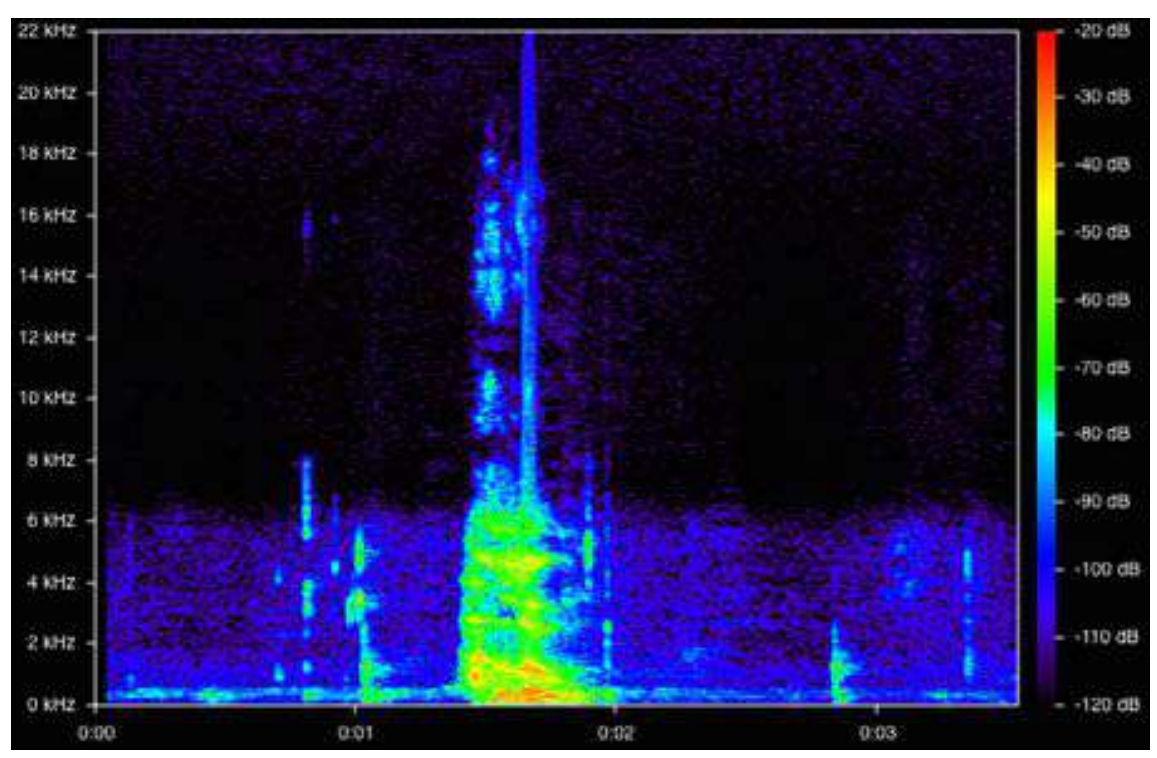

Figure 7.3: Spectrogram for "Hello", produced using Spek (`htpp://spek.cc`) showing rainbow colour scheme.

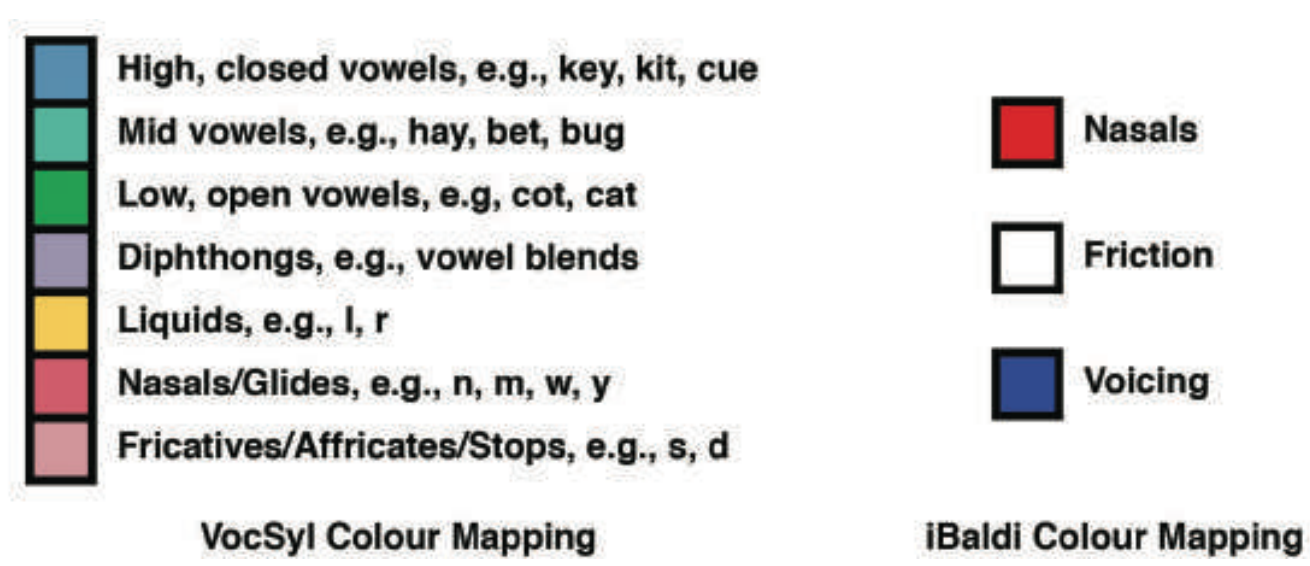

Figure 7.4: Left: Phonetic colour mapping for VocSyl implementation. Right: Colour mapping for iBaldi implementation.

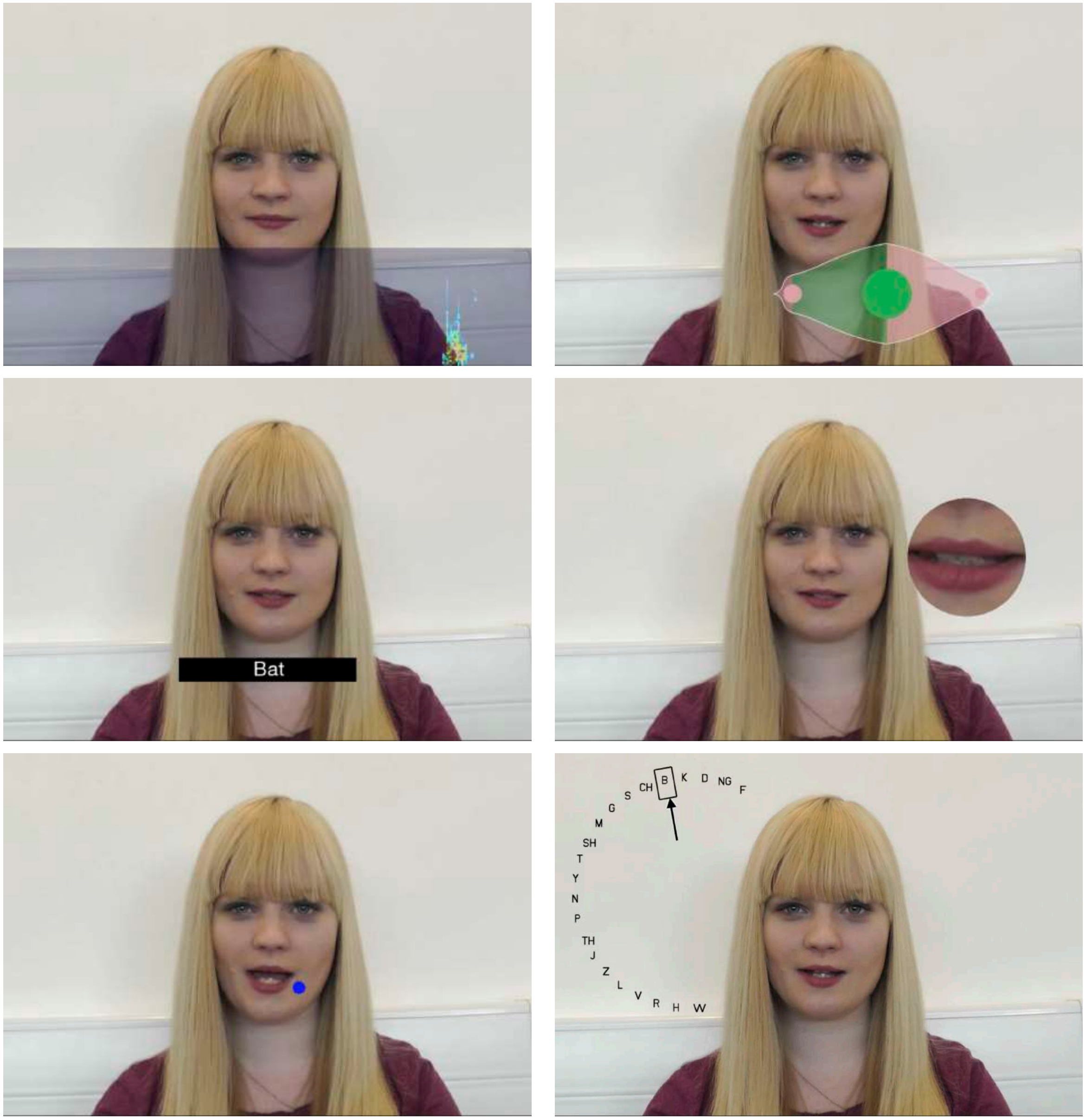

Figure 7.5: Each visualisation shown for the word ‘Bat’, First Row: Spectrogram (left) and VocSyl (right), Second Row: Captions (left) and Lip Magnification (right), Third Row: iBaldi (left) and PhonemeViz (right).

### 7.4.3 Viseme groups, Word List & Task

The words for the evaluation were chosen by looking at the phoneme to viseme table by Lucey et al. [90]. Three words were chosen for each viseme group that were similar apart from the initial consonant phoneme. In total the evaluation used four groups /p/,/t/,/k/,/ch/ with three (/p/, /t/ and /k/) being repeated, albeit with different words. A native Scottish female speaker was recorded (in 1080p) from the shoulders up and 1.5m away. The speaker was recorded saying each word from Table 7.2 three times. This ensured subtle variations in speech, which would be the case in day-to-day conversation, as well as reducing how familiar participants would become with each video. For each technique, participants watched a sequence of these videos (without audio) in which the speaker uttered words from one group. The participant was told to press the spacebar when they identified the speaker saying the target word (indicated by an * in Table 7.2).

| Viseme | Word 1 | Word 2 | Word 3 |
|---|---|---|---|
| /p/ | /P/æ/t/<br>**Pat** | /M/æ/t/<br>**Mat** | /B/æ/t/<br>**Bat*** |
| /t/ | /S/ʌ/n/<br>**Sun** | /D/ʌ/n/<br>**Done*** | /T/ʌ/n/<br>**Tonne** |
| /k/ | /K/ɪ/l/<br>**Kill** | /G/ɪ/l/<br>**Gill*** | /N/ɪ/l/<br>**Nil** |
| /ch/ | /t/ʃɪ/l/<br>**Chill** | /ʃ/ɪ/l/<br>**Shill** | /dʒ/ɪ/l/<br>**Jill*** |
| /p/ | /B/æ/n/d/<br>**Banned** | /M/æ/n/d/<br>**Manned** | /P/æ/n/d/<br>**Panned*** |
| /k/ | /L/ʌɪ/t/<br>**Light** | /N/ʌɪ/t/<br>**Night** | /K/ʌɪ/t/<br>**Kite*** |
| /t/ | /Z/əʊ/n/<br>**Zone*** | /T/əʊ/n/<br>**Tone** | /S/əʊ/n/<br>**Sewn** |

Table 7.2: Experimental words and their pronunciation, with the corresponding viseme group in the first column. Target words identified with (*).

### 7.4.4 Participants

I recruited 14 participants from a local university (mean = 32.71 years, SD = 14.44, male = 11). Participants were compensated for their time with a £5 Amazon gift voucher. Participants were over 18 years-old and had good to excellent typical or corrected vision.

#### Demographic Questionnaire

The questionnaire included 13 questions across two sections (shown in Appendix D.3). The first section contained six questions that were used to gather basic demographic information; age, sex, corrected visual acuity, handedness, highest level of education, level of computer literacy.

The second section contained seven questions to gauge participants' experience of speechreading:

1) "When a person's mouth is not visible (e.g., it is covered or they turn their back to you) do you find yourself thinking: 'If I could see their lips it would help me to understand what they are saying'?"
2) "When a person lowers their voice (e.g., to whisper) do you find that you look at their lips more?"
3) "When somebody is across the room talking (e.g., giving a presentation) do you find that you want to look at their mouth more to determine what they are saying?"
4) "When talking over the phone do you feel that any confusion over certain words or letters being spoke (e.g. somebody telling you their postcode) would be avoided if you could see the other person's lips?"
5) "When talking to a person behind a screen (e.g., somebody at the bank or a bus driver) do you look more at their lips to help you understand what has been said?"
6) "When talking to a person behind a screen (e.g., somebody at the bank or a bus driver) do you look more at their lips to help you understand what has been said?"
7) "How often do you rely on lipreading in your daily life?"

Participants reported their reliance on speechreading in daily life (Q7) using a scale where 1 indicated never and 7 almost always. Six participants chose 5, two participants indicated 3, three participants reported 2 and three participants chose 1. The mean rating was 3.21 (SD = 1.71). Participants then responded 'yes' or 'no' to questions 1-6 that asked them if they resorted to Speechreading in certain situations (Q2, Q3, Q5) or if they felt a better understanding would be reached if the other persons lips were visible (Q1, Q4, Q6). The percentage of yes responses are as follows: Q1 - 50%, Q2 - 35.71%, Q3 - 57.14%, Q4 - 71.43%, Q5 - 42.86%, Q6 - 64.29%.

One participant answered ‘no’ to all of the questions (the same participant who responded with 1 to the initial question). All other participants answered with at least one ‘yes’, while two participants answered ‘yes’ to all of the questions.

### 7.4.5 Speechreading Proficiency Test (SPT)

To evaluate participants’ speechreading ability I required a speechreading proficiency test. However, no test has yet been generally accepted for use in the UK [102]. Although there are many existing speechreading tests (outlined in Appendix A of Mohammed’s Thesis [102]), they were not suitable for this evaluation for four main reasons. First, the majority of the speechreading tests in the literature do not provide access to the full training material because the material was either originally supplied on a format that is now out of date (e.g., 8mm film, video laserdisc, or performed live) or the material was never publicly available [38]. Second, each test uses a different type of material ranging from phonemes and ‘nonsense syllables’ to sentences, questions, or passages from books [38]. In this evaluation, I wanted to assess each visualisation’s ability to help participants determine words within a viseme group, something that few tests are focused on evaluating. Third, within the material there can be use of location specific language (e.g., USA, Denmark, Australia [38]) that could influence results because participants may be unfamiliar with certain terms and colloquialisms. Finally, the speakers in the material often had region specific dialects, which could also influence the results.

Therefore, I assessed participants’ differences in speechreading proficiency using a custom-built Speechreading Proficiency Test (SPT). My SPT consisted of 40 words chosen from a list of 14,735 words that were taken from demographic spoken material (e.g., conversations). The list was part of the larger British National Corpus (BNC) and was downloaded from (`http://www.kilgarriff.co.uk/bnc-readme.html`). Ten words were selected from each quarter of the list, which was ordered by frequency of occurrence in speech, thus the first group of words were more common than the fourth group of words. In addition to frequency, the number of syllables was also considered (e.g., all words in the first group had one syllable).

To create the video material for the SPT, a second native Scottish female speaker sat 1.5m away from the camera and was recorded (in 1080p) from the shoulders up saying each word (as shown in Figure 7.6). Afterwords the audio was removed. I identified that trying to speech-read individual words without any conversational context was too challenging. To overcome this, participants were provided with a sheet with four groups of ten words, randomly ordered (as shown in Appendix D.5). Participants viewed each word twice and were in control of moving to the next word. This provided participants with enough time to write down the video number next

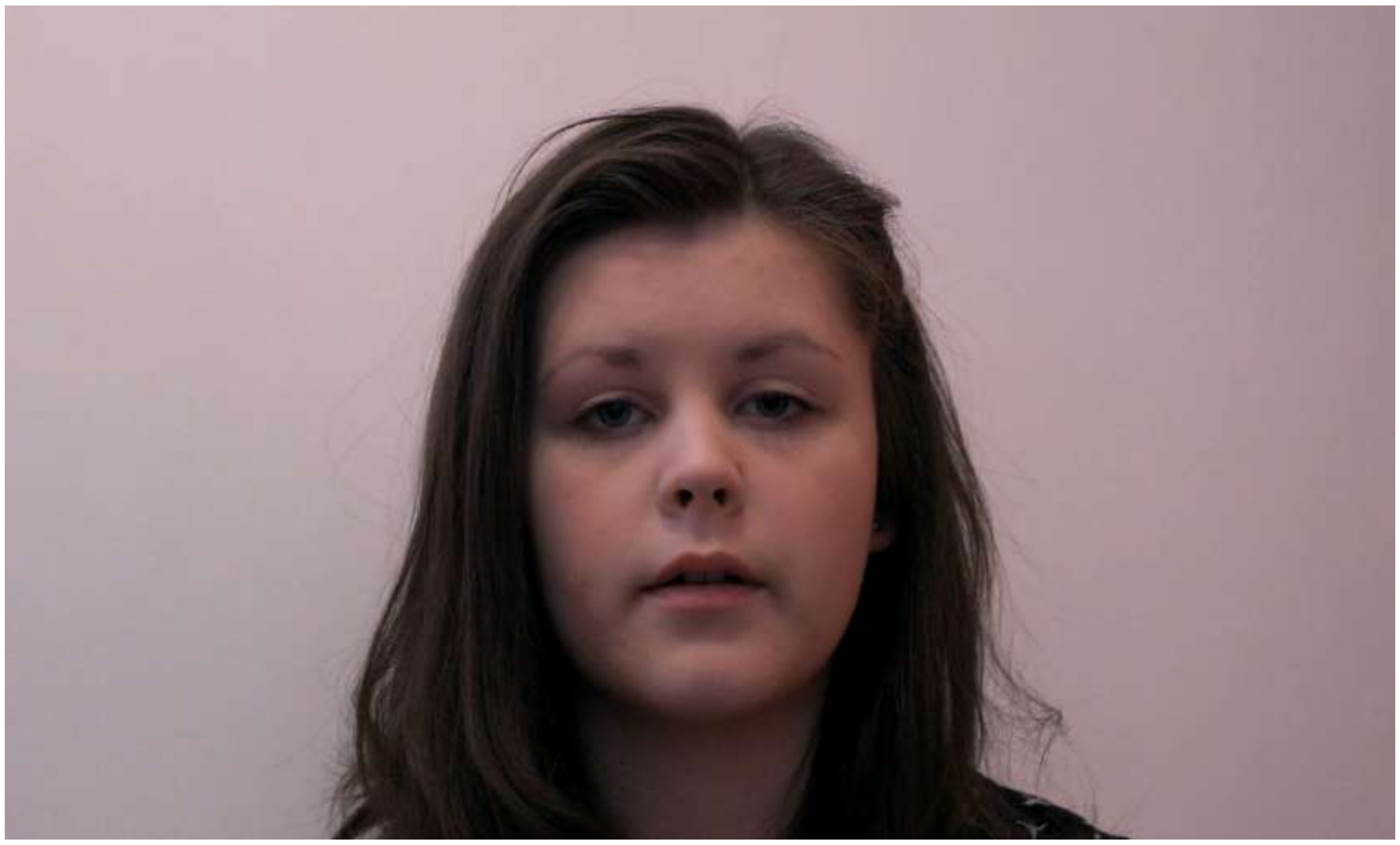

Figure 7.6: Example of the video material used for the Speechreading Proficiency Test.

to the word on the sheet they thought was being spoken.

### Participants' Speechreading

The SPT was marked out of 40 (one point was awarded for each correct word) and I calculated a percentage score for each participant. The minimum SPT score was 15% correct and the maximum SPT score was 100% correct (mean SPT score = 74.29, SD = 22.78). These results indicate that I had a diverse set of participants in terms of their experience with speechreading.

## 7.4.6 Results

### Session x Technique

For each trial, participants' responses were recorded (whether they did or did not hit the spacebar) as well as what the appropriate response should have been (they should or should not have hit the spacebar). The first 9 trials were excluded from analysis as these were considered additional training. I then calculated the $F_1$ Scores for each condition per participant using their precision and recall values calculated from their responses. The $F_1$ score was the dependent variable, while the independent variables were the session and technique used. A 2x7 RM-ANOVA was carried out to investigate whether there was any interaction between session and technique used.

I expected to see a positive learning effect between session one and session two. The assumption of sphericity was violated ($p < .05$) for both effects (session and technique) and the interaction between session and technique, however this was likely due to PhonemeViz having no variance (no participants made any errors with PhonemeViz), so I continued with the originally-planned tests.

There was no significant main effect of session, $F(1.00, 13.00) = .88$, $p = .37$, $\eta_p^2 = .06$, however, there was a significant main effect of technique, $F(2.86, 37.13) = 37.57$, $p < .001$, $\eta_p^2 = .74$. There was no significant interaction effect between session and technique used, $F(3.32, 43.10) = 1.73$, $p = .17$, $\eta_p^2 = .12$.

There is a confirmed main effect for technique, however, my prediction for a difference over session was false. This could be due to the still relatively short period of time given to participants to familiarise themselves with the techniques and they require more practice before their $F_1$ scores would significantly increase compared to those from the first session.

## Session Two Results

Since there was no significant effect found across sessions, I chose to regard the first session as training and focus the analysis on the data from the second session. I conducted a one-way RM-ANOVA across the seven conditions (None, Spectrogram, VocSyl, Captions, Lip Magnification, iBaldi, PhonemeViz). The assumption of sphericity was violated, $\chi^2(20) = 79.41$, $p < .001$, therefore the Greenhouse-Geisser correction ($\varepsilon = .42$) was used. The results indicate that $F_1$ scores were significantly affected by the visualisation technique used, $F(2.55, 33.09) = 29.20$, $p < .001$, $\eta_p^2 = .69$. Mean $F_1$ scores are summarised in Table 7.3.

| **Technique** | **Mean $F_1$ score** | **s.e.** |
|---|---|---|
| None | 0.44 | 0.03 |
| Spectrogram | 0.36 | 0.07 |
| VocSyl | 0.55 | 0.09 |
| Captions | 0.99 | 0.01 |
| Lip Magnification | 0.37 | 0.04 |
| iBaldi | 0.42 | 0.07 |
| PhonemeViz | 1.00 | 0.00 |

Table 7.3: Mean $F_1$ scores $\pm$s.e. for each technique.

A pairwise comparison using Bonferroni-corrected *t*-tests (21 in total with alpha level = .002) was used to investigate the performance of each visualisation technique as well as the None condition. Individually comparing None, Spectrogram, VocSyl, Lip Magnification and iBaldi with all other techniques only revealed a significant difference when the other techniques were Captions and PhonemeViz (both of which achieved near-perfect $F_1$ scores).

These results indicate that four of the techniques (Spectrogram, VocSyl, Lip Magnification and iBaldi) were not significantly better than None (the mean $F_1$ scores for Spectrogram, Lip Magnification, and iBaldi actually fell below the mean $F_1$ score for the None condition). Captions and PhonemeViz both significantly increased participants' $F_1$ scores, with PhonemeViz allowing participants to achieve 100% correct identification of the words being spoken.

## Task Load Index

I performed Friedman tests on each of the six measures of the NASA-TLX and this was followed with Bonferroni-corrected Wilcoxon signed-rank tests (alpha = .008) comparing each technique with PhonemeViz when there was a significant main effect. Results are summarised in Figure 7.7.

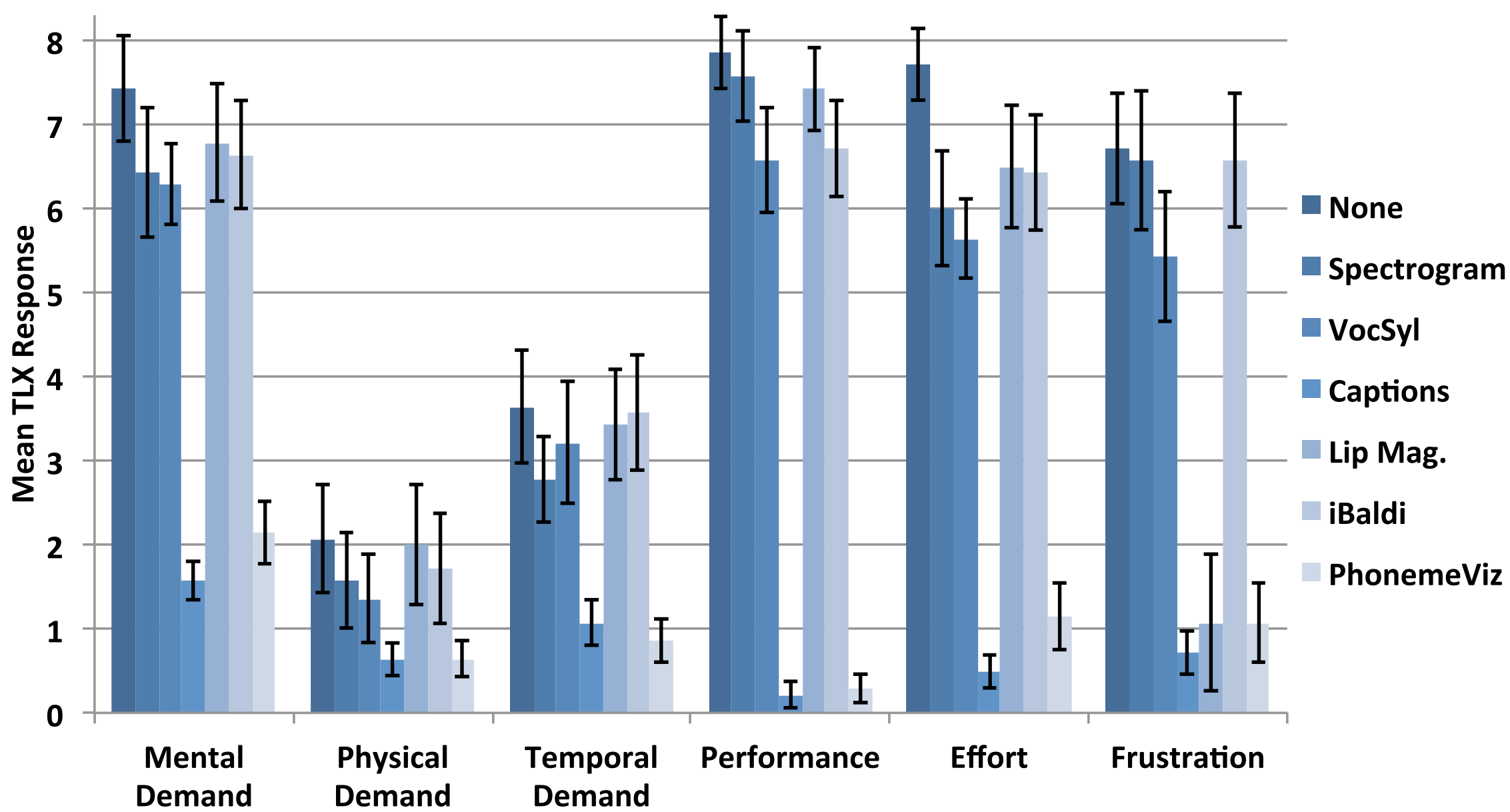

Figure 7.7: Mean TLX score ±s.e. for None and the six techniques.

There was a significant main effect ($p < .001$) for Mental Demand ($\chi^2(6) = 53.04$), Physical Demand ($\chi^2(6) = 21.58$), Temporal Demand ($\chi^2(6) = 37.11$), Performance ($\chi^2(6) = 57.53$), Effort ($\chi^2(6) = 59.13$) and Frustration ($\chi^2(6) = 53.45$). Post-hoc tests showed that PhonemeViz had significantly lower perceived Mental Demand, Temporal Demand, Effort and Frustration and better perceived Performance against all other conditions apart from Captions. Post-hoc tests for physical demand revealed that PhonemeViz was not significantly different from the other conditions.

These findings show that both PhonemeViz and Captions resulted in a much better perceived task load index in terms of Mental Demand, Temporal demand, Performance, Effort, and Frustration compared with the remaining visualisation techniques and None.

## Closing Questionnaire

The participants were first asked to rank each technique in order of preference. A value of 1 indicated they would most definitely use the technique again and 7 indicated they were least likely to use the technique again. I calculated an overall ranking for each technique by calculating its mean score and then ranked those means from lowest to highest. These results are shown in Table 7.4.

| **Technique** | **Mean Rank** | **s.e.** |
|---|---|---|
| None | 5.36 | 0.56 |
| Spectrogram | 5.50 | 0.31 |
| VocSyl | 4.36 | 0.31 |
| Captions | 1.21 | 0.31 |
| Lip Magnification | 4.64 | 0.33 |
| iBaldi | 5.00 | 0.39 |
| PhonemeViz | 1.93 | 0.13 |

Table 7.4: Mean rank with ±s.e. for each condition (1=most likely to use, 7=least likely to use).

The rankings showed that participants preferred Captions followed by PhonemeViz. Interestingly, there is a quite a difference (2.43) between the mean ranking for PhonemeViz and VocSyl, which was ranked third. The subjective rankings further support the $F_1$ data in showing how much of an improvement Captions and PhonemeViz provided.

Participants were asked to explain the reason behind their ranking order. A summary for None and each technique was formed from those responses. Overall, None was found to be difficult with no additional information present. However, one participant said they preferred it to any technique relying on shape and colour. For Spectrogram the consensus was that it provided unclear assistance, it was difficult to understand and distinguish clear differences between words. One participant felt you would need to be an expert in order to use Spectrogram well (which is supported by the literature [65]).

Although VocSyl was considered confusing, one person felt that it was the best technique after Captions and PhonemeViz. This is supported by both mean $F_1$ scores and the mean ranking. Captions were identified as being straightforward in that the whole word is given. Lip Magnification was considered distracting. A couple of participants wanted to see more facial features rather than just the lips.

For iBaldi, comments suggested it was distracting and the visualisation occurred too quickly making it difficult to focus on what was being spoken. The colours were also difficult to comprehend, however one participant felt that with additional training, iBaldi could be more useful. Finally, PhonemeViz was considered clear and easy to understand with little time or mental effort required. One participant suggested that it would be difficult to identify words that begin with the same letter, which for the current version of PhonemeViz is true, however I discuss extensions to PhonemeViz later to help address this.

Finally, participants were asked why they ranked their chosen technique the best. Only three of the techniques were ranked as best by participants (Captions, PhonemeViz, and None). The most popular technique was Captions (11 out of 14 participants put it as their first choice). From the written feedback, I identified that participants felt Captions was simple to use, it was familiar, there was little effort required, and they were confident using it. PhonemeViz was ranked best by two participants. P1 wrote: *"...it let me identify the words that the person was speaking quickest."* and P2 explained: *"I could look at her lips and keep my peripheral vision on the cue, for double confirmation of what was being said."* (this was echoed by P3, P6 and P12). The final top ranking was for None and it was given by just one participant. P12 noted that *"...it was more natural and didn't require me to do two kinds of thinking at the same time."*

# 7.5 Discussion

## Summary & Explanation of Results

The study uncovered three main results:

1) PhonemeViz enabled participants to achieve perfect $F_1$ scores (100% accuracy, 0% errors), which was significantly higher than all other techniques except Captions. In addition to this, none of the remaining techniques performed significantly differently than having no technique at all, indicating that they did not offer much assistance in the study's speechreading task.
2) When using PhonemeViz, participants reported lower Mental Demand, Temporal Demand, Effort, and Frustration, as well as higher Performance, than all of the other techniques except Captions.

3) Participants rated PhonemeViz as their second-most preferred option when asked if they would want to use the technique again (Captions were ranked first). There was a sizeable gap in average rankings between PhonemeViz and the third-best technique (VocSyl).

I attribute PhonemeViz's strong $F_1$ score performance to two main factors. First, I designed PhonemeViz to be easily learned, with no colour coding and simple visual feedback. As a result, participants were able to quickly ascertain how the visualisation works, and as a result were able to use it very effectively. The first day results support this, with PhonemeViz achieving nearly 100% $F_1$ scores in the first session as well.

Second, PhonemeViz facilitates the speechreader's natural tendency to be able to focus on the lips and face of the speaker by placing the visualisation mechanism in the user's peripheral vision. Several of the participants commented positively on this aspect of PhonemeViz. By incorporating the three design principals identified in Section 7.3.1 (Persistence, Short Training, and Peripheral Position), PhonemeViz required less from participants than most of the other techniques, so they rated PhonemeViz highly in the TLX responses. PhonemeViz did not require as much attention or focus, and supported participants in the study tasks. This resulted in participants experiencing low mental demand, temporal demand, effort, and frustration when using the technique, in addition to higher perceived performance.

Finally, the large gap between PhonemeViz and the third-best rated technique (VocSyl) again reflect the strengths of PhonemeViz over the other techniques. However, here as in the other results, Captions either outperformed PhonemeViz or there was no difference between the two techniques. I will now discuss why I think this was the case, and what other advantages PhonemeViz might offer over Captions that were not explored in this evaluation.

### 7.5.1 Captions and PhonemeViz

Captions repeatedly did as well as PhonemeViz or better. There are several possible reasons for this. First, PhonemeViz is an entirely new technique, so participants will have never seen it before the study. Participants' previous experience with Captions (which many participants are likely to have seen before, but I did not explicitly ask this in the demographics questionnaire) was likely to make participants feel more at ease and comfortable with captions – they were able to fully predict how it would function, and rest in the confidence that Captions allowed them to complete the study task correctly.

Second, due to PhonemeViz's peripheral projection of the first phoneme of each test word, participants may have felt torn between attending to the visualisation and attending to the speaker to gather the rest of the word (the part of each test word that was not visualised by PhonemeViz). Captions presented the entire test word to the participant, so they could completely disregard the speaker if they so wanted to, which may also explain the better subjective experience overall, leading to stronger TLX responses and rankings (which is what I found).

Third, although designed carefully, the layout of phoneme characters within PhonemeViz might have not been as easy to learn as I hoped, and this could have decreased participants' subjective ratings and rankings for PhonemeViz. Perhaps a purely alphabetical layout with some alternative visual presentation to help disambiguate proximate viseme members (e.g., /m/ and /p/) would make PhonemeViz easier to use.

Captions are processing and human-resource intensive, whereas PhonemeViz can be fully automated through existing phoneme classifiers. Of course, classifiers rarely function at 100% accuracy, but PhonemeViz has the built-in redundancy of allowing speechreaders to continue to focus on the speaker's face (thereby not inhibiting the speechreaders pre-existing speechreading skills). Unlike Captions, PhonemeViz should be able to be fully automated.

Producing captions has inherent delay. As human languages are context sensitive, the correct identification of a word must always happen after the word is uttered (prediction aside), and sometime cannot be determined until the entire sentence containing the word has been spoken. On the contrary, phoneme classifiers function on relatively small samples of sound, so can result in phoneme classifications much more quickly than automatic-speech recognition systems can produce speech-to-text output. Unlike Captions, PhonemeViz should be able to operate in near real time.

## 7.5.2 Limitations and Extensions

The results of the comparative evaluation show that PhonemeViz allowed all participants to achieve 100%-word recognition (showing successful disambiguation), and participants rated the technique well in subjective and qualitative feedback. This demonstrates that visualising phonemes can improve visual only speechreading in constrained word recognition tasks. However, there are four limitations with the evaluation of PhonemeViz.

First, in this evaluation the participants were not individuals who relied on speechreading for communication, therefore this preliminary study gives no direct insights into the performance of speechreaders. However, from the results of the Speechreading Proficiency Test, I demonstrated that participants had a variety of speechreading ability. Furthermore, none of the participants self-reported having hearing loss. However, people who speechread are typically individuals who loose their hearing after they have acquired a language and therefore these results still give some insight into potential performance levels of PhonemeViz when used by individuals with similar language and lexical abilities.

Second, the evaluation of PhonemeViz used videos of isolated words, presented with pauses between each stimulus video, with no audio volume – thus, the task in this study was different than someone speechreading in typical conversational speech, in which the individual could also use context clues about the topic and meaning of the conversation. Therefore, the results do not indicate if visualising phonemes would result in better comprehension with sentences or during natural conversation. Furthermore, I evaluated PhonemeViz using pre-recorded videos, as this allowed the evaluation of each visualisation technique in isolation from any automated detection software needed to detect elements of speech, phonemes, and words. At this point I have demonstrated that PhonemeViz works well with pre-recorded video, so could be used in such situations (e.g., news broadcasts, TV programs). However, for use in the smart glasses concept discussed earlier, this would require further evaluation.

Third, in this study, PhonemeViz only shows consonant phonemes. As part of expanding the set of phonemes, we will revisit how to distribute and highlight the phonemic character representations in the periphery, to ensure we do not lose the strengths demonstrated. Currently, PhonemeViz disambiguates visemes at the start of word, however for homophenous words like 'fifty' and 'fifteen', the confusing viseme occurs later in the word. To supplement this, I will need to revisit the 'Persistence' property (as information will need to be shown at a faster rate).

Finally, the results do not demonstrate if the visualisation detracted from the participants' ability to speech-read, as I do not know to what extent the participants were splitting their attention between looking at the face and the visualisation.

## Generalisations & Extensions

PhonemeViz is based on identifying phonetic elements of a person's speech, and not identifying the words that someone says. The 48 typical phonemes listed in Table 2.2 are for English, but many other languages carry a similar set of phonemes (e.g., French [16], German [31], Korean [41], and Japanese [72] each have distinct viseme-phoneme mappings). As a result, PhonemeViz can be directly extended to work in other languages; all that is needed is a phoneme classifier for the desired language, and a phoneme to character mapping to be used in the 'phoneme meter' portion of PhonemeViz. As phoneme classifiers for different languages are developed, phoneme meter character mappings for each language could be developed simultaneously, allowing speechreaders the world over to benefit from PhonemeViz.

As discussed in Chapter 6, questionnaire participants reported watching TV as method of practice, but reported that there is a lack of feedback when watching without subtitles, and too much reliance on the subtitles when watching with them turned on. To support speechreading practice during watching television, PhonemeViz could be superimposed onto video content (in a similar manner to subtitles) as a form of speechreading practice or as a way to access media that reinforces their speechreading skills.

## 7.6 Conclusion

The findings from the speechreading student questionnaire presented in Chapter 6 suggest that even with extensive training, sometimes speechreading does not provide enough information. Participants also reported challenges likely to be caused by the presence of visemes [54]. Four out of seven of the tutors interviewed in Chapter 4 highlighted the value of initial-letter fingerspelling during teaching as it helps to demonstrate when words are visually similar on the lips, due to being classed under the same viseme category. Even though initial-letter fingerspelling is a powerful teaching technique, it requires the speaker to know how to fingerspell for successful use outside of classes.

Smart glasses, such as the Google Glass, Microsoft Hololens, and the Epson Moverio glasses, project light over the user's vision and enhances how users see the real world [143], but also allow for the incorporation of virtual elements into the physical world. Using such devices, it may be possible for us to augment the speechreading process with additional information. This would allow for decreasing the challenge posed by visemes and also results in no additional learning for the other conversation partner when conversing with a person who is speechreading.

In this chapter, I presented an SAT called *PhonemeViz* – a phoneme visualisation technique designed to allow speechreaders to use their peripheral vision to attend to a text-based visualisation to help them disambiguate confusing viseme-to-phoneme mappings that occur at the start of words. In a comparative evaluation, I found that PhonemeViz resulted in participants achieving 100% word recognition (showing successful disambiguation), and PhonemeViz was well-received in subjective and qualitative feedback.

In the future, I will continue to expand PhonemeViz's capabilities in four distinct ways. First, I will look to expand the phonemes presented in PhonemeViz's visualisation to include all potentially-confusing viseme-to-phoneme mappings, not just consonants.

Second, as part of expanding the set of phonemes, I will revisit how I distribute and highlight the phonemic character representations in the periphery to ensure I do not lose the strengths demonstrated here as I expand the phoneme set.

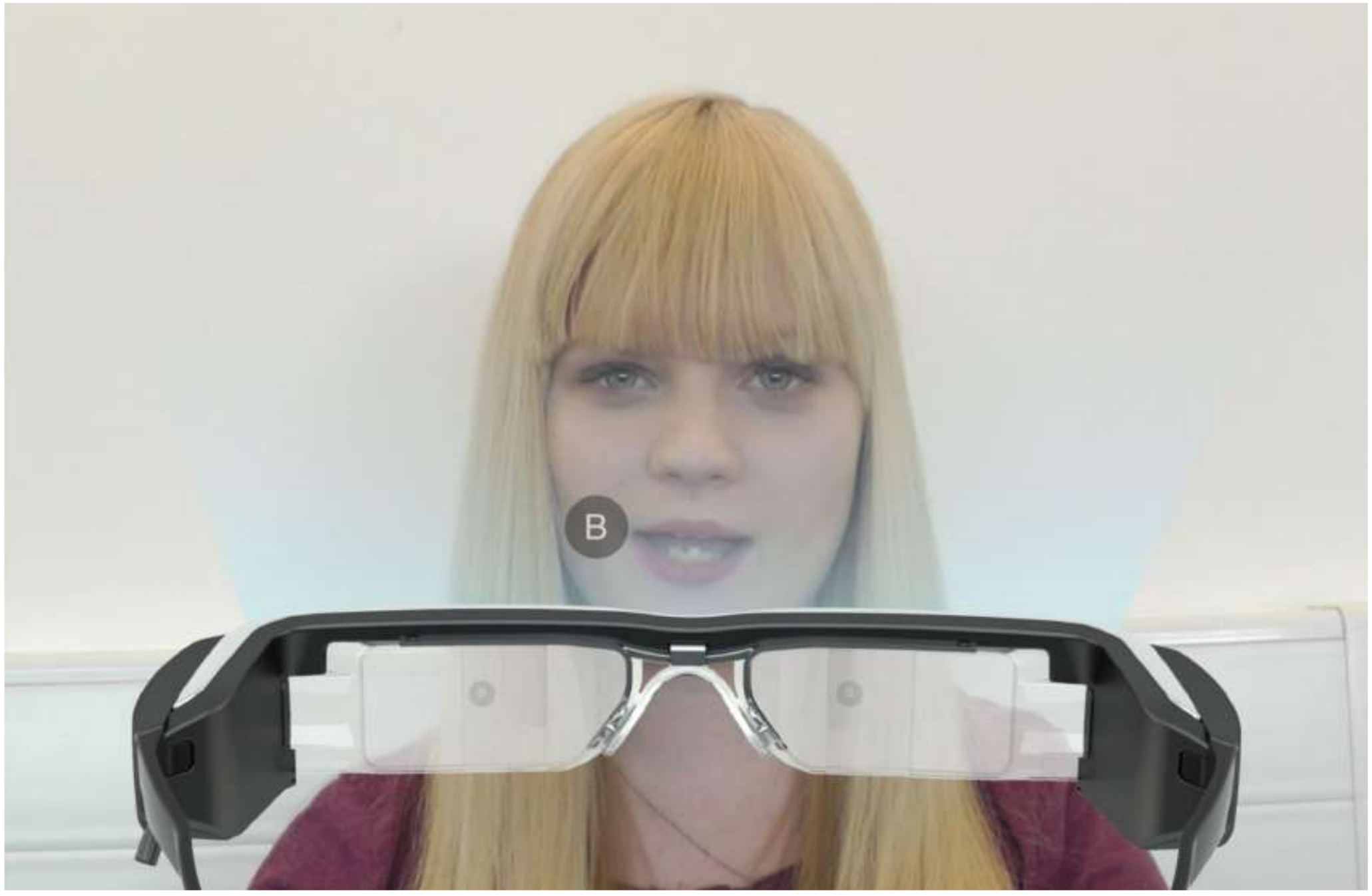

Figure 7.8: Mockup of what PhonemeViz could look like when viewed through Epson Moverio glasses (`http://www.epson.com/moverio`) for 'Bat'.

Third, I want to allow PhonemeViz to provide disambiguations for confusing viseme-to-phoneme mapping wherever they occur in a word – this might also include concatenated visemes, which often arise when people are speaking rapidly [129]. To facilitate this, I am revisiting how I provide persistence in PhonemeViz, and will experiment with progressive transparency to suitably increase the information provided by the visualisation.

Fourth, I plan to test PhonemeViz in actual conversations, which will require integration of phoneme detection software. Fortunately, existing open source phoneme detection software is available (`cmusphinx.sourceforge.net/wiki/about`). By extending PhonemeViz in these four directions, I hope to carry the strengths I present here in the lab-based study of PhonemeViz to real world conversations and then deploy on commodity Head Mounted Displays (HMD) (as shown in 7.8).

Although the evaluation of PhonemeViz demonstrates its potential to help support speechreaders during conversation, the automatic speech recognition systems needed to generate PhonemeViz's visualisations are not currently accurate enough. This is discussed in more detail in Chapter 9. Therefore, in the next chapter I describe the development of a new SAT called MirrorMirror which is a mobile application that focuses on improving speechreading practice.

# 8

# MirrorMirror

## 8.1 Introduction

This chapter presents a Speechreading Acquisition Tool (SAT) called MirrorMirror. MirrorMirror is an Android application designed using the framework presented in Chapter 5, which provides a variable (low to high) amount of hybrid information. MirrorMirror was primarily inspired by the mirror training technique that was highlighted by seven speechreading tutors during the interviews presented in Chapter 4. The rationale behind the design of MirrorMirror was further informed by comments from tutors during the interviews, along with findings from the student questionnaire presented in Chapter 6. The motivation and rationale behind the design of MirrorMirror is given first, followed by a case-study based evaluation with three speechreading students.

## 8.2 Motivation

The findings from the questionnaire results presented in Chapter 6 report that 67% of questionnaire participants have been in speechreading classes for over two years, yet less than a third (32%) of participants rated their speechreading ability as 'Good' and nobody rated it 'Excellent', suggesting that speechreading practice needs to be improved. During the interviews with speechreading tutors presented in Chapter 4, tutors reported that they think students practice outside of class by using techniques from class. The questionnaire findings suggest that this is only partially correct;

the 66% of participants who practiced at home, did so primarily by observing speakers in daily life or on television.

Participants reported a high frequency of watching television (with subtitles on and off) to practice speechreading. Together with the reported use of observation, it could be argued that these techniques are used to provide the speechreader with an endless supply of practice material. However, with television and observation, it is difficult for speechreaders to assess whether they are understanding the speaker correctly. Along with current SATs, neither TV nor observation allow for targeted practice around the challenges or situations that participants reported most negatively impact their speechreading ability.

In Chapter 4, spechreading tutors reported that mirror practice plays a key role in speechreading training, and is also recommended by Action on Hearing Loss [4, 106] for practice, as it may develop visual cue integration skills needed during speechreading [7]. In Chapter 6, Participants report that mirror practice allowed them to learn lipshapes, compare them with others, and show visual differences between words. However, they dislike watching themselves, that they have full knowledge of what they are saying, that they over-emphasise words, and that mirror practice was not akin to genuine speechreading. This resulted in a low frequency of usage by participants (57% of those who answered reported ‘Never’).

The quality of speechreading practice outside of class is of limited value because: 1) current SATs (e.g., lipreading.org, lipreadingpractice.co.uk, ConversationMadeEasy [130] and DAVID [123]) have limited content, 2) TV and observation provide limited feedback, and 3) mirror practice provides too much feedback. To address this, I designed a mobile application called MirrorMirror that allows speechreaders to practice lipshapes and words by recording videos of people they frequently talk to. MirrorMirror provides a multiple choice quiz game where the user selects the word they think the speaker has spoken. MirrorMirror provides feedback on whether they are speechreading correctly, and allows them to target specific challenges and situations.

## 8.3 Implementation

MirrorMirror was implemented as an Android application. I chose Android as the target platform as there are a variety of inexpensive Android tablets on the market that could be used for the evaluation. MirrorMirror is primarily designed for tablet based displays, but could also be adapted to smaller and larger screens. MirrorMirror’s visual design follows Google’s Material Design guide (`https://material.io/guidelines/`). The application is targeted for API

Level 25 with a minimum SDK of 16.

### 8.3.1 Data Storage

As reccomended by the android developer guidelines[a], I started to design MirrorMirror by first expressing the information model required on an entity-relationship (ER) diagram (as shown in Figure 8.1). To implement the information model I used SQLite[b]. MirrorMirror saves recorded videos within the application's internal directory and to protect the privacy of speakers, the videos are not available (to other applications or third parties) outside of the application.

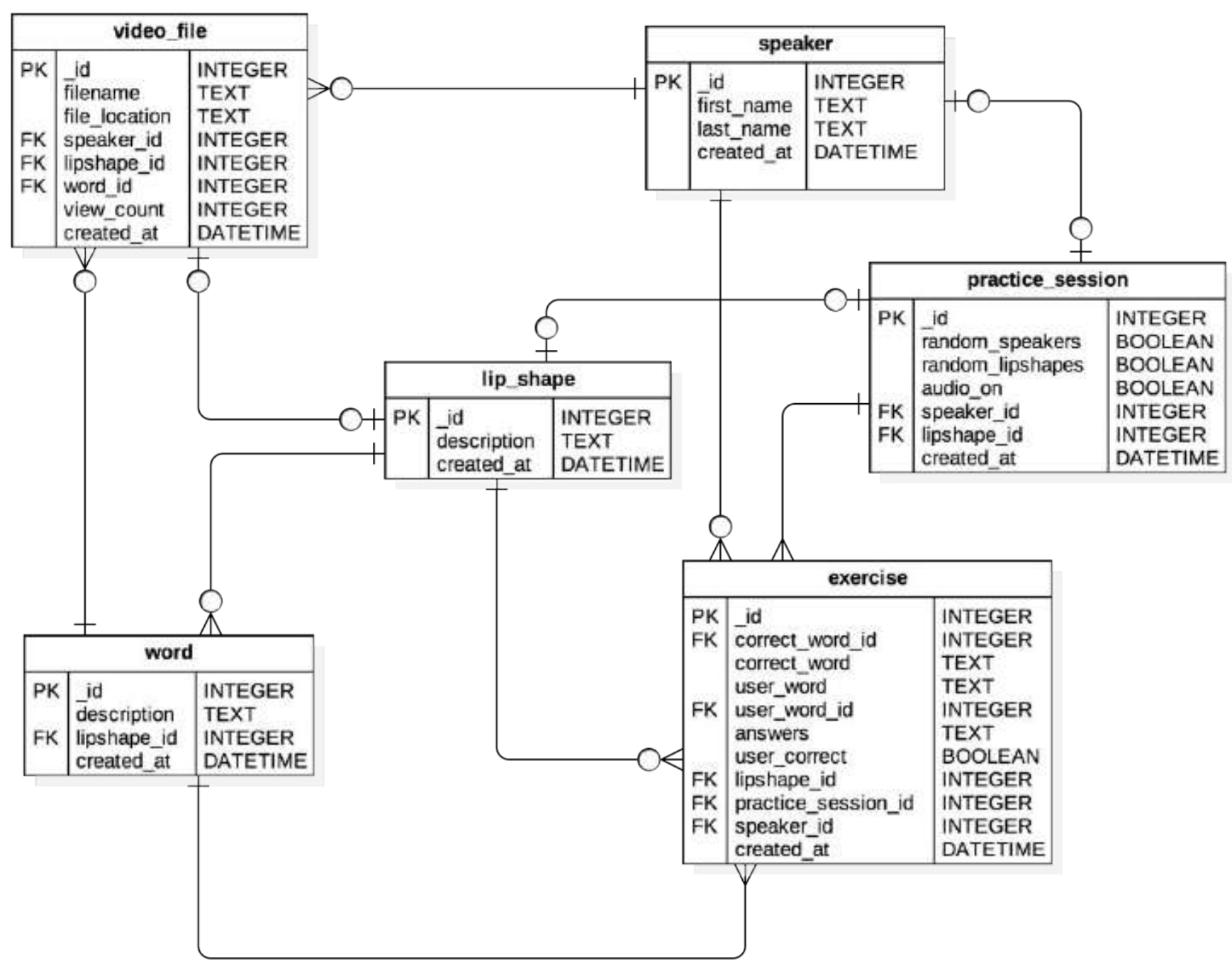

Figure 8.1: Entity-relationship diagram for MirrorMirror, showing each of the database tables along with their attributes and data types.

### 8.3.2 Screen List

Using the information model, I then defined the context necessary to enable users to discover, view and act upon the data within MirrorMirror by determining the set of 'screens' necessary. In

[a]https://developer.android.com/training/design-navigation/screen-planning.html
[b]https://www.sqlite.org/

MirrorMirror, the primary goal is to enable users to view, save and practice with videos. Below is a list of 'screens' that cover these use cases:

- 'Home' or 'launchpad' screen for accessing videos and capturing videos
- List of speakers
- Speaker view
- List of lipshapes
- List of words within a lipshape
- Word list for a lipshape
- List of videos
- Video view
- Capture video 'screen'
- Lipshape practice session 'screen'
- Word practice session 'screen'
- Help view

Using the screen list above, I then defined the directed relationships between screens; where an arrow from screen A to screen B implies that screen B should be directly reachable from screen A. The screen map for MirrorMirror is shown in Figure 8.2.

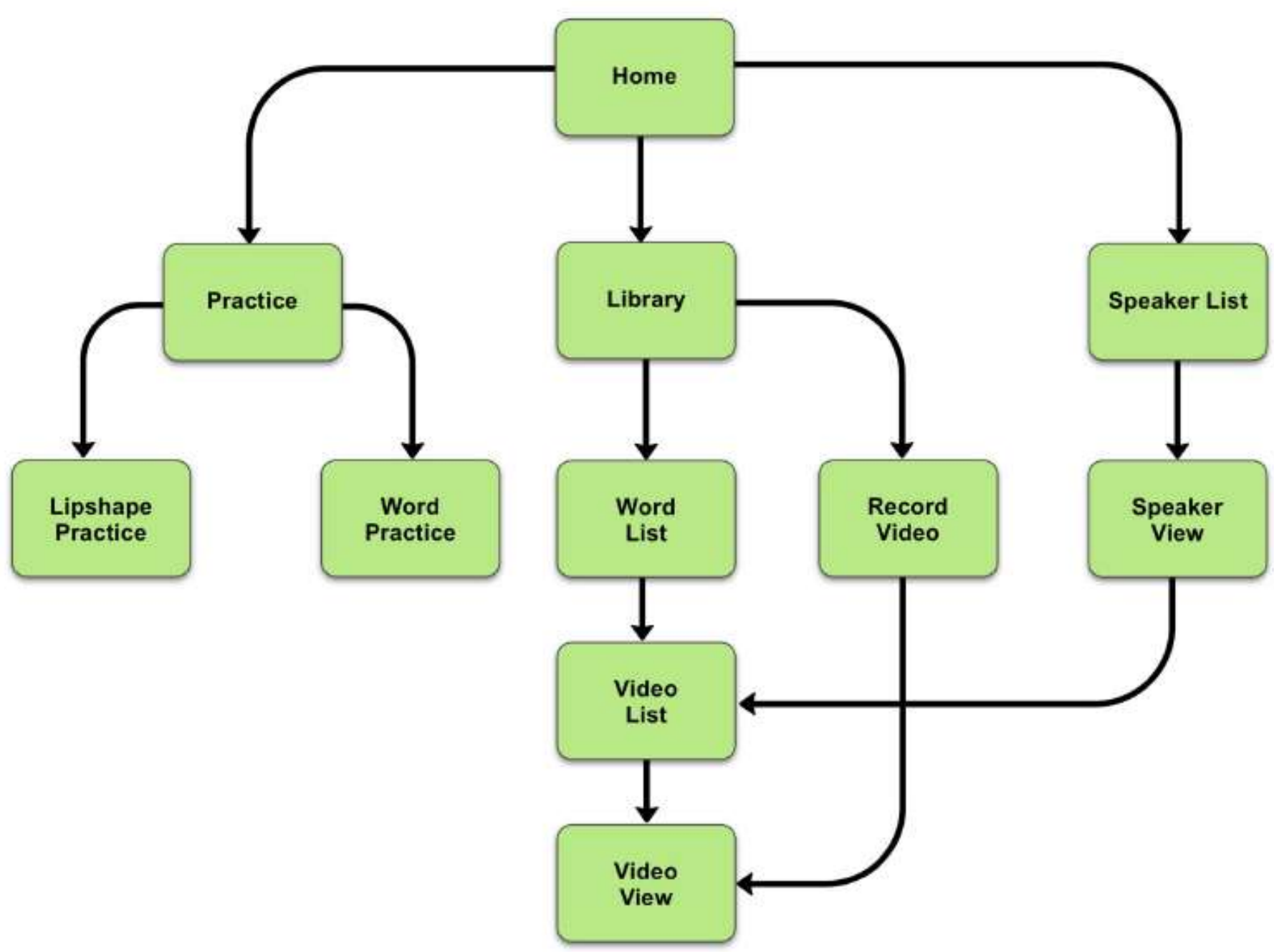

Figure 8.2: MirrorMirror's initial 'screen map', which shows all of the 'screens' and their relationships.

### 8.3.3 External Libraries

To allow users to record videos, a library that provides more control over video capture settings compared to the default Android activity was required. To achieve this, I used the "LandscapeVideoCamera" library developed by Jeroen Mols (`https://github.Com/JeroenMols/LandscapeVideoCamera`). This library is open source and offers a capture activity with granular controls over video quality, storage location and file size, and it can also restrict recording to landscape orientation.

## 8.4 Application Features

### 8.4.1 Navigation

MirrorMirror uses tab-based navigation as this is a popular solution for lateral navigation in Android applications. Android design guidelines (`https://material.io/guidelines/patterns/navigation.html#navigation-patterns`) state that tabs allow users to easily move between a small number of section-related screens, and because MirrorMirror has three main navigation areas ("Library", "Speakers" and "Practice"), tabs are the most effective navigation solution. Users switch tabs by either tapping on the tab name or swiping left to right.

### 8.4.2 'Library' Tab

#### Lipshape Library

The lipshape library is initially displayed on the 'Library' tab and is implemented as an Android ListView. Each lipshape is displayed as a row in the list with a title (displaying the lipshape) and subtitle (showing the number of words contained within that lipshape) as shown in Figure 8.3. Tapping on a lipshape item loads the word library for that lipshape as shown in Figure 8.4.

#### Word Library

The word library displays the word list for a particular lipshape. Each word is a row in the list with a title (displaying the word) and subtitle (showing the number of videos recorded with it) as shown in Figure 8.4. Tapping on a word loads a collection view of videos. To return to the lipshape library, the user can press the device back button, or the back arrow on the 'Lipshape' title strip.

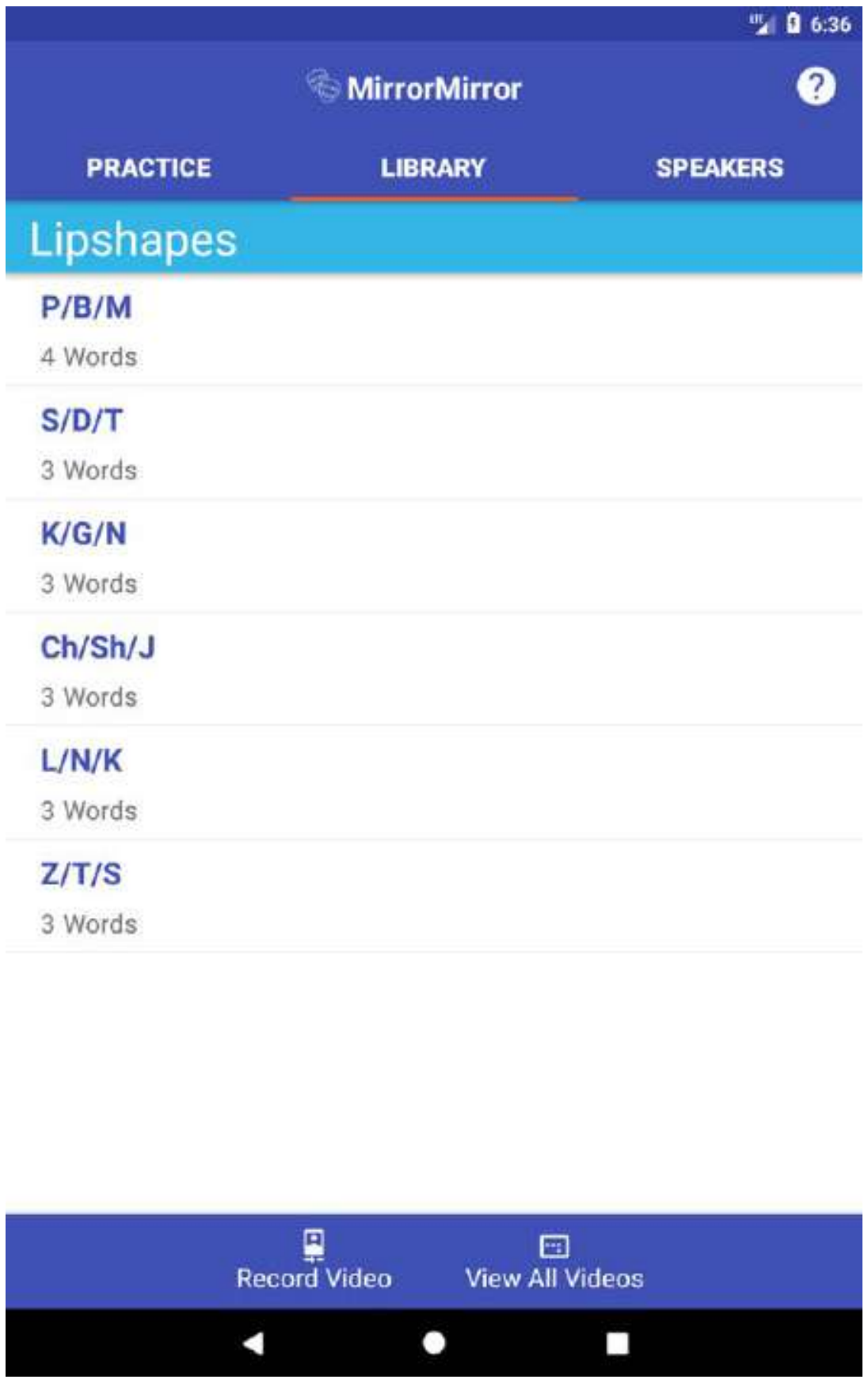

Figure 8.3: Screenshot of the 'Library' tab displaying the lipshape ListView. Each lipshape is displayed as a row in the ListView with a title displaying the lipshape and subtitle showing the number of words contained within that lipshape. Tapping on an individual item loads a sub list of words as shown in Figure 8.4.

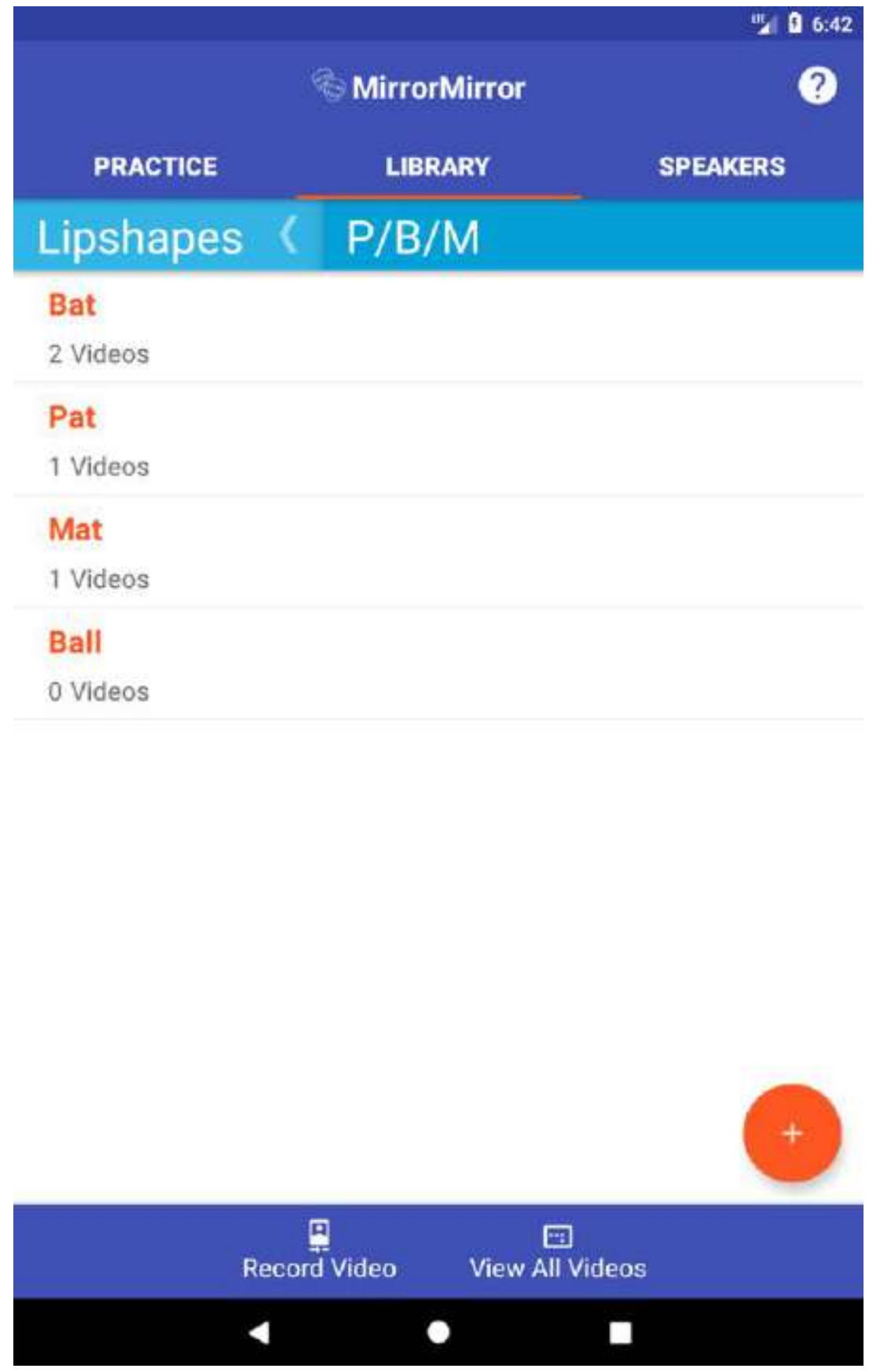

Figure 8.4: Screenshot of the 'Library' tab displaying the word ListView of lipshape "P/B/M". Each word is an item in the ListView and has a title displaying the word and subtitle that shows the number of videos available for it. Tapping on an individual item displays a collection containing those videos. Users can add new words to this lipshape by tapping on the '+' button in the right hand bottom corner.

### Adding Words

When in the word library, the add new word activity is launched by tapping on the '+' button in the right hand bottom corner (as shown in Figure 8.4). The add new word activity displays a TextView (as shown in Figure 8.5 and when a user taps the submit button on the keyboard the word is added to the lipshape it was launched from.

### Deleting Words

When in the word library, users can delete words from the current lipshape by long pressing on a word row, which launches a dialog window as shown in Figure 8.6. When a user deletes a word, all videos of that word are deleted as well.

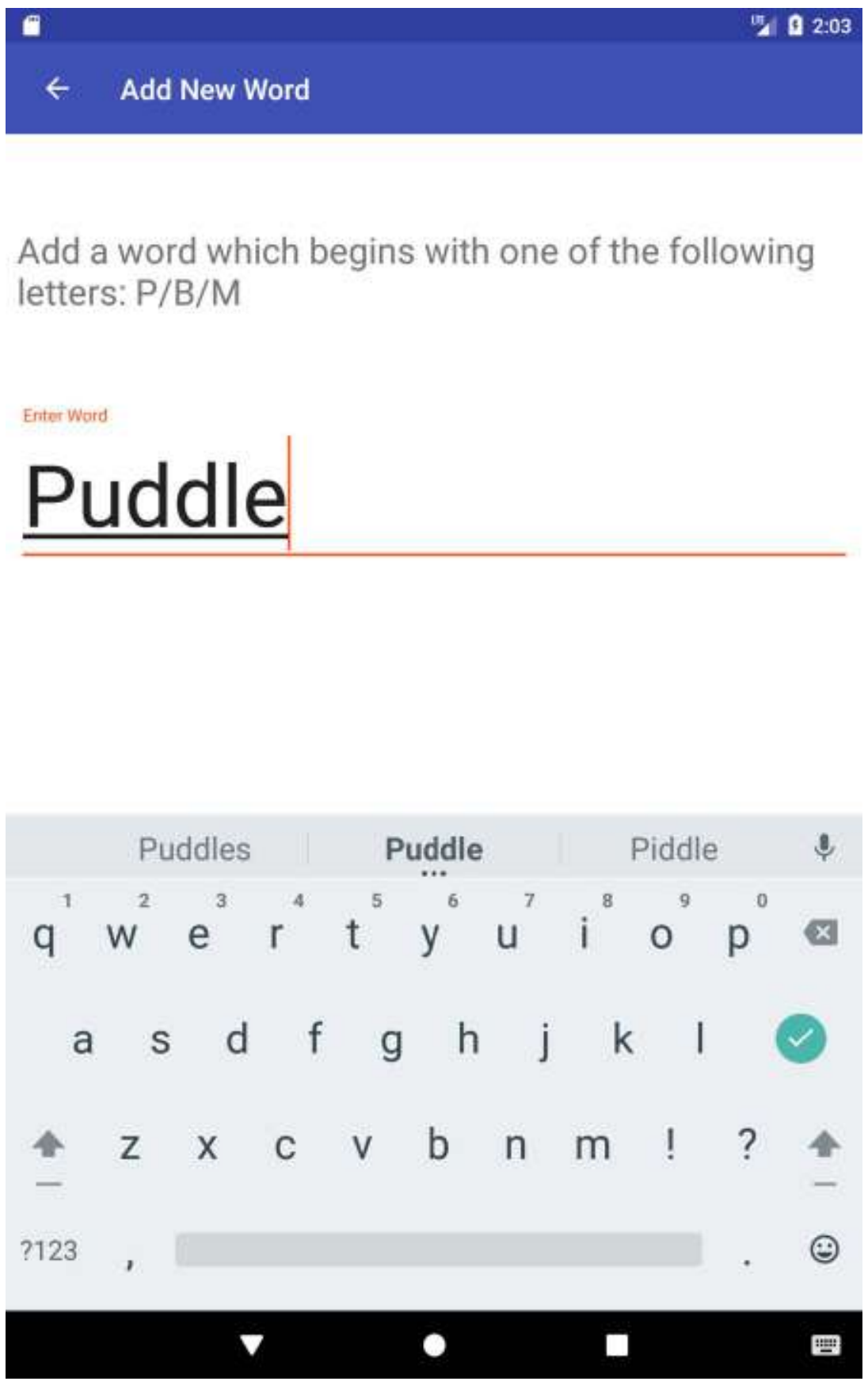

Figure 8.5: Screenshot of the 'Add word activity shown here with a user entering the word 'Puddle' for lipshape 'P/B/M'. When the user hits enter the word is added as long as it passes validation (e.g., must be a word, word must start with a capital letter, and the word must start with one of the phonemes of the lipshape).

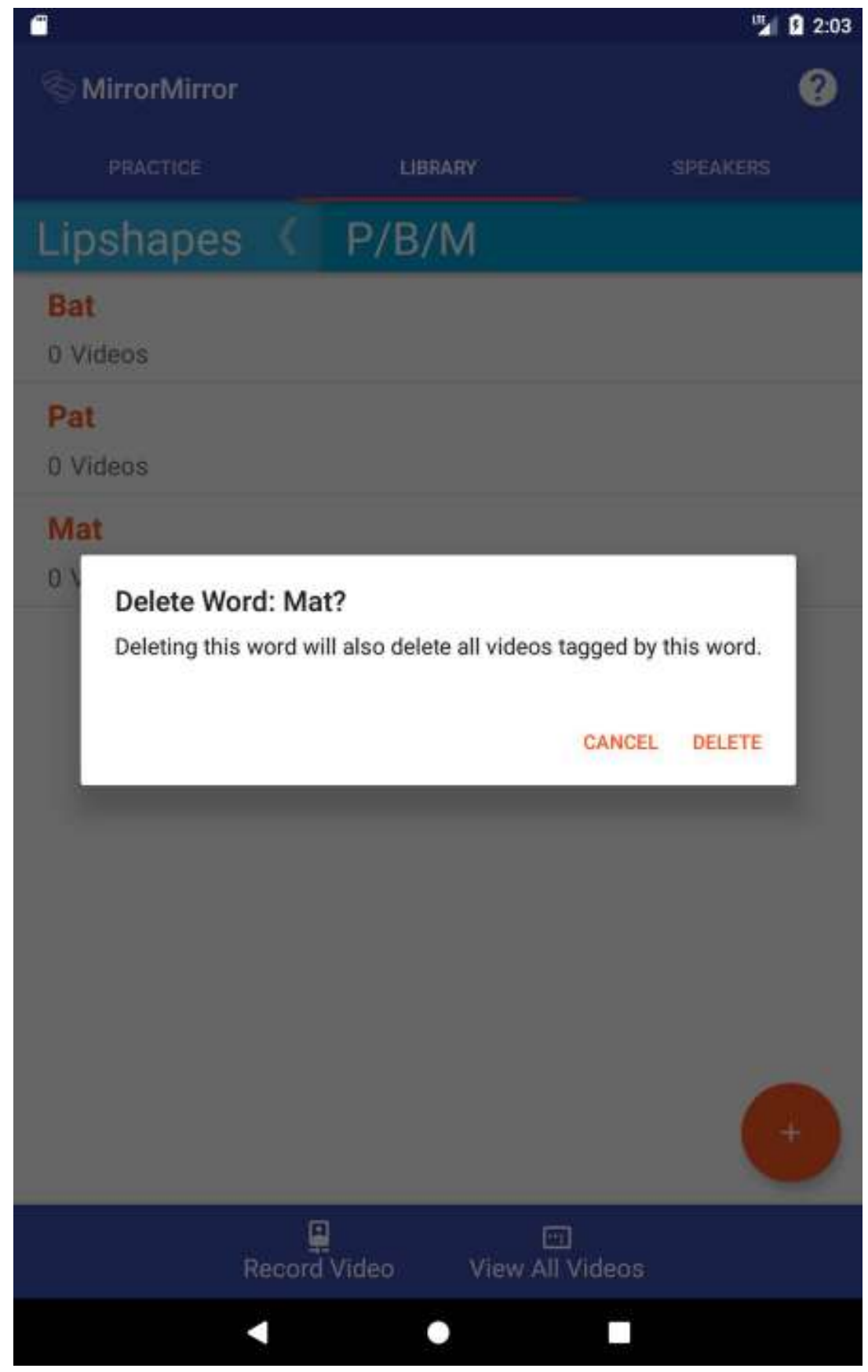

Figure 8.6: Screenshot of the Delete word dialog shown here for 'Mat' that appears after the user long presses on an item in a word ListView. If the user taps 'Delete', the word is deleted (along with its associated videos) and if the user taps 'Cancel' the dialog is dismissed.

### 8.4.3 Recording Videos

Users can record videos by tapping on the “Record Video” button on the bottom navigation drawer of the ‘Library’ tab as shown in Figure 8.3. This launches the Record Video activity, which prompts the user to select a word and the speaker who is going to be saying the word as shown in Figure 8.7. When the “Record Video” button is tapped, the video capture activity is launched as shown in Figure 8.9. To begin recording, the user taps the red circle, and the user is presented with the word they are to speak along with a timer showing the duration of the recording as shown in Figure 8.10. The icon in the lower right hand corner flips from front camera to back camera. To stop the video recording, the user taps the red circle again, and is then asked if they wish to save or discard the video as shown in Figure 8.11. Once the video has been saved, the user is then shown a VideoView.

### 8.4.4 Video Library

The video library is accessed via the bottom navigation drawer of the ‘Library’ tab as shown in Figure 8.3. The video library displays a collection of videos and the user can swipe left or right to select a video as shown in Figure 8.8. Tapping on the play button or the thumbnail plays the video in a fullscreen view. Underneath the video, the word, lipshape, and speaker is displayed. By tapping the edit video button, the user can edit the lipshape, word or speaker of the video, or delete the video.

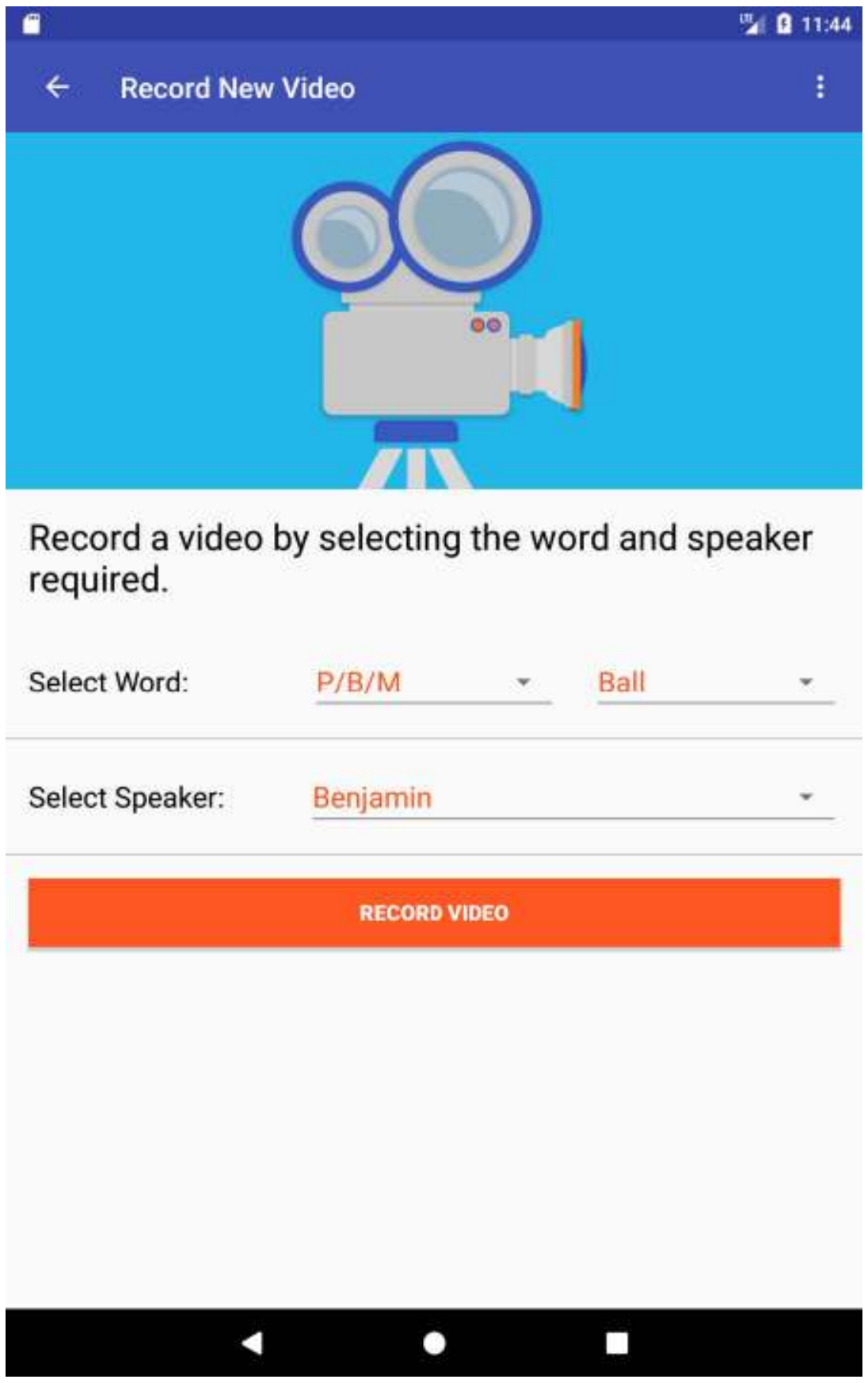

Figure 8.7: Screenshot of the Record Video activity, which displays three dropdown lists: lipshapes, words, and speakers. Tapping on Record Video starts the video capture activity as shown in Figure 8.9.

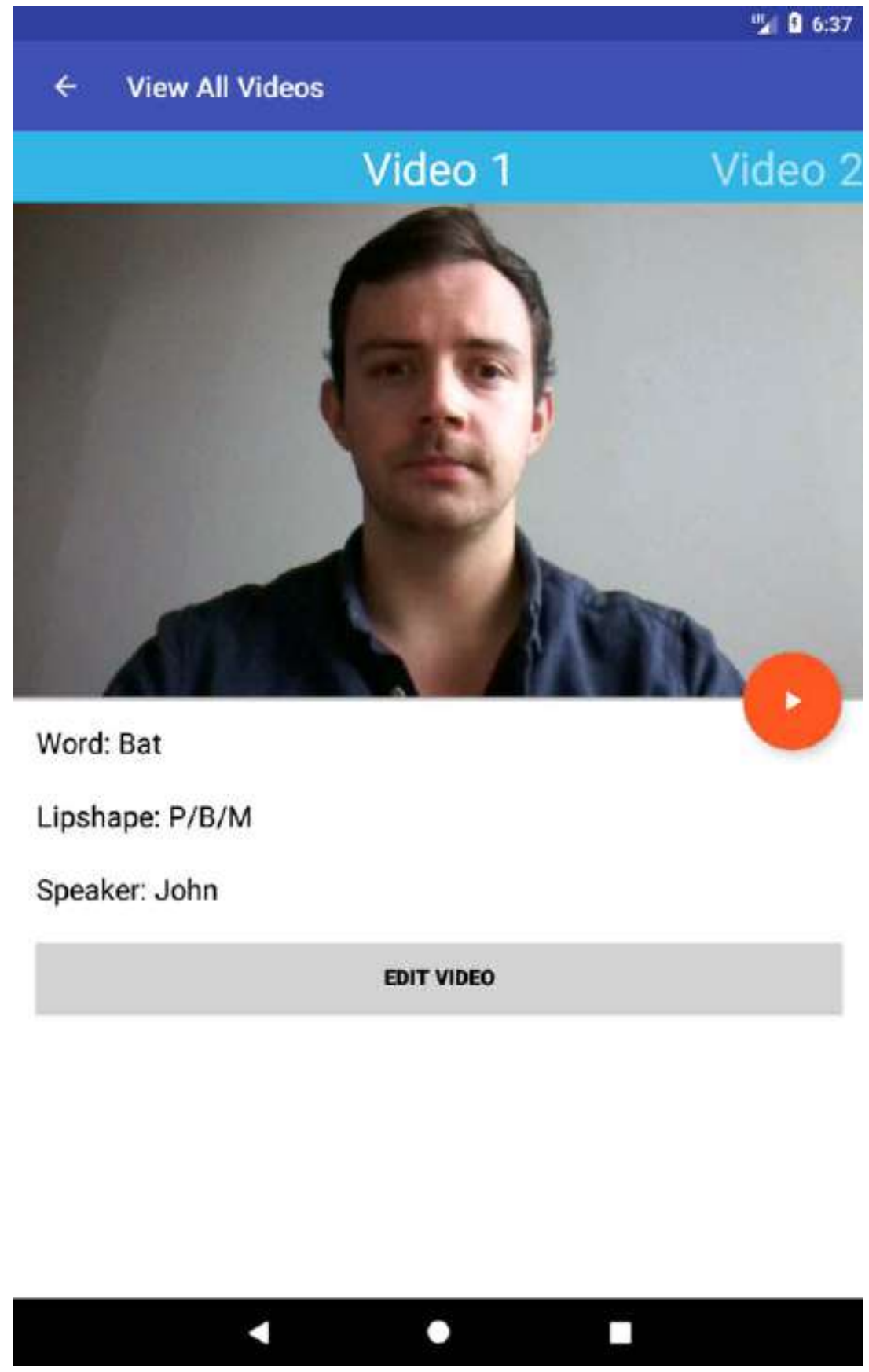

Figure 8.8: Screenshot of the Video Library activity that displays a collection of VideoViews in a ViewPager. The user can swipe left and right to select a video object. The VideoView displays a thumbnail of the video. Tapping on the play button or the thumbnail plays the video in a fullscreen view. Underneath the video thumbnail the word, lipshape, and speaker of the video are shown. By tapping the 'Edit Video' button, the user can edit how the video is tagged or delete the video.

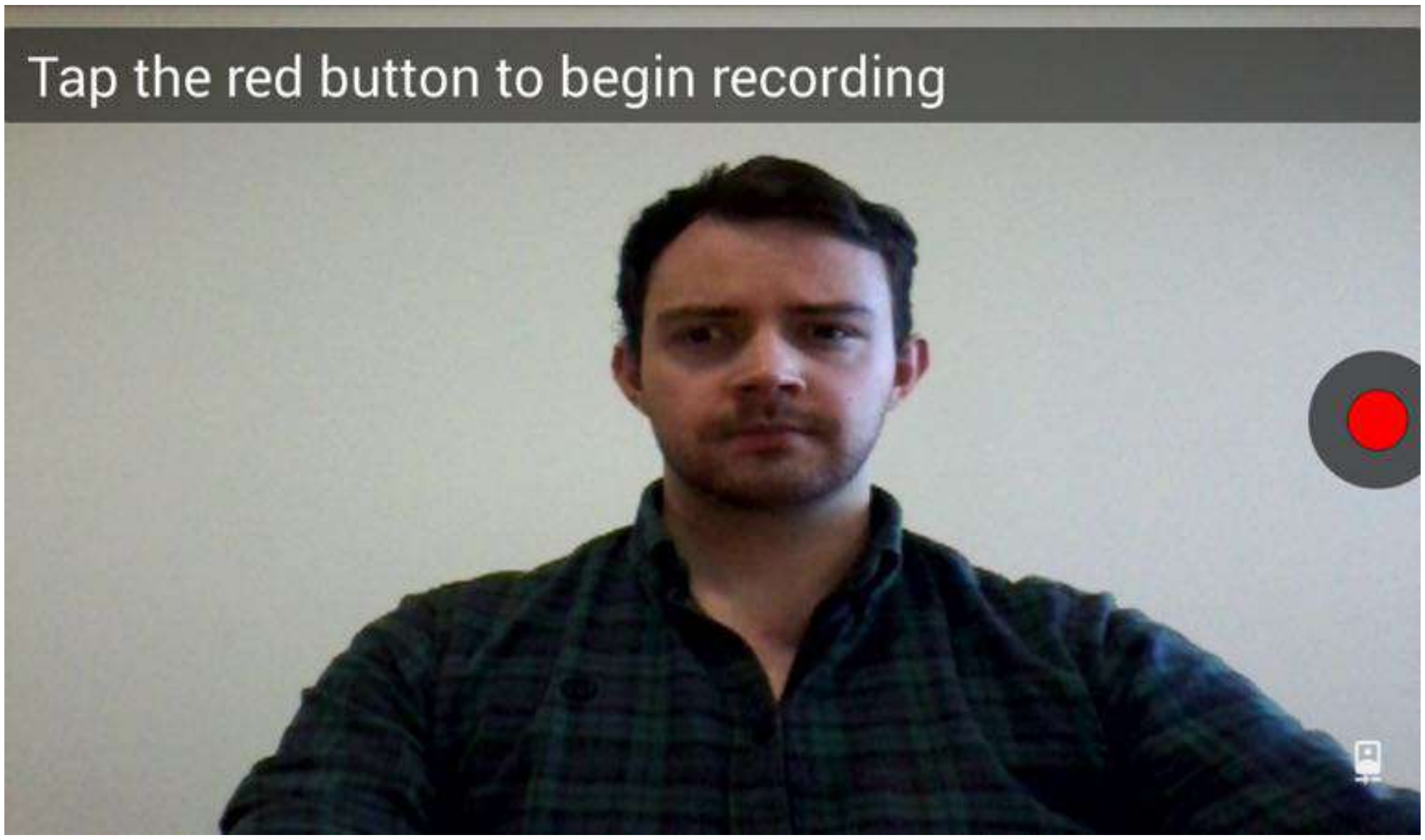

Figure 8.9: Screenshot of the Video Recorder activity. When the user taps the red circle, recording begins. The icon in the lower right hand corner toggles between the front-facing and rear facing cameras.

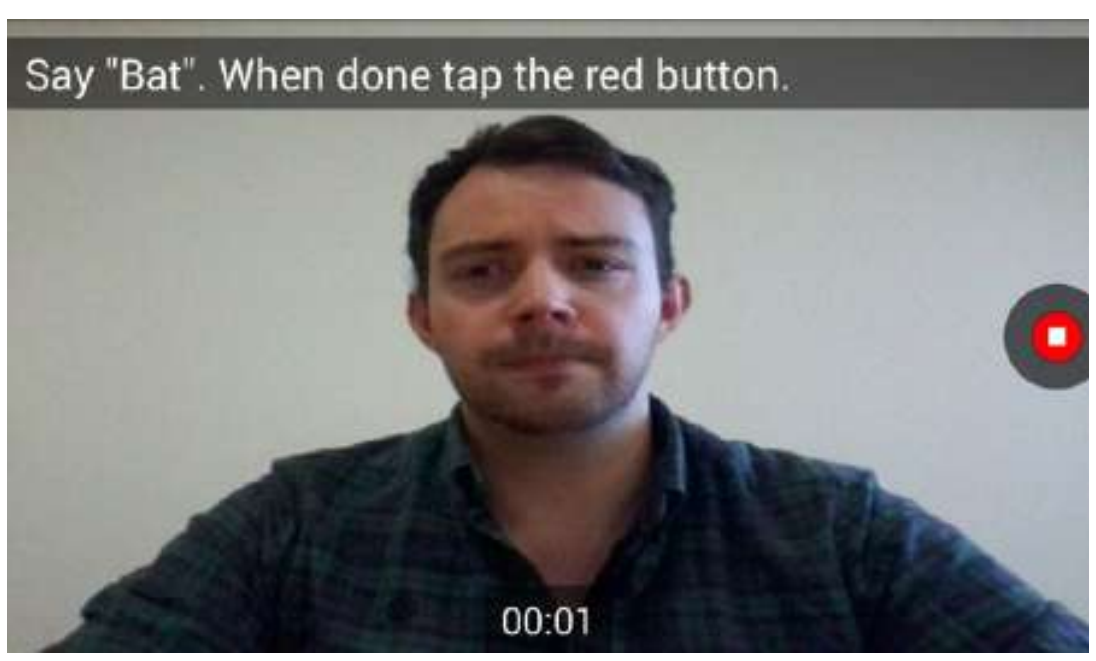

Figure 8.10: Screenshot of the Video Recorder activity in progress, the word chosen for the speaker to say is displayed in an overlay. Tapping the red button stops the video recorder.

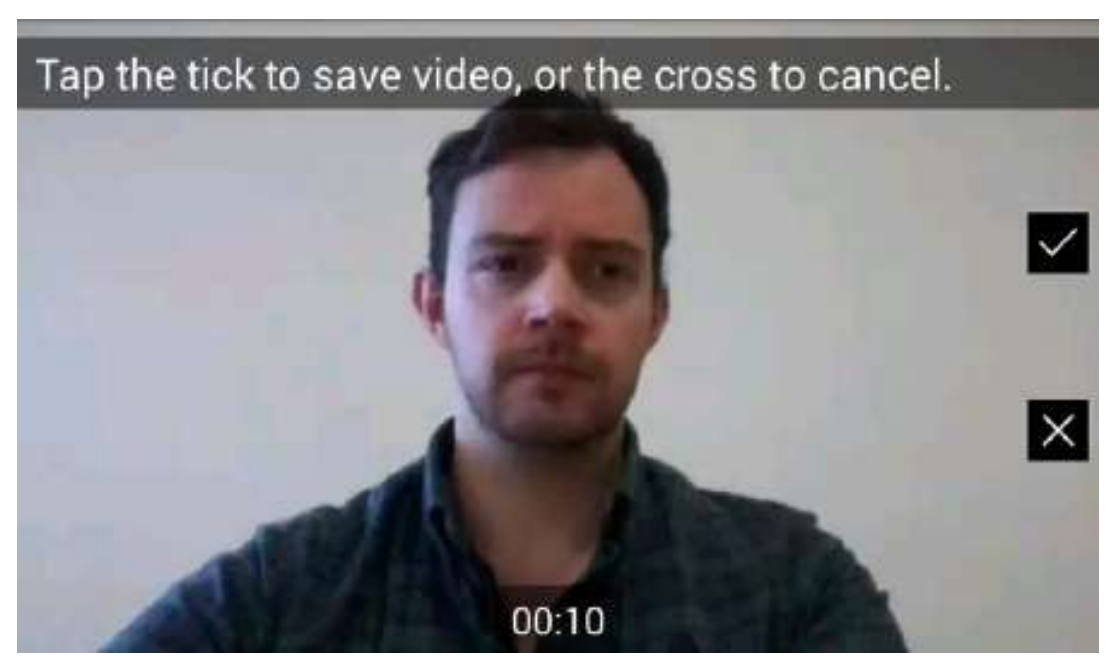

Figure 8.11: Screenshot of the Video Recorder activity confirmation screen, the user has to accept the video by pressing the tick or reject it by pressing the cross. Tapping the Android back button also cancels the video saving.

### 8.4.5 'Speaker' Tab

The speaker tab displays the collection of speakers as an Android ListView. Each speaker is displayed as a row in the ListView with a title showing the name of the speaker and subtitle showing the number of videos available with this speaker, as shown in Figure 8.12.

#### Speaker View

Tapping on an individual item in the speaker list loads a speaker view as shown in Figure 8.13. The speaker view displays the full name of the speaker and the number of videos available for that speaker. Tapping on 'View All Videos' opens a video library for this speaker as shown in Figure 8.8. Tapping on 'Edit Speaker' displays a button to delete the speaker. When a speaker is deleted, all videos of that speaker are also deleted.

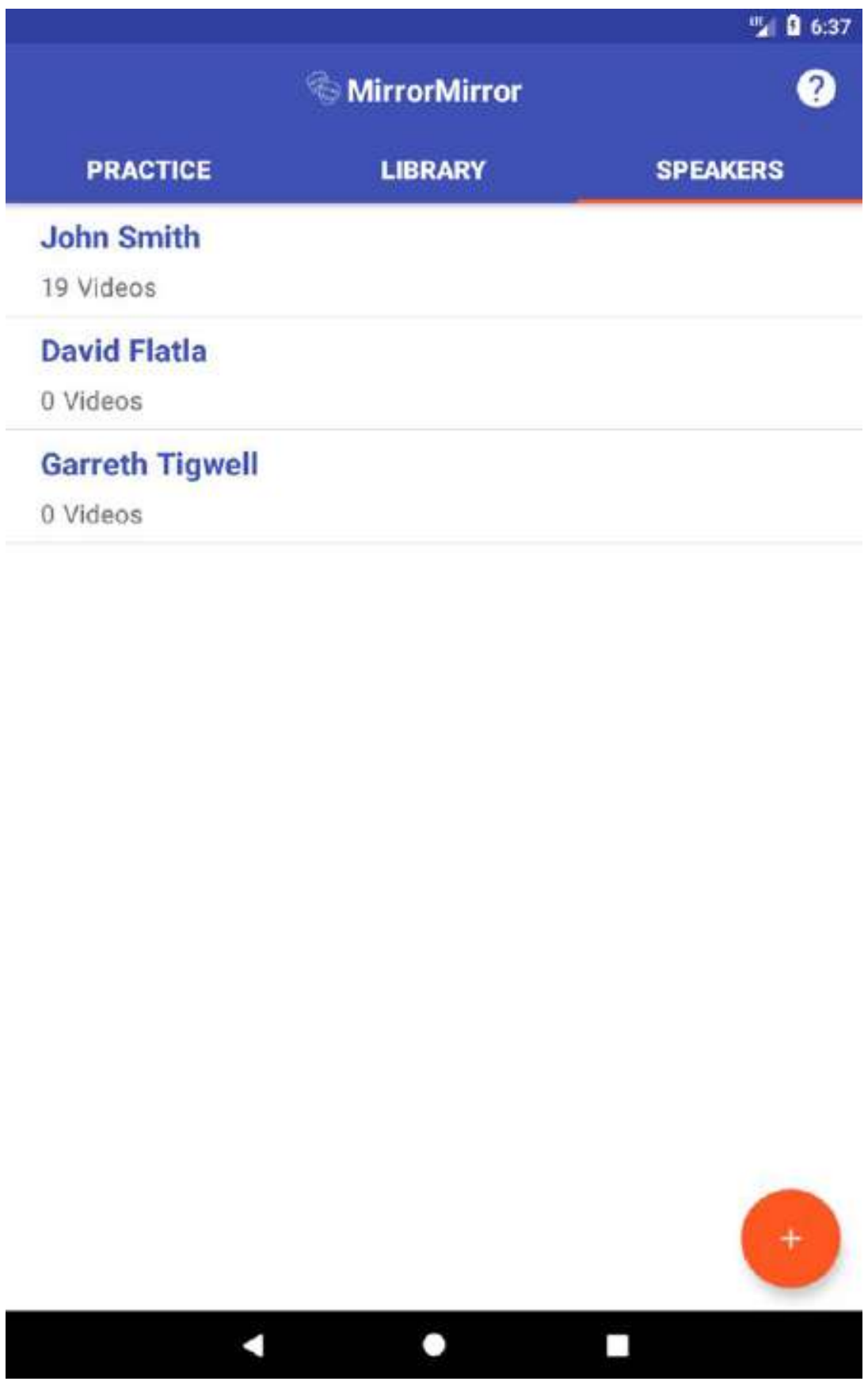

Figure 8.12: Screenshot of the 'Speaker' tab displaying a ListView of speakers. Each speaker is an item in the ListView and has their full name as a title and subtitle showing the number of videos available. Tapping on a speaker loads a detail view as shown in Figure 8.13.

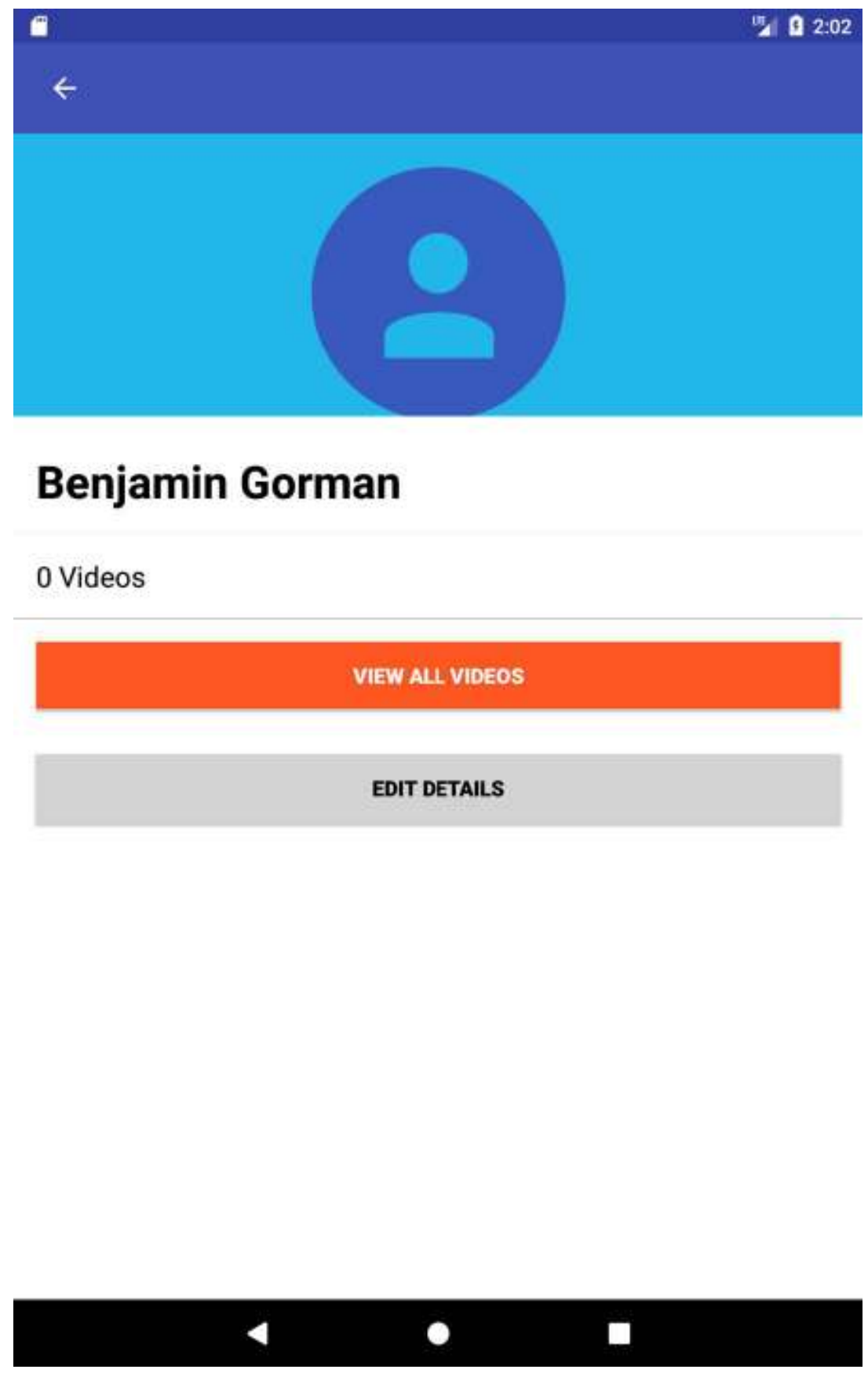

Figure 8.13: Screenshot of the speaker view displaying the full name of the speaker and the number of videos available, tapping on 'View All Videos' opens a video library for this speaker as shown in Figure 8.8. Tapping on 'Edit speaker' displays a button to delete the speaker.

### Adding a speaker

The user can add a speaker by pressing on the '+' button anchored at the bottom of the list as shown in Figure 8.12. The 'Add Speaker' activity has a field for the first name and last name as shown in Figure 8.14. When the users taps the submit button, there is a dialog box that informs the speaker about the research project and asks them if he/she consents to the use of their data and videos as shown in Figure 8.15.

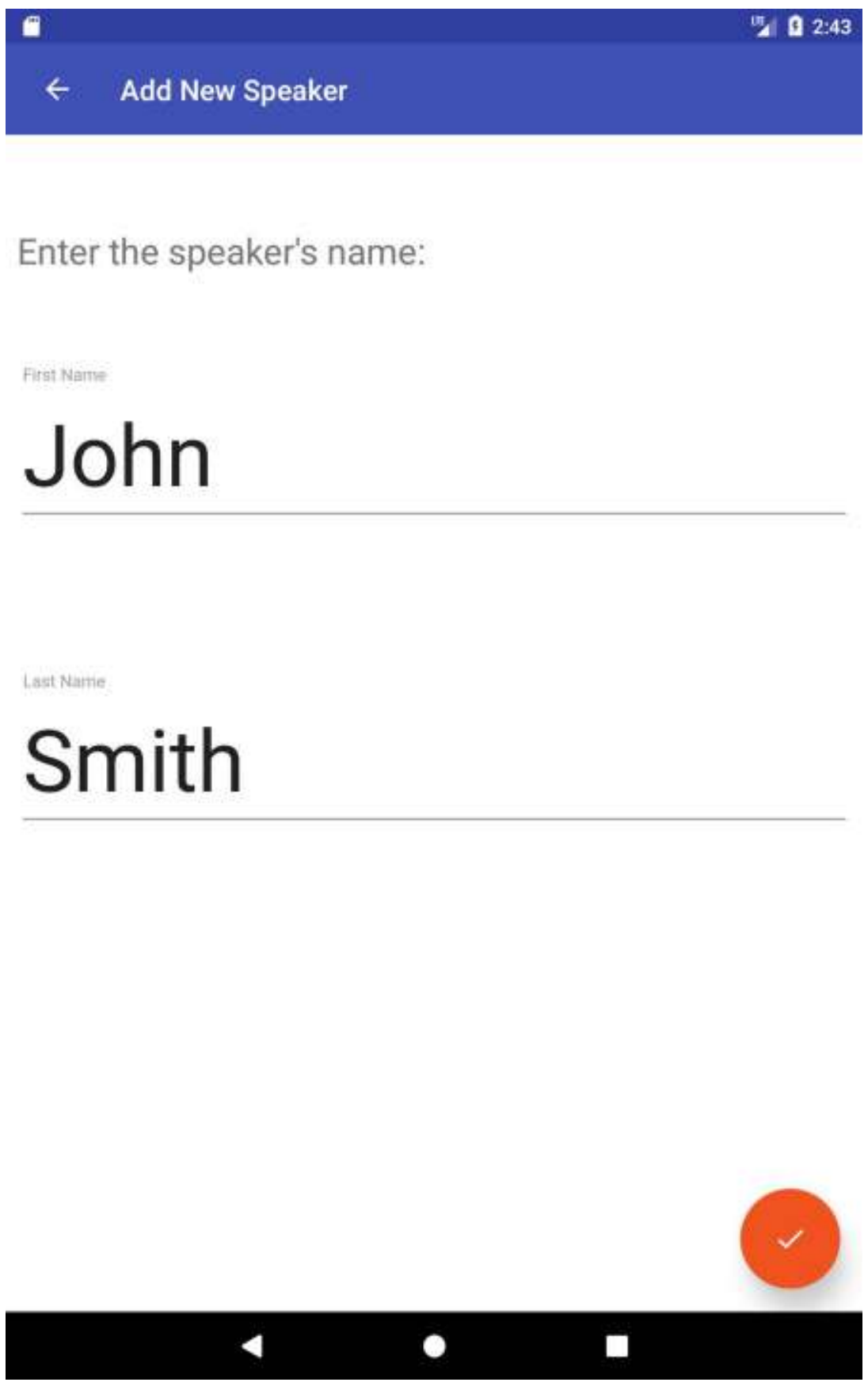

Figure 8.14: Screenshot of the Add Speaker activity shown here with a user entering the name 'John Smith'. When the user taps on the submit button, the dialog box shown in Figure 8.15 is displayed.

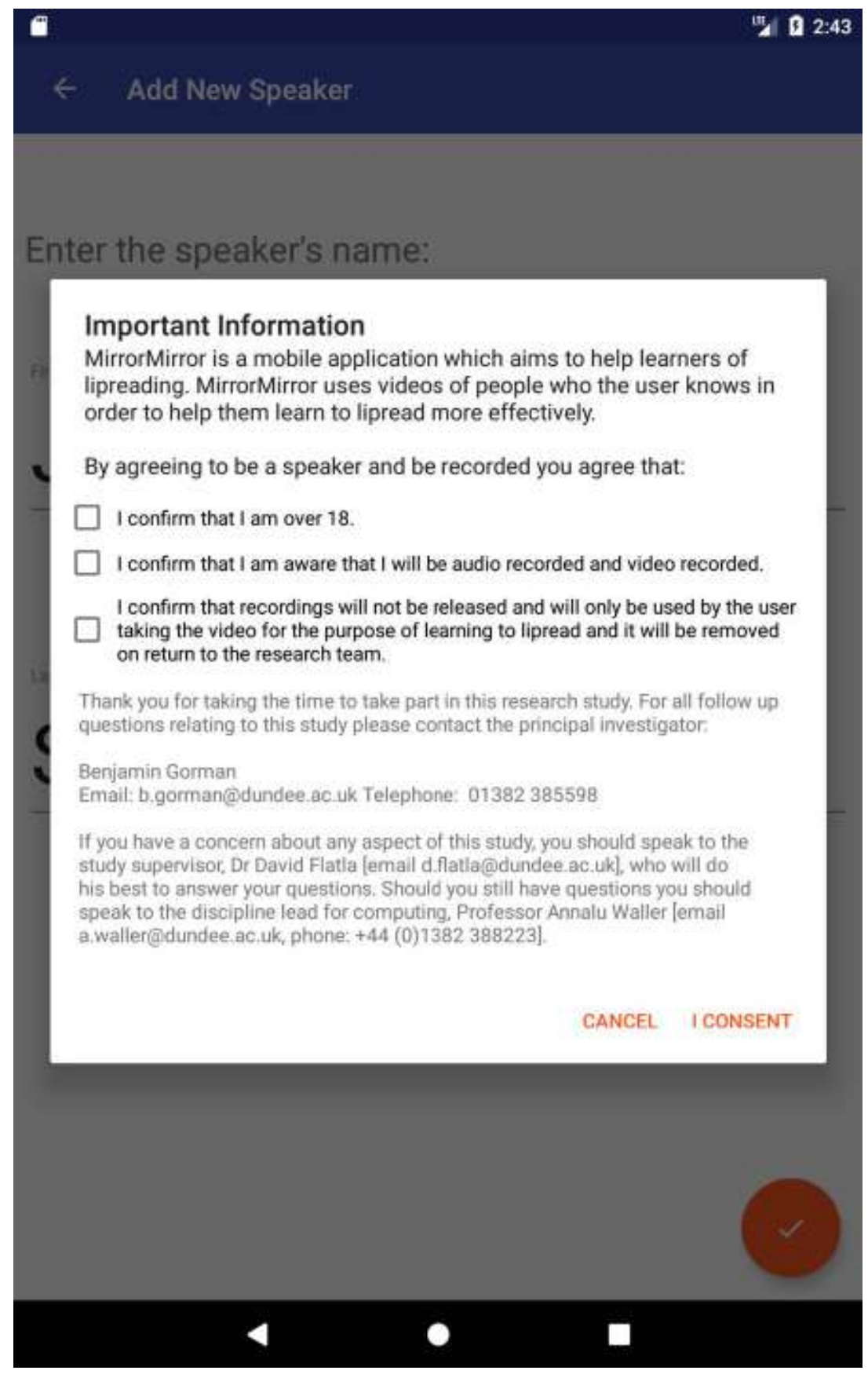

Figure 8.15: Screenshot of the consent dialog that is displayed after submitting the 'Add Speaker' form shown in Figure 8.14. If the user does not accept all checkboxes, adding the speaker is cancelled.

### 8.4.6 'Practice' Tab

There are two practice modes available in the 'Practice' tab as shown in Figure 8.16: 'Lipshape Practice' and 'Word Practice'.

#### Lipshape Practice

Lipshape Practice is a multiple choice quiz game where the user selects the word they think the speaker has spoken in the video. Lipshape practice chooses a random video from a selected lipshape, and two random words are presented along with the correct word as shown in Figure 8.18. The user selects an answer and is given feedback whether they are correct before the next video is presented. Lipshape practice shows a minimum of one trial and a maximum of ten trials.

##### Lipshape Practice Setup

The lipshape practice setup activity allows for setting parameters for the practice session as shown in Figure 8.17. The user can select the lipshape they wish to practice on, or they can practice on all lipshapes. When 'All Lipshapes' is selected, the answers can be from any other lipshape, which make the challenge easier (because there is greater variety in the multiple choice options). The user can also choose videos from a specific speaker or from all speakers. Finally the user can select to have audio on or off. Once the user has selected the parameters, they press the play button to begin the session. If there are not enough videos in the library for the session an error message will appear.

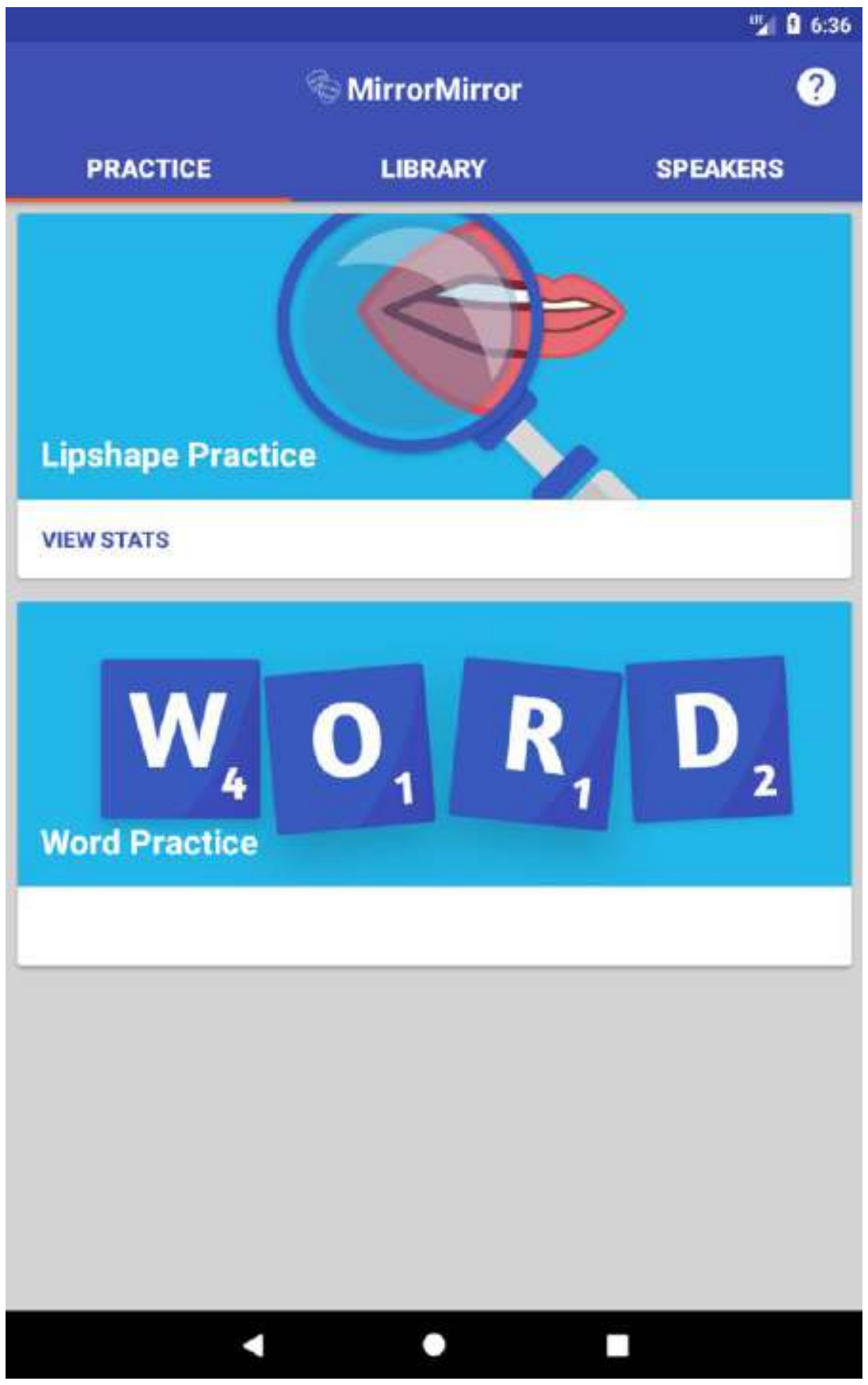

Figure 8.16: Screenshot of the 'Practice' tab showing CardViews for 'Lipshape Practice' and 'Word Practice'. On the 'Lipshape Practice' card there is a button for the 'View Stats' activity.

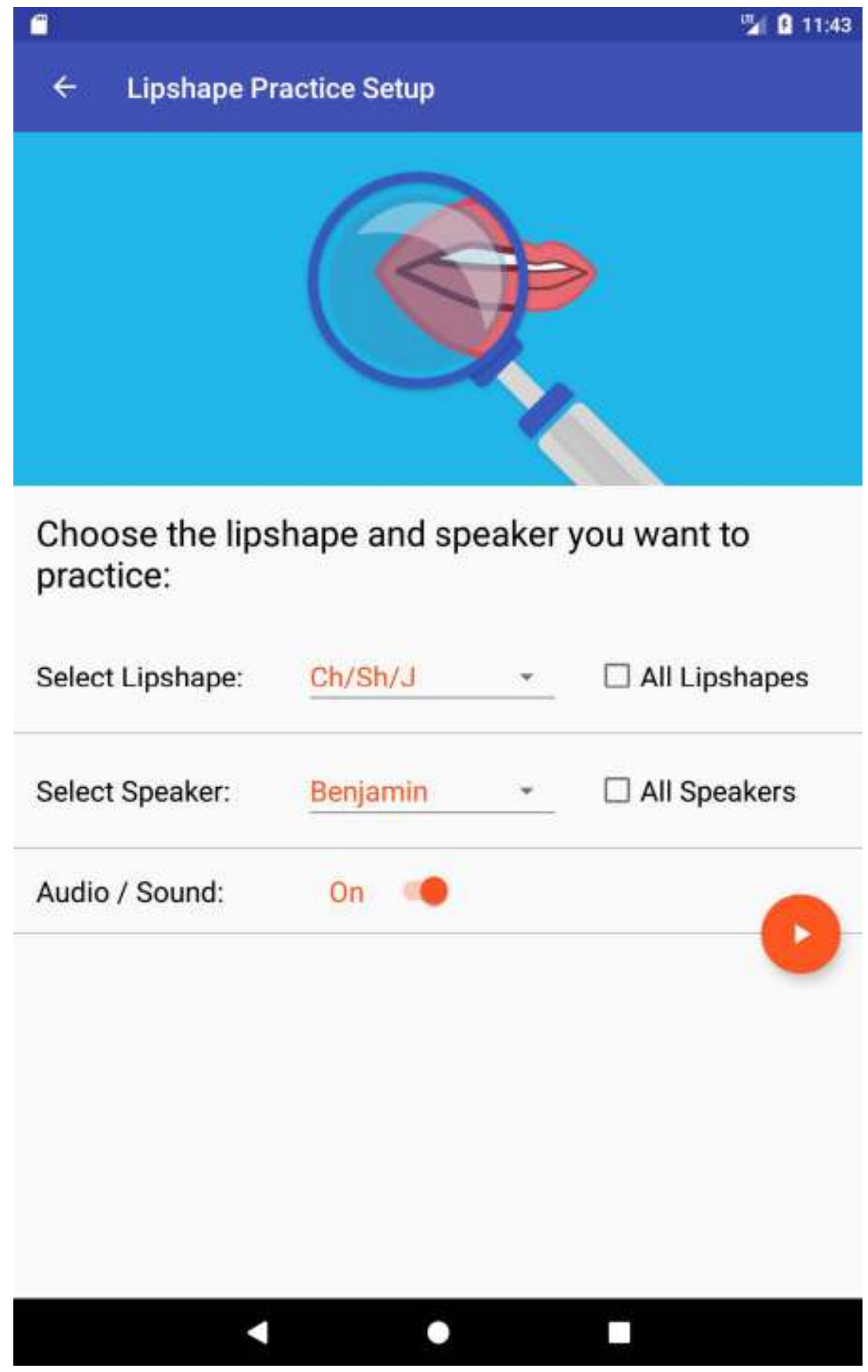

Figure 8.17: Screenshot of the 'Lipshape Practice' setup activity, which has three settings: 1) a dropdown list of lipshapes and a checkbox for 'All Lipshapes', 2) a dropdown list of speakers and a checkbox for 'All Speakers', and 3) an audio slider with a text label displaying if audio is on or off. Tapping on the play button begins the 'Lipshape Practice' activity.

### Lipshape Practice Session View

The ‘Lipshape Practice’ session view displays the trial video in a VideoView as shown in Figure 8.18. The video plays automatically and can be replayed by pressing the play button. A progress bar and numerical indicator displays the trial number and the progress through the practice session. At the bottom of the video are three buttons displaying three words, one of which is the correct answer. When the user taps on a word they are shown a ‘correct’ or ‘incorrect’ message (as shown in Figure 8.19 and Figure 8.20 respectively).

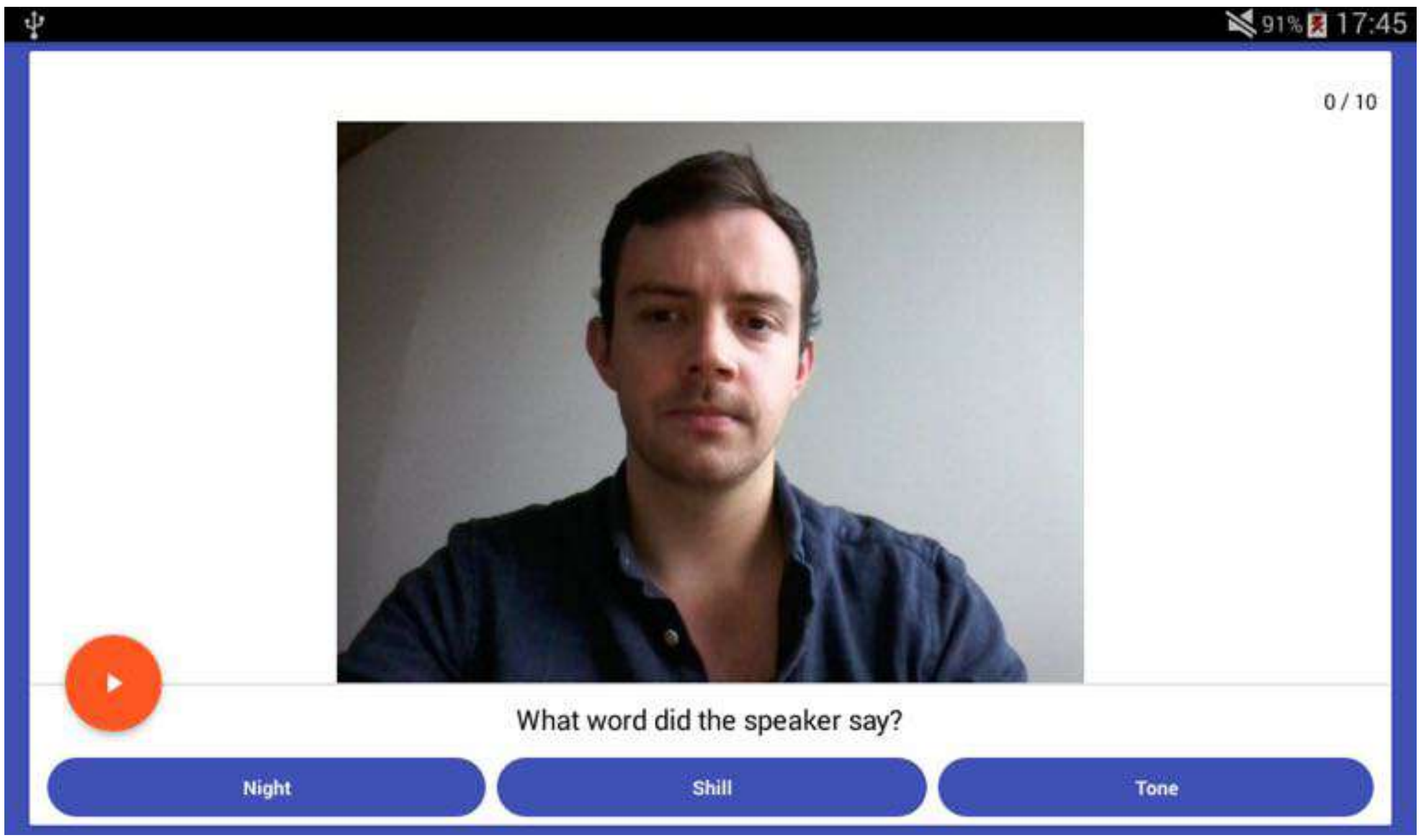

Figure 8.18: Screenshot of the ‘Lipshape Practice’ session that displays the current video in a VideoView. The video plays automatically and can be replayed by pressing the play button. A progress bar and numerical indicator displays the video number and the progress through the practice session. At the bottom of the video are three buttons displaying three words, one of which is the correct answer.

### Lipshape Practice Results

The ‘Lipshape Practice’ ResultView displays a table of the videos that lists each video plus the correct answer, the user’s answer, and the result as shown in Figure 8.21.

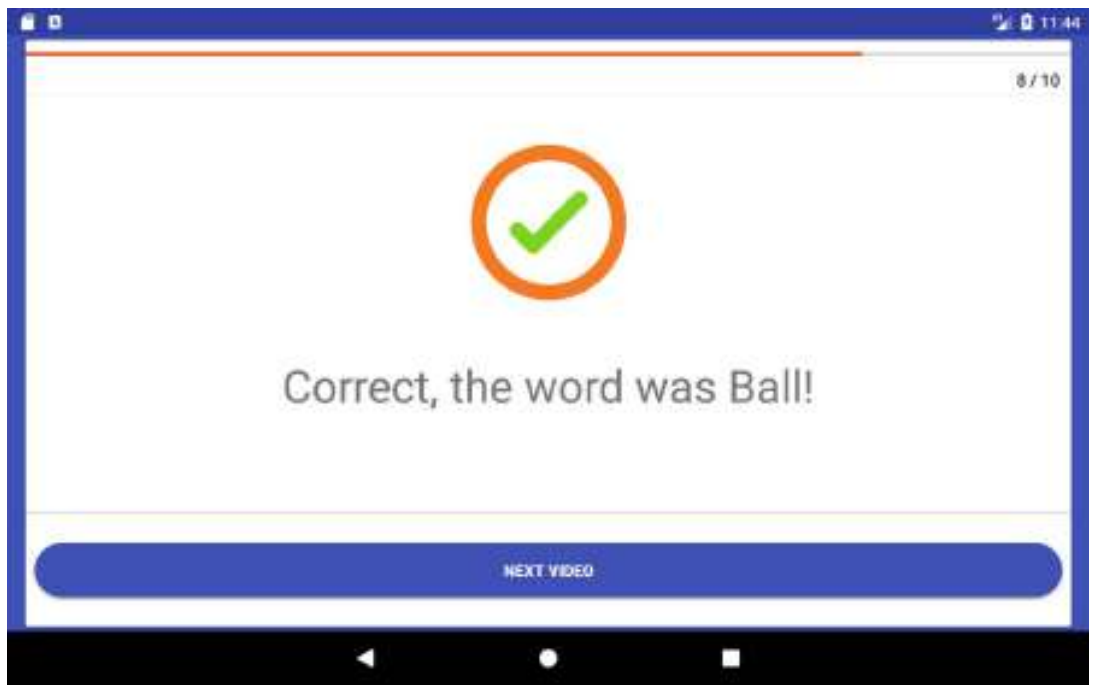

Figure 8.19: Screenshot of the ‘Lipshape Practice’ activity result card for a correct response.

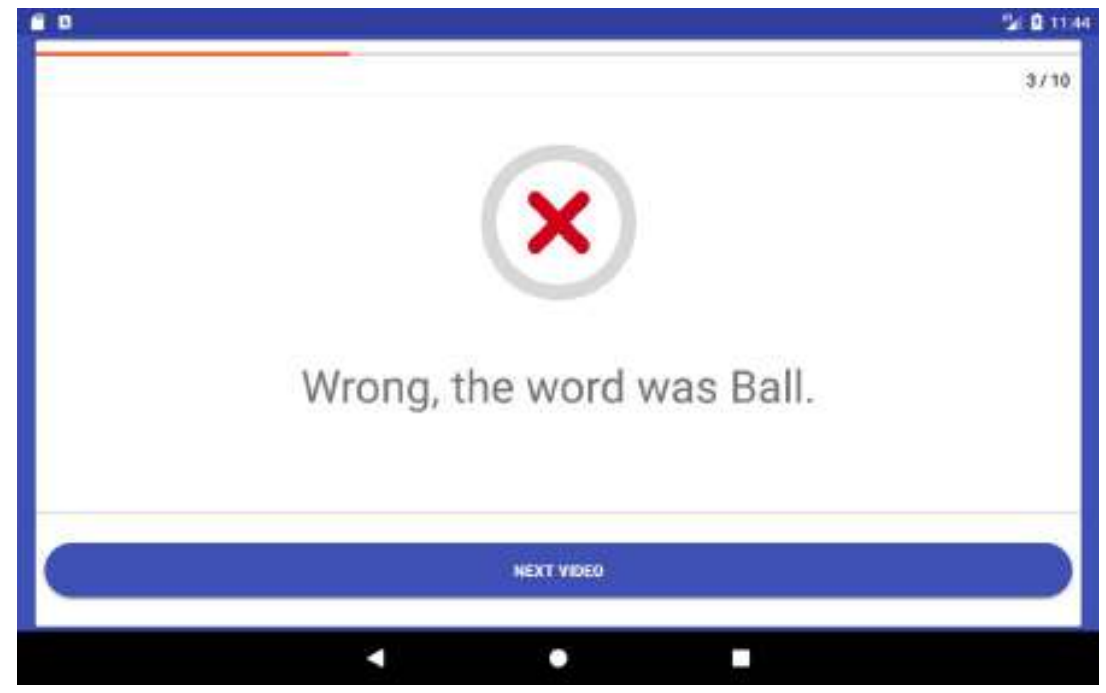

Figure 8.20: Screenshot of the ‘Lipshape Practice’ activity result card for a incorrect response.

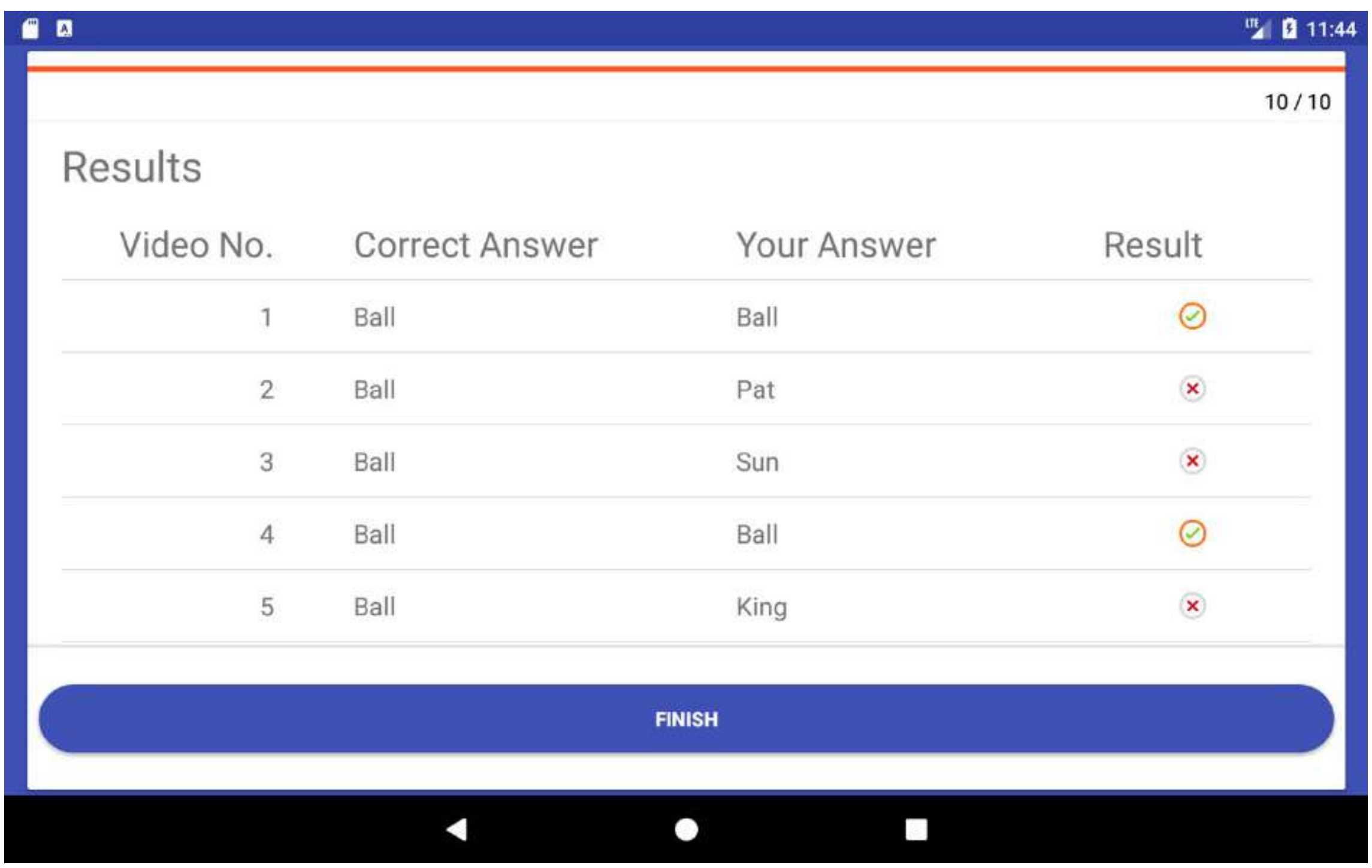

Figure 8.21: Screenshot of the ‘Lipshape Practice’ ResultView that displays a table of the videos, showing the correct answer, the user’s answer, and the result.

### View Statistics View

The user can view details of their previous Lipshape Practice sessions by tapping on the 'View Stats' button under the Lipshape Practice CardView as shown in Figure 8.16. The 'View Statistics' screen displays the date, speaker(s), lipshapes(s), audio status and the results for each session as shown in Figure 8.22.

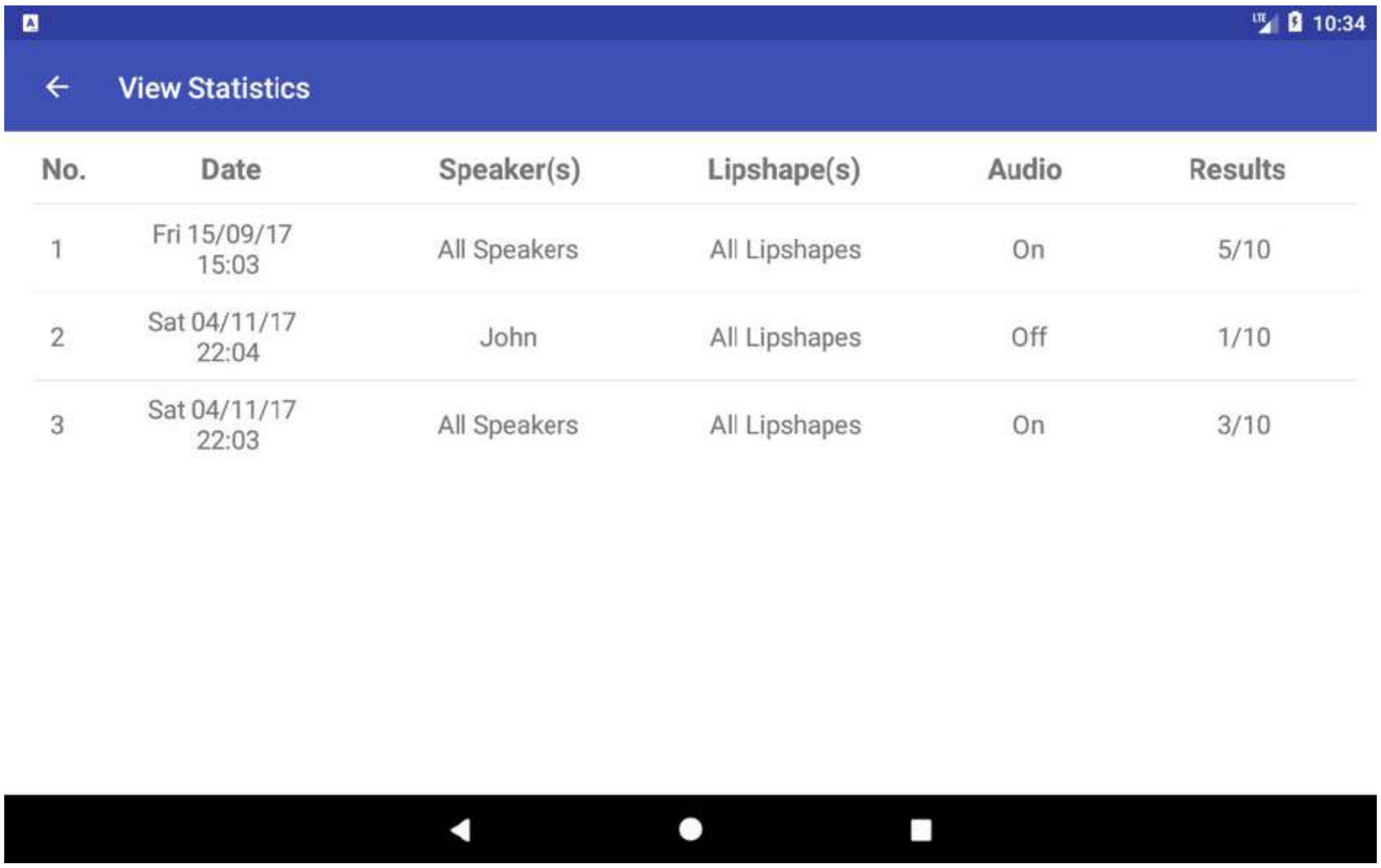

Figure 8.22: Screenshot of the 'View Stats' view that displays a table previous lipshape practice sessions.

## Word Practice

Word Practice allows for all videos from a selected word to be played in sequence, which allows for a quicker and more focused practice session than the user scrolling through their video library.

### Word Practice Setup

The user can select the word and if they wish to practice with a specific speaker or all speakers as shown in Figure 8.23. The user can also choose to have the audio for the videos on or off.

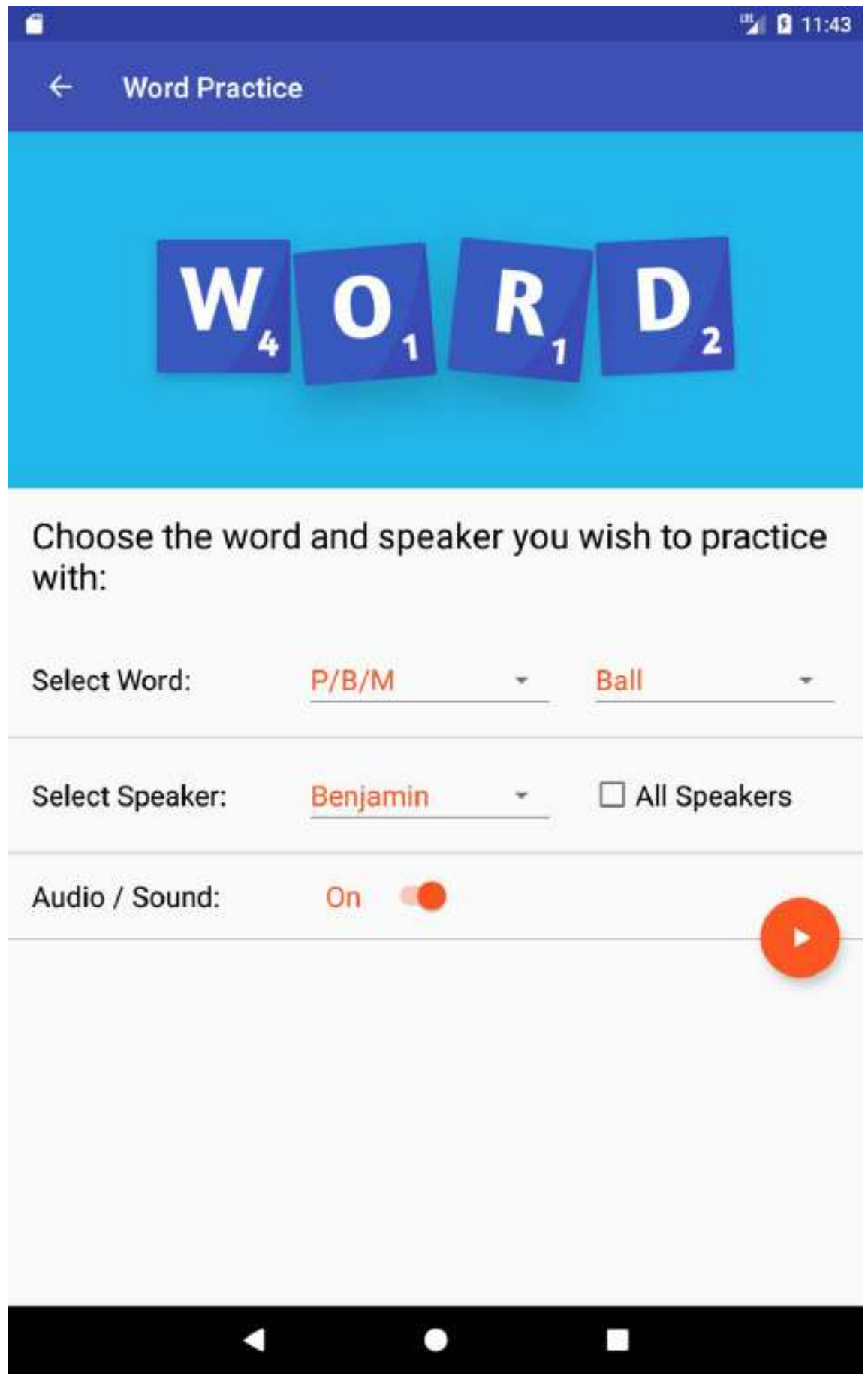

Figure 8.23: Screenshot of the 'Word Practice' setup activity, which has three options: 1) a dropdown list for lipshapes and words, 2) a dropdown list of speakers and a checkbox for 'All Speakers' and 3) an audio slider with a text label displaying if audio is on or off. Tapping on the play button begins the 'Word Practice' activity.

### Word Practice Session View

The 'Word Practice' session view displays the current video in a VideoView as shown in Figure 8.24. The video plays automatically and can be replayed as many times as desired by pressing the play button. A progress bar and numerical indicator displays the video number and the progress through the practice session. At the bottom of the video is a button that allows the user to proceed to the next video.

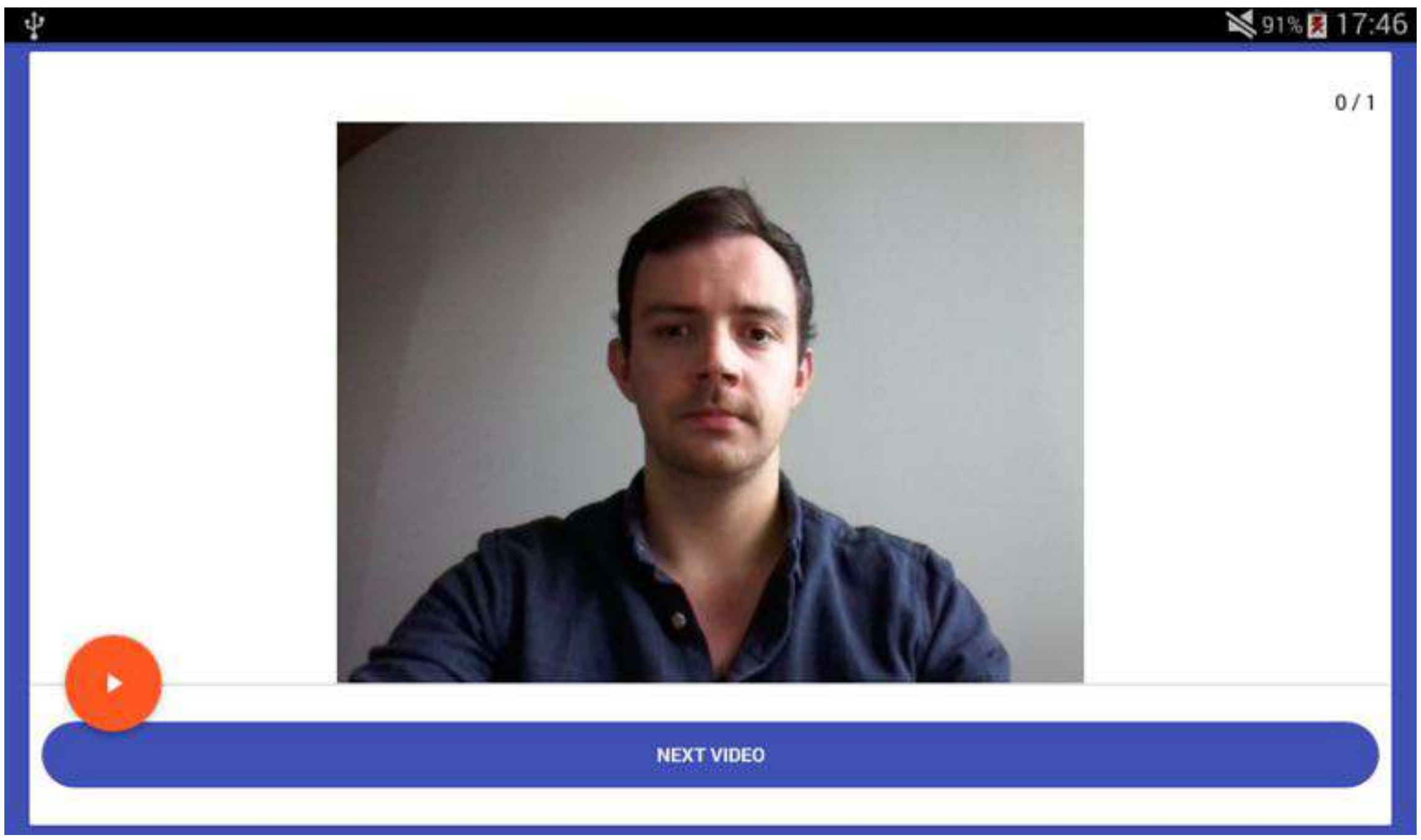

Figure 8.24: Screenshot of the ‘Word Practice’ activity, which displays the current video in a VideoView. The video plays automatically and can be replayed by pressing the play button. A progress bar and numerical indicator display the video number and the progress through the practice session. At the bottom of the video is a button that allows the user to proceed to the next video.

## 8.5 Case Study Evaluation

To evaluate MirrorMirror, I decided to conduct case studies with speechreading students from the classes taught by the speechreading tutors who took part in the interviews presented in Chapter 4. Case study research involves "intensive study of a single unit for the purpose of understanding a larger class of (similar) units…observed at a single point in time or over some delimited period of time" [63]. As such, these case studies provide an opportunity to gain a deeper understanding of how MirrorMirror could be used by speechreading students compared to a lab based study as it provides a more realistic understanding of how MirrorMirror would actually be used.

### 8.5.1 Procedure, Apparatus & Design

The case study evaluation of MirrorMirror was comprised of three stages: 1) a briefing, initial questionnaire, and tutorial session, 2) a week-long in-the-wild-deployment, and 3) a post-deployment discussion session.

#### Stage 1: Briefing and Tutorial Session

I met with each participant and explained the information sheet (shown in Appendix E.5). The participant was then asked to sign the consent form before completing the pre-deployment questionnaire.

##### Pre-deployment questionnaire

The questionnaire had two sections (shown in Appendix E.3). The first section contained nine questions and was used to gather basic demographic information; age, sex, highest level of education, level of computer literacy and details about each participant's hearing. The second section was used to understand participants' daily experience of speechreading and contained four questions: 1) "Please rate your lipreading ability", 2) " How long have you been in lipreading classes?", 3) "Do you practice lipreading outside of classes?", '4) 'If yes, how do you practice lipreading at home?". Finally, participants were also asked: "Do you own a mobile device?".

##### Tutorial Session

During this session, I introduced and explained each feature of MirrorMirror. To help participants remember how to use MirrorMirror I also provided a printed copy of the tutorial.

## Stage 2: In-The-Wild-Deployment

I supplied a mobile device with MirrorMirror pre-installed to participants and asked them to use MirrorMirror for daily speechreading practice for one week. At this stage, MirrorMirror included six typical lipshapes that are practiced in lipreading classes (P/B/M, S/D/T, K/G/N, Ch/Sh/J, L/N/K and Z/T/S) [106]. I added three words to each lipshape (e.g., Pat, Bat, Mat for P/B/M), and a video of each word spoken by me, totalling 18 videos. During the deployment, MirrorMirror recorded details of each lipshape practice session, including the date, time and results of each trial.

#### Device

Participants were supplied with a Samsung Galaxy Tab 3 (T210R, White, Wi-Fi) tablet that was rooted and running a slim ROM of Android KitKat. The Samsung Galaxy Tab 3 has a 7 inch 1024 x 600 pixel screen, a 1.3-megapixel front facing camera, and a 3 -megapixel rear facing camera.

#### Task List

During the course of the one week in-the-wild deployment, the participants were asked to complete the following tasks:

- “Add at least three new words to each lipshape.”
- “At a minimum, we ask you to try and practice at least three lipshape per day using the ‘Lipshape Practice’ feature.”
- “Add at least three new speakers to your library (speakers can be family, friends, colleagues, anyone else you see on a regular basis e.g., coffee shop worker, newsagent etc).”
- “Record at least one video for each lip shape for each new speaker.”

## Stage 3: Post-deployment Discussion and Results Gathering

After the one week in-the-wild deployment, I met with each participant to discuss MirrorMirror in a questionnaire-based structured interview (questionnaire shown in Appendix E.3), which was audio recorded for later transcription. The participant was then debriefed about the purpose of the study and given an opportunity to ask any final questions. After this session, I took back the device and downloaded the usage statistics, before removing the app and all participant data from the device (as required by the research ethics review board).

### 8.5.2 Participants

All participants had to be above the age of 18, and be currently enrolled in a speechreading class. Speechreading tutors were contacted through existing contacts, and were asked to pass on details of the study to their students. Students were then asked to contact the researcher via email if they were interested in taking part. As participants were expected to have limited hearing, tasks for stage 1 and stage 3 were conducted in a location that was chosen by the participant that suited their hearing needs. Additionally, all study material for stage 1 and 3 were presented in written form, as well as verbally. I recruited three participants (mean = 67.66 years, SD = 11.84, two males). The participants' backgrounds, hearing loss history, and speechreading experience were varied:

**P1, (Male, 74)**

P1 is a retired teacher and his highest education level is university. He self-reported having moderate-to-severe hearing loss for 40 years due to ageing[c]. He wears one in-the-ear (ITE) hearing aid in his right ear. He has been in speechreading classes for over a year and rates his speechreading ability as 'Fair'. He reported that he practices outside of class by watching television, in particular he said "Usually BBC news bulletins". He owns a Samsung Android phone and rates his computer literacy as "Good".

**P2, (Female, 54)**

P2 is a librarian and her highest education level is college. She self-reported having profound hearing loss for an unknown amount of time, with an unknown cause. She wears one cochlear implant in her left ear. She has been in speechreading classes for two years and rates her speechreading ability as 'Fair'. She reported that she does not practice outside of class. She owns an iPhone and rates her computer literacy as "Good".

**P3, (Male, 75)**

P3 is a retired medical physicist and his highest education level is university. He self-reported having severe hearing loss for 20 years due to exposure to loud noise. He wears behind-the-ear hearing aids in each ear. He has been in speechreading classes for two years and rates his speechreading ability as 'Good'. He reported that he does practice outside of class by speechreading as often as he can. He owns an Android tablet, and rates his computer literacy as "Excellent".

[c]Although early-onset age-related hearing loss (as evident with P1) is rare [62], it is not uknown [86].

### 8.5.3 Task Results

#### Speakers

All participants added multiple speakers into their speaker library, as requested in the task list. The participants mainly added close family members as speakers, which was expected given the brief period of the evaluation. **P1** added two speakers (his wife and his son), **P2** added four speakers (herself, her husband, her daughter, and a colleague), and **P3** added two speakers (himself and his wife).

#### Words

All participants added new words into their word library, as requested in the task list. P1 added 19 new words, P2 added 18 new words, and P3 added 36 new words. The words were mostly added evenly across lipshapes (as shown in Table 8.1). The words typically followed the pattern set up in the examples where the words were similar aside from the initial letter or syllable (e.g., P2 added 'Patter', 'Batter', and 'Matter'). Some of the words added, such as 'Norman' and 'George' by P1 may hold additional meaning (e.g., friends or relative names).

| **Lipshape** | **P1** | **P2** | **P3** |
|---|---|---|---|
| P/B/M | 3 | 3 | 6 |
| S/D/T | 3 | 3 | 6 |
| K/G/N | 4 | 3 | 6 |
| Ch/Sh/J | 3 | 3 | 6 |
| L/N/K | 3 | 3 | 6 |
| Z/T/S | 3 | 3 | 6 |

Table 8.1: Number of words added to each lipshape by each participant.

### Videos

All participants added new videos into their video library, as requested in the task list. However, the participants varied greatly in the number of videos they chose to add to their library. P1 explained in his interview that he found the pre-populated videos to be quite useful, therefore this could explain the smaller number of videos he chose add. P1 was also the only participant to not record videos of himself using MirrorMirror. **P1** added 19 new videos, **P2** added 145 new videos, and **P3** added 71 new videos (shown in Tables 8.2, 8.3, and 8.4 respectively).

| Speaker | No. Videos Recorded |
|---|---|
| P1 | 0 |
| Author | 1* |
| Wife | 12 |
| Son | 6 |
| **Total** | 19 |

Table 8.2: Number of videos recorded of each speaker during the evaluation by P1. *This video was recorded during the tutorial session.

| Speaker | No. Videos Recorded |
|---|---|
| P2 | 36 |
| Daughter | 36 |
| Husband | 36 |
| Colleague | 37 |
| **Total** | 145 |

Table 8.3: Number of videos recorded of each speaker during the evaluation by P2.

| Speaker | No. Videos Recorded |
|---|---|
| P3 | 54 |
| Wife | 17 |
| **Total** | 71 |

Table 8.4: Number of videos recorded of each speaker during the evaluation by P3.

### Lipshape Practice Sessions

Overall, participants made great use of the Lipshape Practice mode, often using it multiple times per day during the evaluation. **P1** completed 14 Lipshape Practice sessions that included 76 individual trials (shown in Figure 8.25). **P2** completed 72 Lipshape Practice sessions that included 706 individual trials (shown in Figure 8.26). **P3** completed 43 Lipshape Practice sessions that included 367 individual trials (shown in Figure 8.27). This data is summarised in Table 8.5.

| Participant | No. Sessions | No. Trials | No. Correct Trials | No. Incorrect Trials |
|---|---|---|---|---|
| P1 | 14 | 76 | 62 | 14 |
| P2 | 72 | 706 | 462 | 244 |
| P3 | 43 | 367 | 234 | 133 |

Table 8.5: Number of Lipshape Practice sessions, trials, correct , and incorrect trials completed by each participant.

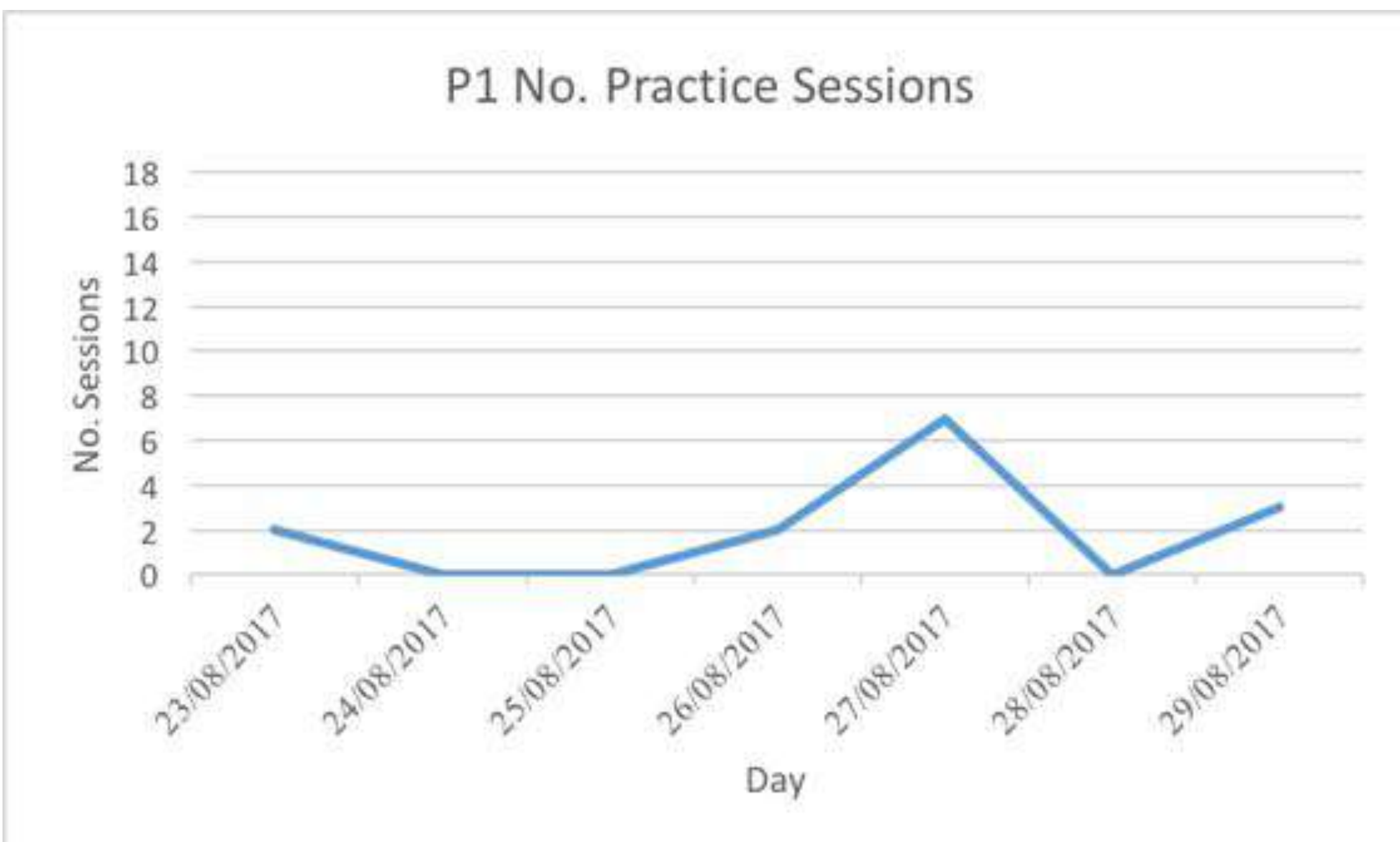

Figure 8.25: Number of practice sessions per day completed by P1.

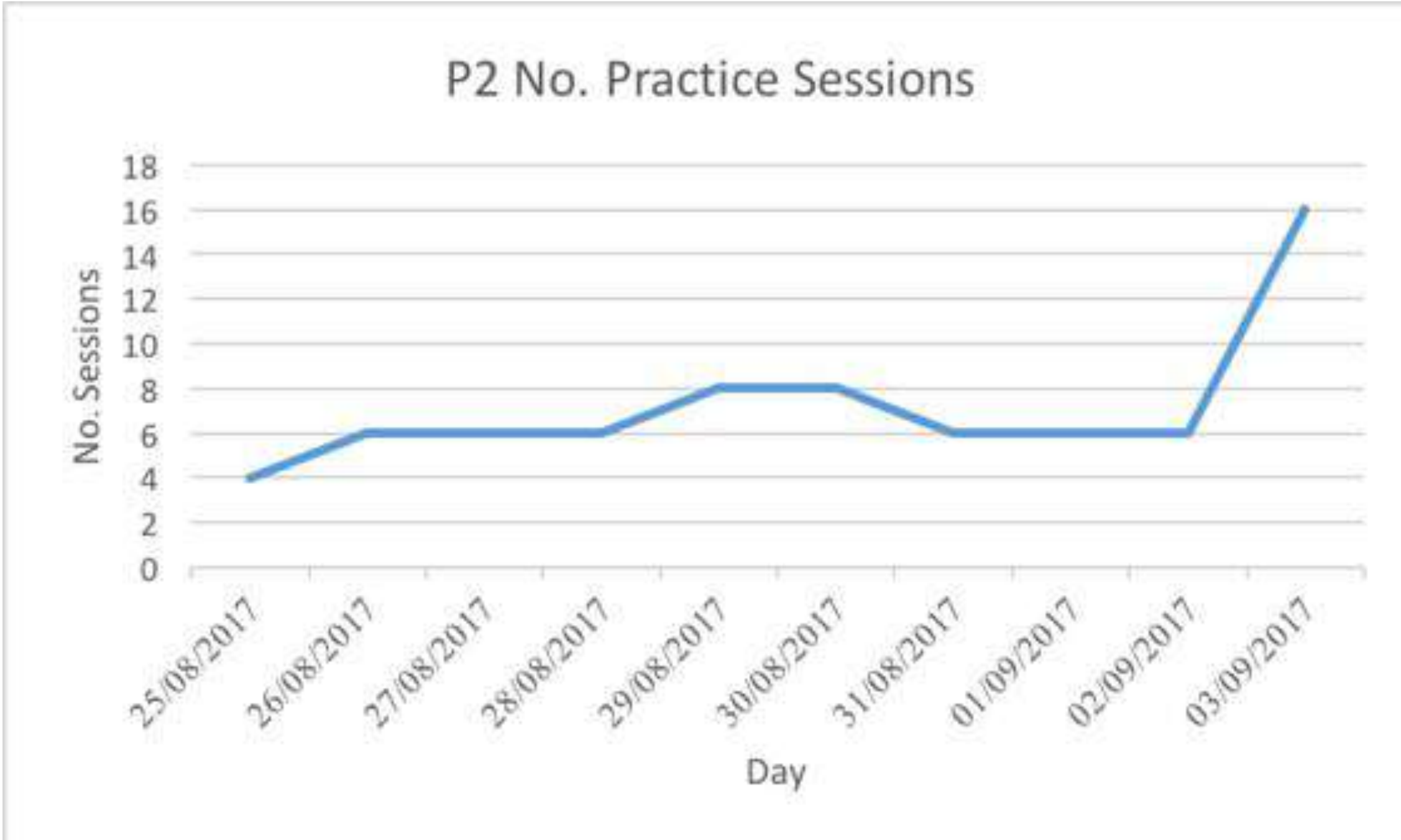

Figure 8.26: Number of practice sessions per day completed by P2.

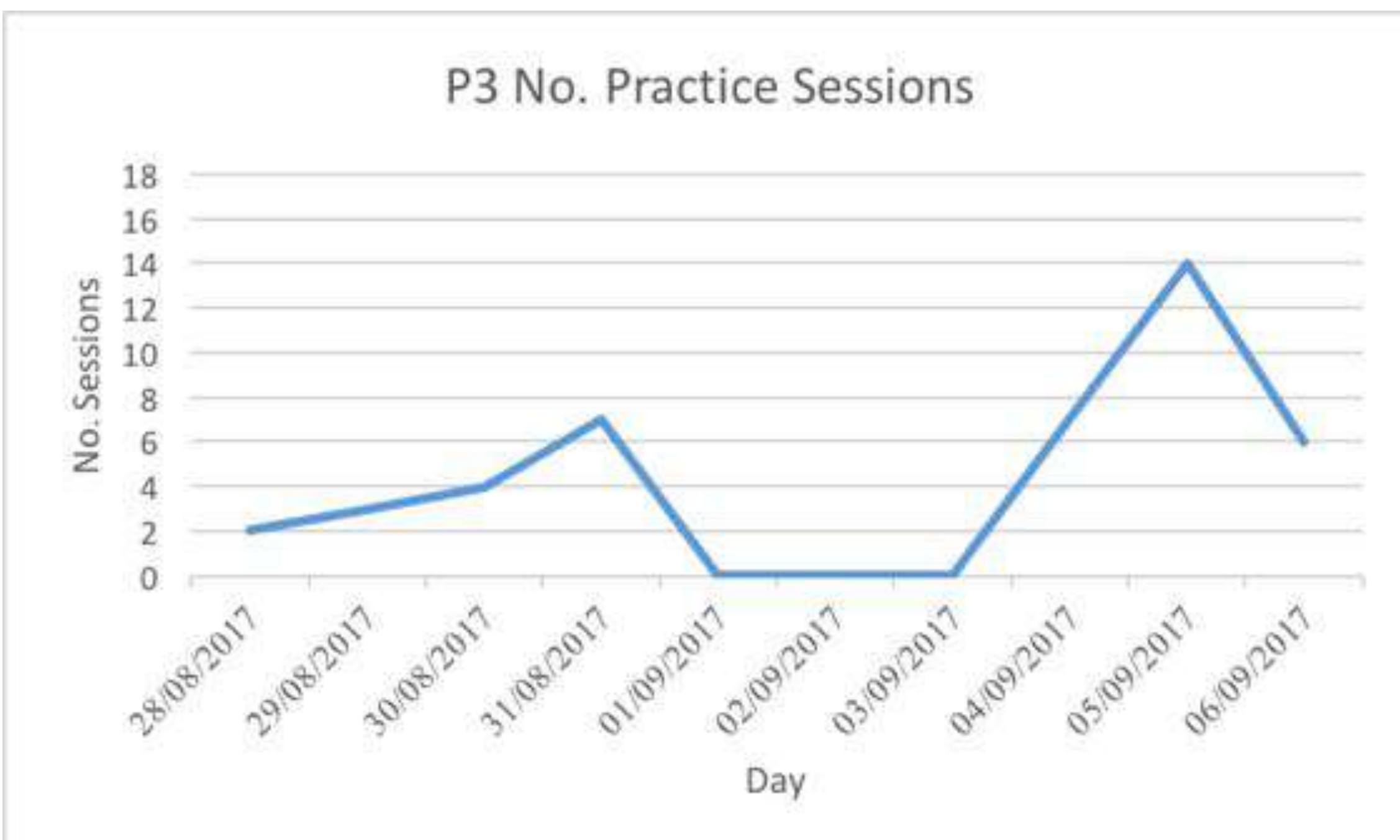

Figure 8.27: Number of practice sessions per day completed by P3.

### 8.5.4 Interview Findings

#### P1

When asked about his overall impressions of MirrorMirror, P1 said that he felt it was *"quite easy, straightforward…and quite helpful"*. The aspect he liked the most was the lipreading practice mode and being able to see *"other peoples faces sort of close up like yourself, my son and my wife"* and that he found it *"interesting to be able to try [and] work out which word was being used"*. When asked how his speakers felt about MirrorMirror he said they *"…thought it was a very good idea"*, and that his son found it easier than his wife adding that they had *"No problem"* with being recorded. When asked if he had practiced speechreading with them before, he said *"no I maybe tried once with my wife…maybe lasted a minute or two but really no I would say basically no"*. These responses indicate that MirrorMirror allowed P1 to practice his speechreading on those closest to him more than he has in the past. However, he did add that when his wife and son recorded videos, they *"both started exaggerating the word[s]"*, indicating preserving naturalness of speech may be an outstanding issue for MirrorMirror.

When asked if he thought his speechreading would be improved using MirrorMirror he said *"Yes…I think so yeah…because I wasn't practicing often enough but that would be a nice easy way to practice particularly if there were videos of people I didn't know how they spoke"*. When asked if he would continue to use MirrorMirror (if he could), he said he would and there was not anything that he did not use or dislike about the application. When asked what the most important aspect of MirrorMirror was he said *"I think the videos, because I don't know your lip movements"*. When asked if he thought that MirrorMirror could be more useful if used on people's lipshapes he was not familiar with, he said that it would be *"Harder but I think more valuable…I don't mean impossibly harder but the lack of familiarity with their speech patterns would be better"*.

#### P2

When asked her overall impressions of MirrorMirror, P2 said that *"Once it was set up, it was nice and easy."* She added that *"It was very interesting, that even after practicing the words…I was still getting them wrong."* She believed this was because some of the words appeared visually similar; *"I don't think I got mat once…because it was too like bat and pat"*. This was mainly with her husband who spoke these words *"very similar."* However, she added that *"It was also interesting seeing my daughter, I found her lipreading quite easy…I think I got more of her ones right than I did my husband."*

When asked what she disliked, P2 found it cumbersome to choose words she wanted to redo, as she had to go into the practice and set all the options again. When recording videos, she said it would be better if you could record a batch of videos at a time, with MirrorMirror remembering you had just recorded a video with a speaker and a certain lip shape. She also said that the angle of the device while recording influences the difficulty of the video, for instance her daughter *"had [the tablet] slightly lower down [so] I could actually see her lip moving sometimes whereas everybody else was much more face on."*. She added that *"I think you probably need to say 'have it at the level of your head' "*. She also said that because the camera on the tablet provided was not centrally located, it can be difficult to record a video because you have to *"reach across the camera to put it on and off"* and that sometimes *"because you are looking at [the camera]"* while recording *"it is quite difficult"* adding however *"once you got used to it, it was fine"*.

She discussed never using mirror practice before, but felt that looking at her own mouth shapes with MirrorMirror was *"As difficult as everybody else"*. On practicing single words, she said that it is *"the most difficult thing for [speechreaders] to do anyway, so if you said you know 'the bat feeds at night' then I know you are not talking about a mat or you are not talking about pat"*. However, she added that practicing *"one word, heightens it, and makes it really obvious that I'm not picking it up."* and that it was *"really good for practice having one word."*

When asked if MirrorMirror could improve awareness in others of her need to speechread, she said that that her colleague was now more *"aware of [P2's] hearing loss but she [was] maybe not aware of the 'non hearing' "* that is required while speechreading and *"was [now] more aware that [speechreading] is really difficult"*.

When asked if she thought her speechreading would be improved using MirrorMirror she said *"Definitely…yesterday, I did the whole lot straight through…and did the worst ones again and…instead of getting four [correct] I was getting five or six."*. She also said that *"I think if there was something in particular that you were going to, an event or something and you knew you were going to be asked certain questions [MirrorMirror] would be really handy."*. Overall, she felt that MirrorMirror *"was nice to get confidence, and there was times when I'm just like saying 'you know I am saying that!"*. When asked if she would continue to use MirrorMirror she said she *"probably would."* but that more default videos would be useful.

## P3

When asked about his overall impressions of MirrorMirror, P3 said that *"[he] found it very interesting."* adding that he *"liked the basic idea of it. It's different from what [his] training in lipreading has been"*. When asked what he liked, he said it was *"interesting with [MirrorMirror]* *that words like parked and packed and I was surprised I could actually tell the difference…I got the 'r'…and that made me think that this was useful"*. When asked what he disliked he said *"nothing fundamental"* but echoed what P2 said, that it was repetitive having to enter the lipshape and speaker when adding multiple videos. He also said that because the camera on the tablet is not centered, it was important how the speaker held the tablet when recording a video and that it would be better if the tablet was on a stand or fixed position. Furthermore, he said that it was *"very important whether a person starts with their mouth closed or open"* because with the mouth open he *"thinks you are opening your mouth to speak as the first syllable…and that is confusing"*. He felt that speakers needed more instruction before taking a video to ensure all videos were consistent because if he knew the speaker was *"starting with [the] mouth closed [he] could lipread"* as normal.

He said that his wife had *"no problem"* with being recorded and that he could *"lipread her from [MirrorMirror] much better than [he] normally lipread her"*. When asked if his wife became more aware of his need to speechread through use of MirrorMirror, he said that *"she may have found it quite instructive because she was watching"* when he was practicing and that he *"was getting them wrong"* even though she could hear the audio. On watching his own videos, he found it *"quite revealing watching [himself] recorded"*; that he did not think he was a very good person to lipread and that he will try to improve this during speechreading classes.

He said that he practiced lipshapes *"always within groups because it's too easy when you do it across groups"* and that *"you do come to learn…well, he never said two of these words"* so sometimes he knows what someone has said without needing to speechread as that speaker had never recorded some of the words presented as multiple choice answers. He said that he did not use the word practice feature as he *"didn't feel like it was practical"* however he did watch individual videos (which is very similiar). He also did not look at his statistics because he felt that he *"didn't feel they would be very encouraging"*. Although earlier in the interview he mentioned that *"the best [he] ever got was 9/10"*. When asked if he thought MirrorMirror would improve his speechreading in the long term, he said he *"thinks it might"* and that he would continue to use MirrorMirror if it had more default videos of other speakers.

## 8.6 Discussion

### 8.6.1 Summary of Findings

Through the postal questionnaire with 59 speechreading students presented in Chapter 6, I identified that students are not currently supported for practice outside of class. Using the findings, I elicited requirements for a new SAT called MirrorMirror that addresses the limitations of current SATs by allowing users to capture and practice with videos of people they frequently speak with. Third, I evaluated MirrorMirror with three speechreading students who felt it enabled them to target their practice on people, words and situations they encounter daily.

### 8.6.2 Limitations

The evaluation of MirrorMirror had a relatively small number of participants that may not be representative of the wider population of people with hearing loss. Recruiting highly specialised participants for evaluations that require face-to-face contact has always been a challenge in accessibility research [120], and this work is no exception. This could be addressed by a deployment of MirrorMirror to a more diverse set of participants that is representative of the wider population of those with hearing loss. However, this should only occur after I extend MirrorMirror to address the feature-specific requests from the participants (e.g., simplified batch video recording, video library sharing – see below) I would then deploy MirrorMirror for longer to a more diverse set of participants.

Although the results indicate that MirrorMirror could improve participants' ability to practice outside of class, I recognise that the evaluation does not evaluate MirrorMirror's effectiveness in improving overall speechreading ability. This would be difficult to show (especially as participants only used it for a week) but I have shown that MirrorMirror has the potential to improve speechreading ability because participants reported that they would use it.

A limitation of MirrorMirror that participants reported was that as speakers capture the videos, sometimes these videos are not the best quality (e.g., bad angle). It is possible to reduce this by including a tutorial within the application for new speakers, informing them how to capture the best possible video. However, when speechreading, it is not always possible to have perfect conditions, as I found from the questionnaire responses on speechreading challenges; although they may increase the difficulty of practice, imperfect videos could actually be beneficial to the user's speechreading acquisition.

A second limitation of MirrorMirror is that the value of practice is directly related to the number of videos captured; users may become familiar with certain videos. This could be addressed by allowing users to share their video libraries with one another so that the number of possible videos to practice on is increased.

### 8.6.3 Generalisations & Extensions

All of the participants discussed that being able to practice sentences with MirrorMirror would be valuable, as it adds context to the practice. MirrorMirror can easily be extended by adding a 'Sentence Practice' mode to the 'Practice' tab. However, this will require revisiting how speakers will record videos, as recording sentences versus words will increase the difficulty for speakers. In addition, MirrorMirror could also be extended through 'Context Practice', in which words and speakers could be tagged with a scenario such as a coffee shop or a doctor's appointment.

MirrorMirror is currently built for English speechreading practice, but by updating the lipshape categories, it could be extended to support other languages. For example, French [16], German [31], and Japanese [72] each have distinct viseme-to-phoneme mappings. In addition, there are many popular ways to learn a foreign language (e.g., Duolingo (*duolingo.com*) or Rosetta Stone (*rosettastone.co.uk*)), however, immersing yourself in a new country is another way to practice [40, 82]. MirrorMirror could be adapted to help people learn by recording speakers of the target language, which could also help with learning pronunciation or region-specific dialects.

## 8.7 Conclusion

Speechreading can help people with hearing loss improve understanding during conversation, but is a challenging skill to acquire. Current Speechreading Acquisition Tools (SATs) are not adaptable to individual student needs, whereas speechreading classes are. Current SATs inflexibility is caused by: 1) a limited selection of content, 2) a limited selection of speakers, and 3) the user not being able to customise the content with the particular words, situations or people they encounter on a daily basis; speechreading classes are based around watching people speak, and are tailored to each student's needs, but currently available SATs are not adaptable.

In this chapter, I used the findings of the postal questionnaire presented in Chapter 6 to inform the requirements for a new SAT called MirrorMirror. MirrorMirror is an Android application that allows students to practice speechreading through recording, watching, and testing their speechreading using videos of people they frequently speak with.

I then described supplying three speechreading students with a tablet running MirrorMirror and asked them to use it for daily practice for one week. Participants willingly engaged with MirrorMirorr, and the findings suggest that through the use of MirrorMirror, speechreaders can effectively target their speechreading practice on people, words and situations they encounter during daily conversations.

9

# Discussion

## 9.1 Introduction

In this chapter, I will discuss implications and extensions of the research work presented in this thesis. This chapter addresses the thesis problem presented in Chapter 1 and explores how the solution presented in this thesis can be improved and extended.

## 9.2 Summary of Contributions

The main contribution of this thesis was the introduction of a novel framework that can support the development of Speechreading Acquisition Tools – a new type of technology designed specifically to improve speechreading acquisition. This framework allows for the development of SATs that reflect the teaching practices of contemporary speechreading classes. Through the development and release of SATs, people with hearing loss will be able to augment their class-based learning, or learn on their own if no suitable classes are available.

This thesis also presented a number of secondary contributions:

1) A critical overview of existing Conversation Aids and related approaches to improving speechreading, framed within the cells of the framework.
2) Novel interview data from seven practicing speechreading tutors with thematic analysis of that data.
3) Novel questionnaire data from a postal survey with 59 students from speechreading classes.

4) A description of the development and evaluation of PhonemeViz, a new SAT in the form of a visualisation that displays a subset of a speaker's spoken phonemes to the speechreader to reduce viseme confusion that occurs at the start of words. The design of PhonemeViz was inspired by the initial fingerspelling technique that was described by speechreading tutors during the interviews.
5) A description of the development and evaluation of MirrorMirror, a new SAT that addresses the limitations of current SATs by allowing users to capture (and practice with) videos of people they frequently speak with. The design of MirrorMirror was inspired by the mirror practice technique that was described by speechreading tutors during the interviews.

## 9.3 Explanation of Contributions

Through thematic analysis of interviews conducted with speechreading tutors, I identified four main themes relevant to the future development of Speechreading Acquisition Tools (SATs): *speechreading as a skill*, limited *access to speechreading*, a broad range of *teaching practices*, and mixed *attitudes to technology*.

Using these themes, I developed a novel framework to help design new SATs that are influenced by the teaching techniques and approaches reported by the tutors. In evaluating the framework, I demonstrated that it can accommodate current teaching techniques (identified by the tutors during the interviews) and existing solutions, as well as be used to design new SATs.

I then conducted a questionnaire with speechreading students to gather data from the student perspective to further enhance the design of new SATs. I identified that students are not currently well supported for practicing outside of class, and that even though 67% of participants had been in speechreading classes for over two years they still felt that their speechreading ability could be improved.

To further evaluate how the framework can support the design of new SATs, I used the findings from the interviews and the questionnaire to influence the development of *PhonemeViz* and *MirrorMirror*. User evaluations of these new SATs showed that using the framework can help design effective tools for speechreading acquisition.

### 9.3.1 Framework

To help design new SATs that are influenced by approaches and techniques used within contemporary speechreading classes, I used findings from thematic analysis of interviews with speechreading tutors. The framework consists of two dimensions (*Type of Skill* and *Amount of Information*), each with three levels (*Analytic*/*Synthetic*/*Hybrid* and *Low*/*Medium*/*High*, respectively).

To evaluate the framework I first used it to categorise every teaching technique identified during the thematic analysis. Second, I used the framework to critically reflect on existing Conversation Aids and SATs. Finally, I used the framework to support the design of three new SATs. By employing the framework in this fashion, I show that it: 1) comprehensively reflects existing speechreading teaching practice, 2) can be used to help understand the strengths and weaknesses of previously-developed solutions, and 3) can be used to identify clear opportunities for the development of new SATs to help improve speechreading skill acquisition.

Through the dissemination and adoption of the framework into the research community and assistive technology commercial sector, I foresee new technology being developed that is grounded in the teaching practices of speechreading tutors, therefore helping improve speechreading acquisition worldwide. Once in the hands of people with hearing loss, SATs will help to enhance their speechreading capabilities, increasing their conversational confidence and reducing their social isolation.

### 9.3.2 PhonemeViz

During speechreading classes, the tutor uses fingerspelling to help students recognise which phonemes within a viseme class are being spoken. Even though fingerspelling is a useful tool in classes, it does not help during natural conversation as people typically do not know how to fingerspell. As a result, participants reported that ‘words looking the same on the lips’ or (confusing visemes [54]) were one of the top challenges for them when speechreading, as there is not enough information through the visual signal to help them disambiguate visemes.

As Head Mounted Displays (HMD) (such as the Google Glass, Microsoft Hololens, and the Epson Moverio) allow extra information to be added to the visual field [143], it may be possible to enhance speechreading with this extra visual information. Inspired by fingerspelling, where each hand shape represents a letter, I anticipated that presenting textual representations of phonemes could benefit the speechreading process.

To explore this, I developed PhonemeViz – a phoneme visualisation technique designed to allow speechreaders to use their peripheral vision to attend to a text-based visualisation to help them disambiguate confusing viseme-to-phoneme mappings that occur at the start of words. In a comparative evaluation, I found that PhonemeViz resulted in participants achieving 100% word recognition (showing successful disambiguation), and PhonemeViz was well-received in subjective and qualitative feedback. The results suggest that PhonemeViz could be deployed on an HMD to improve understanding during speechreading by reducing the challenges caused by visemes.

These results demonstrate that presenting textual representations of initial consonant phonemes (a *medium amount of information*) results in a high level of accuracy in a visual-only speechreading task. This accuracy level suggests that the framework helped to design an effective SAT that was inspired by a teaching technique used within contemporary speechreading classes.

### 9.3.3 MirrorMirror

During the interviews, tutors reported that mirror practice plays a key role in speechreading training and is also recommended by Action on Hearing Loss [4, 106] for practice, as it may develop visual cue integration skills needed during speechreading [7].

However, questionnaire participants reported a low frequency of using mirror practice (57% of those who answered said 'Never'). This was explained by a number of factors: 1) they disliked watching themselves, 2) they have full knowledge of what they are saying, 3) they would over-emphasise words, and 4) that mirror practice was not akin to genuine speechreading. However, participants reported that they liked how mirror practice allowed them to learn lipshapes, compare them with others, and show visual differences between words.

Currently SATs that focus on assisting the practice of speechreading have three limitations: 1) a limited selection of content, 2) a limited selection of speakers, and 3) the user cannot customise the content with particular words, situations or people they encounter on a daily basis. To address this, I introduced a mobile application called MirrorMirror that allows speechreaders to practice lipshapes and words by recording videos of people they frequently talk to. MirrorMirror provides a multiple choice quiz game in which the user selects the word they think the speaker has spoken. MirrorMirror provides feedback on whether they are speechreading correctly, and allows them to target specific challenges and situations.

To evaluate MirrorMirror, I supplied three speechreading students with a tablet running MirrorMirror and asked them to use it for daily practice for one week. Participants willingly engaged with the application, and the findings suggest that through the use of MirrorMirror, speechreaders would be able to augment their class-based learning, or support learning on their own if no suitable classes are available. This evaluation showed that participants felt that MirrorMirror allowed them to practice their speechreading more effectively by allowing them to control the *amount of hybrid information* that was presented. This suggests that the framework helped to design an effective SAT for speechreading practice, which was inspired by a technique used by all speechreading tutors who took part in the interviews – mirror practice.

## 9.4 Is the thesis problem solved?

In Section 1.1, the problem to be addressed in this thesis was presented: *Existing tools designed to improve speechreading acquisition are not effective because they have not been designed within the context of contemporary speechreading lessons or practice*.

In this thesis, I demonstrated that the development of a framework influenced by findings from interviews with speechreading tutors allows for the development of SATs (such as PhonemeViz and MirrorMirror) that are effective in supporting the practice and use of speechreading.

The evaluation of two SATs (PhonemeViz and MirrorMirror) designed using the framework, show that it can successfully influence the development, while still including insight from speechreading students. However, the framework, along with both PhonemeViz, and MirrorMirror all have possible extensions that I will now briefly discuss.

## 9.5 Extensions to the Framework

During the evaluation of the framework I fitted existing teaching techniques into the framework dimensions. Although I fitted each teaching technique that was identified during the interviews, it is possible that there are other techniques used by other tutors that also fall within the framework. Therefore, it is necessary to conduct a larger evaluation of the framework's coverage using a greater number of speechreading tutors from other countries.

Although I have demonstrated the framework has allowed me to develop two examples of SATs that are effective, it would be necessary for an evaluation of the framework's influence to be conducted with AT researchers with no prior knowledge of speechreading or hearing loss. If novices can design an effective SAT that helps people acquire aspects of speechreading, then the impact of the framework would be demonstrated.

Furthermore, I aimed to make the framework as accessible as possible, but it still requires an individual who wants to build a new SAT to read about how each teaching technique and previously designed tool fits within the framework dimensions. The development of an online interactive tool that explains the framework would allow an individual to quickly become aware of the related work in this area of research. It would also allow for the upkeep of fitting new techniques and SATs into the framework, giving a broader representation of the space of SATs.

The framework could also be used in courses such as the SCCTL[a] and those run by ATLA[b], to provide a new way to teach each of the approaches and techniques used within speechreading classes to new speechreading tutors.

Furthermore, during the interviews with speechreading tutors, they reported that creating material for speechreading classes can take a large amount of time and effort (e.g., they reported that they would often create custom material from topics in the news). An interactive tool that takes news articles as input and then performs content analysis could be designed. Speechreading tutors would then use the framework to select the area they wanted to practice (e.g, low amount of analytic information and medium amount of synthetic information) and the tool could produce material with those constraints for the speechreading classes.

## 9.6 Extensions to PhonemeViz

In Chapter 7, I discussed how PhonemeViz could be extended by adding support for different languages and extended to help with visemes that occur within words rather than just initial phonemes. PhonemeViz would also benefit from an evaluation on its effectiveness within connected speech. Some further extensions to PhonemeViz are discussed below.

### 9.6.1 Subtitles Alternative

Typically, subtitles are a verbatim copy of what is spoken by the speaker, and are presented as text on black bars, centred at the bottom of the television screen. However, participants from the interviews and questionnaire reported that subtitles do not allow effective speechreading practice. Participants also reported that turning the subtitles off was too difficult, and that they would either read the text or watch the face.

Brown et al. [33] investigated the creation of 'dynamic subtitles', that presented text in varying positions (such as closer to an actor's face) according to the underlying video content. In an evaluation with participants with hearing loss, they found that their approach improved the overall viewing experience. Furthermore, results from eye-tracking data suggested that participants watching dynamic subtitles were closer to the baseline of participants watching without subtitles versus those watching with subtitles. PhonemeViz could be adapted and used with this ‘dynamic subtitles’ approach to allow viewers who speechread, the opportunity to practice their speechreading while they watch video content.

[a]The Scottish Course to Train Tutors Of Lipreading - `http://www.scotlipreading.org.uk`
[b]The Association of Teachers of Lipreading to Adults - `http://atlalipreading.org.uk`

### 9.6.2 PhonemeViz on a Head Mounted Display (HMD)

As discussed in Chapter 7, the ultimate goal of PhonemeViz is to deploy it using a Head Mounted Display (HMD). First, although I designed PhonemeViz with regards to this eventuality, the evaluation process did not evaluate whether PhonemeViz used on an HMD would introduce other problems such as background occlusion. Therefore, a broader evaluation of PhonemeViz on an HMD could demonstrate its value in a broad range of situations.

Second, evaluating PhonemeViz on an HMD would also eventually require an Automatic Speech Recognition (ASR) system in order to provide the information required to generate the visualisation. There has recently been significant increases in ASR performance rates, although they still remain susceptible to noise [60]. However many HMDs are fitted with cameras and could therefore make use of the visual speech information to improve accuracy rates [15, 115]. When the visual signal is used in combination with the acoustic signal, this is referred to as audio-visual-speech recognition (AVSR) [116].

For instance, Google have used an AVSR system called ‘Watch, Listen, Attend and Spell’ (WLAS) that was trained using hours of recored video footage from the British Broadcasting Corporation (BBC) and could transcribe the content with around 47% accuracy [42]. Furthermore, an AVSR system called LipNet [12] can achieve 93.4% accuracy on the GRID audiovisual sentence corpus [44]. Therefore, in the near future it might be possible to use AVSR systems to provide information for PhonemeViz using cameras on the HMD.

However, to evaluate how feasible PhonemeViz was with current ASR systems, I evaluated an open source ASR system[c] using a corpus of pre-recorded audiovisual sentences to see if PhonemeViz could be built with current ASR systems.

#### How feasible is PhonemeViz using Automatic Speech Recognition?

For the feasibility evaluation, I used a set of 84 sentences from a database of pre-recorded audiovisual sentences (CUNY sentences [27]) spoken by an American female speaker, which were provided through the supplementary materials of a paper written by Alteri et al. [8]. The sentences were judged in the original report by three independent judges and deemed to appropriately match to every day conversation. Within the 84 sentences there are a variety of statements, questions and commands. There are seven sentences of each length varying continuously from from 3 to 14 words.

[c] Pocketsphinx – `https://cmusphinx.github.io/`

I manually keyframed each sentence video (using the system described in Section 7.4.2) to provide an accurate transcription of what the speaker said. Each sentence video was then automatically transcribed by passing it through an implementation of the Pocketsphinx speech recogniser[d]. The audio was converted to 16KHz, 16bit Mono (single channel) Little-Endian files. Noise was removed from the files using Audacity[e] and then I converted each file into the .raw format used by Pocketsphinx.

I calculated errors in the automatic transcriptions using the Levenshtein distance. The Levenshtein distance is obtained by finding the cheapest way (fewest steps) to transform one string into another. Transformations are the one-step operations of (single-phoneme) insertion, deletion and substitution and cost one unit each. Before calculating the errors, I made each transcription lowercase in order to not penalise for capitalisation errors. I then normalised the Levenshtein error scores against the number of characters within each sentence.

The average word error was 3.1 units where character error was 0.28 units. Although these are large errors (especially when they do not account for the speech recogniser being offline, which would also introduce delays), PhonemeViz do not require the whole word to be correct, only that the initial Consonant phoneme is correct. Across the 84 sentences, 64% of the initial phonemes were correct. While promising this includes vowel phonemes and does not account for the delay that could be present due to wrongly recognised words being longer than expected (which would result in the visualisation being out of sync with the words being spoken). Finally, this evaluation used pre-recorded sentences and the speech recogniser was not performing in real-time.

In summary, although currently available open source speech recognition systems are not able to be used to produce accurate PhonemeViz visualisations, it is possible that the accuracy of future AVSR systems would allow them to be used for PhonemeViz.

## 9.7 Extensions to MirrorMirror

In Chapter 8, I discussed how MirrorMirror could be extended by adding support for practicing sentences and context. It would also be beneficial to conduct a larger evaluation with MirrorMirror using participants from two speechreading classes. Using a between subject design, it would be possible to evaluate MirrorMirror's ability to improve speechreading acquisition longterm. Some additional extensions to MirrorMirror are discussed below.

[d]CMUSphinx – 'Open Source Speech Recognition Toolkit'. https://cmusphinx.github.io/
[e]Audacity – Open source audio software. http://www.audacityteam.org/

### 9.7.1 Video Sharing

For the evaluation of MirrorMirror, I recorded 18 videos of myself speaking sample words for each lipshape. All of the participants reported that having these additional videos along with the videos that they recorded was beneficial. A current limitation of MirrorMirror is that the amount of practice possible is directly related to the number of videos captured; users may become familiar with certain videos.

To solve this limitation, MirrorMirror could be extended by allowing participants to share and upload videos of themselves and their speakers (if the speaker consented) to a cloud based storage service. MirrorMirror could then allow users to download videos of speakers with certain speechreading challenges such as different accents, speaking rates, and facial distractions (e.g., moustaches, beards). This would allow users to increase the effectiveness of their practice by allowing them to practice speechreading on a larger variety of speakers. This could also result in the words spoken in the videos being added to the users' word library, which would increase the challenge of practice, as there would be more potential answers within the lipreading practice sessions.

### 9.7.2 AVSR Training Datasets

As discussed earlier, research is currently investigating how to integrate the visual speech signal into speech recognition systems to improve accuracy rates [115]. These systems are typically trained on a corpus of audiovisual speech data [15]. For instance, Google's 'Watch, Listen, Attend and Spell' (WLAS) system is trained on footage of news content from the BBC [42]. However, the reported high accuracy (47%) of this system may be because it was trained and then evaluated on this highly-specific scenario (e.g., high quality footage of a trained speaker with good lighting and good audio). Other systems such as LipNet [12] also report a high level of accuracy, but again are using a high quality corpus of videos that were recorded specifically to train AVSR systems [44].

If these AVSR systems are to be used in realistic situations (e.g., to read the lips of someone speaking into a voice interface in a car from an odd angle with a high degree of noise) it is necessary to train the system with a less-than-perfect corpus of video data. However, creating such a corpus that includes a large variety of words would be labour intensive because each video needs to be segmented and then labelled correctly. The video libraries produced by each user of MirrorMirror results in an audiovisual corpus that has a large amount of speakers, saying a variety of words recorded in real-world situations. Therefore, these corpuses could be leveraged (with each speaker's informed consent) to train AVSR systems.

### 9.7.3 Person-Specfic-Viseme-Models (PSVMs)

Currently, Phonemeviz shows all of the initial consonant phonemes within each viseme class. However, it may not be necessary to supply the speechreader with redundant information in cases where they can speechread individual speech movements. For instance, even thought p/b/m is viseme class, a speaker may have a very noticeable /p/ versus /m/ speech movement.

Flatla [55] introduces the concept of Situation-Specific Modelling (SSM), which is a process where the user's performance on a sample of a task is captured and then this data is used to build a model that represents the user's abilities. Flatla introduced SSM in the context of modelling a user's colour differentiation abilities, however this approach could also be used to generate a model of a speechreader's ability to disambiguate phonemes on a per speaker basis.

Taking a user's responses to a particular speaker's videos during lipshape practice sessions in MirroMirror would provide the data necessary to generate such a model. This model could be called a Person-Specfic-Viseme-Model (PSVM) – a model that represents a user's ability to disambiguate phonemes within viseme classes for particular speakers. This model could then be used to provide more effective information for visualisations such as PhonemeViz.

## 9.8 Implications for Practitioners

The findings of this thesis demonstrates the impact that speechreading can have on the lives of people with hearing loss. Speechreading classes are currently the most accessible way for people to learn how to speechread, however teaching within classes could still be improved. From the analysis of teaching techniques presented in Chapter 4, there are large differences across classes in how students are taught.

The first important implication of this thesis for practitioners, is derived from the diversity of teaching techniques used within classes. This diversity suggests that even though speechreading tutors are trained on courses such as the Scottish Course to Train Tutors of Lipreading (SCTTL) [f], they use their own personal speechreading, and teaching experience to inform their teaching

[f] http://www.scotlipreading.org.uk/index.php/classes/

practice.

Therefore, perhaps speechreading tutors should take part in frequent training workshops. The attendees of these workshops should comprise of speechreading tutors, SCTTL instructors, and a selection of speechreading students. The goal would be for tutors and students to share insights from their classes with each other to uncover which techniques are most effective. Finally, the SCTTL instructors could use information from across workshops to inform future course development so that new speechreading tutors are teaching with the latest techniques.

The second important implication of this thesis for practitioners, is derived from the interview and questionnaire findings presented in Chapter 4, and 6 respectively. These findings demonstrate that learning to speechread is a very individual process. Therefore, speechreading tutors should look into different ways in which they can support their students individually within their classes.

## 9.9 Implications for Developers

Developers of assistive technology can use the findings in this thesis to help inspire new SATs. However, developers should be aware that fitting a technology into the framework presented in Chapter 5 does not guarantee that it will be useful for speechreading acquisition. Although the framework provides guidance for the design of new SATs, any new SAT still needs to be evaluated using speechreaders. If developers are aiming to use speech data within their designs, the use of Wizard of Oz [68] evaluations can be used to identify problems with conveying this type of information during the speechreading process as shown in Chapter 7.

For developers designing new SATs it is also important to consider how their new designs will be accepted by speechreaders. This will in part be helped by evaluating SATs with speechreaders, however it could be useful to demonstrate designs to speechreading tutors and also gather their insight. Finally, technology should be designed to support the speechreading process rather than replace it entirely. This approach is favourable as it supports their skill acquisition rather than having the user rely on the technology.

## 9.10 Implications for Future Research

A limitation of the work presented in the thesis is that all participants taught, spoke, and speechread in English. Therefore, future research should investigate speechreading practice outside of English speaking countries. Although there is information on how visemes exist in other languages, there is little research into how speechreading is taught in other languages.

A second limitation of the work presented in this thesis is that it focuses on adult speechreading. All speechreading students who took part in this research were over the age of 18 and therefore there is no insight into how speechreading is taught to children and how we could design technology to support acquisition of speechreading at at early age.

# 10

# Conclusion and Future Work

At least 360 million people worldwide have disabling hearing loss that frequently causes difficulties in day-to-day conversations. Traditional technology (e.g., hearing aids) often fails to offer enough value, has low adoption rates, and can result in social stigma.

People with hearing loss find that speechreading can help to improve understanding during conversation. Speechreading can be described as a special case of audio-visual speech recognition where emphasis is placed on the visible, rather than on the audible, speech information. Speechreading helps to improve conversational confidence (thereby reducing social isolation), enhance employability, and improve educational outcomes.

However, speechreading is a skill that takes considerable practice and training to acquire. Publicly-funded speechreading classes are sometimes provided, and have been shown to improve speechreading acquisition. However, classes are only provided in a handful of countries around the world. Existing tools have been designed to help improve speechreading acquisition, but are often not effective because they have not been designed within the context of contemporary speechreading lessons or practice. In general, these previous tools are not helpful to acquiring speechreading because their designs were not influenced or based on how speechreading is currently taught.

To address this, in this thesis I presented a novel speechreading acquisition framework that can be used to design Speechreading Acquisition Tools (SATs) – a new type of technology to improve speechreading acquisition. I used thematic analysis of interviews conducted with speechreading tutors to identify and organise the dimensions of the framework.

In evaluating the framework, I demonstrated that it can accommodate current teaching techniques (identified by the tutors during the interviews) and existing solutions, as well as be used to design new SATs. I then conducted a questionnaire with speechreading students to gather data from the student perspective which can further enhance the design of new SATs.

To further evaluate the effectiveness of the framework, I developed two new SATs (*PhonemeViz* and *MirrorMirror*) that were influenced by findings from the speechreading tutor interviews and the speechreading student questionnaire.

*PhonemeViz* was inspired by the initial fingerspelling technique that was highlighted by four speechreading tutors during the interviews presented in Chapter 4. Fingerspelling is used by tutors to indicate the initial letter of a word within a viseme class during a speech movement exercise. Findings from the student questionnaire showed that participants find that words and phonemes looking the same on the lips present one of the biggest challenges while speechreading. Therefore, PhonemeViz is a visualisation that is positioned at the side of a speaker's face, beginning at the forehead and ending at the chin and presents textual representations of consonant speech sounds in a semi-circular arrangement, with an arrow beginning from the centre of this semi-circle pointing at the last spoken initial consonant speech sound (phoneme) to provide persistence. This design is intended to enable a speechreader to focus on the speaker's eyes and lip movements (as in traditional speechreading), while also monitoring changes in PhonemeViz's state using their peripheral vision to help disambiguate phonemes with a viseme class.

I evaluated PhonemeViz with 14 participants against five existing visualisation techniques (plus a no visualisation control condition) in a lab-based user study. The results demonstrated that PhonemeViz allowed participants to achieve 100% word recognition (showing successful disambiguation), and PhonemeViz was well-received in subjective and qualitative feedback.

*MirrorMirror* was initially conceptualised during the third step of the framework evaluation. MirrorMirror is a new SAT in the form of a mobile application that allows students to practice their speechreading by recording and watching videos of people they frequently speak with. The design of MirrorMirror was inspired by the mirror training technique that was highlighted by seven speechreading tutors during the interviews presented in Chapter 4. MirrorMirror's design was further informed by the positive and negative aspects of mirror training as reported by participants of the student questionnaire presented in Chapter 6.

I evaluated MirrorMirror through case studies with three speechreading students. The case study evaluation of MirrorMirror was comprised of three stages: 1) a briefing, initial questionnaire, and tutorial session, 2) a week-long in-the-wild-deployment, and 3) a post-deployment discussion session. The findings demonstrated that MirrorMirror enabled participants to effectively target their speechreading practice on people, words and situations they encounter during daily conversations.

The findings from the evaluation of these two new SATs demonstrate that using the framework can help design effective tools for speechreading acquisition.

## 10.1 Contributions

The central contribution of this thesis is the development of a novel framework that can be used to develop Speechreading Acquisition Tools (SATs) – a new type of technology designed specifically to improve speechreading acquisition. Through the development and release of SATs, people with hearing loss will be able to augment their class-based learning, or learn on their own if no suitable classes are available.

This thesis also presented a number of secondary contributions:

1) A critical overview of current Conversation Aids, and related approaches to improving speechreading framed within the framework.
2) Presentation of novel interview data from seven practicing speechreading tutors and thematic analysis of that data.
3) Presentation of novel questionnaire data from a postal survey with 59 students from speechreading classes.
4) I introduce PhonemeViz, a new SAT in the form of a visualisation that displays a subset of a speaker's spoken phonemes to the speechreader to reduce viseme confusion which occurs at the start of words.
5) I introduce MirrorMirror, a new SAT that addresses the limitations of current SATs by allowing users to capture and practice with videos of people they frequently speak with.

## 10.2 Future Work

In the future, I will pursue a number of the extensions to the Framework, PhonemeViz and MirrorMirror as discussed in Chapter 9.

*Framework*: First, I will look to interview speechreading tutors from outside Scotland in order to gather more teaching techniques used within speechreading classes. Second, I will employ other researchers to use the framework to design new SATs and document how they utilise the framework during their development process. Finally, I will design an interactive online tool that helps to help inform individuals who wish to use the framework about the techniques and tools that are fitted within each cell.

*PhonemeViz*: First, I will look to evaluate PhonemeViz with sentences and then in connected speech. Second, I will evaluate PhonemeViz within the context of commodity Head Mounted Displays. Third, I will investigate how PhonemeViz could be used as a subtitle substitute within video content for speechreaders.

*MirrorMirror*: First, I will add sentence and context practice session modes. Second, I will develop the video library sharing system, to help increase the amount of videos that are available to practice with.

## 10.3 Closing Remarks

Speechreading is a vital communication technique for people with all severities of hearing loss. Speechreading classes are taught using approaches and techniques that result in the acquisition of this vital skill. However, there are not enough classes to satisfy demand, and due to an ageing population this demand will rise. Although there have been attempts to develop tools to help with the acquisition of speechreading, these tools often try to help in ways that are not effective.

In this thesis, I developed a framework to help develop more effective tools for speechreading acquisition by designing them within the context of contemporary speechreading classes. It is my hope that through use of the framework, many more effective tools will be designed to support speechreading acquisition to support people who need to rely on speechreading during conversations.

# A

# Ethical Approval Forms

This appendix contains the letters of approval from the University Teaching and Research Ethics committee.

## A.1 Speechreading Tutor Interviews

UNIVERSITY OF DUNDEE

Computing, School of Science and Engineering

*Head of Discipline: Professor Emanuele Trucco*

**Ethics Committee**
*Convener*
Professor Annalu Waller
*Administrator*
Mrs Kathleen Cummins

27/11/15

Benjamin Gorman
Computing
School of Science and Engineering
University of Dundee

Dear Benjamin

***Full title of study:*** **Improving Speechreading**

***SoCEC reference number:*** **AC-012**

Thank you for submitting an ethics application on 8 November. Your application has been reviewed by the Ethics committee.

**Ethical issues arising from the proposed study**

You have indicated that there are no significant ethical issues arising from this project. The Ethics Committee has approved this study.

**Conditions of approval**

By submitting an application to the Ethics Committee you confirm that you have read and understand the University of Dundee Guidelines for Ethical Practices in Research and the School of Computing Code of Practice for Research involving Human Participants and undertake to abide by these guidelines. Permission is therefore granted for you to proceed with the study.

Please inform the committee of any change in project methodology which may have ethical implications.

Best wishes for your research,

Yours sincerely

Annalu

Annalu Waller MBCS MIPEM
Professor
Convener: Computing Ethics Committee

Administrator: Mrs Kathleen Cummins
*email* SoC-EthicsMembers@dundee.ac.uk *telephone* 01382 386532

School of Science and Engineering
UNIVERSITY OF DUNDEE Dundee DD1 4HN Scotland UK *t* +44 (0)1382 388085
www.computing.dundee.ac.uk

## A.2 Speechreading Student Questionnaire

UNIVERSITY OF DUNDEE

School of Science and Engineering

*Head of Discipline:(Computing) Professor Annalu Waller*

**Ethics Committee**
*Convener*
Professor Annalu Waller
*Administrator*
Mrs Kathleen Cummins

02/12/16

Benjamin Gorman
Computing
School of Science and Engineering
University of Dundee

Dear Benjamin

***Full title of study:*** Improving Lipreading

***SoCEC reference number:*** 16-011

Thank you for submitting an ethics application on 11 November. Your application has been reviewed by the Ethics committee.

**Ethical issues arising from the proposed study**

You have indicated that there are no significant ethical issues arising from this project. The Ethics Committee has approved this study.

**Conditions of approval**

By submitting an application to the Ethics Committee you confirm that you have read and understand the University of Dundee Guidelines for Ethical Practices in Research and the School of Science of Engineering Code of Practice for Research involving Human Participants and undertake to abide by these guidelines. Permission is therefore granted for you to proceed with the study.

Please inform the committee of any change in project methodology which may have ethical implications.

Best wishes for your research,

Yours sincerely

Annalu Waller MBCS MIPEM
Professor
Convener: School of Science and Engineering Ethics Committee

Administrator: Mrs Kathleen Cummins
*email* SSE-Ethics@dundee.ac.uk *telephone* 01382 386532

School of Science and Engineering
UNIVERSITY OF DUNDEE Dundee DD1 4HN Scotland UK *t* +44 (0)1382 388085
www.computing.dundee.ac.uk

# A.3 PhonemeViz Evaluation

UNIVERSITY OF DUNDEE

School of Computing

*Dean*
Professor Janet Hughes

Ethics Committee
*Convener*
Professor Annalu Waller

*Administrator*
Mrs Kathleen Cummins

14/04/15

Ben Gorman
School of Computing
University of Dundee

Dear Ben

***Full title of study:*** Viseme visualisations

***SoCEC reference number:*** 15-006

Thank you for submitting an ethics application on 3 April. Your application has been reviewed by the Ethics committee.

**Ethical issues arising from the proposed study**

You have indicated that there are no significant ethical issues arising from this project. The Ethics Committee has approved this study.

**Conditions of approval**

By submitting an application to the Ethics Committee you confirm that you have read and understand the University of Dundee Guidelines for Ethical Practices in Research and the School of Computing Code of Practice for Research involving Human Participants and undertake to abide by these guidelines. Permission is therefore granted for you to proceed with the study.

Please inform the committee of any change in project methodology which may have ethical implications.

Best wishes for your research,

Yours sincerely

Annalu

Annalu Waller MBCS MIPEM
Professor
Convener: School of Computing Ethics Committee

Administrator: Mrs Kathleen Cummins
*email* SoC-EthicsMembers@dundee.ac.uk *telephone* 01382 386532

School of Computing
UNIVERSITY OF DUNDEE

College of Art, Science and Engineering
UNIVERSITY OF DUNDEE Dundee DD1 4HN Scotland UK *t* +44 (0)1382 384145 *f* +44 (0)1382 385509
www.computing.dundee.ac.uk

# A.4 MirrorMirror Evaluation

UNIVERSITY OF DUNDEE

School of Science and Engineering

*Head of Discipline:(Computing) Professor Annalu Waller*

**Ethics Committee**

*Convener*
Professor Annalu Waller

*Administrator*
Mrs Kathleen Cummins

6 July 2017

Ben Gorman
School of Science and Engineering
University of Dundee

Dear Ben

***Full title of study:*** MirrorMirror: A mobile app to teach lipreading

***SoCEC reference number: 17-007***

Thank you for submitting an ethics application on 13 June. Your application has been reviewed by the Ethics committee.

**Ethical issues arising from the proposed study**

You have indicated that there are no significant ethical issues arising from this project. The Ethics Committee has approved this study.

**Conditions of approval**

By submitting an application to the Ethics Committee you confirm that you have read and understand the University of Dundee Guidelines for Ethical Practices in Research and the School of Science of Engineering Code of Practice for Research involving Human Participants and undertake to abide by these guidelines. Permission is therefore granted for you to proceed with the study.

Please inform the committee of any change in project methodology which may have ethical implications.

Best wishes for your research,

Yours sincerely

Annalu

Annalu Waller MBCS MIPEM
Professor
Convener: School of Science and Engineering Ethics Committee

Administrator: Mrs Kathleen Cummins
*email* SSE-Ethics@dundee.ac.uk *telephone* 01382 386532

School of Science and Engineering
UNIVERSITY OF DUNDEE Dundee DD1 4HN Scotland UK *t* +44 (0)1382 388085
www.computing.dundee.ac.uk

B

# Study Material For Speechreading Tutors Interviews

## B.1 Introdution

This appendix contains study material used during the speechreading tutor interviews presented in Chapter 4.

## B.2 Audio Release Form

**IMPROVING SPEECHREADING**

**Audio Consent Form**

Thank you for taking part in the Improving Speechreading project. You are to take part in a one to one interview with a member of the research team and if you consent, this session will be audio recorded.

We ask that you complete the section below and return this form to us.

Please complete the following and sign below:

| | YES | NO |
|---|---|---|
| I confirm that I am aware the event will be audio recorded. | | |
| I confirm that the recording will not be released and will only be used by the research team to produce an anonymised transcript. | | |

____________________ ____________________ ________

Name Signature Date

Improving Speechreading Audio Release Form
March 2016

# B.3 Consent Form

**IMPROVING LIPREADING**

## Consent Form

Please Initial

1. I confirm that I have read and understand the information sheet for the above study. I have had the opportunity to consider the information, ask questions and have had these answered satisfactorily.

2. I understand that my participation is voluntary and that I am free to withdraw from the study at any time without giving any reason and without penalty.

3. I understand that individuals may look at data collected during the study from the research where it is relevant to my taking part in this research. I give permission for these individuals to have access to my data.

4. I agree to take part in the above study.

5. Your personal information will be kept confidential. No reference will be made to your identity in publications or other documents.

____________________________________________________________

Name of participant    Signature    Date

____________________________________________________________

Name of Researcher taking Consent    Signature    Date

(When completed: 1 for participant; 1 for researcher site file)

Improving Speechreading - Consent Form - Version 1, March 2016

## B.4 Debriefing

IMPROVING LIPREADING

### Debrief Statement

Thank you for taking the time to take part in this research study. Lipreading's effectiveness can be limited due to the confusion caused by visemes. A viseme is any of several speech sounds in which the position of the face and mouth look the same.

In this study we were looking to gather information on how lipreading is currently taught in an effort to gain information on how technology can support the way it is taught and if we can support people while they lipread.

Thank you for taking the time to take part in this research study. For all follow up questions relating to this study please contact the

Principal Investigator:

**Benjamin Gorman**
Email: b.gorman@dundee.ac.uk
Telephone: 01382 385598

Improving Speechreading

Debriefing Statement - Version 1, March 2016

# B.5 Demographics Questionnaire

**IMPROVING LIPREADING** **ParticipantID:**

**Demographics Form**

- **All Questions Are Optional.**
- **Your personal information will be kept confidential. No reference will be made to your identity in publications or other documents.**
- **Your participation is voluntary and that I am free to withdraw from the study at any time without giving any reason and without penalty.**

1. Age:

_________

2. Sex:

**M** **F** **Other**

3. Highest level of education:

**Other** **High School** **College** **University**

5. Please rate your level of computer literacy:

**Excellent** **Good** **Fair** **Poor**

6. How many years have you taught lipreading/speechreading?

7 a). Do you know sign language?

7 b). How many years have you known sign language?

9. Do you have a hearing loss?
If yes, please give details:

## B.6 Information Sheet

PARTICIPANT INFORMATION SHEET

## Improving Speechreading

We would like to invite you to take part in our research study. Before you decide if you wish to take part, we would like you to understand why the research is being undertaken and what it will involve. A member of the research team will go through the information sheet with you and answer any questions you have.

### What is the 'Improving Speechreading' study?

In everyday conversation, people with typical vision and hearing subconsciously use information from a speaker's lips and face to understand what they are saying. People who can speechread (more commonly known as lipreading) are more skilled at extracting this information.

However, speech reading is limited in that many parts of a word share the same mouth and lip shape and thus are impossible to distinguish from visual information alone.

In this study, we are looking to gain an insight into how speechreading/lipreading is currently taught and how technology could help individuals learning to lipread and also support them during lipreading.

### Do I have to take part in the study?

It is up to you to decide to join the study. We will describe the study and go through this information sheet with you. If you agree to take part, we will then ask you to sign a consent form. You will be given copies of these forms to keep.

You will then be provided time to ask any questions you may have of the researchers. Please also feel free to ask questions at any time during the interview.

### What happens if I wish to withdraw from the study?

You are free to withdraw at any time, without giving a reason and without penalty. Any data that has already been gathered from you will also be discarded.

Computing
at the University of Dundee

### What will I have to do?

You will be seen by a researcher (Benjamin Gorman) at the School of Computing, University of Dundee, one to one interview based session. The researcher will begin by asking you for some general information about yourself (e.g., age) in the form of a demographic questionnaire. All questions are optional.

You will then begin the one-to-one interview with the researcher, which we estimate will take about an hour. You will help inform the researcher around how lipreading is currently taught. You can decline to answer any question during the interview.

If you consent, the interview will be audio recorded in order for the researcher to later transcribe and gather more data from the interview. This audio recording will be kept confidential to the research team.

### What are the possible disadvantages and risks of taking part?

There are no risks associated with this study and we hope that the task will be enjoyable. The timing and location of sessions will be discussed with you.

### What are the possible benefits of taking part?

In our experience, people enjoy taking part in research as they are helping to develop new technology. Your involvement will help us understand how visualisations can be designed to support lipreading, which can potentially help people with hearing loss in the real world.

### What happens at the end of the study?

The analysis of the data will be completed by October 2016. The results of this study may be published in academic journals and presented at academic conferences. If you would like to know the outcome of the study, I will send you a copy of the study report by October 2016.

### What if there is a problem?

If you have a concern about any aspect of this study, you should speak to the study supervisor, Dr David Flatla [email d.flatla@dundee.ac.uk], who will do his best to answer your questions.

Computing
at the University of Dundee

### Will my information be kept confidential?

Yes. We will follow ethical and legal practice and all information about you will be handled in confidence. To ensure anonymity, personal records will only be available to the research team for the duration of the study and will not be kept together with the results or be presented in the report. If your data is used for publications, no reference to your identity will be made.

### Who has reviewed this study?

Computing at the University of Dundee's Ethics Committee, which has responsibility for scrutinising all proposals for non-clinical research on humans has examined the proposal and has raised no objections from the point of view of ethics.

### Who can I contact in connection with this research?

This research is part of an on-going research project directed by Benjamin Gorman. He is a PhD Student in the School of Computing at the Univeristy of Dundee.

Please feel free to contact him about the study. His contact details are:

**Benjamin Gorman**
**Email:** b.gorman@dundee.ac.uk
**Telephone: 01382 385598**

**Thank you for taking the time to read this information sheet and for considering taking part in this study.**

Computing
at the University of Dundee

## B.7 Interview Guide

IMPROVING LIPREADING

## Discussion Guide

- **All Questions Are Optional.**
- **Your personal information will be kept confidential. No reference will be made to your identity in publications or other documents.**
- **Your participation is voluntary and you are free to withdraw from the study at any time without giving any reason and without penalty.**

**Background**

1. Why did you decide to become a lipreading tutor?
2. How did you learn to become a lipreading tutor?
3. How long have you been a lipreading tutor?
4. How many classes do you teach on an average week?
5. How long do people typically spend in a class - months, years etc?
6. When starting, does everyone have roughly the same level of lipreading experience?
7. Do you assess each individual to determine what stage of lipreading they are at?
8. How do you get rid of bad habits from self-teaching?
    1. Are these difficult to correct?
9. Why do people tend to stop coming to a class?
    1. Do they just not need classes after a while?

**General Teaching**

10. Do you use voice when giving instructions?
11. Do you use voice when giving training material?
12. Do you encourage students to wear their hearing aids during a class?
13. Where do you get your material from?
    1. Is it from a textbook, from the Scottish course to teach lipreaders?
14. What do you consider to be the basic teachable unit?
    1. Word, Syllable, Phoneme, Vowel/Consonant
15. Is the majority of your material story based/sentence based/word based or movement based?
16. Do you classify sounds based on their visible appearance or how they are produced by the mouth/lips?
17. Do you use syllable drills?
18. Do you teach kinaesthetic awareness? (e.g. using a hand on throat to feel voicing)
19. Do you use mirror practice?
20. Do you find it better to teach students to recognize the individual sounds or is it better for the students to get a general idea of what is being said?

21. What is a lesson focused around?
    1. Context or a particular movement/word?

**Learning Info**

22. What do students learn in their first month?
23. Is there any particular area that the majority of your students find difficult when they are first learning?
24. What would a general outline of an average lesson look like in the...
    1. ...first year?
    2. …second year?
    3. …third year & beyond?

**Further Learning Information**

25. How do people keep learning at home?
    1. Homework, mirror practice?
26. Do you show any videos?
    1. Are they subtitled?
27. Do you teach strategies?
    1. (Where to sit, where a speaker sits, to avoid shadows, etc)
28. Do you use animations?
    1. Animated graphics/pictorial representations of lips
29. Do you teach lip shapes?
    1. How do you teach the lip shapes?
30. Do you teach how to read gestures?
31. Do you teach how to read facial expressions?
32. What are the key elements to effective lipreading?

**Assessment**

33. How do you know as a teacher that they are getting better?

**Future**

34. What could be improved with lipreading?
35. Has anyone in your classes ever expressed interest over making lipreading easier?
36. Do you know of any assistive technologies for lipreading?
37. Do you know of any computer based lipreading training tools?
38. Do you find that captioned videos help lipreaders when watching television?
39. Do you think people who are looking into learning lipreading would benefit from additional technology to help them?

C

# Study Material For Student Questionnaires

## C.1 Introduction

This appendix contains study material used during the student questionnaire presented in Chapter 6.

## C.2 Questionnaire

**Lipreading Questionnaire** **ParticipantID** __________

- **All Questions Are Optional.**
- **Your personal information will be kept confidential. No reference will be made to your identity in publications or other documents.**
- **Your participation is voluntary and you are free to withdraw from the study at any time without giving any reason and without penalty.**

☐ *I have read the accompanying information sheet. By completing this questionnaire, I agree to my data being used for the research project.*

**General**

1. Age: __________

2. Sex (Circle one):

| M | F | Other |
|---|---|---|

3. Highest level of education (Circle one):

| High School | College | University | Other |
|---|---|---|---|

4. Please rate your level of computer literacy (Circle one):

| Excellent | Good | Fair | Poor |
|---|---|---|---|

5. Do you have a hearing loss? (Circle one):

| Yes | No | Don't know |
|---|---|---|

If yes:

A) How long have you had a hearing loss

B) How would you describe your hearing loss (Circle one):

| | | | |
|---|---|---|---|
| Mild | Moderate | Severe | Profound |

C) Cause, if known (Tick all that apply)

- ☐ Hearing loss present at birth (congenital)
- ☐ Exposure to loud noise
- ☐ Head trauma
- ☐ Virus/Disease
- ☐ Aging

Other:

D) Do you use:

- ☐ Cochlear Implant(s)
- ☐ Hearing Aid(s)

Other:

**Lipreading**

1. Please rate your lipreading ability (Circle one):

| | | | |
|---|---|---|---|
| Excellent | Good | Fair | Poor |

2. How long have you been in lipreading classes?

3. Do you practice lipreading outside of class? (Circle one):

| | |
|---|---|
| Yes | No |

A). If yes, please describe how you practice:

4. Do you use mirror practice outside of class? (Circle one):

| | |
|---|---|
| Yes | No |

**If yes:**

A). How often do you use mirror practice at home? (Circle one):

| | | | | | |
|---|---|---|---|---|---|
| Never | 1-2 times per Year | 1-2 times per Month | Once a Week | 2-3 Times a Week | Daily |

B) What do you **like** about mirror practice?

C) What do you **dislike** about mirror practice?

5. Do you use videos to practice lipreading outside of class? (Circle one):

| | |
|---|---|
| Yes | No |

**If yes:**

A) How often do you use videos or watch television to practice lipreading outside of class **with subtitles turned on**? (Circle one):

| | | | | | |
|---|---|---|---|---|---|
| Never | 1-2 times per Year | 1-2 times per Month | Once a Week | 2-3 Times a Week | Daily |

B) How often do you use videos or watch television to practice lipreading outside of class with **subtitles turned off**? (Circle one):

| | | | | | |
|---|---|---|---|---|---|
| Never | 1-2 times per Year | 1-2 times per Month | Once a Week | 2-3 Times a Week | Daily |

6. In what situations do you find lipreading challenging?
(Tick all that apply)

- ☐ Home
- ☐ Group Conversations
- ☐ Dentist
- ☐ Doctors
- ☐ Shopping
- ☐ Opticians
- ☐ Coffee Shops
- ☐ Restaurants
- ☐ Transport (Bus/Taxi/Train/Plane)

Other:

7. What do you find challenging when lipreading?
(Tick all that apply)

- ☐ Words looking the same on the lips
- ☐ People talking quickly
- ☐ People covering their mouths
- ☐ People turning away from you
- ☐ Accents
- ☐ Beards/Facial hair
- ☐ Quiet Speakers
- ☐ Concentration
- ☐ Fatigue

Other:

8: Do you rehearse/anticipate possible phrases or words that you may have to lipread before being in a situation?

| Yes | No |
|---|---|

A) If yes, please describe how:

9. Do you own a mobile device?
(e.g., iPhone/Android smartphone, iPad/Tablet, Amazon Kindle Fire.) If yes, please give details:

## C.3 Student Information Sheet

PARTICIPANT INFORMATION SHEET

## Improving Lipreading

We would like to invite you to take part in our research study. Before you decide if you wish to take part, we would like you to understand why the research is being undertaken and what it will involve.

### What is the 'Improving Lipreading' study?

In this study, we are looking to gain an insight into how lipreading is currently taught and how technology could help individuals learning lipreading.

### Do I have to take part in the study?

It is up to you to decide to join the study.

### What happens if I wish to withdraw from the study?

You are free to withdraw at any time, without giving a reason and without penalty. Any data that has already been gathered from you will also be discarded.

### What will I have to do?

Attached to this sheet is a questionnaire, which asks you some questions about yourself and your experience with lipreading and technology. If you agree to join the study, please fill in this questionnaire. All questions are optional.

Once you have completed your questionnaire, place it into the envelope provided, seal the envelope, and then hand this back to the tutor of your class. If you wish to, you can also post this directly back to the University.

### What are the possible disadvantages and risks of taking part?

There are no risks associated with this study and we hope that the task will be enjoyable. The timing and location of sessions will be discussed with you.

### What are the possible benefits of taking part?

In our experience, people enjoy taking part in research as they are helping to develop new

University of Dundee

Computing at the University of Dundee

1

technology. Your involvement will help us understand how technology can be designed to support lipreading, which can potentially help people with hearing loss in the real world.

### What happens at the end of the study?

The analysis of the data will be completed by October 2017. The results of this study may be published in academic journals and presented at academic conferences. If you would like to know the outcome of the study, I will send you a copy of the study report by October 2017.

### What if there is a problem?

The University of Dundee School of Science and Engineering's Research Ethics Committee, which has responsibility for scrutinising all proposals for non-clinical research on humans within the School has examined the proposal and has raised no objections from the point of view of ethics. If you have a concern about any aspect of this study, you should speak to the study supervisor, Dr David Flatla [email d.flatla@dundee.ac.uk], who will do his best to answer your questions.

### Will my information be kept confidential?

Yes. We will follow ethical and legal practice and all information about you will be handled in confidence. To ensure anonymity, personal records will only be available to the research team for the duration of the study and will not be kept together with the results or be presented in the report. If your data is used for publications, no reference to your identity will be made.

### Who has reviewed this study?

Computing at the University of Dundee's Ethics Committee, which has responsibility for scrutinising all proposals for non-clinical research on humans has examined the proposal and has raised no objections from the point of view of ethics.

### Who can I contact in connection with this research?

This research is part of an on-going research project directed by Benjamin Gorman. He is a PhD Student in the School of Computing at the University of Dundee. Please feel free to contact him about the study. His contact details are:

**Benjamin Gorman**
**Email:** b.gorman@dundee.ac.uk
**Telephone: 01382 385598**

**Thank you for taking the time to read this information sheet and for considering taking part in this study.**

UNIVERSITY OF DUNDEE    Computing at the University of Dundee

## C.4 Tutor Information Sheet

**TUTOR INFORMATION SHEET**

## Improving Lipreading

We would like to invite your students to take part in our research study. Before you decide if you wish to help facilitate them taking part, we would like you to understand why the research is being undertaken and what it will involve.

### What is the ‘Improving Lipreading’ study?

In this study, we are looking to gain an insight into how lipreading is currently taught and how technology could help individuals learning lipreading.

### Do I have to take part in the study?

It is up to you to decide to help us distribute the study. It is up to the student if they wish to take part. Please make it clear that they do not have to take part. It is their choice.

### What happens if I wish to withdraw from the study?

You are free to withdraw at any time, without giving a reason and without penalty. Any data that has already been gathered from you will also be discarded.

### What will I have to do?

In this information pack you will find Questionnaire Forms, and Information Sheets such as this stapled to the front. You will also find individual envelopes. Please distribute these to your students during or before a class is due to begin. Potential participants must be above the age of 18. Additionally, potential participants must be literate, as the investigator will not be present to read the information sheet and questionnaire to the potential participant.

Once students have completed the questionnaire they should place it in a provided envelope and seal the envelope. Once all forms have been completed students may either post the survey themselves directly or envelopes can be collected and placed in the larger envelope provided. You will find stamps and a pre-addressed label inside. Please send this back to the university at your convenience.

### What are the possible disadvantages and risks of taking part?

There are no risks associated with this study and we hope that the task will be enjoyable.

UNIVERSITY OF DUNDEE    Computing at the University of Dundee    1

## What are the possible benefits of taking part?

In our experience, people enjoy taking part in research as they are helping to develop new technology. Your involvement will help us understand how technology can be designed to support lipreading, which can potentially help people with hearing loss in the real world.

## What happens at the end of the study?

The analysis of the data will be completed by October 2017. The results of this study may be published in academic journals and presented at academic conferences. If you would like to know the outcome of the study, I will send you a copy of the study report by October 2017.

## What if there is a problem?

The University of Dundee School of Science and Engineering's Research Ethics Committee, which has responsibility for scrutinising all proposals for non-clinical research on humans within the School has examined the proposal and has raised no objections from the point of view of ethics. If you have a concern about any aspect of this study, you should speak to the study supervisor, Dr David Flatla [email d.flatla@dundee.ac.uk], who will do his best to answer your questions.

## Will my information be kept confidential?

Yes. We will follow ethical and legal practice and all information about you will be handled in confidence. To ensure anonymity, personal records will only be available to the research team for the duration of the study and will not be kept together with the results or be presented in the report. If your data is used for publications, no reference to your identity will be made.

## Who has reviewed this study?

Computing at the University of Dundee's Ethics Committee, which has responsibility for scrutinising all proposals for non-clinical research on humans has examined the proposal and has raised no objections from the point of view of ethics.

## Who can I contact in connection with this research?

This research is part of an on-going research project directed by Benjamin Gorman. He is a PhD Student in the School of Computing at the University of Dundee. Please feel free to contact him about the study. His contact details are:

**Benjamin Gorman**
**Email:** b.gorman@dundee.ac.uk
**Telephone: 01382 385598**

**Thank you for taking the time to read this information sheet and considering taking part in this study.**

UNIVERSITY OF DUNDEE

Computing
at the University of Dundee

2

D

# Study Material For PhonemeViz

This appendix contains study material used during PhonemeViz study presented in Chapter 7.

# D.1 Consent Form

Viseme Visualisations Evaluation

Please Initial

1. I confirm that I have read and understand the information sheet for the above study. I have had the opportunity to consider the information, ask questions and have had these answered satisfactorily.

2. I understand that my participation is voluntary and that I am free to withdraw from the study at any time without giving any reason and without penalty.

3. I understand that individuals may look at data collected during the study from the research where it is relevant to my taking part in this research. I give permission for these individuals to have access to my data.

4. I agree to take part in the above study.

________________________________________________

Name of participant          Signature          Date

________________________________________________

Name of Researcher taking Consent          Signature          Date

(When completed: 1 for participant; 1 for researcher site file
- No reference will be made to your identity in publications or other documents)

Viseme Visualisations Evaluation

Viseme Visualisations Consent Form - Version 1, April 2015

## D.2 Debriefing Form

Debriefing Statement

## Viseme Visualisations

Day 1

Thank you for taking the time to take part in this research study. Speechreading's effectiveness can be limited due to the confusion caused by visemes. A viseme is any of several speech sounds in which the position of the face and mouth look the same.

In this study we were evaluating how different visualisation techniques can be used to overcome the issues caused by visemes. The purpose of today's session was to familiarise you with each technique. The next session will follow a similar pattern as the main part of today's session, and afterwards we will ask some general questions about each technique.

Thank you for taking the time to take part in this research study. For all follow up questions relating to this study please contact the Principal Investigator:

Benjamin Gorman
Email: b.gorman@dundee.ac.uk
Telephone: 01382 385598

Viseme Visualisations

Debriefing Statement - Version 1, April 2015

D. Study Material For PhonemeViz

Debriefing Statement

Viseme Visualisations

Day 2

Thank you for taking the time to take part in this research study. Speechreading's effectiveness can be limited due to the confusion caused by visemes. A viseme is any of several speech sounds in which the position of the face and mouth look the same.

In this study we were evaluating how different visualization techniques can be used to overcome the issues caused by visemes. Within the visualizations our solution was titled PhonemeViz and the other techniques were previously established in academic literature. Therefore, the evaluation was to determine whether our solution offered benefit over the others.

Thank you for taking the time to take part in this research study. For all follow up questions relating to this study please contact the Principal Investigator:

Benjamin Gorman
Email: b.gorman@dundee.ac.uk
Telephone: 01382 385598

Viseme Visualisations

Debriefing Statement - Version 1, April 2015

## D.3 Demographics Questionnaire

## Viseme Visualisation Demographics Questionnaire

All questions are optional.

1. **ParticipantNo**

2. **Age**

3. **Sex**

   Male
   Female
   Other

4. **Handedness**

   Left
   Right
   Ambidextrous

5. **Please rate your corrected visual acuity:**

   Excellent
   Good
   Fair
   Poor

6. **Highest level of education:**

   Other
   High School
   College
   University

7. **Please rate your level of computer literacy:**

   Excellent
   Good
   Fair
   Poor

8. **When a person's mouth is not visible (e.g., it is covered or they turn their back to you) do you find yourself thinking: "If I could see their lips it would help me to understand what they are saying"?**

   Yes

   No

9. **When a person lowers their voice (e.g., to whisper) do you find that you look at their lips more?**

   Yes

   No

10. **When somebody is across the room talking (e.g., giving a presentation) do you find that you want to look at their mouth more to determine what they are saying?**

    Yes

    No

11. **When talking over the phone do you feel that any confusion over certain words or letters being spoke (e.g. Somebody telling you their postcode) would be avoided if you could see the other person's lips?**

    Yes

    No

12. **When talking to a person behind a screen (e.g., somebody at the bank or a bus driver) do you look more at their lips to help you understand what has been said?**

    Yes

    No

13. **Do you often mishear lyrics until you see somebody perform the song?**

    Yes

    No

14. **How often do you rely on lipreading in your daily life?**

| | 1 | 2 | 3 | 4 | 5 | 6 | 7 | |
|---|---|---|---|---|---|---|---|---|
| Never | | | | | | | | Always |

# D.4 Closing Questionnaire

**Closing Questionnaire** **Participant ID:** ___________

1. Please rank the visualization techniques in order of preference (1 indicating "most definitely use again" and 7 indicating "least likely to use again")

| Technique | Rank |
|---|---|
| Spectrogram | |
| Caption | |
| AlphaViz | |
| VocSyl | |
| None | |
| Lip Magnification | |
| iBaldi | |

2. Please explain your reasoning behind the ranking order

3. Please explain why you ranked the technique at position 1.

## D.5 Information Sheet

INFORMATION SHEET

## Viseme Visualisations

We would like to invite you to take part in our research study. Before you decide if you wish to take part, we would like you to understand why the research is being undertaken and what it will involve. A member of the research team will go through the information sheet with you and answer any questions you have.

### What is the 'Viseme Visualisations' study?

A phoneme is a basic unit of a language's phonology. Phonemes are combined to form meaningful units such as words.

In everyday conversation, people with typical vision and hearing subconsciously use information from a speaker's lips and face to aid comprehension. People can to some extent deduce what phoneme has been produced based on visual cues, even if the sound is unavailable or degraded (e.g., by background noise). People who can speechread (more commonly known as lipreading) are more skilled at extracting this information.

However, speech reading is limited in that many phonemes share the same mouth and lip shape (known as a viseme) and thus are impossible to distinguish from visual information alone. Sounds whose place of articulation is deep inside the mouth or throat are not detectable, such as some consonants and most gestures of the tongue. Voiced and unvoiced pairs look identical, such as [p] and [b], [k] and [g], [t] and [d], [f] and [v], and [s] and [z]; likewise for nasalisation (e.g. [m] vs. [b]). It has been estimated that only 30% to 40% of sounds in the English language are distinguishable from sight alone.

With recent advancements in mobile devices, augmented reality and high-resolution display technology we can overlay information directly onto the visual field of a user. Algorithms have previously been demonstrated which can convert speech audio into phonemes units.

In this study, we are evaluating a number of phoneme/speech visualisation techniques to determine the degree to which they assist speech reading.

### Do I have to take part in the study?

It is up to you to decide to join the study. We will describe the study and go through this information sheet with you. If you agree to take part, we will then ask you to sign a consent form. You will be given copies of these forms to keep. You will then be provided time to ask any questions you may have of the researchers. Please also feel free to ask questions at any time during the study.

### What happens if I wish to withdraw from the study?

You are free to withdraw at any time, without giving a reason and without penalty. Any data that has already been gathered from you will also be discarded.

### What will I have to do?

You will be seen by a researcher (Benjamin Gorman) at the School of Computing, University of Dundee, in a lab-based user study session. The researcher will begin by asking you for some general information about yourself (e.g., age) in the form of a demographic questionnaire. All questions are optional.

You will then begin the study session, which we estimate will take about 45 minutes. You will help evaluate several different visualisations which have been designed to help support speech reading. For each visualisation, you will watch a video in which a speaker will say a list of selected words; those that commonly cause confusion due to visemes. The audio will be turned off. You will be required to press a button when you hear a particular word, using the visualisations to support your answers. More instructions will be given during the experiment.

### What are the possible disadvantages and risks of taking part?

There are no risks associated with this study and we hope that the task will be enjoyable.

The timing and location of sessions will be discussed with you. The first session will require an hour and fifteen minutes and the second session an hour.

### What are the possible benefits of taking part?

In our experience, people enjoy taking part in research as they are helping to develop new technology. Your involvement will help us understand how sound localisation can be improved, which can potentially help people with impaired hearing in the real world.

### What happens at the end of the study?

The analysis of the data will be completed by October 2015. The results of this study may be published in academic journals and presented at academic conferences. If you would like to know the outcome of the study, I will send you a copy of the study report by October 2015.

### What if there is a problem?

If you have a concern about any aspect of this study, you should speak to the study supervisor, Dr David Flatla [email d.flatla@dundee.ac.uk], who will do his best to answer your questions. If

Viseme Visualisatons

Participant Information Sheet - Version 1, April, 2015

you remain unhappy and wish to complain formally, you can do this by speaking to Dr Janet Hughes, Dean and Head of School of Computing, University of Dundee [phone: 01382 385195 or email jhughes@computing.dundee.ac.uk].

### Will my information be kept confidential?

Yes. We will follow ethical and legal practice and all information about you will be handled in confidence. To ensure anonymity, personal records will only be available to the research team for the duration of the study and will not be kept together with the results or be presented in the report. If your data is used for publications, no reference to your identity will be made.

### Who has reviewed this study?

The School of Computing's Ethics Committee, which has responsibility for scrutinising all proposals for non-clinical research on humans at the University of Dundee's School of Computing, has examined the proposal and has raised no objections from the point of view of ethics.

### Who can I contact in connection with this research?

This research is part of an on-going research project directed by Benjamin Gorman.
He is a PhD Student in the School of Computing at the Univeristy of Dundee.

His contact details are:

Benjamin Gorman
Email: b.gorman@dundee.ac.uk
Telephone: 01382 385598

Please feel free to contact him about the study.

### Thank you.

Thank you for taking the time to read this information sheet and for considering taking part in this study.

Viseme Visualisatons

Participant Information Sheet - Version 1, April, 2015

# D.6 Video Release Form

## VIDEO RELEASE FORM

Thank you for taking part in the Viseme Visualisation project. You will be required to read aloud a list of words and this session will be video recorded. The video or photographs may be shown online in publications or at academic conferences. Your participation is voluntary.

We ask that you complete the section below and return this form to us. You may change your mind at a later date to withdraw the points below that you have consented to. Let us know and we will ask you to complete a new form.

Please complete the following and sign below:

| | YES | NO |
|---|---|---|
| I confirm that I am aware the event will be video recorded. | | |
| I confirm that I give permission for video or photographs of myself to be disseminated in publications or at academic conferences. | | |
| I confirm that I give permission for video/photograph of myself to be used for research and teaching purposes relevant to the subject matter. | | |

____________________ ____________________ ________

Name Signature Date

Viseme Visualisation Video Release Form
April 2015

# D.7 Proficiency Test Words

**Speechreading Proficiency Test** **Participant ID:**___________________

| | | | |
|---|---|---|---|
| 1 | It | 21 | Salvation |
| 2 | You | 22 | Pinched |
| 3 | The | 23 | Necklace |
| 4 | And | 24 | Machinery |
| 5 | She | 25 | Involved |
| 6 | Far | 26 | Failure |
| 7 | Left | 27 | Apology |
| 8 | Rest | 28 | Booking |
| 9 | Use | 29 | Loose |
| 10 | Sign | 30 | Gather |
| 11 | Burnt | 31 | Surviving |
| 12 | Vehicle | 32 | Market |
| 13 | Riding | 33 | Federation |
| 14 | Goodbye | 34 | Entertaining |
| 15 | Entrance | 35 | Distribution |
| 16 | Mouse | 36 | Operating |
| 17 | Kicked | 37 | Librarian |
| 18 | Spread | 38 | Kidney |
| 19 | Outside | 39 | Horizon |
| 20 | Crying | 40 | Comparative |

# D.8 Proficiency Test Response Sheet

**Speechreading Proficiency Test** **Participant ID:____________________**

| | | | |
|---|---|---|---|
| 1 | | 21 | |
| 2 | | 22 | |
| 3 | | 23 | |
| 4 | | 24 | |
| 5 | | 25 | |
| 6 | | 26 | |
| 7 | | 27 | |
| 8 | | 28 | |
| 9 | | 29 | |
| 10 | | 30 | |
| 11 | | 31 | |
| 12 | | 32 | |
| 13 | | 33 | |
| 14 | | 34 | |
| 15 | | 35 | |
| 16 | | 36 | |
| 17 | | 37 | |
| 18 | | 38 | |
| 19 | | 39 | |
| 20 | | 40 | |

# D.9 Evaluation Words

| | Viseme | Word 1 | Word 2 | Word 3 |
|---|---|---|---|---|
| **1.** | /p/ | Pat | Mat | Bat |
| **2.** | /t/ | Sun | Done | Tonne |
| **3.** | /k/ | Kill | Gill | Nil |
| **4.** | /ch/ | Chill | Shill | Jill |
| **5.** | /p/ | Banned | Manned | Panned |
| **6.** | /k/ | Light | Night | Kite |
| **7.** | /t/ | Zone | Tone | Sewn |
| ***Training words*** | */f/* | *Fan* | *Van* | |

E

# Study Material For MirrorMirror

This appendix contains material used during the MirrorMirror study presented in Chapter 8.

# E.1 Consent Form

**MIRROR MIRROR**

**Consent Form**

Please Initial

1. I confirm that I have read and understand the information sheet for the above study. I have had the opportunity to consider the information, ask questions and have had these answered satisfactorily.

2. I understand that my participation is voluntary and that I am free to withdraw from the study at any time without giving any reason and without penalty.

3. I understand that individuals may look at data collected during the study from the research where it is relevant to my taking part in this research. I give permission for these individuals to have access to my data.

4. I agree to take part in the above study.

5. Your personal information will be kept confidential. No reference will be made to your identity in publications or other documents.

______________________________________________

Name of participant Signature Date

______________________________________________

Name of Researcher taking Consent Signature Date

(When completed: 1 for participant; 1 for researcher site file)

Mirror Mirror - Consent Form - Version 1, June 2017

## E.2 Debriefing

MIRROR MIRROR

### Debrief Statement

Thank you for taking the time to take part in this research study.

In this project we have developed a mobile application called MirrorMirror. MirrorMirror is essentially a brain training game centred around speechreading. MirrorMirror allows speechreaders to capture and store videos of speech movements and words in what we call a "lipshape library". Users can gather coded videos of friends and family, allowing them to practice speechreading on those they speak with most. The repository of videos overcomes the 'full knowledge' limitation of current mirror practice.

We were looking to answer the following questions:

Does MirrorMirror help with speechreading acquisition?
Does MirrorMirror raise awareness of Speechreading?

Thank you for taking the time to take part in this research study. For all follow up questions relating to this study please contact the

Principal Investigator:

**Benjamin Gorman**
Email: b.gorman@dundee.ac.uk
Telephone: 01382 385598

### What if there is a problem?

The University of Dundee School of Science and Engineering's Research Ethics Committee, which has responsibility for scrutinising all proposals for non-clinical research on humans within the School has examined the proposal and has raised no objections from the point of view of ethics. If you have a concern about any aspect of this study, you should speak to the study supervisor, Dr David Flatla [email d.flatla@dundee.ac.uk], who will do his best to answer your questions.

Should you still have questions you should speak to the discipline lead for computing, Professor Annalu Waller [email a.waller@dundee.ac.uk, phone: +44 (0)1382 388223].

Mirror Mirror

Debriefing Statement - Version 1, June 2017

## E.3 Pre-deploment Questionnaire

**MirrorMirror Questionnaire** **ParticipantID** __________

- **All Questions Are Optional.**
- **Your personal information will be kept confidential. No reference will be made to your identity in publications or other documents.**
- **Your participation is voluntary and you are free to withdraw from the study at any time without giving any reason and without penalty.**

☐ *I have read the accompanying information sheet. By completing this questionnaire, I agree to my data being used for the research project.*

**General**

1. Age: __________

2. Sex (Circle one):

| M | F | Other |
|---|---|---|

3. Highest level of education (Circle one):

| High School | College | University | Other |
|---|---|---|---|

4. Please rate your level of computer literacy (Circle one):

| Excellent | Good | Fair | Poor |
|---|---|---|---|

5. Do you have a hearing loss? (Circle one):

| Yes | No | Don't know |
|---|---|---|

If yes:

A) How long have you had a hearing loss

B) How would you describe your hearing loss (Circle one):

| Mild | Moderate | Severe | Profound |
|---|---|---|---|

C) Cause, if known (Tick all that apply)

- ☐ Hearing loss present at birth (congenital)
- ☐ Exposure to loud noise
- ☐ Head trauma
- ☐ Virus/Disease
- ☐ Aging

Other:

D) Do you use:

- ☐ Cochlear Implant(s)
- ☐ Hearing Aid(s)

Other:

E. Study Material For MirrorMirror

**Lipreading**

1. Please rate your lipreading ability (Circle one):

| Excellent | Good | Fair | Poor |
|---|---|---|---|

2. How long have you been in lipreading classes?

3. Do you practice lipreading outside of class? (Circle one):

| Yes | No |
|---|---|

A). If yes, please describe how you practice:

4. Do you own a mobile device?
(e.g., iPhone/Android smartphone, iPad/Tablet, Amazon Kindle Fire.) If yes, please give details:

## E.4 Post Deployment Discussion Guide

MIRROR MIRROR

### Discussion Guide

- **All Questions Are Optional.**
- **Your personal information will be kept confidential. No reference will be made to your identity in publications or other documents.**
- **Your participation is voluntary and you are free to withdraw from the study at any time without giving any reason and without penalty.**

**Overall**

1. **What are your overall impressions of MirrorMirror?**
2. **What do you like about MirrorMirror?**
3. **What do you dislike about MirrorMirror?**

**Speakers**

1. **How did your speakers feel about MirrorMirror?**
2. **How did they feel about being recorded?**
3. **Did they know you had a hearing loss/needed to lipread?**
4. **Did it raise their awareness of your hearing loss or need to lipread?**

**Lipreading**

1. **Do you think MirrorMirror would improve your lipreading?**
2. **If it was available to you, would you continue to use MirrorMirror?**
3. **What would make MirrorMirror better?**

Computing
at the University of Dundee

## E.5 Information Sheet

**PARTICIPANT INFORMATION SHEET**

## MirrorMirror App

We would like to invite you to take part in our research study. Before you decide if you wish to take part, we would like you to understand why the research is being undertaken and what it will involve.

### What is the 'MirrorMirror App' study?

In this study, we are looking to evaluate a mobile application called MirrorMirror.

MirrorMirror is a mobile application which aims to help learners of lipreading. In this three stage evaluation you will 1) Meet with a researcher who will introduce the application to you, and ask some questions about your yourself and your background, 2) Use the application for a week on a loaned tablet and perform some tasks. 3) Meet with a researcher and discuss how you used the application and your thoughts on the application.

### Who can take part in the study?

We are looking for participants who are above 18, have a hearing loss and are currently taking lipreading classes.

### Do I have to take part in the study?

It is up to you to decide to join the study.

### What happens if I wish to withdraw from the study?

You are free to withdraw at any time, without giving a reason and without penalty. Any data that has already been gathered from you will also be discarded.

### What will I have to do?

We are evaluating a mobile app called MirrorMirror. To evaluate the app we have three stages

1) A briefing and tutorial session 2) a week long in-the-wild-deployment. 3) A post-deployment discussion session.

**MirrorMirror Tutorial session**
You will meet the researcher who will describe the study to you. If you agree you will sign the consent form. The researcher will then ask you to complete a short questionnaire with some

UNIVERSITY OF DUNDEE    Computing at the University of Dundee

information about you and your lipreading background. The researcher will then begin the tutorial session for MirrorMirror and detail the features it has. Finally, the researcher will discuss the task list for the deployment.

**In-The-Wild-Deployment**

You will be given an Android Tablet for a week and asked to use MirrorMirror for daily lipreading practice. During the week we will ask you to complete a set of tasks detailed in the attached task sheet. MirrorMirror will retain usage statistics in the background and these are saved internally to the device.

**Post-interview questionnaire**

At the end of the week (or sometime after) you will meet with the researcher for a follow up session. You will be asked to complete the closing questionnaire and this discussion may be audio recorded for later transcription. If so you will be asked to complete an audio release form. The audio will only be used by the researcher to transcribe into an anonymised form.

Finally, the researcher will take back the device and offload the usage statistics and then remove the app and all participant data from the device.

## What are the possible disadvantages and risks of taking part?

There are no risks associated with this study and we hope that the task will be enjoyable. The timing and location of sessions will be discussed with you.

## What are the possible benefits of taking part?

In our experience, people enjoy taking part in research as they are helping to develop new technology. Your involvement will help us understand how technology can be designed to support lipreading, which can potentially help people with hearing loss in the real world.

## What happens at the end of the study?

The analysis of the data will be completed by October 2017. The results of this study may be published in academic journals and presented at academic conferences. If you would like to know the outcome of the study, I will send you a copy of the study report by October 2017.

## What if there is a problem?

The University of Dundee School of Science and Engineering's Research Ethics Committee, which has responsibility for scrutinising all proposals for non-clinical research on humans within the School has examined the proposal and has raised no objections from the point of view of ethics. If you have a concern about any aspect of this study, you should speak to the study supervisor, Dr David Flatla [email d.flatla@dundee.ac.uk], who will do his best to answer your questions.

Should you still have questions you should speak to the discipline lead for computing, Professor Annalu Waller [email a.waller@dundee.ac.uk, phone: +44 (0)1382 388223].

### Will my information be kept confidential?

Yes. We will follow ethical and legal practice and all information about you will be handled in confidence. To ensure anonymity, personal records will only be available to the research team for the duration of the study and will not be kept together with the results or be presented in the report. If your data is used for publications, no reference to your identity will be made.

### Who has reviewed this study?

Computing at the University of Dundee's Ethics Committee, which has responsibility for scrutinising all proposals for non-clinical research on humans has examined the proposal and has raised no objections from the point of view of ethics.

### Who can I contact in connection with this research?

This research is part of an on-going research project directed by Benjamin Gorman. He is a PhD Student in the School of Computing at the University of Dundee. Please feel free to contact him about the study. His contact details are:

**Benjamin Gorman**
**Email:** b.gorman@dundee.ac.uk
**Telephone: 01382 385598**

**Thank you for taking the time to read this information sheet and for considering taking part in this study.**

UNIVERSITY OF DUNDEE

Computing at the University of Dundee

3

## E.6 Audio Release Form

MIRROR MIRROR

**Audio Consent Form**

Thank you for taking part in the MirrorMirror project. You are to take part in a one to one interview with a member of the research team and if you consent, this session will be audio recorded.

We ask that you complete the section below and return this form to us.

Please complete the following and sign below:

| | YES | NO |
|---|---|---|
| I confirm that I am aware the event will be audio recorded. | | |
| I confirm that the recording will not be released and will only be used by the research team to produce an anonymised transcript. | | |

____________________ Name
____________________ Signature
________ Date

Mirror Mirror Audio Release Form
June 2017

# E.7 Task List

MIRROR MIRROR

## Task List

- **Your personal information will be kept confidential. No reference will be made to your identity in publications or other documents.**
- **Your participation is voluntary and you are free to withdraw from the study at any time without giving any reason and without penalty.**

**We would like you to use MirrorMirror for daily lipreading practice for the week you have been allocated a tablet. Below is a list of tasks we would like you to complete**

**Before you begin, please:**
Add at least 3 new words to each Lip Shape

**Daily Lipreading Practice:**
At a minimum we ask you to try and practice at least 3 lip shapes per day using the "Lipshape Practice" feature.

To begin with you can use the videos that are preloaded onto your tablet.

**During the week:**

- Add at least 3 new speakers to your library
- (Speakers could be family, friends, colleagues, anyone else you see on a regular basis e.g., coffee shop worker, newsagent etc.)
- Record at least 1 video for each lip shape for each new speaker